\documentclass[11pt]{article}
\pdfoutput=1
\usepackage{graphicx,epsfig}
\usepackage{color}
\usepackage[colorlinks,citecolor=red,linkcolor=blue]{hyperref}
\usepackage{slashed}

\usepackage{epsfig,float,amsmath,amsfonts,amssymb,graphicx,bm,bbm,array,dcolumn,mathabx}
\usepackage{subfigure}

\usepackage[left=2.5cm, right=2.5cm, top=2cm]{geometry}
\usepackage{amsmath, amssymb, graphics, setspace}
\newcommand{\mathsym}[1]{{}}
\newcommand{\unicode}[1]{{}}

\DeclareMathOperator{\im}{Im}

\newcommand{\bigO}{\mathcal{O}}

\def\beq{\begin{equation}}
\def\eeq{\end{equation}}
\def\bea{\begin{eqnarray}}
\def\eea{\end{eqnarray}}
\def\bmat{\begin{pmatrix}}
\def\emat{\end{pmatrix}}

\def\eD{e_D}
\def\eU{e_U}
\def\eDB{\bar{e}_D}
\def\eUB{\bar{e}_U}
\def\eQ{e_Q}
\def\eQB{\bar{e}_Q}
\def\nXz{n_\Chi^{(0)}}
\def\nX{n_\Chi}
\def\nU{n_\Up}
\def\nD{n_\Dp}
\def\nUB{n_{\bar\Up}}
\def\nDB{n_{\bar\Dp}}
\def\nDU{n_{\Delta\Up}}
\def\nDD{n_{\Delta\Dp}}
\def\nB{n_B}

\def\muhh{\hat{\hat\mu}}

\def\muhhB{\hat{\hat\mu}_B}

\def\lrarr{\leftrightarrow}

\def\ssty{\scriptstyle}
\def\sssty{\scriptscriptstyle}

\newcommand{ \slashchar }[1]{\setbox0=\hbox{$#1$}   
   \dimen0=\wd0                                     
   \setbox1=\hbox{/} \dimen1=\wd1                   
   \ifdim\dimen0>\dimen1                            
      \rlap{\hbox to \dimen0{\hfil/\hfil}}          
      #1                                            
   \else                                            
      \rlap{\hbox to \dimen1{\hfil$#1$\hfil}}       
      /                                             
   \fi}                                             %
\def\pslash{\slashchar{p}}
\def\qslash{\slashchar{q}}
\def\qoslash{\slashchar{q}_1}
\def\qtslash{\slashchar{q}_2}
\def\qthslash{\slashchar{q}_3}
\def\qfslash{\slashchar{q}_4}

\def\to{\rightarrow}
\def\Chi{{\cal X}}
\def\Dp{{D}}
\def\Up{{U}}

\def\LIntVV{{\cal L}^{\rm VV}_{\rm int}}

\def\Umat{{\cal U}}

\def\PV{{\rm PV}}

\def\AsymB{{\cal A}_B}
\def\AsymBh{\hat{\cal A}_B}
\def\AsymBsig{{\cal A}_B^{(\sigma)}}
\def\AsymBsigh{\hat{\cal A}_B^{(\sigma)}}
\def\AsymBsign{{{\cal A}_B^{n(\sigma)}}}
\def\AsymBsignh{{\hat{\cal A}_B^{n(\sigma)}}}
\def\AsymBP{{\cal A}_B^\prime}
\def\AsymBPh{\hat{\cal A}_B^\prime}

\def\AsymBPsigh{\hat{\cal A}_B^{\prime(\sigma)}}

\def\ampA{{\cal A}}

\def\ampALOn{{{\cal A}^n_0}}

\def\ampANLOn{{{\cal A}^n_1}}
\def\ampAhatNLOn{{\hat{{\cal A}}^n_1}}
\def\ampAcLOn{{{\cal A}^{c\, n}_0}}

\def\ampM{{\cal M}}

\def\ampMPz{\ampM'_0}
\def\ampMPzt{\widetilde\ampM'_0}

\def\IBBbklnmh{\hat{\cal I}^{\scriptscriptstyle (\bar B B)}_{k\lambda\, nm}}
\def\IBBbolooh{\hat{\cal I}^{\scriptscriptstyle (\bar B B)}_{1\lambda\, 11}}
\def\IBBbtlooh{\hat{\cal I}^{\scriptscriptstyle (\bar B B)}_{2\lambda\, 11}}
\def\IBBboltth{\hat{\cal I}^{\scriptscriptstyle (\bar B B)}_{1\lambda\, 22}}
\def\IBBbtltth{\hat{\cal I}^{\scriptscriptstyle (\bar B B)}_{2\lambda\, 22}}

\def\IBBbTAklnmh{\langle \hat{\cal I}^{\scriptscriptstyle (\bar B B)}_{k\lambda\, nm}\rangle }
\def\IBBbTAolooh{\langle \hat{\cal I}^{\scriptscriptstyle (\bar B B)}_{1\lambda\, 11}\rangle }
\def\IBBbTAtlooh{\langle \hat{\cal I}^{\scriptscriptstyle (\bar B B)}_{2\lambda\, 11}\rangle }
\def\IBBbTAoltth{\langle \hat{\cal I}^{\scriptscriptstyle (\bar B B)}_{1\lambda\, 22}\rangle }
\def\IBBbTAtltth{\langle \hat{\cal I}^{\scriptscriptstyle (\bar B B)}_{2\lambda\, 22}\rangle }

\def\Smat{{\cal S}}
\def\Tmat{{\cal T}}
\def\ampM{{\cal M}}

\def\ampMsigSq{|{\cal M}^{(\sigma)}|^2}

\def\ampMPsigSq{|{\cal M}^{'(\sigma)}|^2}

\def\GtV{(\widetilde{G}_V)}
\def\GtVB{{\widebar{\widetilde{G}}_V}}
\def\GV{{G_V}}
\def\GVn{{G_V^n}}
\def\ghLn{\hat{g}_{L_n}}

\def\ghRn{\hat{g}_{R_n}}

\def\GVm{{G_V^m}}
\def\GZ{{G_\Lambda}}
\def\GZn{{G_\Lambda^n}}
\def\GZm{{G_\Lambda^m}}
\def\GVB{{\phantom{\,}\bar{G}_V\phantom{\,\!}}}
\def\GVBn{{\phantom{\,}\bar{G}_V^n\phantom{\,\!}}}
\def\GVBm{{\phantom{\,}\bar{G}_V^m\phantom{\,\!}}}
\def\GZB{{\phantom{\,}\bar{G}_\Lambda\phantom{\,\!}}}
\def\GZBn{{\phantom{\,}\bar{G}_\Lambda^n\phantom{\,\!}}}
\def\GZBm{{\phantom{\,}\bar{G}_\Lambda^m\phantom{\,\!}}}

\def\rMLtxt{M_\chi/\Lambda}

\def\DGmzoh{\Delta\hat\Gamma_{01}}

\def\sigzP{\sigma_0'}
\def\sigzPh{\hat\sigma_0'}

\def\GmzPTA{\langle \Gamma'_0\rangle }
\def\GmzPTAh{\langle \hat\Gamma'_0\rangle }
\def\alzPh{\hat\alpha'_0}

\def\nspc{\!\!\!\!}

\def\GmTA{\langle \Gamma\rangle }

\def\GmBTA{\langle \bar\Gamma\rangle }
\def\GmzTA{\langle \Gamma_0\rangle }

\def\GmzPTA{\langle \Gamma'_0\rangle }

\def\GmzPTAh{\langle \hat\Gamma'_0\rangle }

\def\GmzSigTA{\langle \Gamma_0^{(\sigma)}\rangle }
\def\GmSigXDBTA{\langle \Gamma_{\chi \bar{D}}^{(\sigma)}\rangle }
\def\GmSigXDTA{\langle \Gamma_{\chi D}^{(\sigma)}\rangle }
\def\GmzSigXDTA{\langle \Gamma_{0\,\chi D}^{(\sigma)}\rangle }
\def\GmzSigXDBTA{\langle \Gamma_{0\,\chi \bar{D}}^{(\sigma)}\rangle }
\def\DGmSigXDTA{\langle \Delta\Gamma_{\chi D}^{(\sigma)}\rangle }
\def\GmSigXUBTA{\langle \Gamma_{\chi \bar{U}}^{(\sigma)}\rangle }
\def\GmSigXUTA{\langle \Gamma_{\chi U}^{(\sigma)}\rangle }
\def\GmSigXQBTA{\langle \Gamma_{\chi \bar{Q}}^{(\sigma)}\rangle }
\def\GmzSigXUTA{\langle \Gamma_{0\,\chi U}^{(\sigma)}\rangle }
\def\GmzSigXUBTA{\langle \Gamma_{0\,\chi \bar{U}}^{(\sigma)}\rangle }
\def\DGmSigXUTA{\langle \Delta\Gamma_{\chi U}^{(\sigma)}\rangle }
\def\DGmSigXQTA{\langle \Delta\Gamma_{\chi Q}^{(\sigma)}\rangle }

\def\GmzSigXQTA{\langle \Gamma_{0\,\chi Q}^{(\sigma)}\rangle }

\def\GmSigXQTA{\langle \Gamma_{\chi Q}^{(\sigma)}\rangle }
\def\GmzPSigTA{\langle {\Gamma'_0}^{\!(\sigma)}\rangle }
\def\GmzPSigUDTA{\langle {\Gamma'_{0\, UD}}^{\nspc\nspc\nspc(\sigma)}\ \rangle }
\def\GmzPSigDDTA{\langle {\Gamma'_{0\, DD}}^{\nspc\nspc\nspc(\sigma)}\ \rangle }
\def\GmzPSigUUTA{\langle {\Gamma'_{0\, UU}}^{\nspc\nspc\nspc(\sigma)}\ \rangle }

\def\DGm{\Delta\Gamma}

\def\DGmTA{\langle \Delta\Gamma\rangle }
\def\DGmzoh{\Delta\hat\Gamma_{01}}

\def\DGmzoTAh{\langle \Delta\hat\Gamma_{01}\rangle }

\def\DGmSigTA{\langle \Delta\Gamma^{(\sigma)}\rangle }

\def\DSigvTA{\langle \Delta\sigma\, v\rangle }

\def\GmTA{\langle \Gamma\rangle }

\def\GmzPTA{\langle \Gamma'_0\rangle }

\def\GmzPTAh{\langle \hat\Gamma'_0\rangle }

\def\GmzSigTA{\langle \Gamma_0^{(\sigma)}\rangle }
\def\GmzPSigTA{\langle {\Gamma'_0}^{\!(\sigma)}\rangle }

\def\Mtn{\tilde{M}_n}
\def\Etn{\tilde{E}_n}
\def\EtnL{\tilde{E}^{\sssty (L)}_n}
\def\Pchi{{\cal P}_\chi}

\def\PchiMSq{\left|{\cal P}_\chi\right|^2_{\sssty (-)}}

\def\Mot{(M^{12})}
\def\MLOot{(M^{12}_0)}
\def\McLOot{(M^{c12}_0)}
\def\MNLOot{(M^{12}_1)}
\def\MNLOoti{M^{12}_{1(i)}}

\def\MotSq{|M^{12}|^2}
\def\MLOotSq{|M^{12}_0|^2}

\def\MhotSq{|\hat{M}^{12}|^2}
\def\Mcot{(M^{c\, 12})}
\def\McotSq{|M^{c\, 12}|^2}
\def\McLOotSq{|M_0^{c\, 12}|^2}
\def\McNLOot{M^{c12}_{1}}

\makeatletter
\def\Lot{%
  \@ifnextchar_%
  {\@Lot}
  {\@latex@warning{Missing argument for \string\Lot}\@Lot_{}}%
}
\def\@Lot_#1{%
  L_{12 {#1}}%
}
\makeatother

\makeatletter
\def\Lhot{%
  \@ifnextchar_%
  {\@Lhot}
  {\@latex@warning{Missing argument for \string\Lhot}\@Lhot_{}}%
}
\def\@Lhot_#1{%
  \hat{L}_{12 {#1}}%
}
\makeatother

\def\Ltf{(L_{34})}
\def\Ltfi{L^{34}_{(i)}}

\def\Lctf{(L^c_{34})}

\def\MLOXU{({{\cal M}_0^{\chi U}})}
\def\MLOXUSq{|{{\cal M}_0^{\chi U\,\!}}|^2}
\def\MhLOXUSq{|{{\cal \hat{M}}_0^{\chi U\,\!}}|^2}
\def\MNLOXU{({{\cal M}_1^{\chi U\,\!}})}
\def\MNLOXUi{({{\cal M}_{1(i)}^{\chi U\,\!}})}               
\def\MLONLOXU{{{\cal M}_{01}^{\chi U\,\!}}}

\def\MLONLOXUnmSt{{{\cal M}_{01\,nm}^{\chi U\,*}}}
\def\MLONLOXUi{({{\cal M}_{01(i)}^{\chi U\,\!}})}
\def\MhLONLOXUi{({{\cal \hat{M}}_{01(i)}^{\chi U\,\!}})}

\def\MhLONLOXUnmi{({\hat{\cal M}_{01(i)}^{\chi U\,\!}})_{nm}}
\def\McLOXU{({{\cal M}_0^{c\, \chi U\,\!}})}
\def\McLOXUSq{|{{\cal M}_0^{c\, \chi U\,\!}}|^2}
\def\McNLOXU{({{\cal M}_1^{c\, \chi U\,\!}})}
\def\McLONLOXU{{{\cal M}_{01}^{c\, \chi U\,\!}}}
\def\McLONLOXUnm{{{\cal M}_{01\,nm}^{c\, \chi U\,\!}}}

\def\Mhzzn{\hat{{\cal M}}^{00}}
\def\IhLOn{{{\cal I}_0^n}}

\makeatletter
\def\IhLOnsig{%
  \@ifnextchar_%
  {\@IhLOnsig}
  {\@latex@warning{Missing argument for \string\IhLOnsig}\@IhLOnsig_{}}%
}
\def\@IhLOnsig_#1{%
  {\cal I}^{n(\sigma)}_{0\, {#1}}%
}
\makeatother

\def\IhLOo{{{\cal I}_0^1}}
\def\IhLOt{{{\cal I}_0^2}}

\def\pnr{{p_n}_r}
\def\pnrh{\hat{p}_{n_r}}

\def\pCM{p_{\scriptscriptstyle CM}}
\def\pCMz{p_{\scriptscriptstyle CM}^0}
\def\pnz{p_n^0}
\def\pnL{p_n^{\scriptscriptstyle (L)}}
\def\pnrL{{p_n}_r^{\scriptscriptstyle \!\!\!\!(L)}}

\def\EnL{E_n^{\scriptscriptstyle (L)}}

\def\kiV{{\bf k}_i}
\def\kir{{k_i}_r}

\def\pnV{{\bf p}_n}

\def\qoV{{\bf q}_1}
\def\qoVSt{{\bf q}^*_1}
\def\qtV{{\bf q}_2}
\def\qtVSt{{\bf q}^*_2}
\def\qUV{{\bf q}_U}
\def\qUVSt{{\bf q}^*_U}
\def\pnV{{\bf p}_n}
\def\pnVSt{{\bf p}^*_n}

\def\kBV{{\bf k}_B}
\def\kBVhat{\!\!\hat{\,\,\bf k}_B}
\def\kBr{{k_B}_r}

\def\kBPV{{\bf k}'_B}

\def\kBPr{{k'_B}_r}

\def\pV{{\bf p}}
\def\pparV{{\bf p}_\parallel}
\def\pparr{{p_\parallel}_r}
\def\pperpV{{\bf p}_\perp}

\def\qor{{q_1}_r}
\def\qorSt{{q_1^*}_{\! r}}

\def\qtrSt{{q_2^*}_{\! r}}

\def\qUr{{q}_{U_{r}}}

\def\qUrt{{{\tilde{q}}_{U_{r}}}}

\def\SSt{{\oasterisk}}

\def\qthr{{q_3}_r}
\def\qthrSt{{q_3^*}_r}
\def\qthrSSt{{q_3}^\SSt_r}

\def\qUPr{{q'_U}_{\! r}}

\def\qUPrSSt{{{{q'}_{\!\! U}}_{\! r}^{\!\! \SSt}}}
\def\EnSSt{{E_n}^{\!\! \SSt}}
\def\EUPSSt{{E'_U}^{\!\! \SSt}}
\def\cUPSSt{{c'_U}^{\!\! \SSt}}
\def\phiUPSSt{{\phi'_U}^{\!\! \SSt}}

\def\qthrSSt{{q_3}_r^{\!\! \SSt}}
\def\EthSSt{{E_3}^{\!\! \SSt}}

\def\EfoSSt{{E_4}^{\!\! \SSt}}

\def\qthV{{\bf q}_3}

\def\qfV{{\bf q}_4}

\def\qUPV{{{\bf q}'_U}}

\def\sigvz{{\sigma_0\, v}}
\def\sigvzn{{\sigma_0^n\, v}}

\def\TAsigv{{\langle  \sigma\, v \rangle }}
\def\TAsigvz{{\langle  \sigma_0\, v \rangle }}

\def\TAsigvzh{{\langle  \hat\sigma_0\, v \rangle }}

\def\ket#1{\left.| #1 \right> }
\def\bra#1{\left<  #1 | \right.}

\def\matel#1#2#3{\left<  #2| #1 | #3 \right> }

\newcounter{partsection}
\newcounter{partsectionnum}[partsection]  
\newcounter{partsubsectionnum}[partsectionnum] 

\renewcommand{\thepartsection}{\Roman{partsection}} 

\newcommand{\custompartsection}[1]{%
    \clearpage
    \thispagestyle{empty} 
    \refstepcounter{partsection} 
    \setcounter{partsectionnum}{0} 
    \setcounter{partsubsectionnum}{0} 
    \vspace*{\fill} 
    \begin{center}
        \LARGE \bfseries Part \thepartsection: #1
    \end{center}
    \vspace*{\fill} 
    \addcontentsline{toc}{section}{Part \thepartsection: #1} 
    \pagestyle{plain} 
}

\begin{document}

\title{
\normalsize{\bf Baryon Asymmetry from the Decay and Scattering of a Majorana Fermion Pair Coupled to Quarks}}

\author{\normalsize{Shrihari Gopalakrishna$^{a,b}$\thanks{shri@imsc.res.in}\ , Rakesh Tibrewala$^c$\thanks{rtibs@lnmiit.ac.in}}
\\
$^a$~\small{Institute of Mathematical Sciences (IMSc), Chennai 600113, India.}\\
$^b$~\small{Homi Bhabha National Institute (HBNI), Anushaktinagar, Mumbai 400094, India.}\\
$^c$~\small{The LNM Institute of Information Technology (LNMIIT), Jaipur 302031, India.}
}

\maketitle

\begin{abstract}
  We compute the baryon asymmetry in decay and scattering processes involving  
  the electromagnetically charge-neutral fermion $\chi$ that carries nonzero baryon number
  and interacts with quark-like fermions $\Up,\Dp$ via a vector-vector dimension-six effective operator,
  in the theory we developed in our earlier work.
  Majorana masses for the $\chi$ break baryon number and split the Dirac fermion $\chi$ 
  into a pair of Majorana fermions $\Chi_n$ with indefinite baryon number. 
  We identify loop amplitudes for $\Chi_n$ decay and scattering processes that are sensitive to the baryon number violation. 
  The phases in the Majorana mass and couplings, in conjunction with the phase from intermediate onshell states,
  lead to $C$ and $CP$ violation in these processes.
  For some representative parameter choices, we numerically compute
  the decay and scattering baryon asymmetries between the process and its conjugate process,  
  and find that the asymmetry generated is very interesting for explaining the baryon asymmetry of the Universe.
\end{abstract}




\section{Introduction}
\label{Intro.SEC}

In any candidate theory, the generation of the baryon asymmetry of the Universe (BAU)
requires that the three Sakharov conditions~\cite{Sakharov:1967dj} be satisfied,
namely,
baryon number violation, $C$ and $CP$ violation, and, departure from thermal equilibrium.
It is generally believed that in the standard model (SM), not enough baryon asymmetry is generated to explain
the observed BAU.
Various theories beyond the standard model (BSM) have been proposed to address this shortcoming of the SM.
For reviews, see for example, Refs.~\cite{Kolb:1990vq,Cline:2006ts}. 
One among them is the BSM effective theory we developed in Ref.~\cite{Gopalakrishna:2022hwk} that has all the ingredients necessary to explain the BAU.

Our theory contains an electromagnetically (EM) charge-neutral fermion $\chi$,
interacting via a vector-vector (VV) dimension-six effective operator
coupling the $\chi$ to an up-type $\Up$ and two identical down-type $\Dp$ quark-like fermions.
The $\Up,\Dp$ each carries baryon number $B=1/3$, like the SM quarks.
We can write the VV interaction schematically as $(1/\Lambda^2)\, (\bar\chi\Gamma^\mu \Up)\,(\bar{\Dp^c}\Gamma_\mu \Dp)$,  
where $\Lambda$ is the cutoff scale.
In this work we identify loop contributions to $\chi$ decay and scattering processes and compute the baryon asymmetry
that result from them. 
In these works we study the VV interaction as opposed to the scalar-scalar (SS) interaction studied in the literature;
see Ref.~\cite{Gopalakrishna:2022hwk} for related references. 

Our motivation in carrying out an effective theory analysis is that it provides an estimate of the baryon asymmetry
generated in a whole class of related ultraviolet (UV) complete theories that have our effective operators
as the low energy limit.
A truly accurate result can only be obtained by calculating in a specific UV completion. 
We take as an example the UV completion given in Ref.~\cite{Gopalakrishna:2022hwk}
that has a propagating state $\xi$ that resolves our effective operators. 

In Ref.~\cite{Gopalakrishna:2022hwk} we showed that our theory has a rich enough structure to potentially satisfy the Sakharov conditions.
Baryon number violation is due to the $\chi$ Majorana mass terms. 
$C$ and $CP$ violation are due to phases in the Majorana mass and couplings.
Considering the $\chi,\Up,\Dp$ in the early Universe, if the interaction rate of the $\chi$ is smaller than the Hubble expansion rate,
the $\chi$ can be out of thermal equilibrium. 

As we demonstrated in Ref.~\cite{Gopalakrishna:2022hwk},
the Majorana mass splits the Dirac Fermion $\chi$ with Dirac mass $M_\chi$
into two nondegenerate Majorana fermions $\Chi_n$ ($n=\{1,2\}$) with masses $M_n$. 
The $\Chi_n$ have indefinite baryon number. 
In our framework, the decay baryon asymmetry is given by the difference in
decay rate for the process $\Gamma(\Chi_n \to \Up\Dp\Dp)$ whose final state has baryon number $B=+1$,
and that for the conjugate process $\Gamma(\Chi_n \to \bar{U}\bar{D}\bar{D})$ whose final state has $B=-1$.
The Majorana (i.e. self-conjugate) nature of the $\Chi_n$ is the reason that the $\Chi_n$ could decay into these decay modes with opposite baryon number. 
Similarly, the scattering baryon asymmetry arises due to the difference in cross sections
$\sigma(\Chi_n \bar{Q} \to QQ)$ and $\sigma(\Chi_n Q \to \bar{Q}\bar{Q})$, where $Q=\{\Up,\Dp\}$, 
and the former process has $\Delta B = +1$ while the latter has $\Delta B = -1$.

We showed in general terms in Ref.~\cite{Gopalakrishna:2022hwk}
that in this effective theory, 
the leading contribution to the baryon asymmetry is from the interference of tree and loop-level amplitudes.
There, we emphasized that nonzero phases in the Majorana masses and/or couplings could act like {\em weak phases},
i.e. flip sign between the process and its conjugate processes, 
and,
also that intermediate $\Up,\Dp$ states going on-shell in the loops leads to a {\em strong phase}, i.e. is the same for the process and its conjugate process.
We showed that both weak and strong phases are essential for the generation of the baryon asymmetry, 
and estimated the size of the baryon asymmetry that could be generated in this effective theory.

In this work, we concretely realize these ideas by identifying specific
one and two loop decay and scattering processes involving the $\Chi_n$ with weak and strong phases, 
and compute the baryon asymmetry generated from them.
We refer to the two-loop diagrams as {\em single operator contributions}
since they require the presence of only the above VV effective operator. 
In addition to the VV operator above, if an operator of the type
$(1/\Lambda^2)\, (\bar\Chi\Gamma^\mu \Up)\,(\bar\Up\Gamma_\mu\Chi)$ is also present in the effective theory,
we show that using both the operators,
we can write one-loop diagrams that we call the {\em multiple operator contributions}
that also yield a baryon asymmetry. 

To compute the total decay and scattering rates and the asymmetry,
we have to perform integrals over phase space and loop kinematical variables. 
These have multiple scales with many variables, and an analytical computation is not feasible.
We therefore resort to a numerical evaluation of these complicated integrals. 
We present our results in a general way without making a particular choice of the scale $\Lambda$ or $M_\chi$, 
but give overall scaling factors involving $\Lambda$ and $r_{M\Lambda}\!\! =\!\! M_\chi/\Lambda$. 
This allows application of our results to different theories with their own preferences for the scale of such new physics.

The results of this work are inputs (as collision terms) into the Boltzmann equations (BE) 
that governs the number densities of the baryons and antibaryons
at high temperature in the early Universe. 
For this purpose, we therefore compute 
the thermal averages of the decay and scattering baryon asymmetries resulting from the processes above. 
Solving the Boltzmann equations with these inputs would let us determine the resulting BAU at late time (today).
The numerical solution of the Boltzmann equations are quite involved and is presented in Ref.~\cite{Gopalakrishna:2024qxk}. 

This paper is organized as follows.
In Sec.~\ref{effOps.SEC} we review the effective theory and the baryon asymmetry generation mechanism from Ref.~\cite{Gopalakrishna:2022hwk},
which we use in this work.  
In Sec.~\ref{ABdec.SEC} we analyze the $\Chi_n$ decay,
starting with the tree-level decay rate in Sec.~\ref{chi2DDUGmLO.SEC},  
computing the single operator decay rate difference in Sec.~\ref{oneOpAsymb.SEC},
and the multiple operator decay rate difference in Sec.~\ref{mixedOpsAsymB.SEC}.
Our computation needs clarity on the subtraction scheme for the $\Chi_n$ 2-point function, which we discuss in Appendix~\ref{Chinm2ptFcn.SEC}. 
In Sec.~\ref{genBAsymScat.SEC} we compute the rate asymmetry from $\Chi_n$ scattering processes,
with the single operator scattering contribution in Sec.~\ref{singOpSig.SEC},
and the multiple operator scattering contribution in Sec.~\ref{multiOpSig.SEC}.
In Sec.~\ref{UDD2UcDcDc.SEC} we compute the rate difference between the $\Delta B = -2$ three-to-three scattering process
and its conjugate process. 
For use in a thermal setting in the early Universe, 
we also compute the thermally averaged decay rate and cross sections of these processes as a function of temperature
using the method we lay down in Appendix~\ref{ThAvgProc.SEC}. 
For computing the late time baryon asymmetry, 
we develop further in Sec.~\ref{nChinBBE.SEC} the Boltzmann equations applicable in the early Universe, 
taking into account the relations imposed by $CPT$ and unitarity on the forward and inverse reaction matrix elements
that we discuss in Appendix~\ref{CPTUniBE.SEC}.
In scattering contributions, the initial state momenta in the thermal spectrum can be
larger than the cutoff scale ($\Lambda$) of the effective theory,
bringing in a dependence on the UV physics, and we discuss our method of handling this in Appendix~\ref{VVintUV.SEC}. 
We offer our conclusions in Sec.~\ref{Concl.SEC}.
We relegate to Part~\ref{techDet.PART} the more detailed technical aspects as supplementary material. 
 
\section{The Effective Theory and the Baryon Asymmetry}
\label{effOps.SEC}

\begin{figure}
\begin{minipage}{6in}
  \centering
  \raisebox{-0.5\height}{\includegraphics[width=0.25\textwidth]{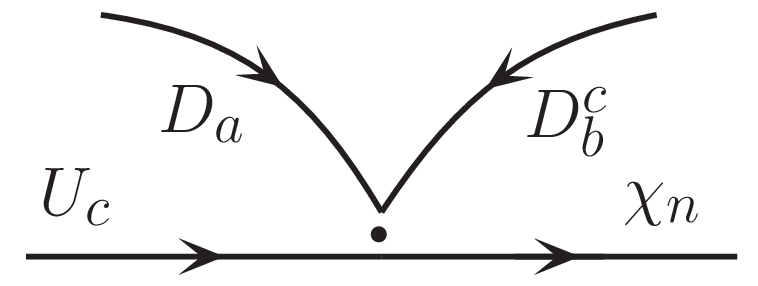}}
  $\phantom{+} \frac{i}{2\Lambda^2} \epsilon^{abc} \widetilde{G}_V^\mu \otimes G^n_{V\mu} $ 
  \hspace*{0.5cm}
  \raisebox{-0.5\height}{\includegraphics[width=0.25\textwidth]{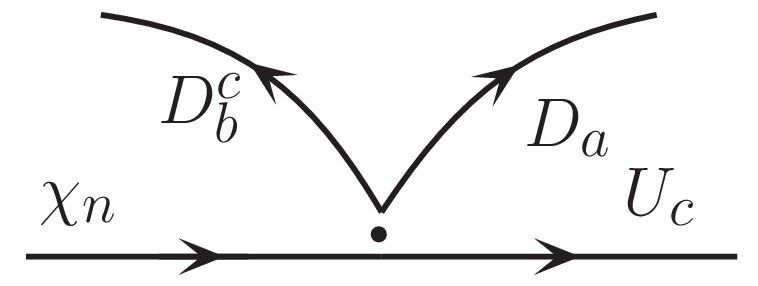}}
  $\phantom{+} \frac{i}{2\Lambda^2} \epsilon^{abc} \bar{\widetilde{G}}_V^\mu \otimes \bar{G}^n_{V\mu} $ 
  \\     \medskip{}
  \raisebox{-0.5\height}{\includegraphics[width=0.25\textwidth]{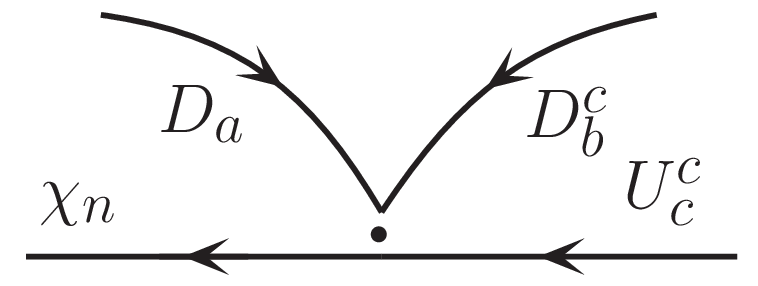}}
  $-\frac{i}{2\Lambda^2} \epsilon^{abc} \widetilde{G}_V^\mu \otimes G^n_{\Lambda\mu} $
  \hspace*{0.5cm}
  \raisebox{-0.5\height}{\includegraphics[width=0.25\textwidth]{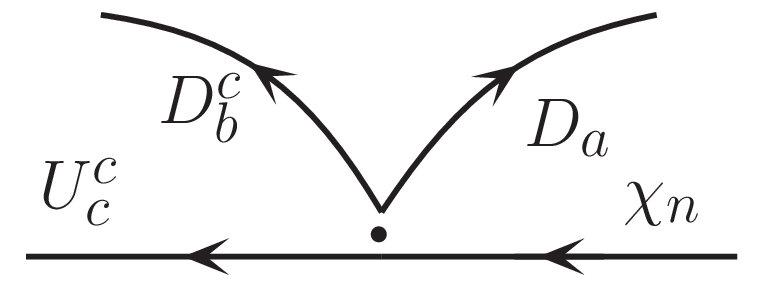}}
  $-\frac{i}{2\Lambda^2} \epsilon^{abc} \bar{\widetilde{G}}_V^\mu \otimes \bar{G}^n_{\Lambda\mu} $
\end{minipage}
\caption{The Feynman rules for the VV interaction, from Ref.~\cite{Gopalakrishna:2022hwk}. 
\label{FeynRules.FIG}}
\end{figure}

In this section we briefly summarize aspects of the effective theory developed in Ref.~\cite{Gopalakrishna:2022hwk} that are essential for our work here.
The theory contains a Dirac (vector-like) fermion $\chi$ 
with a Dirac mass $M_\chi$ and also Majorana masses $\tilde{M}_{L,R}$ that split the $\chi$
into two Majorana fermions $\Chi_n$ (with $n=\{1,2\}$)
with mass eigenvalues $M_n$.
The Dirac fermion $\chi$ can be assigned baryon number $B\!\! =\!\! +1$,
and the Majorana masses break baryon number leading to the Majorana fermions $\Chi_n$ having indefinite baryon number. 
We focus on the VV interaction\footnote{In Ref.~\cite{Gopalakrishna:2022hwk},
we showed that the scalar-scalar (SS) interaction involving two same $\Dp$ flavors cannot be written down.
We solidify this claim further at the matrix element level
(cf. Appendix~\ref{DDM12.SEC}). 
}
dimension-six effective operator that couples each Majorana state $\Chi_n$ 
to the quark-like fermions $\Up$ and two same flavor $\Dp$ as
\beq
\LIntVV = \frac{\epsilon^{abc}}{2 \Lambda^2} \left\{
\left[  \overline{\Dp^c_b} \, \widetilde{G}_V^\mu \Dp_a \right] \ \left[ \bar{\Chi}_n \, G^n_{V \mu} \Up_c \right]  +
\left[  \overline{\Dp}_a \, \bar{\widetilde{G}}_V^\mu \Dp^c_b \right] \ \left[ \bar{\Up}_c \, \bar{G}^n_{V \mu} \Chi_n \right]
\right\}  \ ,
\label{LIntVVMB.EQ}
\eeq
where
$G^n_{V\mu} \equiv \gamma_\mu \left( g_L\, \Umat^*_{1n} P_L + g_R\, \Umat_{2n}  P_R \right) $,
$\bar{G}^n_{V\mu} = \gamma_\mu \left( g_L^* \, \Umat_{1n} P_L + g_R^* \, \Umat^*_{2n}  P_R \right) $, 
$\widetilde{G}_V^\mu = \bar{\widetilde{G}}_V^\mu = \gamma^\mu \tilde{g}$,
with the $\Umat$ being the $2\times 2$ unitary matrix that takes us to the $\Chi_n$ mass basis as worked out in Ref.~\cite{Gopalakrishna:2022hwk}. 
We define, for future use, the combinations $\ghLn \equiv g_L\, \Umat^*_{1n}$ and $\ghRn \equiv g_R\, \Umat_{2n}$.
An equivalent form with charge-conjugated fields is derived in Ref.~\cite{Gopalakrishna:2022hwk},
and is given in terms of
$G^n_{\Lambda \mu} = G^n_{V \mu}|_{(P_L \leftrightarrow P_R)}$ and $\bar{G}^n_{\Lambda \mu} =  \bar{G}^n_{V \mu}|_{(P_L \leftrightarrow P_R)}$
(i.e. $G_\Lambda$ is the corresponding $G_V$ with $P_L$ and $P_R$ interchanged).
The Feynman rules given in Ref.~\cite{Gopalakrishna:2022hwk} are reproduced in Fig.~\ref{FeynRules.FIG}. 
The couplings are shown in the form $(...)\otimes (...)$, where the first factor is for the $\Dp^c - \Dp$ fermion line,
and the second factor for the $\Chi_n - \Up$ line.
On the fields, the superscript $c$ denotes the charge conjugate, and the subscript $\{a,b,c\}$ denotes the color index,
with the color indices contracted using the $\epsilon^{abc}$ as shown.
The direction of the arrow denotes the flow of fermion number (also baryon number),
and even though the Majorana fermions $\Chi_n$ have indefinite baryon number due to their Majorana mass,
we show an arrow on the $\Chi_n$ line indicating the flow direction in the Dirac limit, i.e. when the Majorana mass is zero.
In this convention, the $\Chi_n$ propagator with clashing arrows indicates a Majorana mass insertion. 

\begin{figure}
\begin{minipage}{6in}
  \centering
  \raisebox{-0.5\height}{\includegraphics[width=0.25\textwidth]{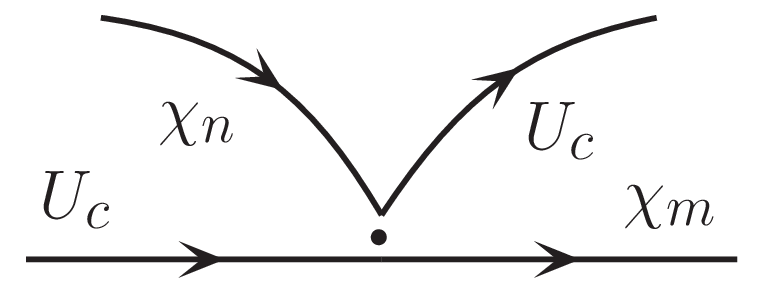}}
  $-\frac{i}{\Lambda^2} \bar{G}_V^{n\mu} \otimes G^m_{V\mu} $ 
  \hspace*{1cm}
  \raisebox{-0.5\height}{\includegraphics[width=0.25\textwidth]{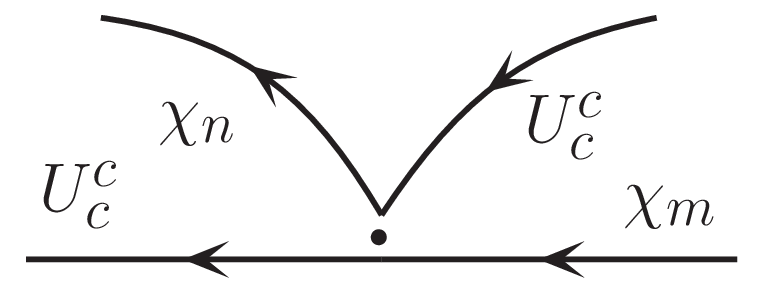}}
  $-\frac{i}{\Lambda^2} G_\Lambda^{n\mu} \otimes \bar{G}^m_{\Lambda\mu} $ 
  \\     \medskip{}
  \raisebox{-0.5\height}{\includegraphics[width=0.25\textwidth]{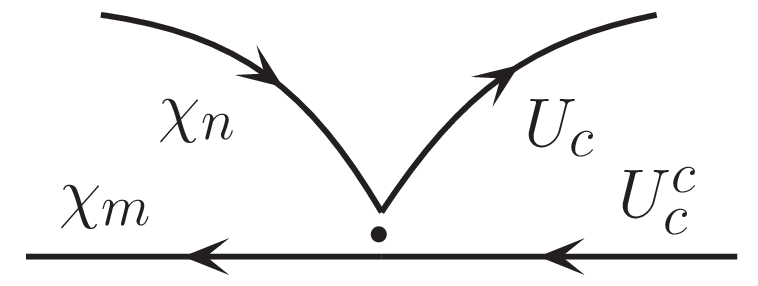}}
  $\frac{i}{\Lambda^2} \bar{G}_V^{n\mu} \otimes G^m_{\Lambda\mu} $ 
  \hspace*{1cm}
  \raisebox{-0.5\height}{\includegraphics[width=0.25\textwidth]{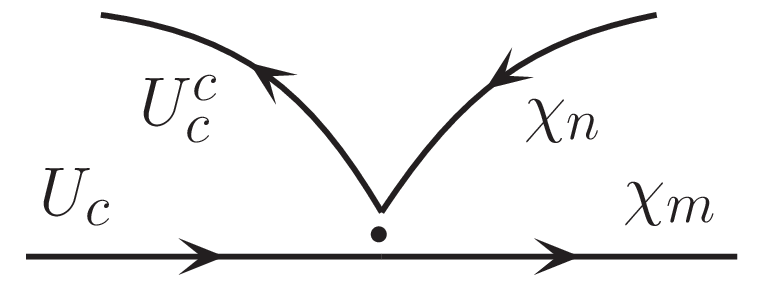}}
  $\frac{i}{\Lambda^2} \bar{G}_\Lambda^{n\mu} \otimes G^m_{V\mu} $ 
  \\     \medskip{}
  \hspace*{3cm}
  \raisebox{-0.5\height}{\includegraphics[width=0.25\textwidth]{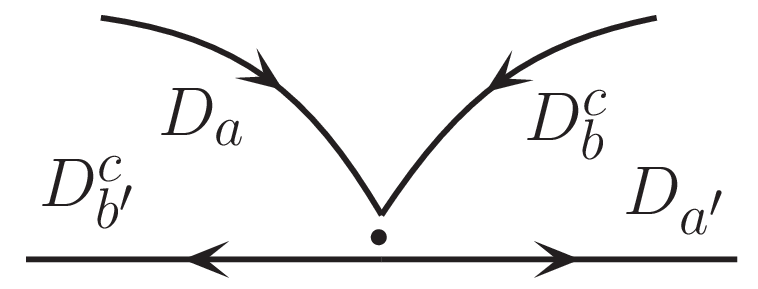}}
  $\frac{i}{2\Lambda^2} (\delta^{aa'} \delta^{bb'} - \delta^{ab'} \delta^{ba'}) \widetilde{G}_V^\mu \otimes \bar{\widetilde{G}}_{V\mu}$
\end{minipage}
\caption{The Feynman rules of the other operators associated with the VV interaction, from Ref.~\cite{Gopalakrishna:2022hwk}.
}
\label{FeynRulesModA.FIG}
\end{figure}

Although our main focus is the effective operator in Eq.~(\ref{LIntVVMB.EQ}),
we also include in our study other associated operators
arising in some ultraviolet (UV) completion examples considered in Ref.~\cite{Gopalakrishna:2022hwk}.
These are
\beq
    {\cal L} \supset - \frac{1}{\Lambda^2} \, [\bar{\Up}_c \bar{G}^n_{V \mu} \Chi_n]\, [\overline{\Chi}_m {G^m_{V}}^\mu \Up_c] 
    - \frac{1}{2 \Lambda^2} (\delta^{aa'} \delta^{bb'}
    - \delta^{ab'} \delta^{ba'}) [\overline{\Dp}^c_b\, \tilde{g} \gamma^\mu \Dp_a]\, [\bar{\Dp}_{a'}\, \tilde{g} \gamma_\mu \Dp^c_{b'}] \ ,
\label{modAops.EQ}    
\eeq
which are self-conjugate.
The Feynman rules for these vertices are reproduced in Fig.~\ref{FeynRulesModA.FIG}. 

\subsection{Computing the baryon asymmetry}
\label{genAB.SEC}

We obtain the baryon asymmetry by following the method described in detail in Ref.~\cite{Gopalakrishna:2022hwk}.
For the $\Chi_n$ decay, we write the decay amplitude as  
$\ampA^n = \ampA_0^n + \ampA_1^n$ in terms of the tree and loop-level amplitudes,
and for the conjugate process as $\ampA^{cn} = \ampA^{cn}_0 + \ampA^{cn}_1$.
We have the decay widths given by $\Gamma^n \propto |\ampA^n|^2$ and $\Gamma^{cn} \propto |\ampA^{cn}|^2$.
The baryon asymmetry arises due to a difference in the decay rates
for the process
$\Gamma^n \equiv \Gamma(\Chi_n \to \Dp \widebar{\Dp^c} \Up)$
and for the conjugate process
$\Gamma^{c n} \equiv \Gamma(\Chi_n \to \Dp^c \widebar{\Dp} \Up^c)$. 
The total decay width for the Majorana $\Chi_n$ is the sum over the process and conjugate process partial widths,
i.e.
$\Gamma_{\Chi_n} = \Gamma^n + \Gamma^{cn} $. 

We find that there is no baryon asymmetry with only the leading order (LO), i.e. tree level, contribution.
When we include loop contributions
a nonzero difference in decay widths between the process and conjugate process could potentially arise,
implying a nonzero baryon asymmetry $\AsymB$.
For a nonzero $\AsymB$ to arise, as explained in detail in Ref.~\cite{Gopalakrishna:2022hwk},
along with a nonzero {\it weak phase} that switches sign in going from the process to the conjugate process,
a {\it strong phase} that does not switch sign must also be present.
The latter can come from loop diagrams with on-shell intermediate states,
giving a discontinuity in the amplitude,
with each cut giving a factor of $i$ (strong phase $\delta =\pi/2$).
Such a discontinuity adds to the loop amplitude $\ampA_1$ the imaginary piece $i\hat\ampA_1$,
which in turn adds to the interference term $\ampA_{01}$ the piece
$i \hat\ampA_{01} = i\hat\ampA_1 \ampA_0^*$,
and we write its contribution to the decay rate as $i\hat\Gamma_{01}$.
Similarly, for the conjugate process, the discontinuity adds the piece $i\hat\Gamma_{01}^c$.
We need to carry only this discontinuity piece for obtaining $\AsymB$ as the other terms are common for the process and conjugate process
and cancel in the difference.
Using the discontinuity, we obtain~\cite{Gopalakrishna:2022hwk}  
\beq
\Delta\hat\Gamma_{01}^n
      = \frac{1}{2 M_n} \int d\Pi_3 \ {\rm Im}(\hat\ampA_{01}^n - \hat\ampA^{nc}_{01})
      = \frac{1}{2 M_n} \int d\Pi_3 \ 2\, {\rm Im}(\hat\ampA_{01}^n) \ , 
\label{DGm01Ahn.EQ}
\eeq
where $d\Pi_3$ is the 3-body phase-space measure, 
and in obtaining the last equality, we have used ${\rm Im}(\hat\ampA_{01} - \hat\ampA^c_{01}) = 2\, {\rm Im}(\hat\ampA_{01})$
that follows if the weak phases are of opposite sign for the process and conjugate process,
which we will show explicitly later is the case in our setup. 
From this, the leading contribution to the baryon asymmetry is
\beq
\AsymB^n \equiv \frac{\Gamma^n - \Gamma^{c n}}{\Gamma^n + \Gamma^{c n}} \approx - \frac{\Delta\hat\Gamma^n_{01}}{\Gamma^n_0} \ ; \quad \AsymB = \sum_{n=1,2} \AsymB^n \ ,
\label{ABDGm01Gm0.EQ}
\eeq
where
the approximation is by taking the denominator at tree-level for the leading contribution to $\AsymB$,
and,
we write the total baryon asymmetry from decay $\AsymB$ as the sum over the contributions $\AsymB^n$ from $n=1,2$.

Proceeding along similar lines to decay, we argue in Ref.~\cite{Gopalakrishna:2022hwk}
that a baryon asymmetry could also be generated in the scattering processes, 
{SC-1}: $\Chi_n(p_n) \bar\Dp(k_i) \to \Dp(q_1) \Up(q_2)$, and 
{SC-2}: $\Chi_n(p_n) \bar\Up(k_i) \to \Dp(q_1) \widebar\Dp^c(q_2)$.
Writing the cross section difference between the SC-1, SC-2 process and its corresponding conjugate process
in terms of the tree-loop interference term
\beq
\Delta\hat\sigma_{01}
= \hat\sigma_{01} - \hat\sigma_{01}^c
= \frac{1}{v} \frac{1}{2E_n 2E_i} \int d\Pi_2 \ {\rm Im}(\hat\ampA^{(\sigma)}_{01} - \hat\ampA^{c\,(\sigma)}_{01}) = 2\,{\rm Im}(\hat\sigma_{01}) \ .
\label{Dsighat.EQ}
\eeq
where $d\Pi_2$ is the 2-body phase space measure,
we define the scattering baryon asymmetry as
\beq
\AsymBsign \equiv \frac{\sigma^{n} - \sigma^{cn}}{\sigma^{n} + \sigma^{cn}}
\approx -\frac{\Delta\hat\sigma_{01}^{n}}{\sigma_0^{n}} \ ;\quad \AsymBsig = \sum_{n=1,2} \AsymBsign \ , 
\label{ABsigDefn.EQ}
\eeq
where $\sigma_0$ is the LO cross section.

This completes a brief summary of the theoretical framework
and the baryon asymmetry generation mechanism put forth in Ref.~\cite{Gopalakrishna:2022hwk}.
In the rest of this paper,
we compute in this theory the baryon asymmetry in $\Chi$ decay and scattering processes and their thermal averages.
We show the expressions we obtain for the matrix elements as Dirac traces.
We evaluate these traces
to yield dot-products of the 4-momenta involved,
but will not present these lengthy expressions as they are not particularly illuminative.
Based on a numerical computation, we present the size of the baryon asymmetry generated in $\Chi$ decay and scattering processes
and study their dependence on the theory parameters. 

\section{The Baryon Asymmetry from $\Chi$ Decays}
\label{ABdec.SEC}

For obtaining the baryon asymmetry
between the decay process $\Chi_n \to \widebar{D^c}DU$ and its conjugate process $\Chi_n \to \widebar{D}D^cU^c$,
as reviewed in Sec.~\ref{genAB.SEC},
we need to compute the interference term between the tree-level and loop-level, $\Delta\hat\Gamma_{01}^n$ (cf. Eq.~(\ref{DGm01Ahn.EQ})). 
We discuss this computation and the baryon asymmetry we obtain in this section
(cf. Appendix~\ref{MatElcomp.SEC}).
We discuss next the tree and loop level contributions to the $\Chi_n$ decay process. 

\subsection{The tree-level decay width}
\label{chi2DDUGmLO.SEC}

\begin{figure}
  \begin{center}
    \includegraphics[width=0.22\textwidth]{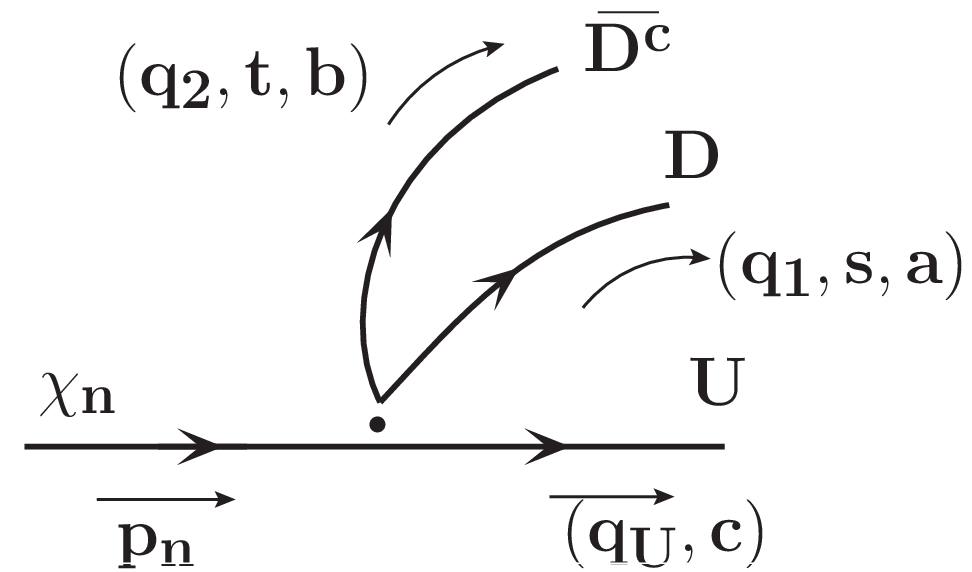}
    \includegraphics[width=0.3\textwidth] {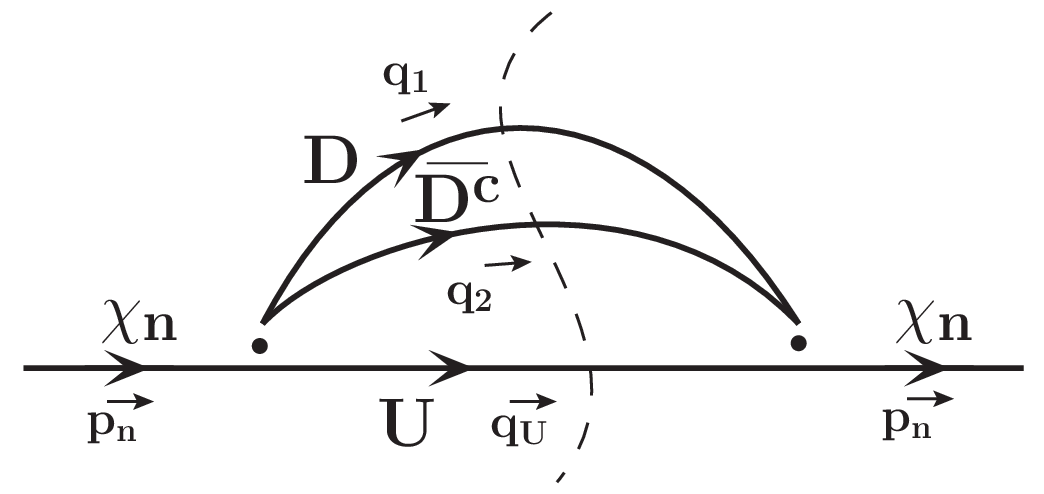}
  \end{center}
\caption{
  The tree-level (LO) $\Chi_n \to \widebar{D^c}DU$ decay process Feynman diagram (left).
  The momentum, spin index, and color index, are shown as $(q,s,a)$ respectively.
The LO $\Gamma_0$ with the dashed lines showing the cut propagators giving 3-body phase-space (right). 
  \label{chi2DDU-LO.FIG}
  }
\end{figure}

In Fig.~\ref{chi2DDU-LO.FIG}~(left) we show the Feynman diagram for the
$\Chi_n \to \widebar{D^c}DU$ decay at LO, i.e. tree-level in the VV effective interaction, giving the decay width $\Gamma_0^n$. 

We write the LO matrix element as
\beq
i \ampALOn = \frac{i}{2 \Lambda^2} C_0 \, \MLOXU^\mu_n \, \MLOot_\mu \ , 
\label{ampA0n.EQ} 
\eeq
where
$\MLOot$ is the $\Dp\Dp$ part, 
$C_0 \equiv \epsilon^{abc}/\sqrt{6}$ is the color-factor
in which the $1/\sqrt{6}$ is from the final state normalization
(cf. Appendix~\ref{DDUstate.SEC}). 
The amplitude $\ampA_0^{cn}$ for the conjugate process $\Chi_n \to \bar{D}D^cU^c$ 
is got from Eq.~(\ref{ampA0n.EQ})
by replacing $\MLOXU_n \to \McLOXU_n$ in which $\GVBn \to \GZn$, and we have $\McLOot = \MLOot$.

We evaluate the 3-body phase-space element equivalently as a cut of a loop diagram as we show in Fig.~\ref{chi2DDU-LO.FIG}~(right).
We thus obtain the LO decay width
\beq
           \Gamma_0^n = \frac{1}{2 M_n} \int d\Pi_3 \ |\ampA_0^n|^2 
           = \frac{1}{2 M_n} \frac{1}{4\Lambda^4} \int d\Pi_3\ (\frac{1}{2}\MLOXUSq)^{\mu\tau}_{nn}\, C_{00}\, (L_{12})_{\mu\tau} \ ,
\label{Gam0-3PSMSq.EQ}
\eeq
where $C_{00} = C_0^2 = \epsilon^{abc} \epsilon^{abc}/(\sqrt{6})^2 = 1$ is the color factor,
the 1/2 in front of the $\MLOXUSq$ is from averaging over the initial state spins, 
and we have 
\beq
(\MLOXUSq)^{\mu\tau}_{nn} = {\rm Tr}[\GVBn^\mu (\pslash_n + M_n) \GVn^\tau (\qslash_U + M_U)] \ ; \quad
(L_{12})_{\mu\tau} = 2 |\tilde{g}|^2 {\rm Tr}[\gamma_\tau (\qslash_1+M) \gamma_\mu (\qslash_2-M) ] \ . 
\eeq
The conjugate process LO decay width $\Gamma_0^{cn}$ is obtained similarly using $|\ampA_0^{cn}|^2$ above.
We can write Eq.~(\ref{Gam0-3PSMSq.EQ}) in terms of the dimensionless $f_{00}^n$ as
(cf. Appendix~\ref{chi2DDUMXULODet.SEC}) 
\beq
\Gamma_0^n = \frac{1}{2} \frac{M_n^5}{4\Lambda^4} f_{00}^n =  \left(\frac{M_\chi^5}{\Lambda^4}\right) \hat\Gamma_0^n \ ; \quad {\rm where}\
    \hat\Gamma_0^n \equiv \frac{1}{8} \left(\frac{M_n}{M_\chi}\right)^5 f_{00}^n \ . \label{Gm0f00.EQ}
\eeq
This also defines a dimensionless rate $\hat\Gamma_0^n$ with the $M_\chi$ and $(\rMLtxt)$ factors removed. 
The $M_n^5$ scaling can be anticipated on dimensional grounds.
As already mentioned above, for obtaining the baryon asymmetry at leading order, we take the dimensionless total width to be $\hat\Gamma_{\Chi_n} \approx 2 \hat\Gamma_0^n$.

We find no rate asymmetry at tree-level, and have to go to higher order in the VV interaction in order to potentially have a nonzero baryon asymmetry. 
We turn next to identifying loop diagrams with this goal in mind.
We first discuss loop contributions involving only the single operator in Eq.~(\ref{LIntVVMB.EQ}),
and following this, with multiple operators including those in Eq.~(\ref{modAops.EQ}).

\subsection{Single operator loop contribution to decay baryon asymmetry}
\label{oneOpAsymb.SEC}

We identify here loop diagrams involving only the operator of Eq.~(\ref{LIntVVMB.EQ})
using the Feynman vertices in Fig.~\ref{FeynRules.FIG}, 
and compute the resulting decay rate difference in the tree-loop interference terms $\Delta\hat\Gamma_{01}$ between the process and conjugate process.

\begin{figure}
  \begin{center}
    \includegraphics[width=0.3\textwidth]{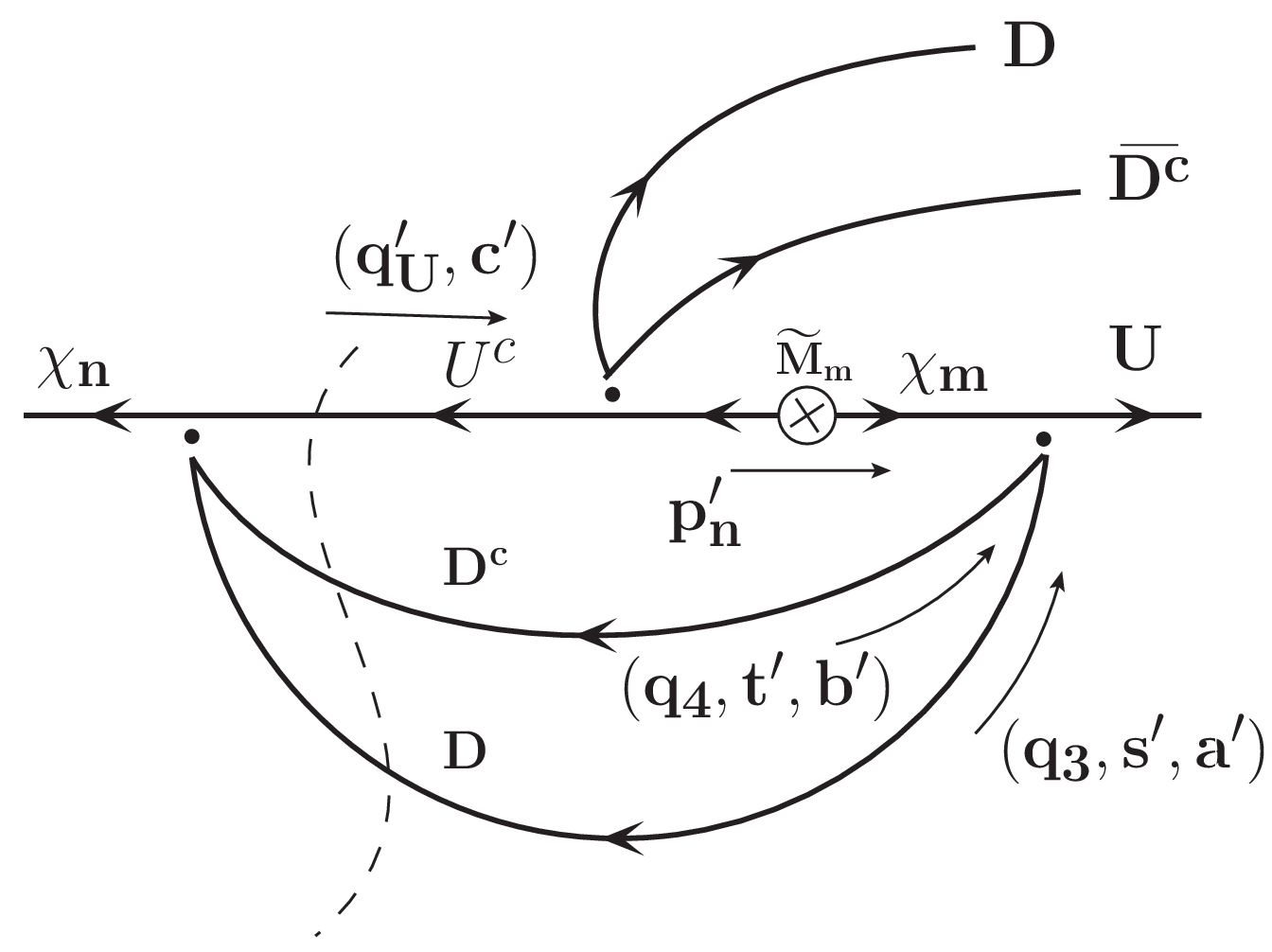}
    \includegraphics[width=0.3\textwidth]{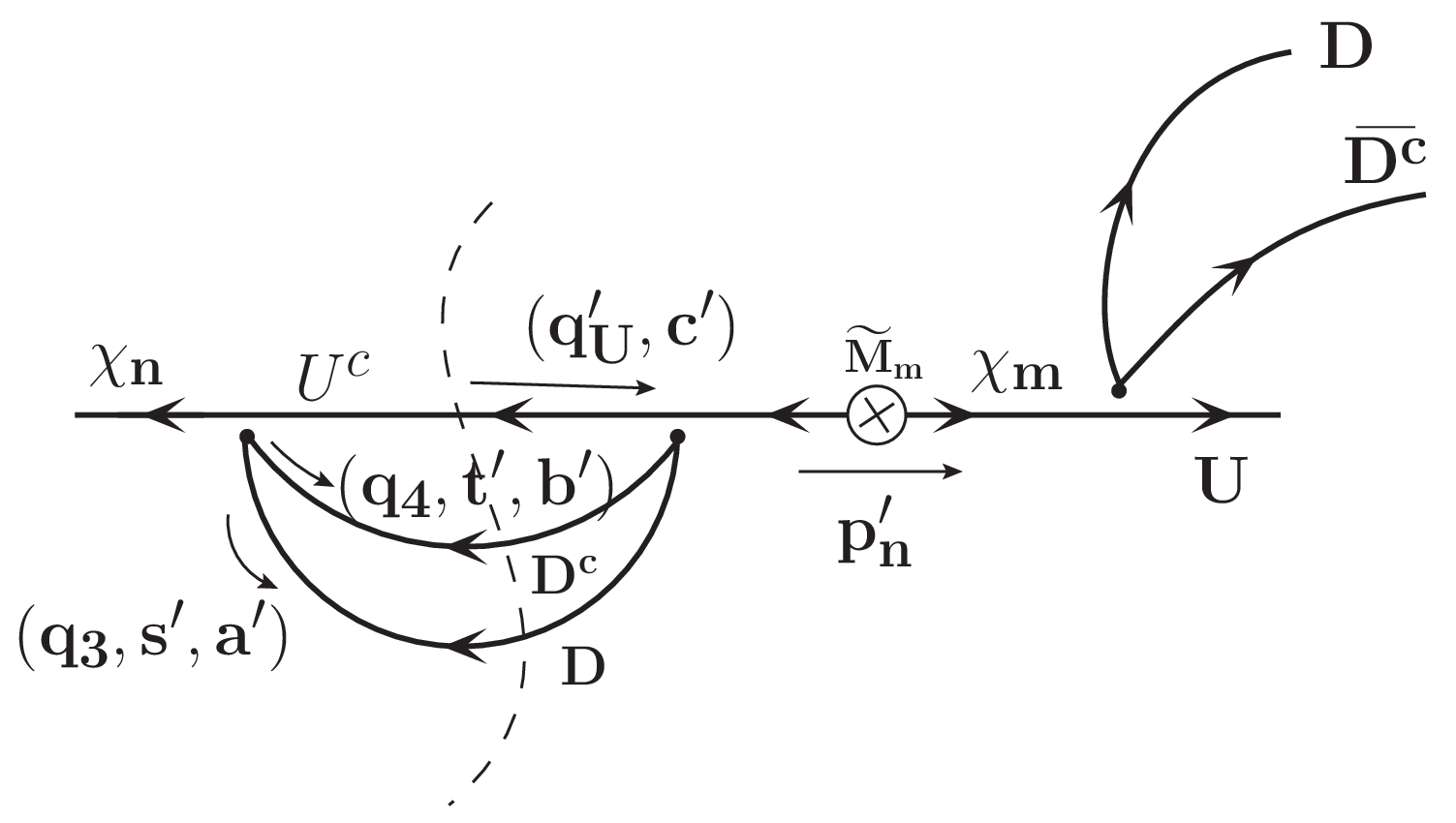}
    \includegraphics[width=0.3\textwidth]{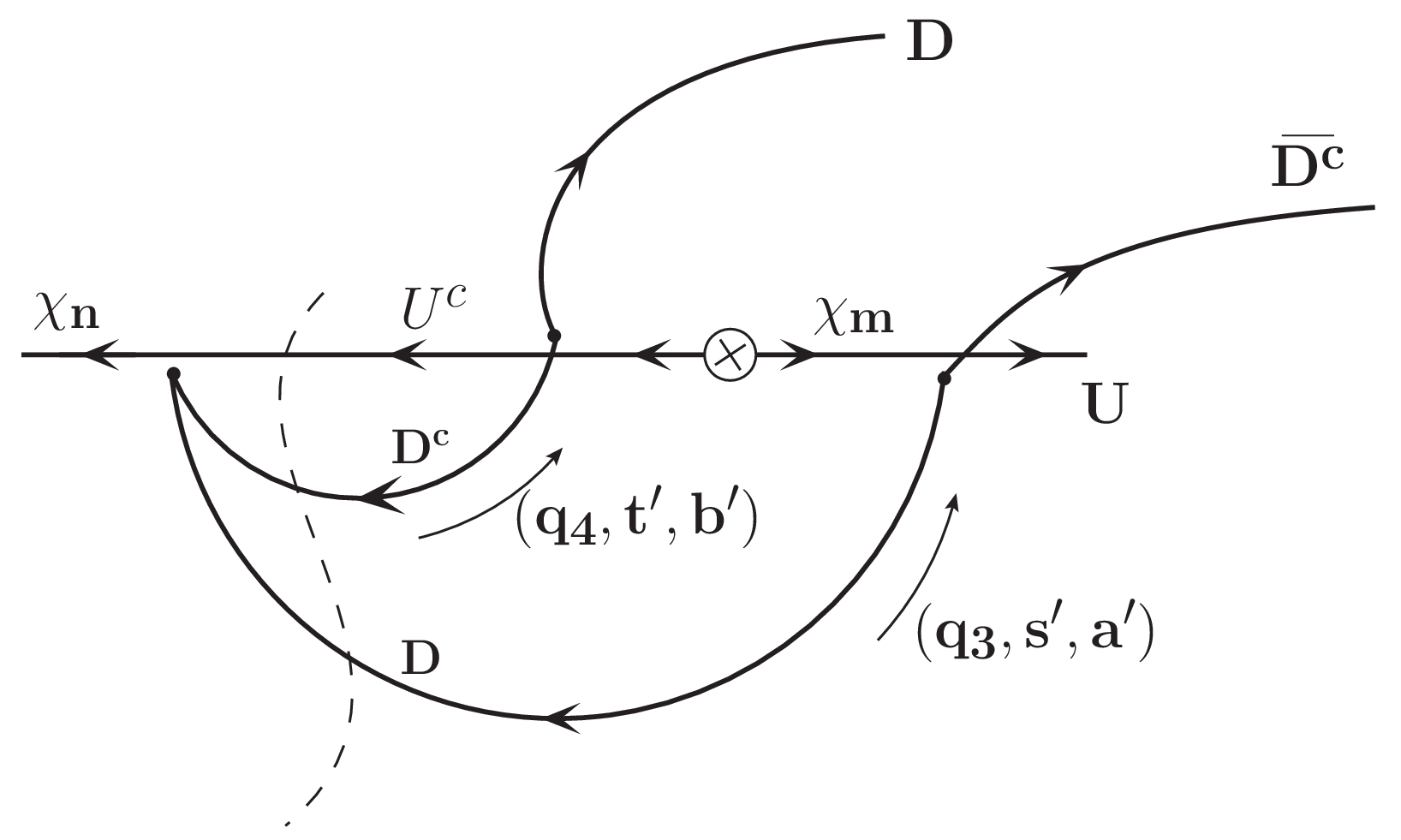}
  \end{center}
\caption{
  The $\Chi_n \to \widebar{D^c}DU$ loop processes A$_1$,~B$_1$,~C$_1$ (left, middle, right).
  The dashed curve shows a cut in the loop, contributing a factor of $i$.
\label{chi2DDU-NLO.FIG}
}
\end{figure}

In Fig.~\ref{chi2DDU-NLO.FIG}, we show the Feynman diagrams contributing to the
$\Chi_n \to \widebar{D^c}DU$ decay at the loop-level, for each $n=\{1,2\}$.
We show three 2-loop processes A$_1$,~B$_1$,~C$_1$ in Fig.~\ref{chi2DDU-NLO.FIG},
while a fourth process D$_1$ differs from C$_1$ by switching the $D$ and $\widebar{D^c}$ in each of the right two vertices
and yields the same contribution as C$_1$ due to them being identical particles, which we therefore do not show explicitly. 
The $\otimes$ shows the Majorana mass insertion $M_m$ on the $\Chi_m$ line (with $m=\{1,2\}$) that violates baryon number.
The clash in the arrows in the $\Chi_m$ line signals the baryon number violation due to the Majorana mass. 

We show in the figure the momenta, spin index, and color index on each fermion line as $(q,s,a)$, or sometimes as $(q,a)$ with the spin label suppressed. 
The final state labels are the same as in Fig.~\ref{chi2DDU-LO.FIG}. 
Momentum conservation implies $q_2 = p_n - q_1 - q_U$, $q_4 = p_n - q_3 - q'_U$, and ${p'_n} = p_n - (q_1 + q_2 + q_3 + q_4)$. 
We are interested in the unpolarized decay width and therefore sum over all spins. 
We carefully handle the fact that the $\widebar{D^c},D$ lead to two identical fermions in the final state. 

We do not show the loop diagrams for the conjugate process $\Chi_n \to \widebar{D}D^cU^c$ in the figure,
but obtain these by switching fields and their charge-conjugated fields, i.e. $\psi \leftrightarrow \psi^c$,
which implies reversing all arrows on the fermion lines.
The VV interaction with the fields conjugated is derived in Ref.~\cite{Gopalakrishna:2022hwk},
and the Feynman vertices are included in Fig.~\ref{FeynRules.FIG}.
As explained in Ref.~\cite{Gopalakrishna:2022hwk}, for the charge conjugated fields,
the fermion-number-flow arrow is to be read in the opposite sense in picking the spinors and the sign of the mass term in the propagator.
The process and conjugate process matrix element expressions have the $u,v$~spinors in the same places,
but only the coupling is changed to the corresponding conjugated forms given in the Feynman rules.
To obtain the difference in the decay width and the baryon asymmetry $\AsymB$,
we compute the difference between the tree-loop interference terms for the process and conjugate process,
and integrate over phase-space.
In our expressions below, we are interested only in the terms that survive when the difference is taken,
and not the common terms that cancel.

The loop contributions in diagrams $(i)$~=~\{A$_1$,~B$_1$,~C$_1$\} in Fig.~\ref{chi2DDU-NLO.FIG} are
\beq
i \ampANLOn = 
\frac{-i}{8 \Lambda^6}
\int d\Pi'_3 \, \sum_{m,(i)}
\frac{{\MNLOXUi}^{\sigma\mu\nu}_{nm}}{[{q'_U}^2 - M_U^2 + i \epsilon] [q_3^2 - M_D^2 + i \epsilon] [q_4^2 - M_D^2 + i \epsilon]} \,
\frac{C_1^{(i)} (\MNLOoti\Ltfi)_{\mu\sigma\nu}}{[{p'_n}^{\!\!(i)\, 2} - M_m^2 + i\epsilon]}  \ , \label{ampANLOn.EQ}
\eeq
where
the $(i)$ on the various factors 
show that they depend on the particular diagram, 
$\MNLOoti$ denotes the $\Dp\Dp$ part of the 3-body final-state phase-space,
$\Ltfi$ stands for the $\Dp\Dp$ loop part,
and
$C_1^{(i)}$ is the color factor. 
$d\Pi'_3$ is the loop integration measure. 
The loop contribution for the conjugate process is obtained by making the change $\MNLOXU \to \McNLOXU$ above with
$\GVB \leftrightarrow \GZ$, and with $\McNLOot, \Lctf$. 

The A$_1$ and C$_1$ diagrams generate a UV finite dimension-6 operator
when computed in terms of the renormalized masses, couplings and wave-functions,
and we think of our expressions being in terms of these renormalized quantities. 
\label{diagBsub.PG}
The B$_1$ diagram (with $p'_n = p_n$), however, has a $\Chi_n \bar\Chi_m$ 2-point function piece,
which being a relevant operator, could potentially be divergent.
This has to be renormalized (subtracted) properly to be consistent with being in the mass basis. 
We explain our renormalization and subtraction scheme for this 2-point function in Appendix~\ref{Chinm2ptFcn.SEC}.
The divergence is only in the hermitian principal value (PV) part of the 2-point function
and is subtracted with mass and wave-function renormalization counterterms.
The subtraction ensures that we leave the $\Chi_n$ masses $M_n$
uncorrected by the loop contributions at the on-shell point, consistent with being in the mass basis.
The discontinuity in the 2-point function gives a non-hermitian (dissipative or absorptive) piece,
which we take without any subtraction as there is no counterterm for this.
For $m\!\!=\!\!n$ with both legs being the same particle and on-shell, this dissipative piece 
yields the LO decay width $\Gamma_0,\Gamma_0^c$ respectively.
For the diagram B contribution, what is picked up is only the discontinuity (in the imaginary part)
of the $\Chi_n \bar\Chi_m$ 2-point function for $m\!\!\neq\!\!n$
with an offshell $\Chi_m$ and the cut on the intermediate $U^cD^c\bar{D}$ giving a strong phase of $i$.   
The discontinuity for $m\!\!=\!\!n$
will not contribute to the baryon asymmetry since it puts $\Chi_m$ also on-shell,
giving an extra $i$
and leaving no strong phase in the amplitude. 
Also, for $m\!\!=\!\! n$, the other possibility of $\Chi_m$ onshell and no $UDD$ cut, 
although has a single factor of $i$ as the strong phase, 
does not survive since this is at the subtraction point at which the counterterm fully cancels it.
Therefore, the discontinuity in the $n\!\!\neq\!\!m$ contribution
that leaves the $\Chi_m$ offshell as an intermediate state at $p_n^2\!\! =\!\! M_n^2$ 
is the only contribution to the baryon asymmetry from the B$_1$ diagram.
The enhancement in the B$_1$ amplitude as $M_m\!\to\!M_n$ for $m\!\neq\!n$ can enhance the baryon asymmetry
and is the idea in {\em resonant baryogenesis};
for recent studies along these lines, but with SS interaction,
see for example Refs.~\cite{Dev:2015uca,Davoudiasl:2015jja} and references therein. 

For $M_n > (2 M_\Dp + M_\Up)$, for any $n$,
the three $\Dp,\Dp,\Up$ intermediate states can all simultaneously go on-shell in the loop diagrams
of Fig.~\ref{chi2DDU-NLO.FIG}, shown as the dashed curve in the diagrams.
The Cutkosky rule (see for example Ref.~\cite{Peskin:1995ev}) prescribes that a cut propagator be replaced as
$1/(k^2 - m^2 + i\epsilon) \to -2\pi i \delta(k^2 - m^2)$. 
Cutting three propagators then gives a factor of
$(-2\pi i)^3 = (2\pi)^3 i = (2\pi)^3 e^{i\pi/2}$, i.e. an extra phase of $\pi/2$,
that is the {\em same} for the process and the conjugate process, acting as a strong phase~\cite{Gopalakrishna:2022hwk}.
Furthermore, for simplicity,
we choose the masses such that the $\Up \to \Chi_m \Dp^c \bar\Dp$ decay is not kinematically allowed
so that a second cut cannot be made in the loop contributions
which can be ensured with $M_m > (M_U - 2 M_D), \forall m$.
If the second cut can be made, its contributions can also be included in the same way.

The discontinuity added by the cut propagators is $i\ampAhatNLOn$, as discussed in Sec.~\ref{genAB.SEC}. 
Replacing the cut denominators by the corresponding $\delta(...)$ functions in Eq.~(\ref{ampANLOn.EQ}), we find $\ampAhatNLOn$ to be
\bea
\ampAhatNLOn = -\frac{1}{8 \Lambda^6}
\int d\Pi'_3 \sum_{m,(i)} {\MNLOXUi}^{\sigma\mu\nu}_{nm}
\frac{C_1^{(i)} (\MNLOot\Ltf)_{\mu\sigma\nu}^{(i)}}{[{p'_n}^{\!\!(i)\, 2} - M_m^2]} \,
\left[ \frac{1}{2} (2\pi)^3 \delta^{\rm cut}_{34U} \right]  \label{ampA1hatn.EQ} \ ,  \\
{\rm where}\ \delta^{\rm cut}_{34U} = \delta(q_3^2 - M_D^2) \delta(q_4^2 - M_D^2) \delta({q'_U}^2 - M_U^2) \ , \nonumber
\eea
where the 1/2 in the last $[...]$ is from the ${\rm Im}(...) = (-i/2){\rm Disc}(...)$.
 Since we take $M_m > (M_U - 2 M_D), \forall m$, as mentioned earlier,
 we cannot have $\Chi_m$ on-shell and therefore we omit the $+i\epsilon$ in the denominator. 
We compute from Eq.~(\ref{ampA1hatn.EQ}) and (the conjugate of) Eq.~(\ref{ampA0n.EQ})
the tree-loop interference term $\hat\ampA_{01}^n = \ampALOn^* \ampAhatNLOn$,
 and take the average over initial-state spins and sum over final-state spins and colors, to get the interference term due to the discontinuity as 
\beq
\hat\ampA_{01}^n = -\frac{1}{(16 \Lambda^8)} 
\int
d\Pi'_3\,
\frac{1}{2}\,
\sum_{m,(i)}
\frac{1}{2}{\MLONLOXUi}^{\sigma\mu\nu\tau}_{nm}
\frac{C_{01}^{(i)} (\MNLOoti\Ltfi)_{\mu\sigma\nu} \MLOot^\dagger_\tau}{[{p'_n}^{\!\!(i)\, 2} - M_m^2]}  \,
\label{ampA01hatn.EQ} \ ,
\eeq
where
the color factor $C_{01}^{(i)} = C_0 C_1^{(i)}$,
the 1/2 in front of $\MLONLOXU$ is from averaging over the initial state spins.
We present the $\Dp$-loop factors later (cf. Eq.~(\ref{LDABCD.EQ})).
For the conjugate process, the analogous factor, $\McLONLOXU$,
is obtained by making the change $\GVB \leftrightarrow \GZ$ in $\MLONLOXU$ above.\footnote{\label{GVB2BZcorr.FN} Including
  the $\overline{\Dp^c}(...)\Dp$ part of the coupling also, the change is more correctly 
  $[\gamma^\mu \otimes \GVB] \to [(-\gamma^\mu) \otimes (-\GZ)]$,
  but can effectively be thought of as $[\gamma^\mu \otimes \GVB] \to [\gamma^\mu \otimes \GZ]$ as stated above.
}

We now make an important observation from the expression for $\MLONLOXU$ and $\McLONLOXU$.
In the limit of the $\Chi_n$ masses $M_{1,2}$ becoming equal, all kinematical factors become independent of $n,m$,
and given that $\AsymB \propto \int \sum_{n,m} \, {\rm Im}(\MLONLOXU - \McLONLOXU)$,
the $n,m$ dependence is now only due to their dependence on
the couplings $\hat{g}_{n,m}$ (and their conjugates) in Eq.~(\ref{ampA01hatn.EQ}).
We can explicitly show that the coupling structure in ${\rm Im}(\MLONLOXU)$ (or in $\McLONLOXU$) when summed over $n,m$
goes to zero in this limit owing to the unitarity of the $\Umat$.
This is expected since in this limit, the two Majorana $\Chi_n$ can be assembled into a single Dirac fermion
with fermion number (baryon number) conserved in the theory,
which therefore implies that ${\rm Im}(\MLONLOXU)$ must go to zero in this limit
since it is a fermion number violating contribution, as evidenced by the clashing arrows in the $\Chi_m$ propagator
in Fig.~\ref{chi2DDU-NLO.FIG}.
In our theory, as discussed in Ref.~\cite{Gopalakrishna:2022hwk}, 
$C$-invariance is present if and only if $\ghLn^* = \ghRn$.
Taking this limit in our expressions leads to a zero baryon asymmetry
(cf. Appendix~\ref{oneOpAsymbDet.SEC}).
Next, $CP$-invariance is present if and only if $\ghLn, \ghRn$ are real, and we find a zero baryon asymmetry in this limit.
Lastly, in the ``Dirac limit'', i.e. the $\tilde{M}_{L,R} = 0$, we find a zero baryon asymmetry.
It is reassuring that our formalism passes these nontrivial checks.

\begin{figure}
  \begin{center}
    \includegraphics[width=0.32\textwidth] {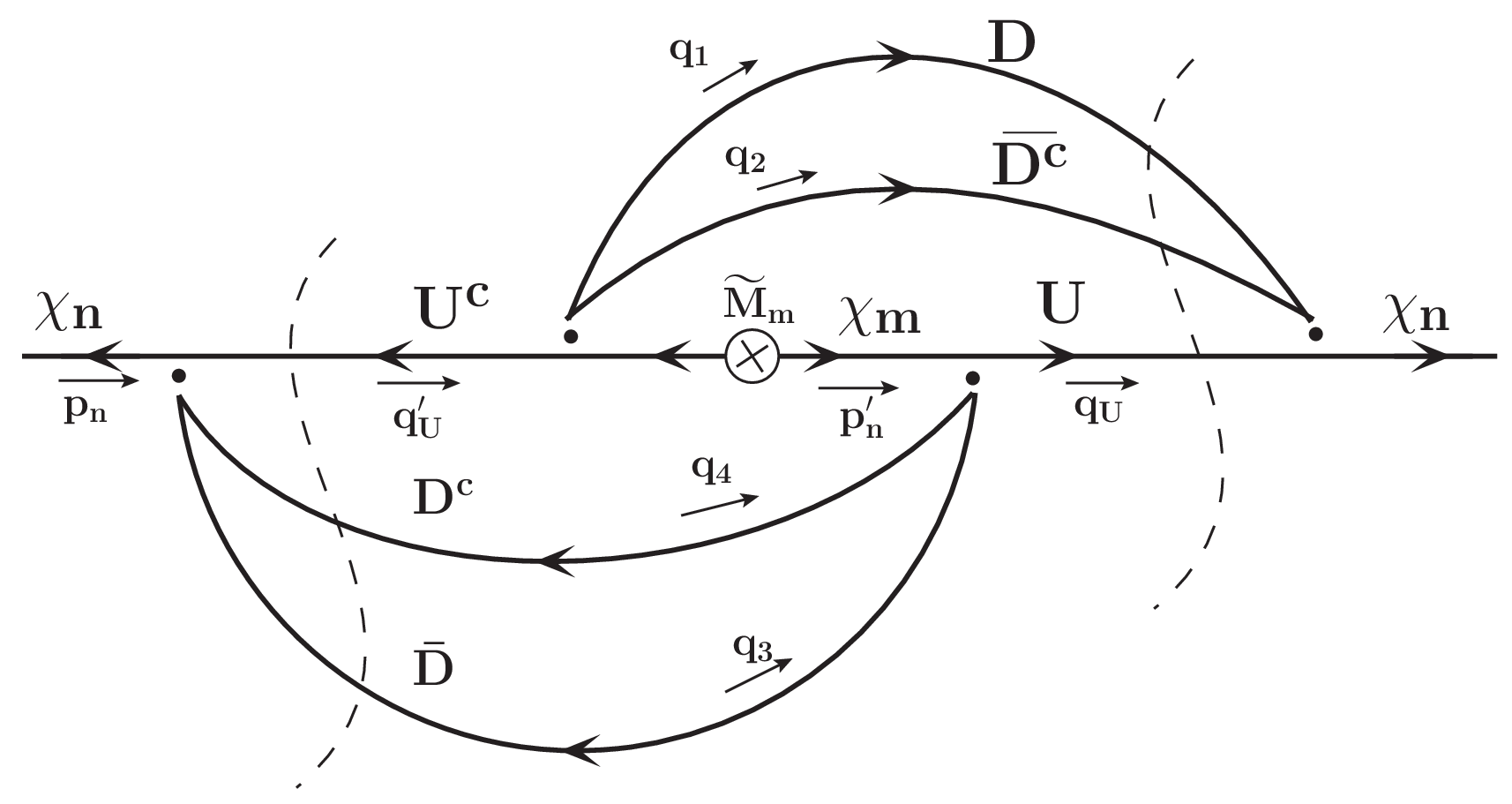}
    \includegraphics[width=0.32\textwidth] {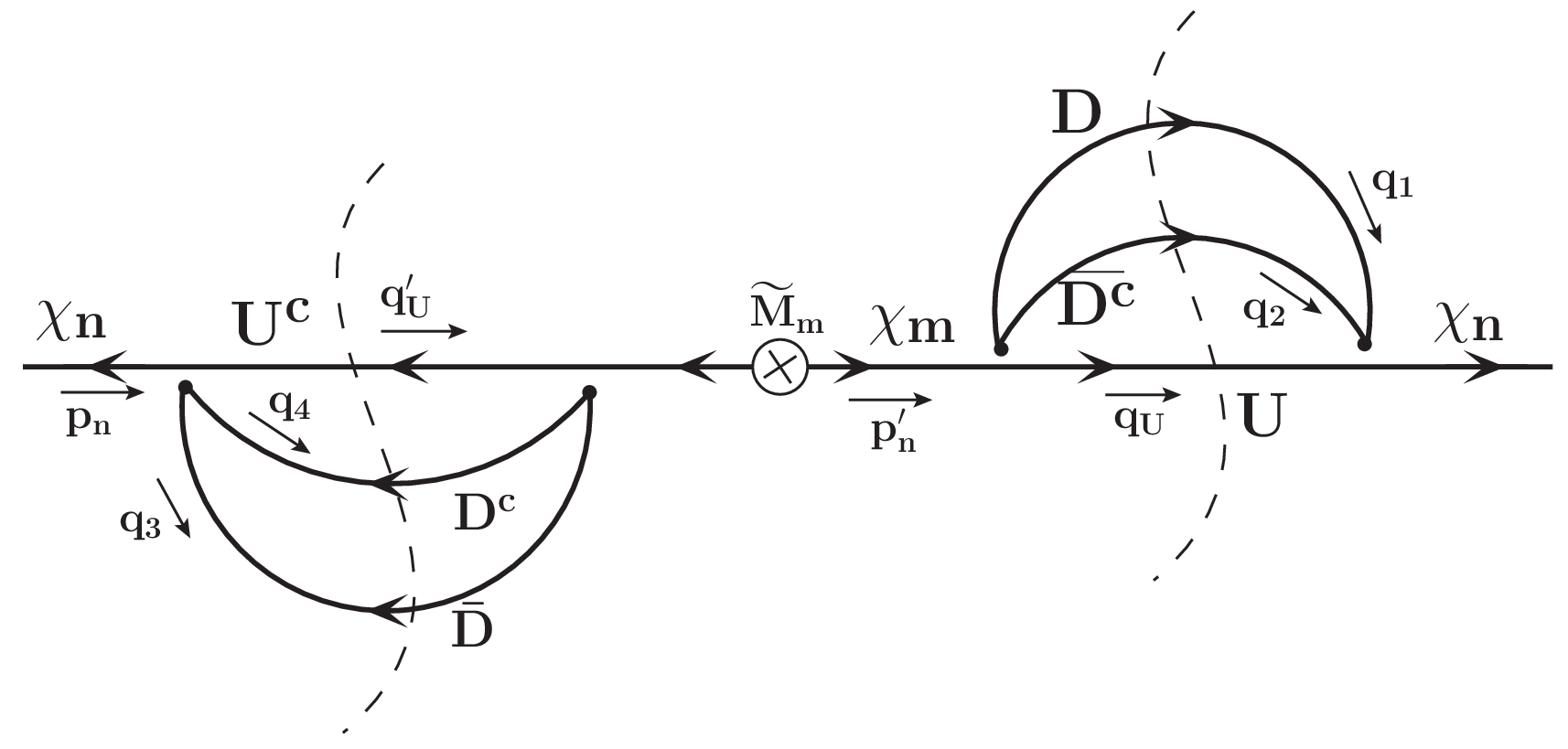}
    \includegraphics[width=0.32\textwidth] {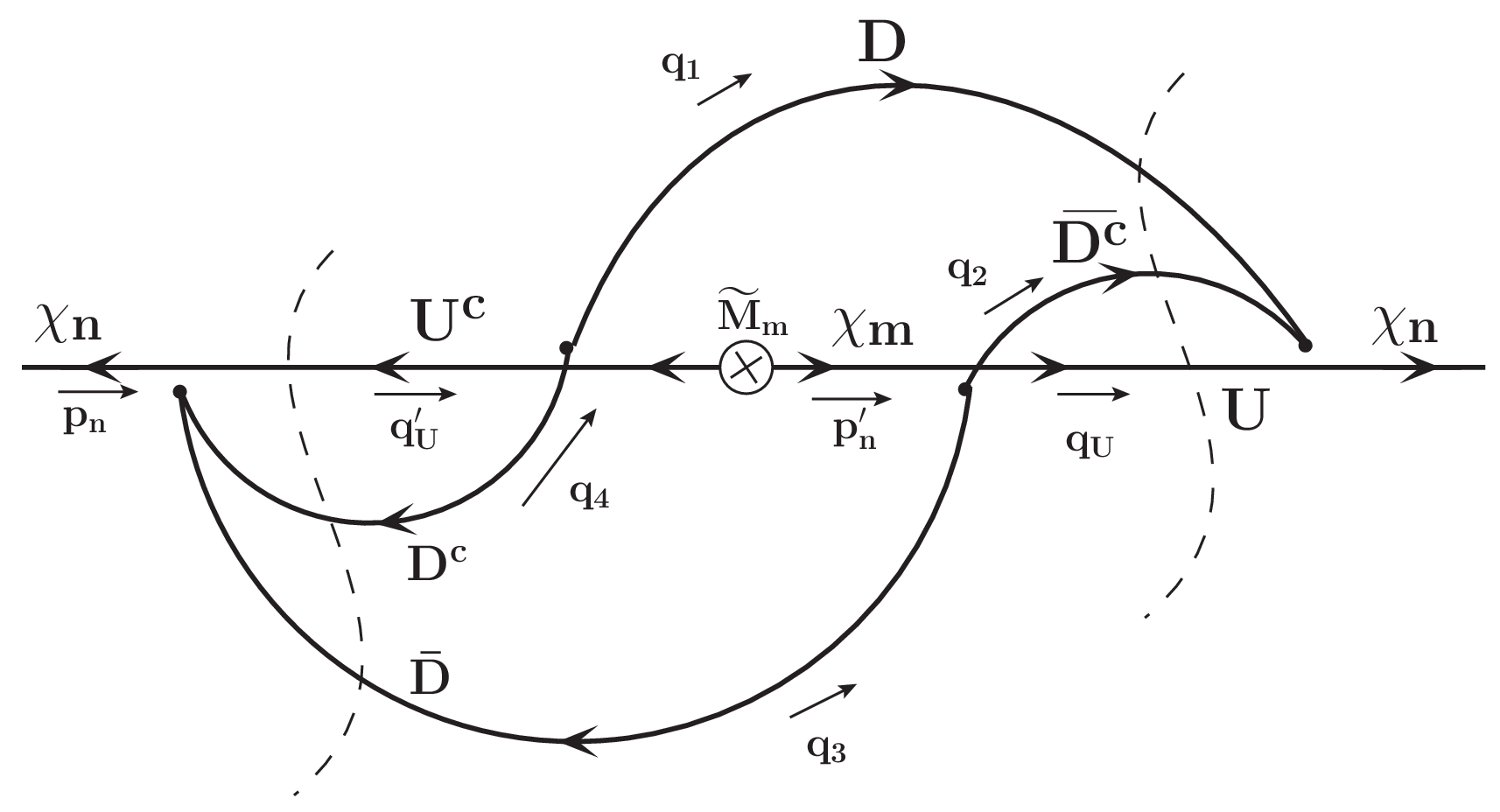}
    \caption{The tree-loop interference term diagrams A, B, C (left, middle, right respectively) required to compute $\Delta\hat\Gamma_{01}$,
      in which the left cuts in each diagram gives the imaginary piece to the amplitude
      while the right cuts yields the 3-body final-state.
      The conjugate process diagrams are obtained by reversing all fermion-line arrows.
      \label{chi2DDU-LONLOCut.FIG}
    }
  \end{center}
\end{figure}
As we did for the tree-level decay rate, 
for the tree-loop interference term $\hat\Gamma_{01}$ also, 
we equivalently write the 3-body phase-space integral as a cut of a loop. 
We thus have the interference term represented by the diagrams in Fig.~\ref{chi2DDU-LONLOCut.FIG}, 
with the phase-space appearing as the right cut in the diagrams,
and yields $\ampANLOn^{(i)}\, \ampALOn^*$
for each loop diagram $(i)$ =\{\!\! A$_1$,~B$_1$,~C$_1$\!\! \} in Fig.~\ref{chi2DDU-NLO.FIG}. 
We now label the corresponding tree-loop interference diagram with $(i)$ =\{\!\! A,~B,~C\!\! \} (left, middle, right, respectively)
in Fig.~\ref{chi2DDU-LONLOCut.FIG}. 
The fourth diagram D,
gives the same contribution as diagram C for the reason mentioned earlier,
and is not shown.
The interference term for the conjugate process $\Gamma^c_{01}$ is obtained by reversing all the arrows on the fermion
lines in the diagram, with all fermion fields conjugated.
The difference between these contributions yields $\Delta\hat\Gamma_{01}$.
The cut shown by the dashed curve on the left gives the discontinuity as noted in Sec.~\ref{genAB.SEC}.
With this cut made, the diagrams in Fig.~\ref{chi2DDU-LONLOCut.FIG} correspond to $\Delta\hat\Gamma_{01}$,
with the final-state 3-body phase-space also included as in Eq.~(\ref{Gam0-3PSMSq.EQ}),
and we have 
\beq
\Delta\hat\Gamma^{n}_{01} = \frac{1}{2M_n} \frac{1}{16 \Lambda^8}
\int
d \Pi_3\, d\Pi'_3\,
\frac{1}{2} \sum_{m,(i)} \, 2\, {\rm Im}\left[\frac{1}{2}{\MLONLOXUi}^{\sigma\mu\nu\tau}_{nm}\right] \frac{C^{(i)}_{01} {(L^{(i)}_D)}_{\mu\tau\nu\sigma}}{({{p'_n}^{\!\!(i)}}^2 - M_m^2)} \ ,
\label{Gm01cuts.EQ} 
\eeq
where the sum over $i=\{A,B,C,D\}$ adds the contributions from diagrams A,~B,~C,~D shown in Fig.~\ref{chi2DDU-LONLOCut.FIG}
(with C and D diagram contributions being equal).
$L^{(i)}_D$ is the $\Dp$-loop part of the matrix element, 
and the superscript $(i)$ on ${p'_n}^{\!\!(i)}$ reminds us that $p'_n$ is different for each of these contributions.
Thus, in Eq.~(\ref{Gm01cuts.EQ}), we exploit the equivalence and side-step
computing the tree and loop contributions separately but rather directly compute the interference term,
as depicted diagrammatically in Fig.~\ref{chi2DDU-LONLOCut.FIG}. 
Computing these we find
(cf. Appendix~\ref{oneOpAsymbDet.SEC})  
\bea
{\MLONLOXUi}^{\sigma\mu\nu\tau}_{nm} &\equiv& [{\MLOXU}^{\tau}_n]^* {\MNLOXU}^{\sigma\mu\nu}_{nm}  \nonumber \\
          &=& (\bar{u}(q_U) \GVBn^\tau u(p_n))^\dagger \bar{u}(q_U) \GVBm^\sigma ({\pslash'_n} + M_m) \GZBm^\mu (\qslash'_U + M_U) \GZn^\nu u(p_n) \nonumber  \\
          &=& {\rm Tr}[ \GVBm^\sigma ({\pslash'_n} + M_m) \GZBm^\mu (\qslash'_U + M_U) \GZn^\nu (\pslash_n + M_n) \GVn^\tau  (\qslash_U + M_U) ] \ , \label{M01XUdefn.EQ}
\eea
and the $\Dp$-loop factors
(cf. Appendix~\ref{LONLOintContr.SEC})
\bea
({L_D^A})^{\mu\tau\nu\sigma} &=& -4 {C_{01}^A} \, {\rm Tr}[{\gamma^\mu} (\qtslash-M_D) {\gamma^\tau} (\qoslash+M_D)]\, {\rm Tr}[{\gamma^\nu} (\qthslash-M_D) {\gamma^\sigma} (\qfslash+M_D)] \ , \nonumber \\
({L_D^B})^{\mu\tau\nu\sigma} &=& -4 {C_{01}^B} \, {\rm Tr}[{\gamma^\tau} (\qoslash+M_D) {\gamma^\sigma} (\qtslash-M_D)]\, {\rm Tr}[{\gamma^\mu} (\qfslash+M_D) {\gamma^\nu} (\qthslash-M_D)]  \ , \nonumber \\
({L_D^{C14}})^{\mu\tau\nu\sigma} &=& {C_{01}^C} \,  {\rm Tr}[{\gamma^\tau} (\qoslash+M_D) {\gamma^\mu} (\qthslash+M_D) {\gamma^\nu} (\qfslash-M_D){\gamma^\sigma} (\qtslash-M_D)]  \ , \nonumber \\ 
({L_D^{C24}})^{\mu\tau\nu\sigma} &=& {C_{01}^C} \, {\rm Tr}[{\gamma^\tau} (\qoslash+M_D) {\gamma^\sigma} (\qfslash+M_D) {\gamma^\nu} (\qthslash-M_D) ({\gamma^\mu} (\qtslash-M_D)]  \ , \nonumber \\
{L_D^{C13}})^{\mu\tau\nu\sigma} &=& {C_{01}^C} \, {\rm Tr}[{\gamma^\tau} (\qoslash+M_D) {\gamma^\mu} (\qfslash+M_D) {\gamma^\nu} (\qthslash-M_D) {\gamma^\sigma} (\qtslash-M_D)]  \ , \nonumber \\ 
({L_D^{C23}})^{\mu\tau\nu\sigma} &=& {C_{01}^C} \, {\rm Tr}[{\gamma^\tau} (\qoslash+M_D) {\gamma^\sigma} (\qthslash+M_D) {\gamma^\nu} (\qfslash-M_D) {\gamma^\mu} (\qtslash-M_D)] \ ,
\label{LDABCD.EQ}
\eea
where the color factors are $C_{01}^{A,C} = 12/(\sqrt{6})^2 = 2$, $C_{01}^B = 36/(\sqrt{6})^2 = 6$. 

We can write the single operator contribution to the decay width difference in terms of the dimensionless $f_{01}$ as 
\bea
\Delta\hat\Gamma_{01}^{n} &=& \frac{1}{2} \frac{M_n^9}{16 \Lambda^8} \sum_m 2 f_{01}^{nm}
   = \left(\frac{M_\chi^9}{\Lambda^8}\right) \left[ \frac{1}{16} \left(\frac{M_n}{M_\chi}\right)^9 \sum_m f_{01}^{nm} \right]  \ , \label{DGm01f01.EQ} 
\eea
where we have a factor of 2 in $\Delta \hat\Gamma_{01}^{n}$ from 2~Im(...) including the imaginary piece from the conjugate process.
Using Eqs.~(\ref{Gm0f00.EQ}) and (\ref{DGm01f01.EQ}) in Eq.~(\ref{ABDGm01Gm0.EQ}) we get 
\bea
\AsymB^n = - \frac{M_n^4}{4\Lambda^4} \frac{\sum_{m} 2 f_{01}^{nm}}{f_{00}^n} =  \left(\frac{M_\chi}{\Lambda}\right)^4 \AsymBh^n \ ; \quad {\rm where}\ 
\AsymBh^n \equiv \left[-\frac{1}{2} \left(\frac{M_n}{M_\chi}\right)^4 \frac{\sum_{m} f_{01}^{nm}}{f_{00}^n} \right] \ .  \label{ABf01f00.EQ} 
\eea 
For the conjugate process, as already discussed, $\McLONLOXU$ is got by $\GVB \leftrightarrow \GZ$ in $\MLONLOXU$
with the $L_D$ unchanged, 
which is $\hat{g}_{\lambda n} \to \hat{g}^*_{\lambda n}$ and $P_L \leftrightarrow P_R$.
The imaginary part of the $\hat{{\cal M}}_{k\, R,L}$ are due to the $\pm \gamma^5$ in $P_{R,L}$,
and therefore we have the property $\hat{{\cal M}}_{k\, R,L} = \hat{{\cal M}}_{k\, L,R}^*$.
This then implies $\McLONLOXUnm = \MLONLOXUnmSt$,
and because $\hat{L}_D$ is real, this again demonstrates explicitly that $\hat\ampA^c_{01} = \hat\ampA_{01}^*$,
\label{phiA01res.PG}
which we used in writing Eq.~(\ref{DGm01Ahn.EQ}).

Next, we choose some example values of masses, couplings and phases, compute the
integrals numerically,
and obtain the $\Delta\hat\Gamma_{01}$ and $\AsymB$.  
We choose these parameters such that the conditions discussed above
for $C$-invariance and $CP$-invariance
do not hold, allowing for a potentially nonzero $\AsymB$.

\subsubsection{Numerical Analysis}
\label{AsymBNumAna.SEC}

In order to illustrate the size of the baryon asymmetry that is generated in our effective theory from the decay of the $\Chi_n$, 
we take a sample parameter-space point ``{\it benchmark point A}'' (BP-A) with values:\\
\indent\indent ${\rm\bf BP\!\!-\!\! A}\! : M_D = 0.25,\ M_U = 0.375,\ M_\chi = 1,\ \tilde{M}_L = 0.1,\ |\tilde{M}_R| = 0.11,\ \phi'_R = -\pi/3 \ , $
\label{BPADefn.PG}\\
\noindent where the masses we show are the values scaled by $M_\chi$.
For the above choices, the $\Chi_n$ mass eigenvalues, are $M_1/M_\chi = 0.91$ and $M_2/M_\chi = 1.09$.
Although our analytical treatment is general, we restrict our numerical analysis to $\tilde{M}_{L,R} \ll M_\chi$, the so-called {\it pseudo-Dirac} limit.
The choice of $M_\Dp, M_\Up$ implicitly assumes that the $\Dp,\Up$ are BSM states since they are a sizable fraction of $M_\chi$.
However, if the $\Dp,\Up$ are SM quarks (cf. Ref~\cite{Gopalakrishna:2022hwk}) and the $M_\Dp,M_\Up$ are much smaller, qualitatively our results should be similar.

We evaluate all the kinematical quantities and the Dirac traces using {\it FeynCalc} (version 9.3.0)~\cite{feynCalcRefs}.
We perform the numerical integrations in {\it Mathematica} (version 13) using the integration technique implemented therein.
The errors quoted in our results below are integration errors which we propagate to $\Gamma_0$, $\Delta\hat\Gamma_{01}$ and $\AsymB$.
We ensure that these errors are better than about 10~\%. 
Next, we show for BP-A the
$\Gamma_{\Chi}$, $\Delta\hat\Gamma_{01}$, and $\AsymB$ values for some sample coupling choices, 
and highlight their dependence on the parameters of the theory. 
 
For the BP-A, with the values of the integrals evaluated numerically, if we additionally take, for example, $\tilde{g}=1$, $g_L=1$, $\phi_L=0$, $g_R = 0$,
we find the following (dimensionless) values:\\ 
\indent $\hat\Gamma_0^{(n=1)} = 1.5\times 10^{-8}$; $\hat\Gamma_0^{(n=2)} = 8.5\times 10^{-7}$;
$\Delta\hat\Gamma_{01}^{(n=2)} = -2.3\times 10^{-12}$; 
$\hat\AsymB^{(n=2)} = 2.7\times 10^{-6}$. \\ 
We omit showing the $\Delta\hat\Gamma_{01}$ and $\hat\AsymB$ values for (n=1) as they are about a factor of 100 smaller.
The $\hat\Gamma_0$ values shown are to be multiplied by $(M_\chi^5/\Lambda^4)$, the $\Delta\hat\Gamma_{01}$ by $(M_\chi^9/\Lambda^8)$, and the $\hat\AsymB$ by $(M_\chi/\Lambda)^4$,
to obtain the actual values.
We can then write for $n\!=\!2$
the actual total decay width $\Gamma_{\Chi_2} \sim 10^{-6}\, \tilde{g}^2 g_{L,R}^2 (M_\chi^5/\Lambda^4)$,
the actual tree-loop interference decay width difference \\ 
$\Delta\hat\Gamma_{01} \sim$ $-10^{-12}\, \tilde{g}^4 g_{L,R}^4 (M_\chi^9/\Lambda^8)$,  
and
the actual baryon asymmetry $\AsymB \sim 10^{-6}\, \tilde{g}^2 g_{L,R}^2 (M_\chi/\Lambda)^4$,
including the couplings scaling.

From the above $\Gamma_0$ and $\Delta\hat\Gamma_{01}$ numbers given for BP-A, $g_L=1,g_R=0,\tilde{g}=1$,
we extract the LO width and tree-loop interference functions parametrized
in Eqs.~(25) and (26) respectively of Ref.~\cite{Gopalakrishna:2022hwk}. 
For $n\!\!=\!\! 2$, we find these to be:   
$\hat{f}_{00} \approx 2.7 \times 10^{-4}$,
and
$\hat{f}_{01} \approx -2.4 \times 10^{-6}$.
We can trace the reason for the suppression in $\hat{f}_{00}$
to a 3-body phase-space suppression leading to a suppressed weight in the phase space measure
due to the 3-momentum magnitude being about ${\cal O}(0.1)$
for the decay kinematics in BP-A. 
Similarly,
we can trace the reason for the suppressed $\hat{f}_{01}$ 
to a product of the suppression factor in the 3-body phase-space as above in the LO integrals,
and another $10^{-3}$ factor from the suppressed weight in the loop-integral measure
due to the cut of the loop-diagram picking out similar sizes of loop-momenta as phase-space momenta,
i.e. suppressed ${\cal O}(0.1)$ 3-momentum magnitudes.
\begin{figure}[h]
  \begin{center}
    \includegraphics[width=0.32\textwidth]{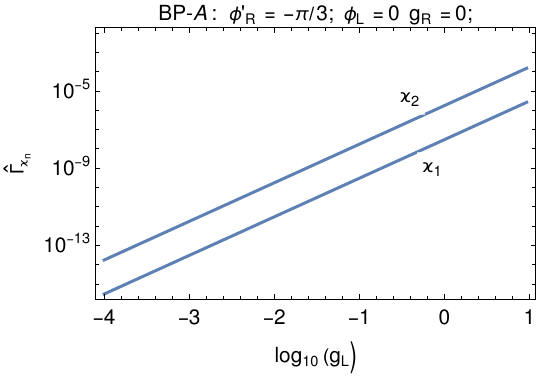}
    \includegraphics[width=0.32\textwidth]{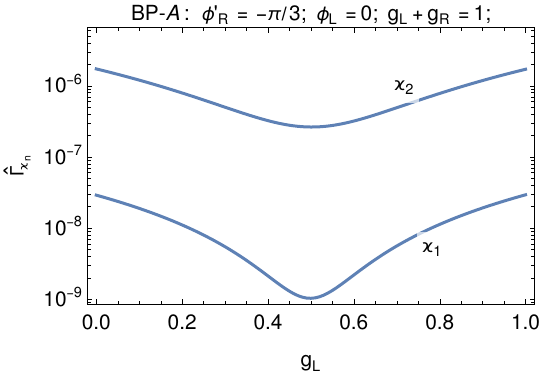}
    \includegraphics[width=0.32\textwidth]{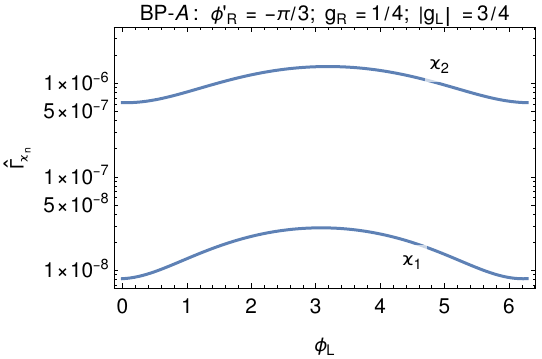}
    \includegraphics[width=0.32\textwidth]{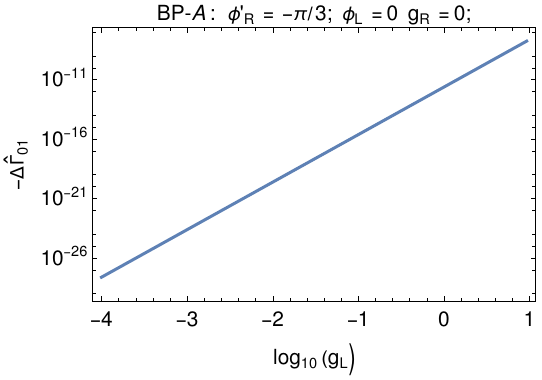}
    \includegraphics[width=0.32\textwidth]{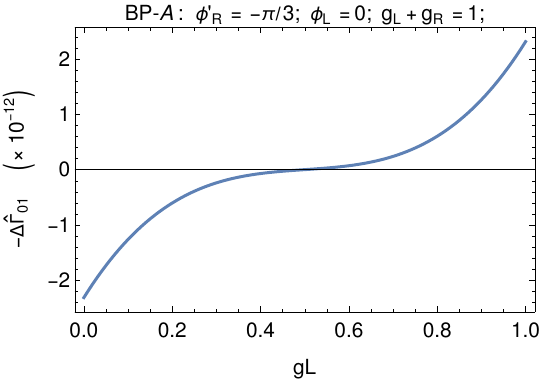}
    \includegraphics[width=0.32\textwidth]{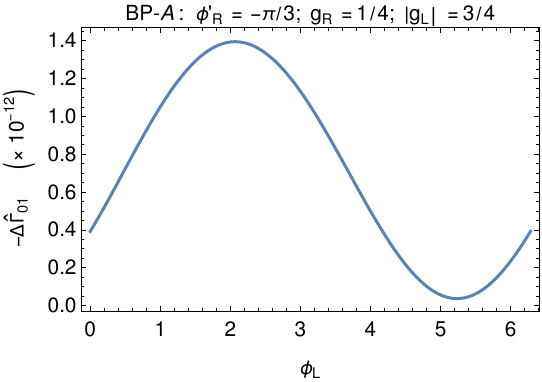}     
    \includegraphics[width=0.32\textwidth]{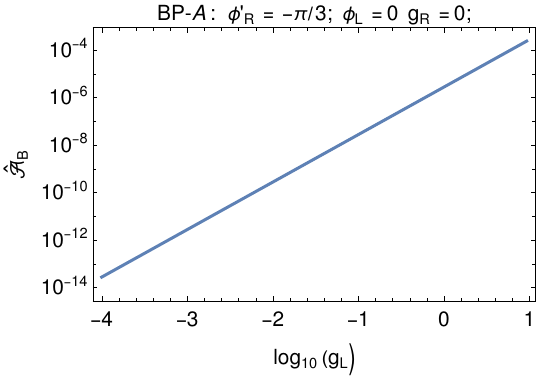}
    \includegraphics[width=0.32\textwidth]{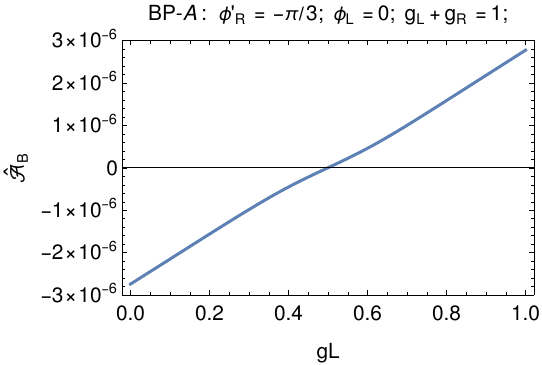}
    \includegraphics[width=0.32\textwidth]{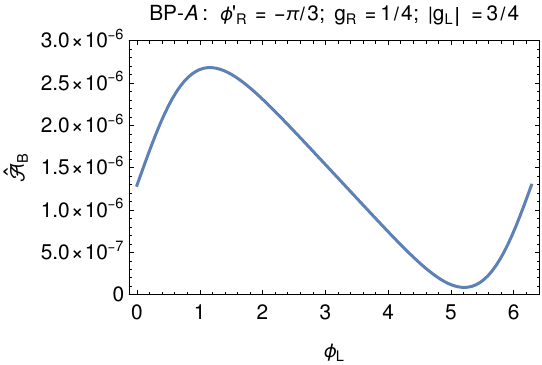} 
  \end{center}
  \caption{
  For BP-A,
  the
 (dimensionless) LO total decay widths
  $\hat\Gamma_{\Chi_n}$ in units of $M_\chi^5/\Lambda^4$ (top row),
  the (negative of) the tree-loop interference decay width difference
  $-\Delta\hat\Gamma_{01}$ in units of $M_\chi^9/\Lambda^8$ (middle row),
  and
  the baryon asymmetry
  $\AsymBh$ in units of $M_\chi^4/\Lambda^4$ (bottom row), 
  for $\tilde{g}=1$, and the following coupling choices: 
  $\phi_L=0$, $g_R = 0$, and $g_L$ varied (left panel);
  $\phi_L=0$, and $g_L, g_R$ varied, subject to $g_L + g_R = 1$ (middle panel);
  $|g_L|=3/4$, $g_R = 1/4$, and $\phi_L$ varied (right panel).
  \label{BPA-GmAsym.FIG}
  }
\end{figure}
In Fig.~\ref{BPA-GmAsym.FIG} we show the dimensionless $\hat\Gamma_{\!\Chi_n}\!\approx\!2 \hat\Gamma_0^n$
(cf. Eq.~(\ref{Gm0f00.EQ})),
the (negative of the) tree-loop interference decay width difference $-\Delta\hat\Gamma_{01}$  for $n\!=\!2$ (cf. Eq.~(\ref{DGm01f01.EQ})), 
and,
$\AsymBh$ (cf. Eq.~(\ref{ABf01f00.EQ})),
    for BP-A, with $\tilde{g}=1$, and the following coupling choices: 
  $\phi_L=0$, $g_R = 0$, and $g_L$ varied;
  $\phi_L=0$, and $g_L, g_R$ varied subject to $g_L + g_R = 1$;
  $|g_L|=3/4$, $g_R = 1/4$, and $\phi_L$ varied.
    As mentioned above, the dimensionless $\hat\Gamma_{\!\Chi_n}$ computed and presented is to be multiplied by $M_\chi^5/\Lambda^4$
    to get the actual LO total decay width,
    the $-\Delta\hat\Gamma_{01}$ by $M_n^9/\Lambda^8$,
and the $\AsymB$ by $M_\chi^4/\Lambda^4$ to get the actual numbers.
We see that the $\Gamma_{\Chi_2}$ is larger than $\Gamma_{\Chi_1}$.
We find that the $\AsymB$ is mainly due to $\Chi_2$ decay and that due to $\Chi_1$ is insignificantly small. 
We recall that the only physical phase in the couplings is in $g_L$, namely $\phi_L$. 
We see that if $g_L$ and $g_R$ are both nonzero, there is a strong dependence of $\AsymB$ on $\phi_L$;
however, although not shown in the figure, we find that if $g_R =0$, there is no variation of $\AsymB$ with $\phi_L$. 

Next, we move away from the BP-A by varying the $|M_R|$ and $\phi'_R$ values,
for which we re-diagonalize the mass matrix and recompute the integrals numerically. 
\begin{figure}
  \begin{center}
    \includegraphics[width=0.24\textwidth]{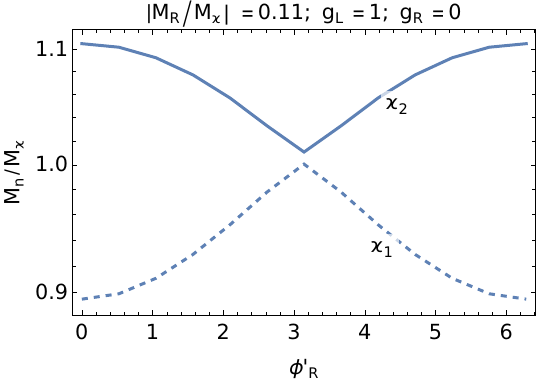}
    \includegraphics[width=0.24\textwidth]{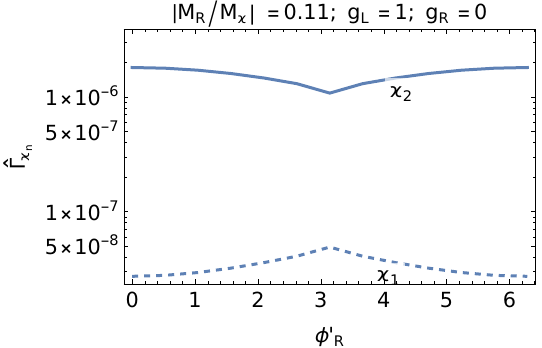}
    \includegraphics[width=0.24\textwidth]{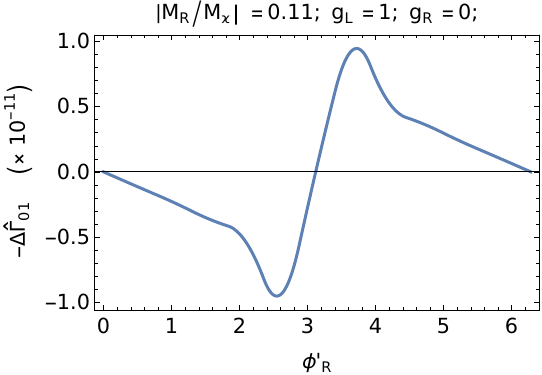}    
    \includegraphics[width=0.24\textwidth]{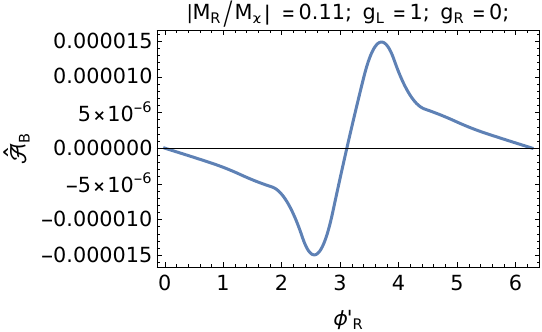}
    \includegraphics[width=0.24\textwidth]{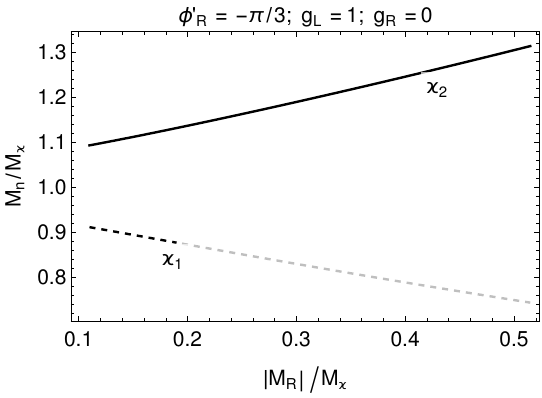}
    \includegraphics[width=0.24\textwidth]{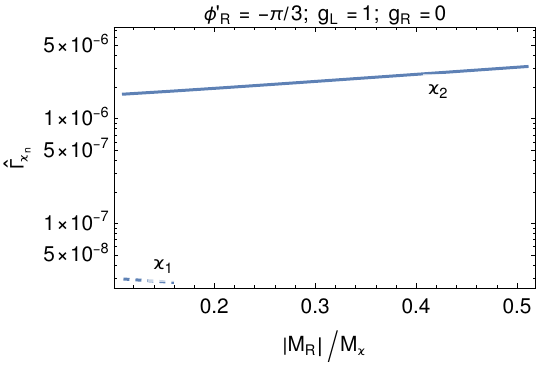}
    \includegraphics[width=0.24\textwidth]{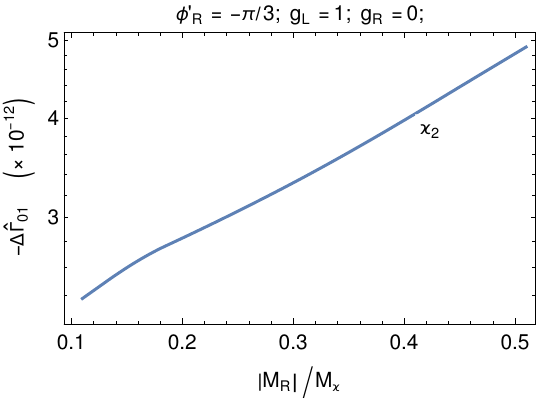}    
    \includegraphics[width=0.24\textwidth]{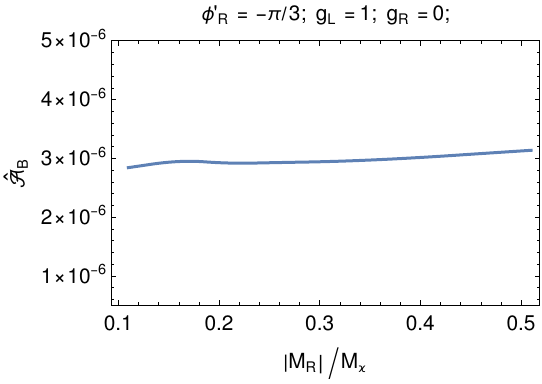}    
  \end{center}
  \caption{
  The $\Chi_n$ mass eigenvalues $M_n/M_\chi$ (first column),
  the $\hat\Gamma_{\Chi_n}$ (second column),
  $-\Delta\hat\Gamma_{01}$ (third column),
  and
  $\AsymBh$ (fourth column),
  with $M_R/M_\chi = 0.11$ and $\phi'_R$ varied (top row), and with $\phi'_R = -\pi/3$ and $|M_R|$ varied (bottom row), 
  with the other parameters as in BP-A, and taking $\tilde{g}=1$, $g_L =1$, $g_R = 0$.
\label{GmAsymMRscan.FIG}
  }
\end{figure}
In Fig.~\ref{GmAsymMRscan.FIG} we show the $\Chi_n$ mass eigenvalues $M_n$, the LO total $\hat\Gamma_{\Chi_n}$,
the (negative of) the tree-loop interference decay width difference $-\Delta\hat\Gamma_{01}$  for $n\!=\!2$, 
and the baryon asymmetry $\AsymB$, 
first with the Majorana mass phase $\phi'_R$ varied, keeping $|M_R|/M_\chi = 0.11$,
and next with the $|M_R|$ varied, keeping $\phi'_R = -\pi/3$;  
the other parameters we take as in BP-A,
and the couplings to be $\tilde{g}=1$, $g_L =1$, $g_R = 0$.
While varying $|\tilde{M}_R|$, we keep $|\tilde{M}_R| > \tilde{M}_L$ so that $M_2 > M_1$ is maintained.
As earlier,
the dimensionless $\Gamma_{\!\Chi_n}$ presented is to be multiplied by $M_\chi^5/\Lambda^4$,
the $-\Delta\hat\Gamma_{01}$ by $M_\chi^9/\Lambda^8$,
and the $\AsymB$ by $M_\chi^4/\Lambda^4$ to get the actual numbers.
As $|\tilde{M}_R|$ increases we see that $M_1$ decreases, and for $M_1 < (2 M_D + M_U)$ the $\Chi_1 \to \Dp\Dp\Up$
decay shuts off kinematically, which is shown as the light-gray part of the line in the $M_1$ vs. $|M_R|$ plot,
and for this region, the $\AsymB$ is only due to $\Chi_2$ decays.
We find the same numerical behavior if we change $\phi'_R \to -\phi'_R$ and interchange $g_L \leftrightarrow g_R$;
in particular the same behavior shown in Figs.~\ref{BPA-GmAsym.FIG}~and~\ref{GmAsymMRscan.FIG} hold for
$\phi'_R = \pi/3$ and $g_L \leftrightarrow g_R$ interchanged. 

We discuss in Appendix~\ref{VVintUV.SEC} the UV state $\xi$ propagator correction resolving the VV effective interaction.
In the decay processes considered here, we expect the effect of these corrections to be minimal
since in decays, the momenta are limited by $M_\chi$, 
and the propagator corrections begin to kick-in for vertex momenta $\gtrsim \Lambda/M_\chi$.
Furthermore, the 3-momentum magnitudes are much lesser than $M_\chi$ given the sizeable final state $M_{\Up,\Dp}$ masses. 
The loop 3-momentum magnitudes are also limited since we enforce the cuts on the $\Up,\Dp$ propagators.  
We perform a numerical computation with the UV propagator correction included and find that indeed the
correction to $\Gamma_{\Chi_n}$, $\Delta\hat\Gamma_{01}$, and $\AsymB$ are smaller than a few percent.
We therefore have only presented our $\Chi$ decay baryon asymmetry results without these UV propagator corrections included. 

We find that diagrams without an arrow clash do not contribute a baryon asymmetry
(cf. Appendix~\ref{noArClSingOp.SEC}).
Next, we explore potential contributions to $\AsymB$ due to adding operators other than the one we have been focusing on.

\subsection{Multiple operator loop contribution to decay baryon asymmetry}
\label{mixedOpsAsymB.SEC}

The UV completion examples in Ref.~\cite{Gopalakrishna:2022hwk} suggest that
associated with the VV interaction of Eq.(\ref{LIntVVMB.EQ}) considered in the previous section, 
other operators such as those in Eq.~(\ref{modAops.EQ}) could also be generated. 
Here we include these other operators also and ask if potentially new contributions to $\AsymB$ could arise,
adding to the baryon asymmetry found in the previous section. 

\begin{figure}
  \begin{center}
    \includegraphics[width=0.25\textwidth]{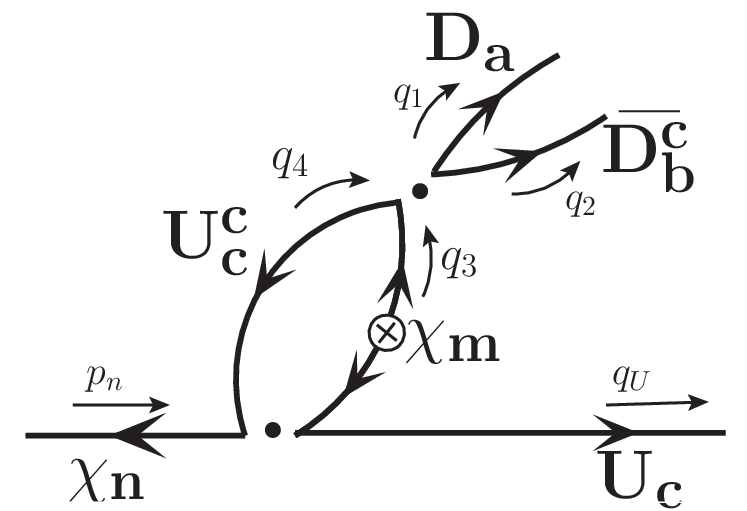}
    \includegraphics[width=0.25\textwidth]{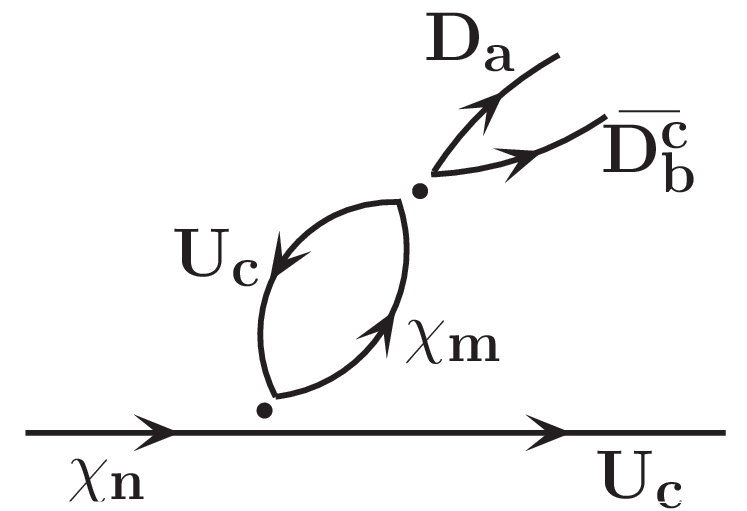}
    \includegraphics[width=0.25\textwidth]{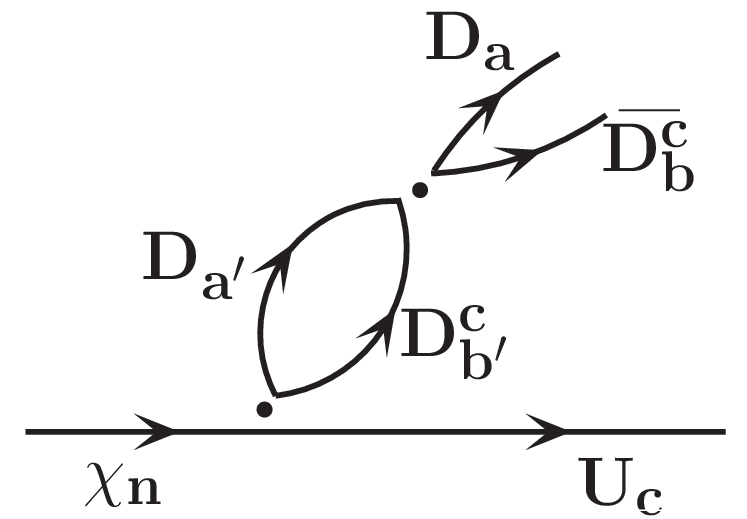}
  \end{center}    
\caption{
  The $\Chi_n \to \widebar{D^c}DU$ loop diagrams A$_1^{\prime(1)}$ (left), A$_1^{\prime(2)}$ (middle), and A$_1^{\prime(3)}$ (right),
  arising from using more than one operator.
\label{chi2DDU-NLO-modA.FIG} }
\end{figure}

We show in Fig.~\ref{chi2DDU-NLO-modA.FIG} such potential new contributions
to the $\Chi_n \to \overline{\Dp^c} \Dp \Up$ decay,
characterized by using two different effective operators
whose Feynman rules are given in Figs.~\ref{FeynRules.FIG}~and~\ref{FeynRulesModA.FIG}.\footnote{
In Ref.~\cite{Yanagida:1980cy}, a GUT UV completion is considered with an SS operator of the type
$[\overline{\Dp} \Gamma \chi]\, [\bar{\chi} \Gamma \Dp]$.
A $VV$ operator with these fields is not in the set we are considering here
(for related details on the UV completion, see Ref.~\cite{Gopalakrishna:2022hwk}), 
but the contribution from analogous operators in our effective theory is that of diagram $A_1^{\prime(1)}$.}
The diagrams for the conjugate mode $\Chi_n \to \overline{\Dp} \Dp^c \Up^c$ are obtained by
reversing the arrows in Fig.~\ref{chi2DDU-NLO-modA.FIG},
i.e. by taking the conjugate of the fields shown.  
We take the independent 1-loop momentum that is integrated over as $q_3$,
and 4-momentum conservation implies $q_4 = p_n - q_3 - q_U$.
We do not show another diagram that involves a correction to the on-shell $\Up$ leg which is subtracted away
when renormalized fields are used.
If the required operators are present in the effective theory such that the one-loop contributions in Fig.~\ref{chi2DDU-NLO-modA.FIG}
can be present,
and if the parameters are favorable as discussed below, 
the baryon asymmetry generated by this could potentially be larger than the two-loop contributions we analyzed in Sec.~\ref{oneOpAsymb.SEC}. 
Using the formalism detailed in Sec.~\ref{genAB.SEC} we compute the decay baryon asymmetry from this mode, which we discuss next. 

We denote the 1-loop amplitudes for the diagrams in Fig.~\ref{chi2DDU-NLO-modA.FIG}
as $\ampA_1^{\prime(1)}$, $\ampA_1^{\prime(2)}$ and $\ampA_1^{\prime(3)}$ respectively, 
and the corresponding loop amplitudes of the conjugate modes as
${\cal A}_1^{\prime c(1)}$, ${\cal A}_1^{\prime c(2)}$, and ${\cal A}_1^{\prime c(3)}$.
For a nonzero rate asymmetry,
in addition to having a weak phase that flips sign between the process and the conjugate process,
we must also have a common strong phase that does not flip sign.
Such a common phase can come from a cut of the loop when the
two intermediate state particles go on-shell.
In diagrams A$^{\prime(1)}$ and A$^{\prime(2)}$, the loop can be cut only for $nm=21$ provided $M_2 > M_1 + 2 M_U$, while in A$^{\prime(3)}$ a cut is always possible. 
We assume here that this condition is satisfied. 

\begin{figure}
  \begin{center}
    \includegraphics[width=0.32\textwidth]{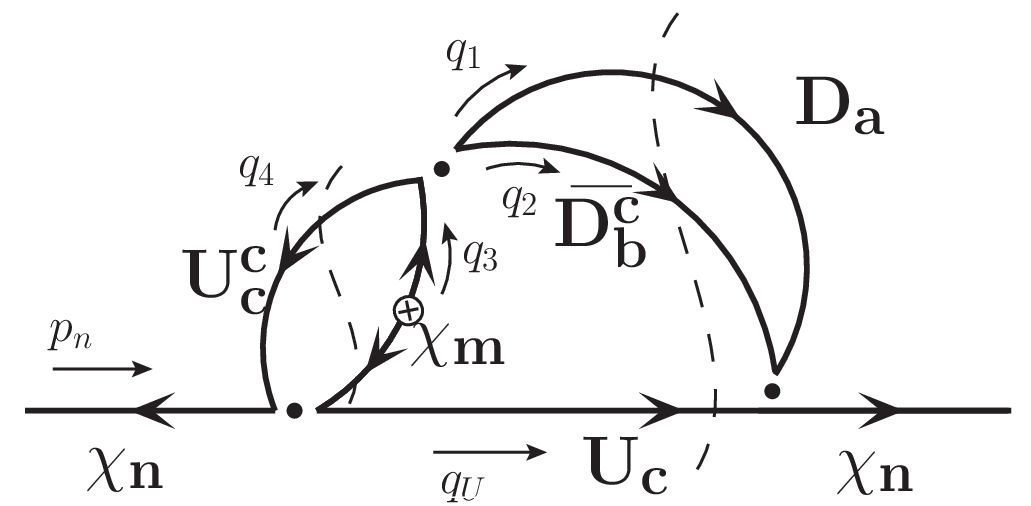}
    \includegraphics[width=0.32\textwidth]{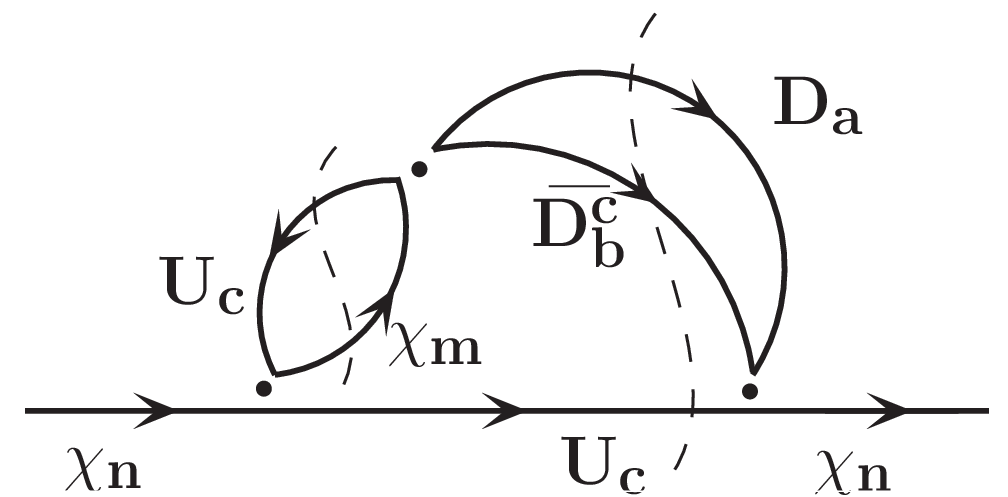}
    \includegraphics[width=0.32\textwidth]{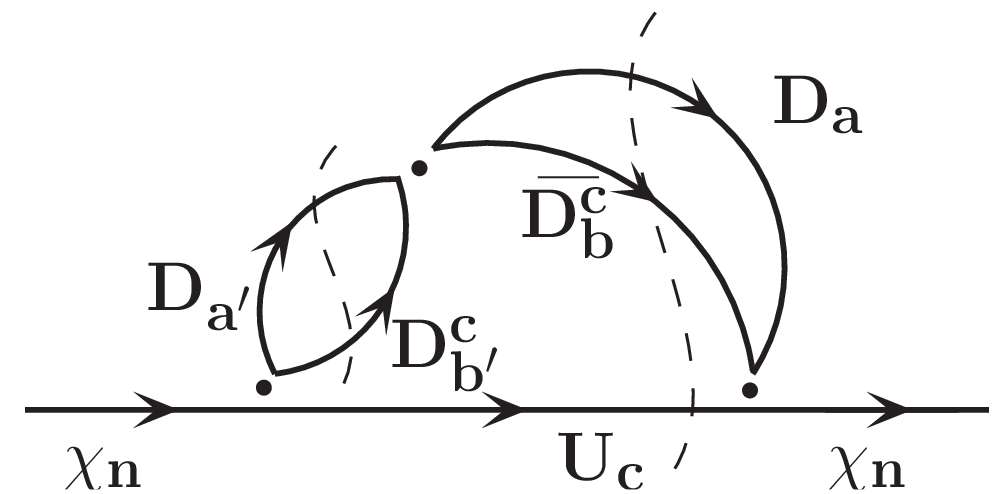}
  \end{center}    
\caption{
  A diagrammatic representation of the tree-loop interference terms
  $\ampA_{01}^{\prime(1)}$ (left), $\ampA_{01}^{\prime(2)}$ (middle), and $\ampA_{01}^{\prime(3)}$ (right),
  corresponding to the contributions in Fig.~\ref{chi2DDU-NLO-modA.FIG}.
  The dashed curves show the cuts.
\label{chi2DDU-LONLO-modA.FIG} }
\end{figure}

We write the tree-loop interference term for the process as
${\cal A}_{01}^{\prime (1),(2),(3)} \equiv {\cal A}_1^{\prime (1),(2),(3)} {\cal A}_0^*$, where ${\cal A}_0$ is the tree amplitude,
and for the conjugate process we correspondingly write ${\cal A}_{01}^{\prime c (1),(2),(3)} \equiv {\cal A}_1^{\prime c (1),(2),(3)} {\cal A}_0^{c*}$. 
The decay rate difference between the process and conjugate process is given by
$\Delta \Gamma\, \propto\, \Delta {\cal A}^{\prime}_{01} \equiv {\cal A}^{\prime}_{01} - {\cal A}^{\prime c}_{01}$,
summed over the processes. 
We include the 3-body phase-space as cuts of a loop integral, 
and show the ${\cal A}_{01}^\prime$ diagrammatically in Fig.~\ref{chi2DDU-LONLO-modA.FIG}.
The dashed curves represent the cuts, with the cut on the left yielding the strong phase and the cut on the right the 3-body phase space.
Computing this with the cuts shown in Fig.~\ref{chi2DDU-LONLO-modA.FIG}~(left) gives   
\bea
    &&\ampA_{01}^{\prime(1)} = -\frac{1}{2} \frac{C_{01}^{\prime (1)}}{(4\Lambda^6)} {\rm Tr}[\GVn_\nu (\qslash_U + M_U) \GVBm^\sigma (\qslash_3 + M_m) \GZBm_\mu (\qslash_4 + M_U) \GZn_\sigma (\pslash_n + M_n)] \cdot \nonumber \\
                           &&\hspace*{9cm} \tilde{g}^2 {\rm Tr}[\gamma^\nu (\qslash_1 + M_D) \gamma^\mu (\qslash_2 - M_D) ] \ , 
\label{A01modA.EQ}
\eea
where the $1/2$ is from averaging over the initial state spins,
and the color-factor $C_{01}^{\prime (1)} = \epsilon_{abc}\epsilon_{abc}/(\sqrt{6})^2 = 1$.\footnote{
We notice that the contribution in Fig.~\ref{chi2DDU-LONLO-modA.FIG}~(left) is very similar to the one in Fig.~\ref{chi2DDU-LONLOCut.FIG}~(left)
when one maps the $q_3q_4$ $\Dp\Dp$ loop and the associated vertices in the former to the left-most vertex in the latter
(corresponding to a heavy UV field exchange).
Consequently, by mapping the momenta appropriately,
we see that Eq.~(\ref{A01modA.EQ}) has the same structure as Eq.~(\ref{M01XUdefn.EQ}) with $L_{12}$ included.
We also notice that the same number of $\GV,\GZ$ couplings are involved in these two contributions.
}
We find no asymmetry from diagrams ${\cal A}_{01}^{\prime (2),(3)}$  
as we could anticipate from there being no arrow clash in these diagrams.

To illustrate the baryon asymmetry arising from diagram $A_1^{\prime(1)}$,
we define another parameter-space point with $M_2 > M_1 + 2 M_U$, namely, ``{\it benchmark point B}'' (BP-B) with values:\\
\indent ${\rm\bf BP\!\!-\!\! B}\! : M_D = 0.05,\ M_U = 0.05,\ M_\chi = 1,\ \tilde{M}_L = 0.1,\ |\tilde{M}_R| = 0.11,\ \phi'_R = -\pi/3 \ ,$
\label{BPBDefn.PG}\\
\noindent where again, the masses we show are the values scaled by $M_\chi$.
In BP-B, the decay mode $\Chi_2 \to \Chi_1 \Up\bar\Up$ is also open (only for $n\!\!=\!\!2$) in addition to the earlier $\Chi_n \to \Up \Dp \overline{\Dp^c}$. 
We compute numerically the partial decay widths to the two modes for $\tilde{g}=1$, $g_L = 1$, $g_R = 0$,
and we find the LO decay widths in units of $(M_\chi^5/\Lambda^4)$ to be 
$\Gamma(\Chi_n \to \Up \Dp \overline{\Dp^c}) = (3.1,7.8)\times 10^{-6}$ for $n\!\!=\!\!(1,2)$,
while 
$\Gamma(\Chi_2 \to \Chi_1 \Up\bar\Up) = 5.3\times 10^{-8}$.
Thus, the branching ratio to the second mode is about $0.3\,\%$.
The suppression is due to the small phase space available for the final-state in the second mode,
the 3-momentum magnitude being of the order $(M_2 - (M_1 + 2M_\Up))/M_2 \approx 0.07$ for BP-B. 

We obtain the baryon asymmetry $\AsymB$ contributed by   
the tree-loop interference term $\ampA_{01}^{\prime(1)}$ depicted in Fig.~\ref{chi2DDU-LONLO-modA.FIG} (left)
following the discussion in Sec.~\ref{genAB.SEC}.
Given that this diagram has an arrow clash, we anticipate a potentially nonzero $\AsymB$. 
We make the cuts shown in the figure and obtain the $\ampA_{01}^{\prime(1)}$. 
We obtain the tree-loop interference term for the conjugate process $\ampA_{01}^{\prime c(1)}$ by interchanging $\GV \leftrightarrow \GZB$.
We compute the difference $\Delta \ampA_{01}^{\prime (1)} = \ampA_{01}^{\prime (1)} - \ampA_{01}^{\prime c(1)} = 2\, {\rm Im}(\ampA_{01}^{\prime (1)})$,
and integrate this after folding in the 3-body phase-space measure $d\Pi_3$ and the one-loop measure $d\Pi'_2$ with the cuts shown
to obtain the decay width difference,
which when written in terms of the scaled (hatted) momenta and masses is 
\beq
\DGmzoh =
\left(\frac{M_\chi^7}{\Lambda^6}\right) \left[ \frac{1}{2} f^{\prime (1)}_{01} \right]
\ ; \quad {\rm where}\ 
     f^{\prime (1)}_{01} \equiv  \left(\frac{M_n}{M_\chi}\right)^7 \int d\hat\Pi_3\, d\hat\Pi_2' \frac{1}{2}\, \Delta \hat\ampA_{01}^{\prime (1)} \ , 
\eeq
where the $M_n^7$ scaling is expected on dimensional grounds.
We then obtain the baryon asymmetry due to the contribution from Fig.~\ref{chi2DDU-LONLO-modA.FIG} (left) using Eq.~(\ref{ABDGm01Gm0.EQ}) as 
\beq
\AsymB^\prime = -\frac{\DGmzoh}{\Gamma^n_0} = \left(\frac{M_\chi}{\Lambda}\right)^2 \AsymBh^\prime \ ; \quad {\rm where}\
    \AsymBh^\prime \equiv \frac{1}{2} \frac{f^{\prime (1)}_{01}}{\hat\Gamma^n_0} \ .
    \eeq
    As already discussed above, for the benchmark point BP-B,
    the BR into the additional decay mode $\Chi_2 \to \Chi_1 \Up\bar\Up$ is suppressed 
    compared to that into the $\Chi_2 \to \overline{\Dp^c} \Dp \Up$ mode, 
    and we therefore take the LO decay width $\Gamma_0$ to be given by the latter (see Eq.~(\ref{Gm0f00.EQ})). 

\begin{figure}
  \begin{center}
    \includegraphics[width=0.45\textwidth]{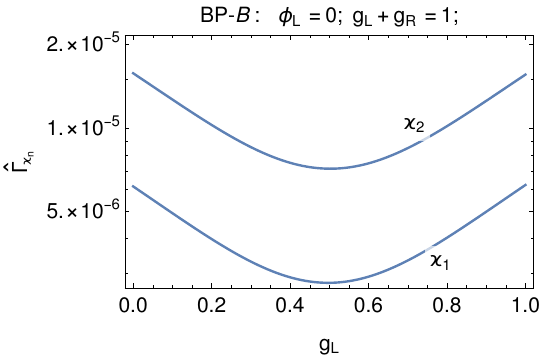}
    \includegraphics[width=0.45\textwidth]{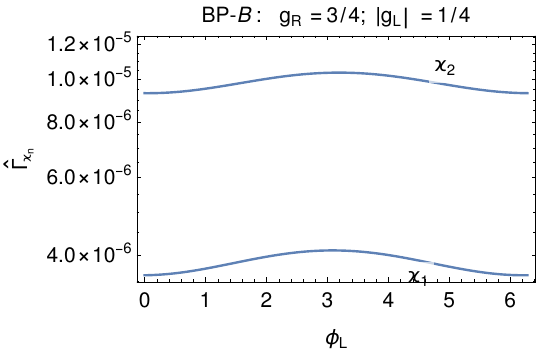}
  \end{center}
  \caption{For BP-B, the LO total $\Chi_n$ decay width $\hat\Gamma_{\Chi_n}$ (in units of $M_\chi^5/\Lambda^4$),
  with $\tilde{g}=1$, $\phi_L=0$, $g_L$ varied with $g_L + g_R = 1$ (left), and, $|g_L|=1/4$, $g_R=3/4$ with $\phi_L$ varied (right). 
  \label{BPB-Gm0.FIG}
  }
\end{figure}

\begin{figure}[t]
  \begin{center}
    \includegraphics[width=0.32\textwidth]{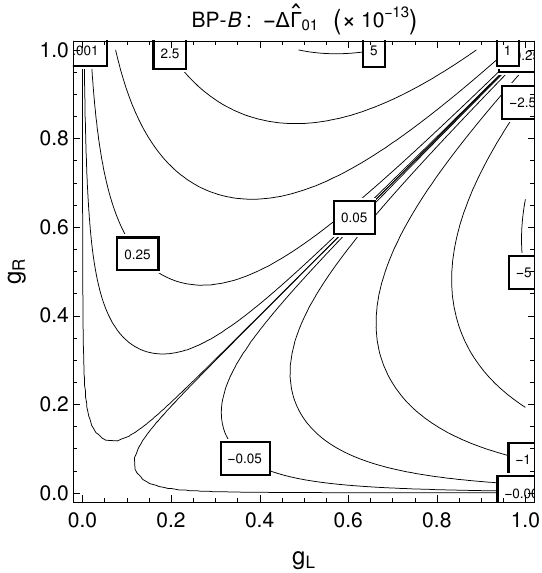}
    \includegraphics[width=0.32\textwidth]{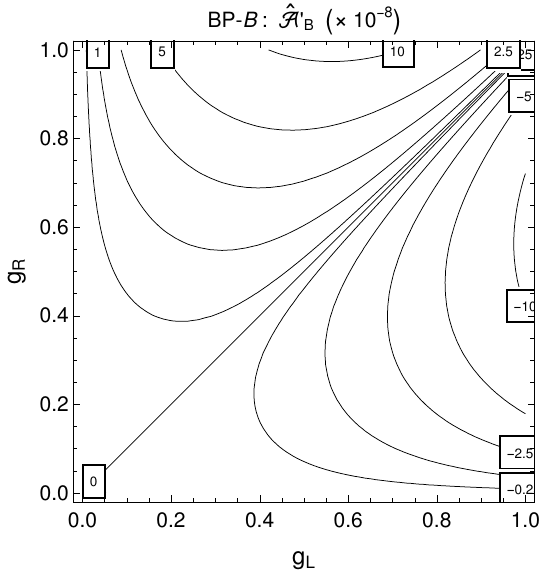} \\
    \includegraphics[width=0.32\textwidth]{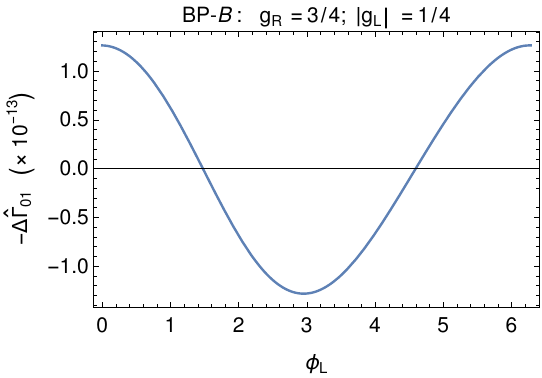} 
    \includegraphics[width=0.32\textwidth]{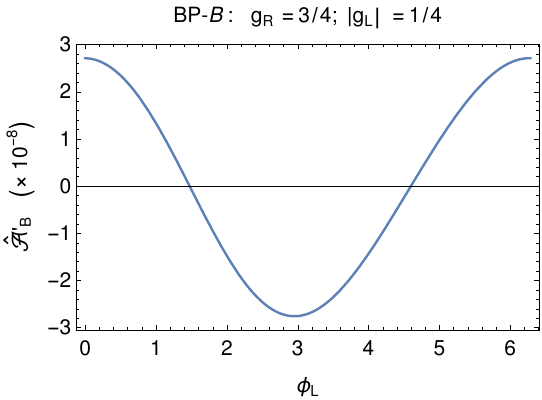} 
  \end{center}
\caption{
  For BP-B,
  the multiple operator $\Chi_n$ decay baryon asymmetry $\AsymBh^\prime$ (in units of $(M_\chi/\Lambda)^2$),
  and the (negative of the) tree-loop interference decay width difference $-\Delta\hat\Gamma_{01}$ (in units of $M_\chi^7/\Lambda^6$), 
  with $\tilde{g}=1$,
  showing contours in the $(g_L,g_R)$ plane with $\phi_L=0$ (top panel); 
  $|g_L|=1/4$, $g_R=3/4$ with $\phi_L$ varied (bottom panel).    
  \label{BPB-Asym.FIG}
  }
\end{figure}

In Fig.~\ref{BPB-Gm0.FIG} we show the total (dimensionless) $\Chi_n$ LO decay width $\hat\Gamma_{\Chi_n} \approx 2 \hat\Gamma_0^n$, 
for BP-B,
on the left as a function of $g_L$ with $g_L + g_R = 1$, and,
on the right as a function of $\phi_L$ for $|g_L|=1/4$, $g_R=3/4$. 
The shown $\hat\Gamma_{\Chi_n}$ values are to be multiplied by $M_\chi^5/\Lambda^4$ to get the actual decay widths.
In Fig.~\ref{BPB-Asym.FIG} we show the resulting $\Delta\hat\Gamma_{01}$, $\AsymBh^\prime$
for various choices of couplings. 
We evaluate the integrals numerically and use them to compute the rate asymmetry. 
For $g_L = 1/4$, $g_R = 3/4$, the decay widths $\hat\Gamma_0^n$ are: $\hat\Gamma_0^1 = 1.8 \times 10^{-6}$ and $\hat\Gamma_0^2 = 4.7\times 10^{-6}$,
which are to be multiplied by $(M_\chi^5/\Lambda^4)$ to get the actual widths.
The tree-loop interference term from $\ampA_{01}^{\prime(1)}$ is $\Delta\hat\Gamma_{01} = -1.3\times 10^{-13}$. 
The resulting baryon asymmetry is $\AsymBh^\prime = 2.7\times 10^{-8}$.
We can thus write for $n\!=\!2$ the actual total decay width $\Gamma_{\Chi_2} \sim 10^{-5}\, \tilde{g}^2 g_{L,R}^2 (M_\chi^5/\Lambda^4)$,
the actual interference term $\Delta\hat\Gamma_{01} \sim - 10^{-13} \tilde{g}^2 g_{L,R}^4 (M_\chi^7/\Lambda^6) $
and the actual baryon asymmetry $\AsymBP \sim 10^{-8}\, g_{L,R}^2 (M_\chi/\Lambda)^2$, including the couplings scaling.

Comparing the $\AsymBh \sim 10^{-6}$ we found from the single operator contribution for BP-A in Sec.~\ref{oneOpAsymb.SEC},
the suppressed $\AsymBPh \sim 10^{-8}$ from the multiple operator contribution for BP-B we find here is surprisingly small
considering the latter is one loop-order lower and is therefore expected to be enhanced.
The reason for the suppression is because we are in the pseudo-Dirac limit with $(M_2 - M_1)/M_2 \sim 0.1$, 
and since we require the $\Chi_1$ to be on-shell in the loop,
the available phase-space and loop momenta are limited by $\qUr \lesssim 0.1$, leading to a suppression in the integrals,
as we detailed above.
The smallness is the result of suppressions due to    
the 3-body phase-space weight, 
the loop measure weight
and,
the dimensionless $\ampA_{01}^{\prime(1)}$  (without the $1/\Lambda^6$) of $\sim 10^{-4}$ including the (imaginary part of) coupling factors.
However, it should be borne in mind that the actual $\AsymB$ has a $M_\chi^4/\Lambda^4$ scaling factor
while $\AsymBP$ has a less suppressed $M_\chi^2/\Lambda^2$ scaling factor due to which the above suppression
may be overcome.
From the above
$\Gamma_{\Chi_2}$ and $\Delta\hat\Gamma_{01}$ numbers given for BP-B, $g_L=1/4,g_R=3/4,\tilde{g}=1$,
we extract the LO width and tree-loop interference functions parametrized
in Eqs.~(25) and (26) respectively of Ref.~\cite{Gopalakrishna:2022hwk}. 
For $n\!\!=\!\! 2$, we find these to be:   
$\hat{f}_{00} \approx 4.3 \times 10^{-4}$,
and
$\hat{f}_{01} \approx -3 \times 10^{-9}$.

In summary,
if the second operator is also present in the effective theory,
and if $M_2 > M_1 + 2 M_U$,
the $\ampA_{01}^{\prime(1)}$ of Fig.~\ref{chi2DDU-LONLO-modA.FIG} (left)
contributes to the baryon asymmetry.  
Comparing this $\Delta\hat\Gamma'_{01}$ contribution to the single operator $\Delta\hat\Gamma_{01}$ contribution of Sec.~\ref{oneOpAsymb.SEC}, we have
$\Delta\hat\Gamma_{01}^\prime/\Delta\hat\Gamma_{01} \sim 2\pi^2 (1/\tilde{g}^2) (\Lambda/M_\chi)^2\, \hat{f}_{01}^\prime/\hat{f}_{01} \sim 0.1\, (1/\tilde{g}^2) (\Lambda/M_\chi)^2$
for our parameter choices.  
Thus, for $M_\chi/\Lambda \lesssim 1/10$, the multiple operator contribution, if present, could dominate over the single operator contribution. 
If any of the above conditions for the presence of the multiple operator contribution are not met,
the single operator baryon asymmetry contribution would remain as the only contribution.
We thermally average the decay width and the width difference so computed
as described in Appendix~\ref{ThAvgProc.SEC}, and obtain the $\GmzTA$ and $\DGmzoTAh$.

\section{The Baryon Asymmetry from $\Chi$ Scattering}
\label{genBAsymScat.SEC}

Here we discuss the baryon asymmetry in
the scattering rates between the $2\to 2$ scattering processes of $\Chi_n$ with $\Up,\Dp$,
and the corresponding ones for $\Chi_n$ with $\Up^c,\Dp^c$,
following up on our discussion in Ref.~\cite{Gopalakrishna:2022hwk}. 
In particular, we consider the difference in scattering cross sections for the process and that for its conjugate process
in each of the channels:
\begin{itemize}
\item[] {SC-1}: $\Chi_n(p_n) \bar\Dp(k_i) \to \Dp(q_1) \Up(q_2)$, 
\item[] {SC-2}: $\Chi_n(p_n) \bar\Up(k_i) \to \Dp(q_1) \widebar\Dp^c(q_2)$,
\end{itemize}
for $n=\{1,2\}$. 
SC-1 includes both $\Chi_n(p_n) \Dp^c(k_i) \to \Dp(q_1) \Up(q_2)$ and $\Chi_n(p_n) \bar\Dp(k_i) \to \widebar{\Dp^c}(q_1) \Up(q_2)$.
The conjugate processes are obtained by interchanging $Q \leftrightarrow Q^c$ for $Q=\{\Up,\Dp\}$. 
We only deal with unpolarized rates and therefore average over initial-state spins and sum over final-state spins.

The cross section for each of the SC-1 and SC-2 channels is
\beq
\sigma^n = \frac{1}{v} \frac{1}{2E_n 2E_i} \int d\Pi_2 \, |{\cal A}^{n(\sigma)}|^2 \ ,
\label{sigmaAmpGen.EQ}
\eeq
where
$|{\cal A}^{n(\sigma)}|^2$ is the scattering matrix element mod-squared,
$v \equiv v_{rel} = |v_n - v_i|$ is the relative velocity between the incoming particles,
we denote the energies of the various particles involved as
$E_n \equiv E_{p_n}$, $E_i \equiv E_{k_i}$, $E_1 \equiv E_{q_1}$, $E_2 \equiv E_{q_2}$, etc.,
and,
the 3-momentum magnitude as ${p}_{r} \equiv \sqrt{{\bf p}\! \cdot\! {\bf p}}$. 
The $d\Pi_2$ is the final-state 2-body phase-space element. 

We obtain the $|{\cal A}^{n(\sigma)}|^2$,
averaged over initial-state spins and summed over final-state spins,
in close analogy with the decay amplitude of Sec.~\ref{ABdec.SEC} as we describe next.
Like the decay amplitude earlier,
the scattering amplitude too can be written as a sum over the tree and loop amplitudes
${\cal A}^{(\sigma)} = {\cal A}_0^{(\sigma)} + {\cal A}_1^{(\sigma)}$ for each $n$.
Using {\it crossing relations}~\cite{Peskin:1995ev}, we can reuse the computation of the decay matrix element of Sec.~\ref{ABdec.SEC}
to obtain the scattering matrix element.
The tree and loop diagrams are similar to those shown in Figs.~\ref{chi2DDU-LO.FIG},~\ref{chi2DDU-NLO.FIG}, and \ref{chi2DDU-NLO-modA.FIG},
but with one $\Up$ or $\Dp$ leg crossed to the initial state to give the scattering processes SC-1 and SC-2. 
In taking the average,
since we sum over the initial state spins, a correspondence to the decay matrix element is maintained,
and for each $n$ we can write the crossing relations 
\bea
&&{\it\rm SC\!-\!1~crossing\!:}\ |{\cal A}^{(\sigma)}(p_n k_i \to q_1 q_2)|^2 = -(1/2)|{\cal A}(p_n \to q_1 q_2 q_U)|^2_{q_2 \to (-k_i);\, q_U \to q_2} \ , \nonumber \\
&&{\it\rm SC\!-\!2~crossing\!:}\ |{\cal A}^{(\sigma)}(p_n k_i \to q_1 q_2)|^2 = -(1/2)|{\cal A}(p_n \to q_1 q_2 q_U)|^2_{q_U \to (-k_i)} \ ,
\label{crRel.EQ}
\eea
where
the 1/2 is due to now an average over the second initial-state particle spins as opposed to just a sum over the spins earlier,
the minus sign arises from the spinor sum changing from $u\bar{u}$ to $v\bar{v}$ for the spinor moved from the final state to the initial state,
and the change in sign of the momenta is because we are crossing an out-going final-state momentum to an incoming initial-state momentum. 

Our intent~\cite{Gopalakrishna:2024qxk} is to apply these results to scattering processes 
happening in the early Universe at temperature $T$.
The initial state momenta sampled from the thermal spectrum are typically of order
$\pnr, \kir \sim T$, and if $T > \Lambda$ is relevant for baryon asymmetry generation, we will be sensitive to the UV physics.
To investigate this effect, we include a correction factor while computing our diagrams,
arising from the effective vertex being resolved by a propagating UV state $\xi$ in the UV completion example,
as discussed in Appendix~\ref{VVintUV.SEC}.
We discuss next, in turn, the single operator and multiple operators scattering contributions to the baryon asymmetry,
with the effect from the UV physics correction factor also brought forth.
Multiple operator scattering contributions involving the SS interaction have been considered, for instance,
in Ref.~\cite{Baldes:2014rda}.

\subsection{Scattering baryon asymmetry from single operator loop contributions}
\label{singOpSig.SEC}

The LO cross section for each of SC-1 and SC-2 from Eq.~(\ref{sigmaAmpGen.EQ}) is
\beq
\sigma^{n}_0\, v = \frac{1}{2E_n 2E_i} \int d\Pi_2\ |\ampA_0^{n(\sigma)}|^2
= \frac{1}{2E_n 2E_i}\, \frac{1}{4 \Lambda^4}\, \int d\Pi_2
    \left[-\frac{1}{4} (\MLOXUSq)^{\mu\tau}_{nn} \, C_{00}\, (\MLOotSq)_{\mu\tau} \right]_{(q_i\ {\rm cross})} \ ,
\eeq
where we obtain the $|\ampA_0^{n(\sigma)}|^2$ using the crossing relations of Eq.~(\ref{crRel.EQ})
in the LO matrix element (cf. Eq.~(\ref{ampALOn.EQ})),
with the ($q_i$~cross) denoting the momentum substitutions shown in the crossing relations for each of SC-1 and SC-2,
and we include the extra (-1/2) as given there. 
We write the LO cross sections equivalently in terms of the dimensionless $f_{00}^{n(\sigma)}$ as
\beq
\sigma^{n}_0\, v = \left(\frac{M_\chi^2}{\Lambda^4}\right) \left[\frac{1}{4} \left(\frac{M_n}{M_\chi}\right)^2 f_{00}^{n(\sigma)} \right] \equiv \left(\frac{M_\chi^2}{\Lambda^4}\right) \hat\sigma^{n}_0\, v \ ,
\label{sig0vf00.EQ}
\eeq
for each of SC-1 and SC-2.
We expect the $M_n^2$ scaling on dimensional grounds.

We are interested in the scattering cross section difference
$\Delta\hat\sigma_{01}$ and asymmetry $\AsymBsig$ between the process and conjugate-process for each of SC-1 and SC-2, 
the conjugate process amplitude being ${\cal A}^{c\,(\sigma)} = {\cal A}_0^{c\,(\sigma)} + {\cal A}_1^{c\,(\sigma)}$.
Again, like in decay, the scattering baryon asymmetry
arises due to the cut in ${\cal A}_1^{(\sigma)}$ (shown as the dashed curve in Fig.~\ref{chi2DDU-NLO.FIG})
giving a discontinuity $(1/2)\,{\rm Disc}({\cal A}_1^{(\sigma)}) = i {\cal \hat{A}}_1^{(\sigma)}$.
For the same reason as in decay, our choice of masses ensures that a second cut through the $\Chi_m$ is not possible.
We denote the interference between the tree ${\cal A}_0^{(\sigma)}$ and the loop ${\cal \hat{A}}_1^{(\sigma)}$
as $\hat\ampA_{01}^{(\sigma)}$,
and analogous to Eq.~(\ref{DGm01Ahn.EQ}), but now integrating over 2-body phase-space, we have
\beq
\Delta\hat\sigma_{01}\, v = \frac{1}{2E_n 2E_i} \int d\Pi_2 \ {\rm Im}(\hat\ampA^{(\sigma)}_{01} - \hat\ampA^{c\,(\sigma)}_{01})
= \frac{1}{2E_n 2E_i} \int d\Pi_2 \ 2\, {\rm Im}(\hat\ampA^{(\sigma)}_{01}) \ .
\label{Dsigv01h.EQ}
\eeq
We obtain the interference term $\Delta\hat\sigma_{01}\, v$ for scattering from the decay interference term Eq.~(\ref{ampA01hatn.EQ})
by using the crossing relation of Eq.~(\ref{crRel.EQ}).

Analogous to the discontinuous piece in decay (cf. Eq.~(\ref{DGm01Ahn.EQ})),
we write the scattering cross section difference between the process and conjugate process
given by the discontinuity in the scattering tree-loop interference term in terms of the dimensionless $f_{01}^{nm(\sigma)}$ 
\beq
 \Delta\hat\sigma_{01}^{n}\, v  = \left(\frac{M_\chi^6}{\Lambda^8}\right) \left[ \frac{1}{16} \left(\frac{M_n}{M_\chi}\right)^6 \, \sum_m 2 f_{01}^{nm(\sigma)} \right] \ ,
\label{Dsig01v.EQ}
\eeq
where the factor of 2 is from equal magnitude but opposite sign contribution from the conjugate process, 
and we expect the $M_n^6$ scaling on dimensional grounds.

Analogous to the decay rate asymmetry of Eq.~(\ref{ABDGm01Gm0.EQ}),
we define a cross section asymmetry which is given by
\beq
\AsymBsign \equiv \frac{\sigma^{n} v - \sigma^{cn} v}{\sigma^{n} v + \sigma^{cn} v}
= -\frac{2 \Delta\hat\sigma_{01}^{n} v}{\sigma^{n} v + \bar\sigma^{n} v}
\approx -\frac{\Delta\hat\sigma_{01}^{n} v}{\sigma_0^{n} v} \ ; \quad
\AsymBsig = \sum_{n=1,2} \AsymBsign \ ,
\label{sigvTADefn.EQ}
\eeq
where
we have a factor of 2 in the numerator from including the $h.c.$ of the interference term. 
For leading order accuracy of the asymmetry we replace the sum in the denominator with the factor of 2 as shown,
and we write the total baryon asymmetry from scattering
$\AsymBsig$ as the sum over the contributions $\AsymBsign$ from $n=1,2$.
We obtain the baryon asymmetry from $\Chi_n$ scattering to be
\beq
\AsymBsign = \left(\frac{M_\chi}{\Lambda}\right)^4 \left[- \frac{1}{4} \left(\frac{M_n}{M_\chi}\right)^4 \sum_{m} \frac{2  f_{01}^{nm(\sigma)} }{ f_{00}^{n(\sigma)} } \right] \ .
\label{ABsigvDefn.EQ}
\eeq

\begin{figure}[h]
  \begin{center}
    \includegraphics[width=0.32\textwidth]{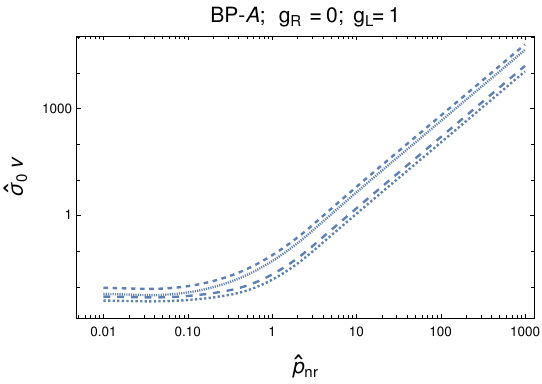}
    \includegraphics[width=0.32\textwidth]{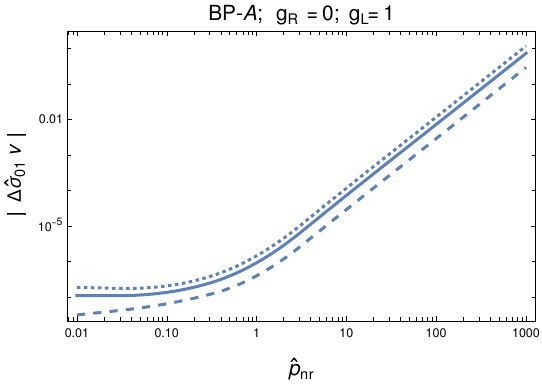}    
    \includegraphics[width=0.32\textwidth]{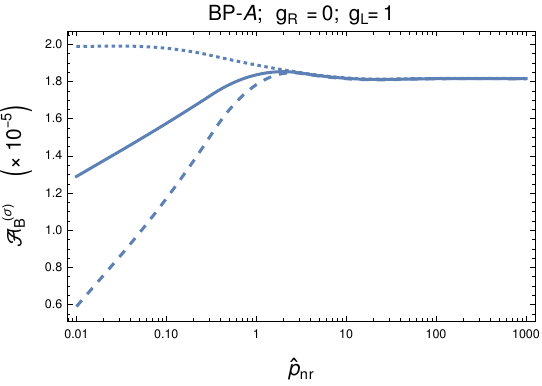}
    \includegraphics[width=0.32\textwidth]{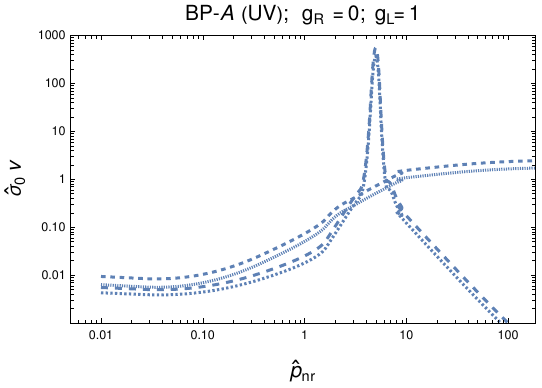}
    \includegraphics[width=0.32\textwidth]{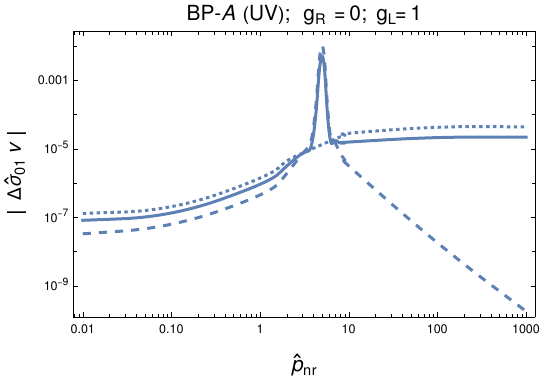}    
    \includegraphics[width=0.32\textwidth]{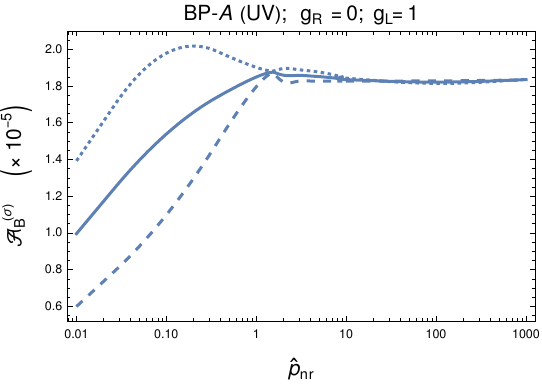}    
    \includegraphics[width=0.32\textwidth]{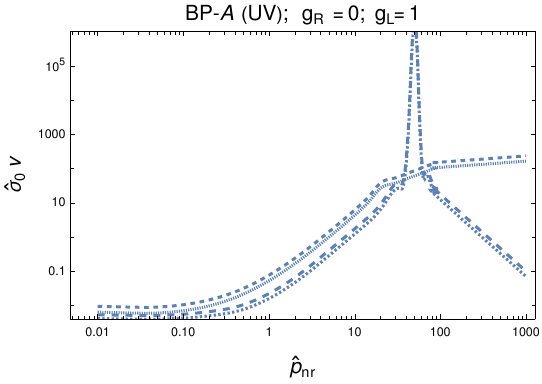}
    \includegraphics[width=0.32\textwidth]{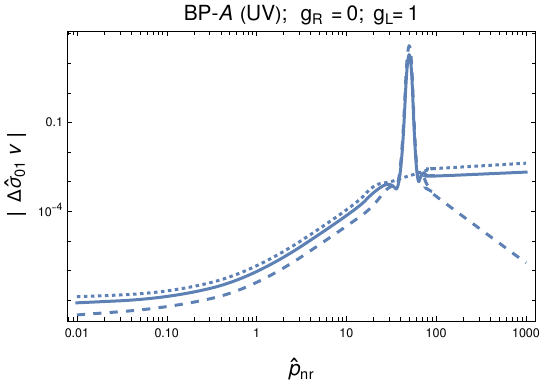} 
    \includegraphics[width=0.32\textwidth]{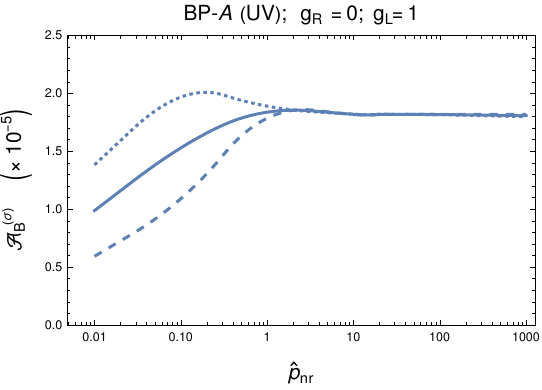}           
  \end{center}    
  \caption{
    For BP-A, in the CM frame,
    with $g_L=1$, $g_R=0$,  
    the $\sigvz$ (in units of $M_\chi^2/\Lambda^4$) (left) 
    for SC-1 n=1 (dotted), SC-1 n=2 (dashed), SC-2 n=1 (long-dotted), SC-2 n=2 (long-dashed), 
    $|\Delta\hat\sigma_{01} v|$ (in units of $M_\chi^6/\Lambda^8$) (middle)
    and
    $\AsymBsig$ (in units of $M_\chi^4/\Lambda^4$) (right)
    summed over $n\!\!=\!\!\{1,2\}$
    for SC-1 (dashed), SC-2 (long-dashed) and the average over SC-1 and SC-2 (solid).
    The topmost row is without UV propagator correction,
    the middle row is with UV propagator correction with $M_\chi/\Lambda\!\! =\!\! 1/10$,
    while the bottom row is with $M_\chi/\Lambda\!\! =\!\! 1/100$, both for $\hat\Gamma_\xi = 1/10$.
    \label{sigABBWpnr.FIG} 
  }
\end{figure}

In Fig.~\ref{sigABBWpnr.FIG} we show as a function of the $\pnrh \equiv \pnr/M_n$, for SC-1 and SC-2 scattering processes,
in the left column the $\sigvzn$ in units of $M_\chi^2/\Lambda^4$,
in the middle column (the absolute value of) the tree-loop interference scattering cross section difference between
the process and conjugate process $|\Delta\hat\sigma_{01} v|$,  
and,
in the right column the baryon asymmetry $\AsymBsig$ summed over n=1,2. 
These are all shown in the CM frame with $\pnr = \kir$ and $c_i \equiv \cos{\theta_i} = -1$,
after integrating over the final-state phase-space angular variables, 
for the benchmark point BP-A, and the couplings $g_L=1$, $g_R=0$.
The top row is with the effective VV interaction,
and we should bear in mind that
one should trust the effective theory results only up to $\pnrh < \Lambda/M_\chi$
and read the top-row plots only up to that for a specific choice of $M_\chi/\Lambda$.
The rise in the cross section we see in the effective theory here is turned around beyond
$\pnrh > \Lambda/M_\chi$ by the UV completion, which we discuss in detail in Appendix~\ref{VVintUV.SEC}. 
As discussed there, we include the UV state $\xi$ propagator correction resolving the VV effective interaction, and
show its effect also in Fig.~\ref{sigABBWpnr.FIG},  
with the middle row for $M_\chi/\Lambda\!\! =\!\! 1/10$ and bottom row for $M_\chi/\Lambda\!\! =\!\! 1/100$, both with $\hat\Gamma_\xi = 1/10$,
where $\hat\Gamma_\xi = \Gamma_\xi/M_\chi$.
As mentioned, we see that for $\pnrh > \Lambda/M_\chi$, the cross section indeed turns around due to the propagating $\xi$ resolving the effective VV interaction. 
We see that the LO SC-1 (t-channel) $\sigvz$ does not fall off for large $\pnrh$ but rather flattens out,
although the growth in $\sigvz$ is not present with the inclusion of the $\xi$ UV propagator correction.
We trace this flattening-out to the fact that the $\xi$ propagator correction becomes independent of $\pnrh$ for $\cos{\theta_1^{CM}}\! =\! -1$
and therefore fails to fall off as $1/\pnrh^2$ at this phase-space point.
In the UV propagator correction, we have included only the real (PV) part of $f_\xi(p)$,
which goes to zero at the position of the pole as discussed in Appendix~\ref{VVintUV.SEC} and seen in Fig.~\ref{fxi00nCh.FIG}. 
The imaginary $\xi$ pole contribution gives the $\xi$ on-shell production and decay,
which we omit in this effective theory analysis viewing it as a UV physics contribution.

Also, the departure from thermal equilibrium required by the Sakharov condition is aided by 
the $\Chi$ being almost decoupled~\cite{Gopalakrishna:2024qxk} from the thermal plasma in the early Universe
due to very weak interactions with $\Up,\Dp$. 
If so, the $\Chi$ is produced mostly when other heavier UV states decay. 
One such example is in the decay of the $\Phi_B$ that is responsible for Majorana mass generation discussed in
Ref.~\cite{Gopalakrishna:2022hwk},
i.e., $\Phi_B \to \Chi\Chi$ at the scale around $\Lambda$.
If such a mechanism is operative,
$\Chi$ scattering is to be considered for $\pnrh \lesssim M_\Phi/M_\chi$ as
there is no appreciable population of $\Chi$ at temperatures above $M_\Phi$.
If $M_\Phi \sim \Lambda$, we should then restrict $\Chi$ scattering to $\pnrh \lesssim \Lambda/M_\chi$. 

As mentioned earlier, eventually, our interest is in the scattering taking place in the early Universe at temperature $T$.
We turn next to computing the thermally averaged cross section difference for use in this setting in Ref.~\cite{Gopalakrishna:2024qxk}. 

\subsubsection{The thermally averaged cross section and baryon asymmetry}
\label{thAvgcsAB.SEC}

For the $2\to 2$ processes SC-1 and SC-2, following the procedure laid-out in Appendix~\ref{ThAvgProc.SEC}, 
we compute the thermally averaged cross section as
\beq
\langle \sigma v\rangle  = \int \frac{d^3 p_n}{(2\pi)^3} \frac{d^3 k_i}{(2\pi)^3} \, \frac{f^{(0)}(E_n)}{n^{(0)}_n} \frac{f^{(0)}(E_i)}{n^{(0)}_i} \, \sigma v \ .
\label{TASC12.EQ}
\eeq
We obtain the thermally averaged $\langle f_{00}^{n(\sigma)}\rangle $
and compute the thermally averaged LO cross sections in analogy with Eq.~(\ref{sig0vf00.EQ}),
for each of SC-1 and SC-2.

Next, we obtain $\langle f_{01}^{nm(\sigma)}\rangle $ 
for each of the SC-1 and SC-2, 
and from these obtain the thermally averaged cross section difference between the process and conjugate process
$\langle  \Delta\hat\sigma_{01}^{n}\, v \rangle $
given by the discontinuity in the scattering tree-loop interference term
analogous to Eq.~(\ref{Dsig01v.EQ}) as 
\beq
\langle  \Delta\hat\sigma_{01}^{n}\, v \rangle  = \left(\frac{M_\chi^6}{\Lambda^8}\right) \left[ \frac{1}{16} \left(\frac{M_n}{M_\chi}\right)^6 \, \sum_m 2 \langle f_{01}^{nm(\sigma)}\rangle  \right] \ .  
\label{TADsig01v.EQ}
\eeq
Also, similar to Eq.~(\ref{ABsigvDefn.EQ}), we define a thermally averaged cross section asymmetry as
\beq
\AsymBsignh \equiv \frac{\langle \sigma^{n} v\rangle  - \langle \sigma^{cn} v\rangle }{\langle \sigma^{n} v\rangle  + \langle \sigma^{cn} v\rangle }
= -\frac{2 \langle \Delta\hat\sigma_{01}^{n} v\rangle }{\langle \sigma^{n} v\rangle  + \langle \bar\sigma^{n} v\rangle }
\approx -\frac{\langle \Delta\hat\sigma_{01}^{n} v\rangle }{\langle \sigma_0^{n} v\rangle } \ ; \quad
\AsymBsigh = \sum_{n=1,2} \AsymBsignh \ .
\label{ABsigvTADefn.EQ}
\eeq
We call attention to the fact that $\AsymBsignh$ is not the thermal average of the earlier $\AsymBsign$. 
The $\AsymBsigh$ is the sum over the contributions $\AsymBsignh$ from $n=1,2$.
We then obtain the thermally averaged baryon asymmetry from $\Chi_n$ scattering to be
\beq
\AsymBsignh = \left(\frac{M_\chi}{\Lambda}\right)^4 \left[- \frac{1}{4} \left(\frac{M_n}{M_\chi}\right)^4 \sum_{m} \frac{2 \langle  f_{01}^{nm(\sigma)} \rangle }{\langle  f_{00}^{n(\sigma)} \rangle } \right] \ .
\eeq

For the above values of the integrals for the benchmark point BP-A, for the choice $g_L=1,g_R=0,\tilde{g}=1$, for example, we find
for $x\!\!=\!\! 1/2$ the LO thermally averaged cross-section for (SC-1,SC-2) 
for $n\!\!=\!\!1$ is $\langle  \sigma_0 v \rangle  = (0.26, 8.8)\, M_\chi^2/\Lambda^4$,
and for $n\!\!=\!\!2$ is $\langle  \sigma_0 v \rangle  = (0.36, 12.4)\, M_\chi^2/\Lambda^4$;
the difference in the interference term in the cross section
for $n\!\!=\!\!2$ is
$\langle \Delta\hat\sigma_{01}\, v\rangle  = (-0.67, -13) \times 10^{-5}\, M_\chi^6/\Lambda^8$; 
and 
the baryon asymmetry from scattering
for $n\!\!=\!\!2$ is $\AsymBsigh = (1.9, 1) \times 10^{-5}\, M_\chi^4/\Lambda^4$. 
We thus can write for
$\hat\Gamma_\xi\!\! =\!\! 1/2$, for $n\!\!=\!\!2$, for (SC-1,SC-2),
$\langle  \sigma_0 v \rangle  = (0.36, 12.4)\, (g^2/\Lambda^2) \, (M_\chi/\Lambda)^2 $,
$\langle \Delta\sigma_{01}\, v\rangle  = (-0.67, -13) \times 10^{-5}\, (g^4/\Lambda^2) \, (M_\chi/\Lambda)^6 $, 
and
$\AsymBsigh = (1.9, 1) \times 10^{-5} \, g^2 \, (M_\chi/\Lambda)^4$, including the couplings scaling.
Again, as in the decay case, 
we find that the imaginary parts of these integrals are zero.

From the above $\langle  \sigma_0 v \rangle $ and $\langle \Delta\hat\sigma_{01}\, v\rangle $ numbers given, for BP-A,
$x\!\!=\!\! 1/2$, with the UV propagator correction factor present with $M_\chi/\Lambda\!\! =\!\! 1/10$ and $\hat\Gamma_\xi = 1/10$,
$g_L=1,g_R=0,\tilde{g}=1$,
we extract the thermal averages of the LO cross section and tree-loop interference functions parametrized
in Eqs.~(31) and (32) respectively of Ref.~\cite{Gopalakrishna:2022hwk}. 
For $n\!\!=\!\! 2$, we find these to be for (SC-1,SC-2):   
$\langle (1/(2\hat{E}_n\, 2\hat{E}_i))\, \hat{f}_{00}^{(\sigma)}\rangle  \approx (2, 66)$,
and
$\langle (1/(2\hat{E}_n\, 2\hat{E}_i))\, \hat{f}_{01}^{(\sigma)}\rangle  \approx (-0.03,-0.6)$. 

\begin{figure}[h]
  \begin{center}
    \includegraphics[width=0.3\textwidth]{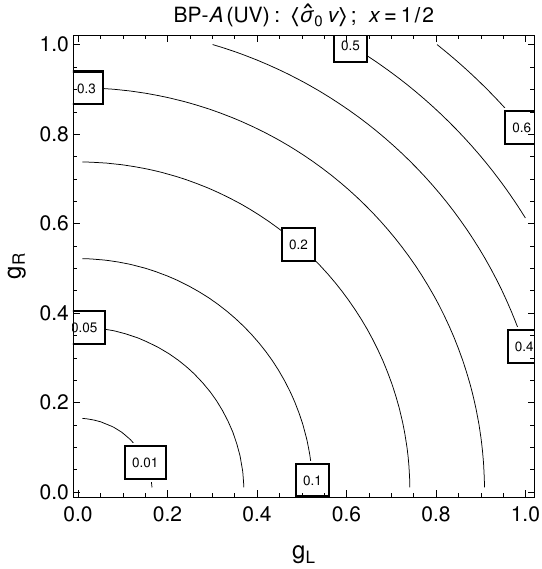} \hspace*{0.25cm}
    \includegraphics[width=0.3\textwidth]{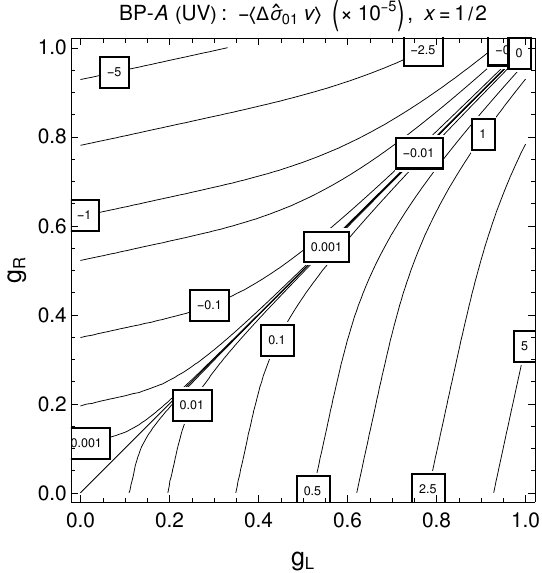} \hspace*{0.25cm}
     \includegraphics[width=0.3\textwidth]{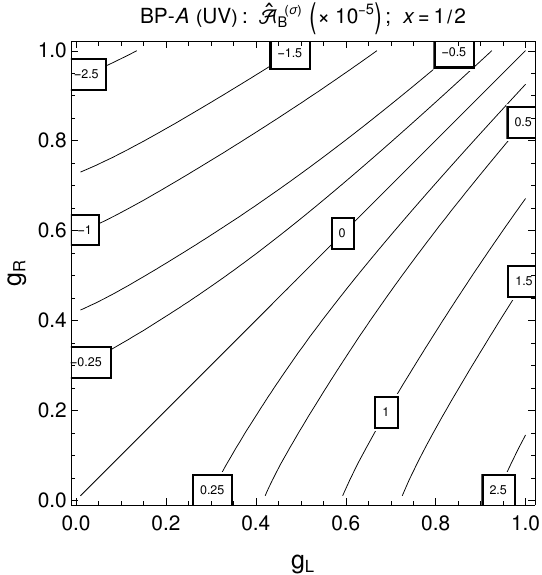} 
     \includegraphics[width=0.3\textwidth]{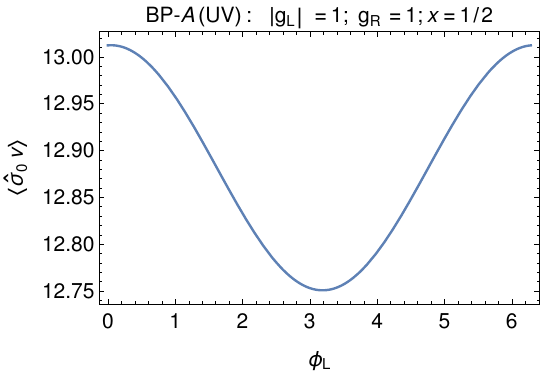} \hspace*{0.25cm}
     \includegraphics[width=0.3\textwidth]{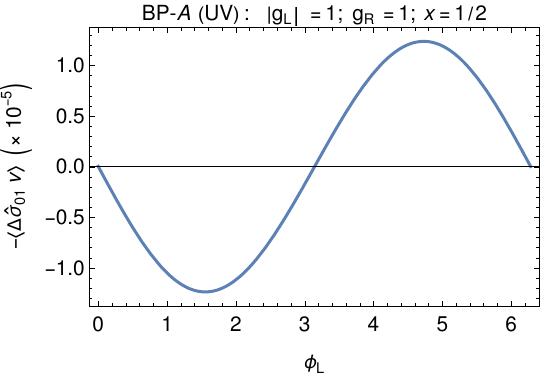} \hspace*{0.25cm}     
     \includegraphics[width=0.3\textwidth]{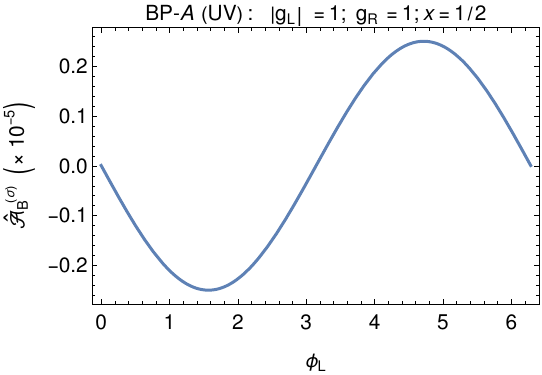}
  \end{center}    
  \caption{
    For BP-A,
    the $\langle  \sigma_0 v \rangle $ (in units of $M_\chi^2/\Lambda^4$)
    averaged over SC-1 and SC-2 for $n\!\!=\!\!2$ (left-panel),
    $-\langle \Delta\hat\sigma_{01} v\rangle $ (in units of $M_\chi^6/\Lambda^8$) averaged over SC-1 and SC-2 for $n\!\!=\!\!2$ (middle-panel),
    and,
    $\AsymBsigh$ (in units of $M_\chi^4/\Lambda^4$)
    from SC-1 and SC-2 for $n\!\!=\!\!2$ (right-panel),
    for $x\!\!=\!\! 1/2$,
    with UV propagator correction included with $M_\chi/\Lambda\!\! =\!\! 1/10$ and $\hat\Gamma_\xi = 1/10$.
    \label{sigvABcoupl.FIG} }
\end{figure}

In Fig.~\ref{sigvABcoupl.FIG} we show, for BP-A, $n\!\!=\!\! 2$,
the LO thermally averaged cross section $\TAsigvzh$ (in units of $M_\chi^2/\Lambda^4$)
averaged over the scattering processes SC-1 and SC-2,
the (negative of the) tree-loop interference scattering cross section difference
$-\langle \Delta\hat\sigma_{01} v\rangle $ (in units of $M_\chi^6/\Lambda^8$)
averaged over SC-1 and SC-2 for $n\!\!=\!\!2$, 
and,
the sum of the baryon asymmetry from scattering $\AsymBsigh$ (in units of $M_\chi^4/\Lambda^4$) from SC-1 and SC-2,
for $x\!\!=\!\! 1/2$.
We choose this $x$ since the $\TAsigvz$ is the largest close to this value, and focus on $n\!\!=\!\!2$ only since the $\AsymBsigh$ is mostly generated for $n\!\!=\!\!2$.
In these plots we highlight the dependence of these quantities on the magnitude of $g_L,g_R$ and the phase of $g_L$ (i.e. $\phi_L$).

\begin{figure}
  \begin{center}
    \includegraphics[width=0.32\textwidth]{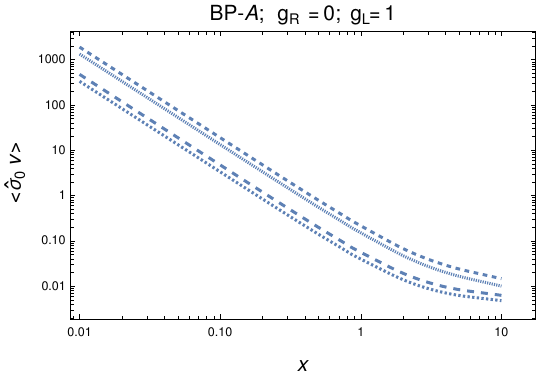}
    \includegraphics[width=0.32\textwidth]{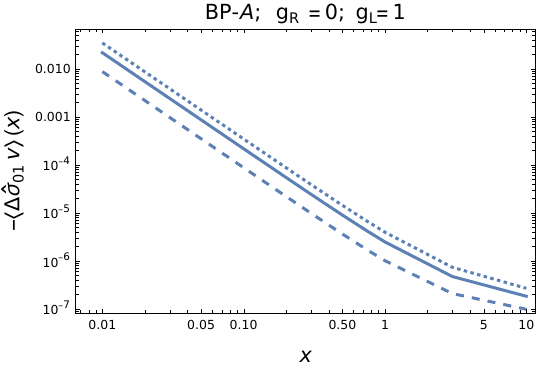}
    \includegraphics[width=0.32\textwidth]{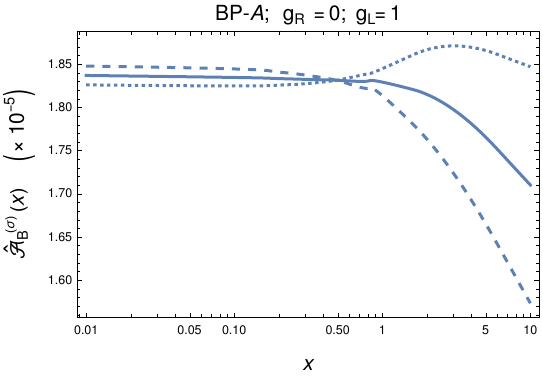}    
    \includegraphics[width=0.32\textwidth]{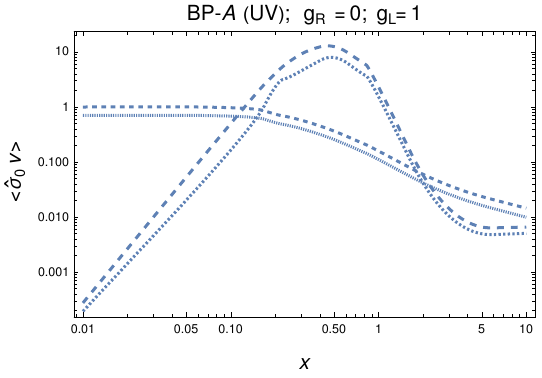}
    \includegraphics[width=0.32\textwidth]{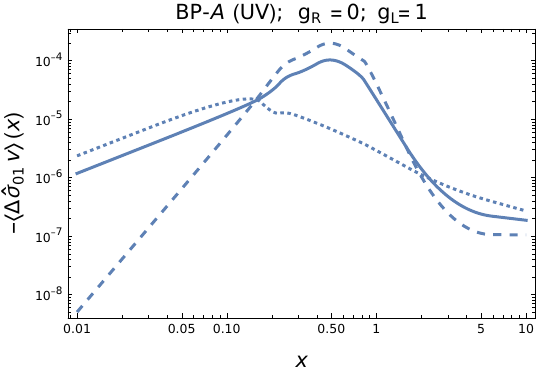}    
    \includegraphics[width=0.32\textwidth]{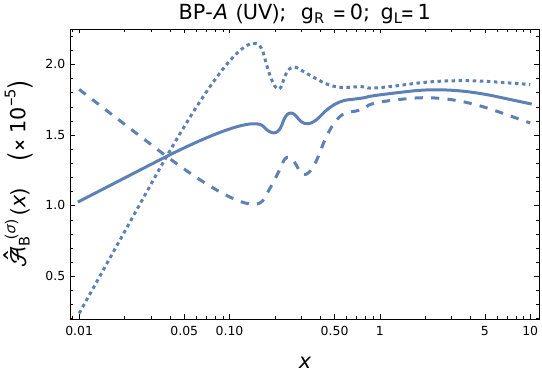}
  \end{center}    
  \caption{
    For BP-A, for $g_L=1$, $g_R=0$,
    the $x\!\!=\!\!M_\chi/T$ dependence of
    $\TAsigvz$ (in units of $M_\chi^2/\Lambda^4$) (left column),
    for SC-1 $n\!\!=\!\! 1$ (dotted), SC-1 n=2 (dashed), SC-2 $n\!\!=\!\! 1$ (long-dotted), SC-2 $n\!\! =\!\! 2$ (long-dashed),
    and,
    for $n\!\! =\!\! 2$
    SC-1 (long-dotted) and SC-2 (long-dashed) 
    the $-\langle \Delta\hat\sigma_{01} v\rangle $ (in units of $M_\chi^6/\Lambda^8$) (middle column)
    and $\AsymBsigh$ (in units of $M_\chi^4/\Lambda^4$) (right column),    
    with their averages over SC-1 and SC-2 (solid).
    The top panel is without UV propagator correction,
    while the bottom panel is with the UV propagator correction
    for $M_\chi/\Lambda\!\! =\!\! 1/10$ and $\hat\Gamma_\xi = 1/10$.
    \label{ABsigh.FIG} }
\end{figure}

In Fig.~\ref{ABsigh.FIG} we show for BP-A, $g_R=0$, $g_L=1$,
the $x$ dependence in the range $x \in (0.01,20)$
$\TAsigvzh$,
and for $n\!\!=\!\!2$, 
the $-\langle \Delta\hat\sigma_{01} v\rangle $
and 
$\AsymBsigh$ generated in the SC-1 and SC-2 scattering processes. 
The plots marked ``UV'' are with the UV propagator correction included with
$M_\chi/\Lambda\!\! =\!\! 1/10$ and $\hat\Gamma_\xi = 1/10$.
To a very good approximation, most of the baryon asymmetry from scattering arises from $n\!\!=\!\!2$,
i.e. from $\Chi_2$ scattering processes, 
as we find the asymmetry generated from n=1 is
about two orders of magnitude smaller than for n=2.

We see that the $\langle \Delta\hat\sigma_{01} v\rangle $ with the UV propagator correction included
is substantial only
in the range $x \in (0.1,10)$
and there is no growth for small $x$ (i.e. for $T > M_\chi$), unlike for the case without the UV correction. 
Regarding the latter case, we should not trust an effective operator analysis for small $x$ (i.e. large $T$)
as the UV state modifications would resolve the effective vertex above the scale $\Lambda$ and soften the growth at small $x$. 
This instructs us to restrict to the above domain while using our baryon asymmetry results.
In this restricted domain, the $\langle \Delta\hat\sigma_{01} v\rangle $ we obtain with and without UV completion
are within an order of magnitude of each other,
being smaller in SC-1 in the effective theory, and bigger in SC-2.
The change is bigger in SC-2, which we expect since the UV state is resonant as it appears in the s-channel.
A truly accurate picture can be obtained only by computing the thermal averages in a specific UV completion
since the thermal spectrum samples UV physics contributions.

\subsection{Scattering baryon asymmetry from multiple operator loop contribution}
\label{multiOpSig.SEC}

We compute here the scattering baryon asymmetry in SC-1 and SC-2 when more than one operator is involved,
obtained from crossing Fig.~\ref{chi2DDU-NLO-modA.FIG}.
We parallel the analysis in Sec.~\ref{singOpSig.SEC} and obtain results here using the crossing relations Eq.~(\ref{crRel.EQ})
in the multiple operator decay contributions in Sec.~\ref{mixedOpsAsymB.SEC}.
The leading contribution to the baryon asymmetry in scattering is in the interference term
between the tree and loop scattering amplitudes, similar to that shown diagrammatically in Fig.~\ref{chi2DDU-LONLO-modA.FIG},
but now crossed. 

For the scattering kinematics, we now have $q_3+q_4 = q_1-k_i$ for SC-1, while $q_3+q_4 = p_n+k_i$ for SC-2.
We have argued (cf. Sec.~\ref{genAB.SEC}) that a nonzero baryon asymmetry
requires a nonzero strong phase coming from a cut in the loop process,
which in the multiple operator contribution of Fig.~\ref{chi2DDU-NLO-modA.FIG} requires $q_3$ and $q_4$ to {\em both} be put onshell.
This is possible only in SC-2 (s-channel kinematics), and not in SC-1 (t-channel kinematics).
We thus infer that a nonzero baryon asymmetry in scattering from the analog of Fig.~\ref{chi2DDU-NLO-modA.FIG} only arises
in SC-2, and we compute this contribution.
We recall that we only need the $nm=21$ contribution for computing the baryon asymmetry. 

For our numerical study, we choose the {\it benchmark point B} (BP-B) mass parameters (see p.\,\pageref{BPBDefn.PG})
that has $M_2 > M_1 + 2 M_U$ allowing the loop to be cut, giving a strong phase. 
Again, for the same reasons explained in the decay case, only the $\ampA_{01}^{\prime(1)}$ leads to a baryon asymmetry.
We write the $\Delta\hat\sigma_{01}\, v$ for the multiple operator scattering contribution for SC-2 in analogy with
Eq.~(\ref{Dsigv01h.EQ}), but now with the matrix element being ${\rm Im}(\hat\ampA^{\prime (\sigma)}_{01})$,
which is obtained from the decay contribution in Eq.~(\ref{A01modA.EQ}) by using the crossing relation of Eq.~(\ref{crRel.EQ}).
We thermally average this and compute for SC-2
the thermally averaged cross section difference $\langle  \Delta\hat\sigma_{01}^{n}\, v \rangle $ between the process
and conjugate process analogous to Eq.~(\ref{TADsig01v.EQ}),
and, the baryon asymmetry from scattering $\AsymBPsigh$ analogous to Eq.~(\ref{ABsigvTADefn.EQ}).
We present the results next.

For illustration, we give here from the multiple operator contribution,
for BP-B, some sample numerical values for the thermally averaged
scattering cross section difference between the process and conjugate process and the baryon asymmetry. 
For the choice $g_L\!\! =\!\! (3/4)\,e^{-\pi/2}$, $g_R\!=\! 1/4$, $\tilde{g}\!\!=\!\! 1$, for example, we find at $x\!\! =\!\! 1/2$,
the LO thermally averaged cross-section
for (SC-1,SC-2) $n\!\!=\!\!1$ is $\langle  \sigma_0 v \rangle \! = \! (0.16, 1) M_\chi^2/\Lambda^4$,
and for $n\!\!=\!\!2$ is $\langle  \sigma_0 v \rangle \! = \! (0.23, 1.45) M_\chi^2/\Lambda^4$.
For SC-2 $n\!\!=\!\!2$, $x\!\! =\!\! 1/2$,
the difference in the interference term in the cross section
$\langle \Delta\hat\sigma_{01}\, v\rangle  \! = \! -5.2\times 10^{-4} M_\chi^4/\Lambda^6$,   
and the scattering baryon asymmetry
is $\AsymBPsigh \! = \! 3.6 \times 10^{-4} M_\chi^2/\Lambda^2$.
We thus can write for $M_\chi/\Lambda \!\! = \!\! 1/10$, for SC-2 $n\!\!=\!\!2$, 
$\langle  \sigma_0 v \rangle \! \sim\! 1.45\times 10^{-2}\, g^2/\Lambda^2$,
$\langle \Delta\hat\sigma_{01}\, v\rangle \! \sim\! -5.2\times 10^{-8} g^4/\Lambda^2  $
and
$\AsymBPsigh\! \sim\! 3.6 \times 10^{-6} \, g^2$, including the couplings scaling. 

From the above $\langle  \sigma_0 v \rangle $ and $\langle \Delta\hat\sigma_{01}\, v\rangle $ numbers
for BP-B, $x\!\! = \!\! 1/2$, $g_L\!\!=\!\!(3/4)\, e^{-i\pi/2}$, $g_R\!\!=\!\!1/4$, $\tilde{g}\!\!=\!\!1$,
with the UV propagator correction factor present, 
we extract the thermal averages of the LO cross section and tree-loop interference functions parametrized
in Eqs.~(31) and (32) of Ref.~\cite{Gopalakrishna:2022hwk}, 
and obtain for SC-2 $n\!\!=\!\! 2$,
$\langle (1/(2\hat{E}_n\, 2\hat{E}_i))\, \hat{f}_{00}^{(\sigma)}\rangle \! \approx\! 12.3$,
and
$\langle (1/(2\hat{E}_n\, 2\hat{E}_i))\, \hat{f}_{01}^{(\sigma)}\rangle \! \approx\! -0.1$,
for the multiple operator scattering contributions. 

\begin{figure}
  \begin{center}
    \includegraphics[width=0.3\textwidth]{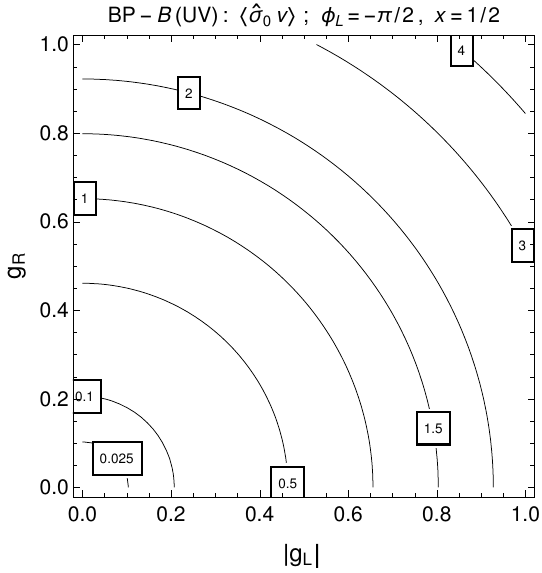}\hspace*{0.25cm}
    \includegraphics[width=0.3\textwidth]{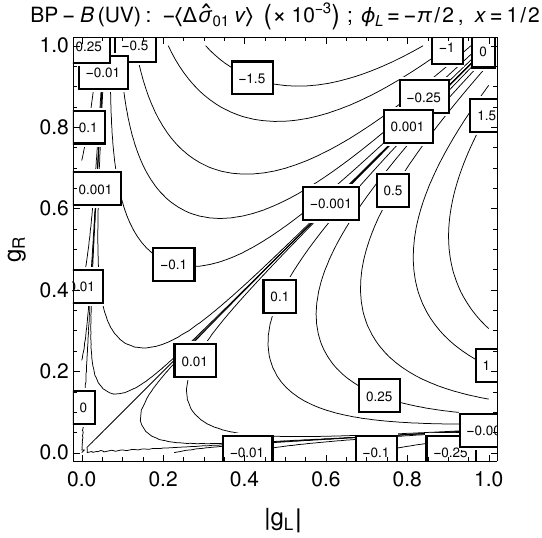} \hspace*{0.25cm}       
    \includegraphics[width=0.3\textwidth]{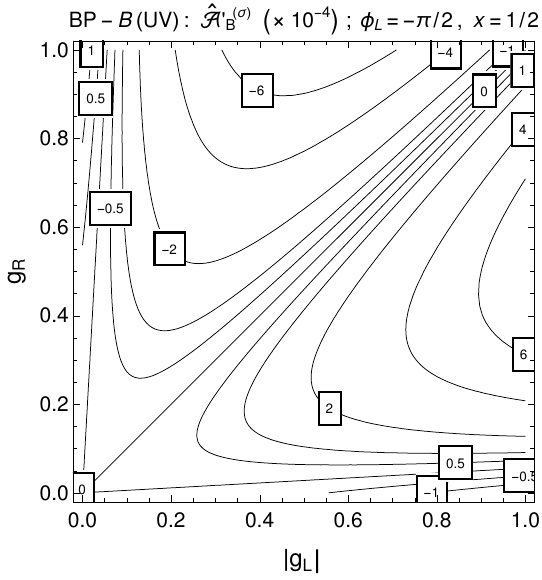}
    \\
    \includegraphics[width=0.3\textwidth]{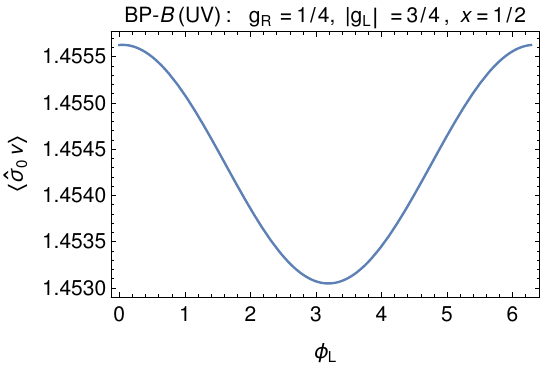}\hspace*{0.25cm}
    \includegraphics[width=0.3\textwidth]{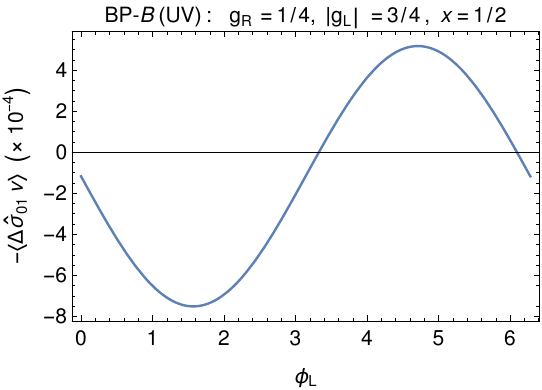} \hspace*{0.25cm}       
    \includegraphics[width=0.3\textwidth]{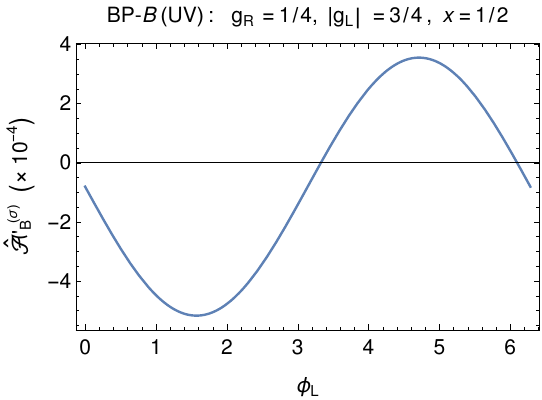}
  \end{center}    
  \caption{For BP-B, SC-2 $n\!\!=\!\!2$, $x\!\!=\!\! 1/2$,
    the
    $\langle  \sigma_0 v \rangle $ (in units of $M_\chi^2/\Lambda^4$) (left-panel),
    $-\langle \Delta\hat\sigma_{01} v\rangle $ (in units of $M_\chi^4/\Lambda^6$) (middle-panel), 
    and $\AsymBPsigh$ (in units of $M_\chi^2/\Lambda^2$) (right-panel)
    from the multiple-operator contributions,
    with the UV propagator corrections included with $M_\chi/\Lambda\!\! =\!\! 1/10$ and $\hat\Gamma_\xi\!\! =\!\! 1/2$.
    \label{sigvABP.FIG} }
\end{figure}

In Fig.~\ref{sigvABP.FIG} we show for BP-B, SC-2 $n\!\!=\!\!2$, $x\!\! =\!\! 1/2$,
the dimensionless LO thermally averaged cross section $\TAsigvzh$ (in units of $M_\chi^2/\Lambda^4$),
the (negative of the) tree-loop interference cross section difference
$-\langle \Delta\hat\sigma_{01} v\rangle $ (in units of $M_\chi^4/\Lambda^6$),
and the baryon asymmetry from scattering $\AsymBPsigh$ (in units of $M_\chi^2/\Lambda^2$).
The UV propagator corrections are included with $M_\chi/\Lambda \!\! = \!\! 1/10$ and $\hat\Gamma_\xi \!\! = \!\! 1/2$.

\begin{figure}
  \begin{center}
    \includegraphics[width=0.32\textwidth]{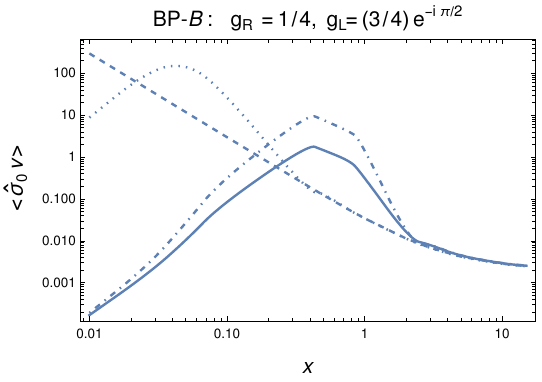}\hspace*{0.25cm}
    \includegraphics[width=0.32\textwidth]{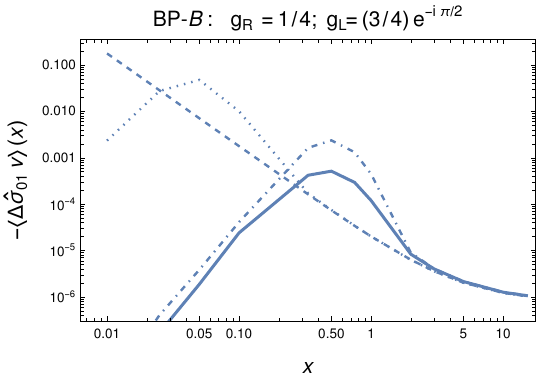}\hspace*{0.25cm}        
    \includegraphics[width=0.32\textwidth]{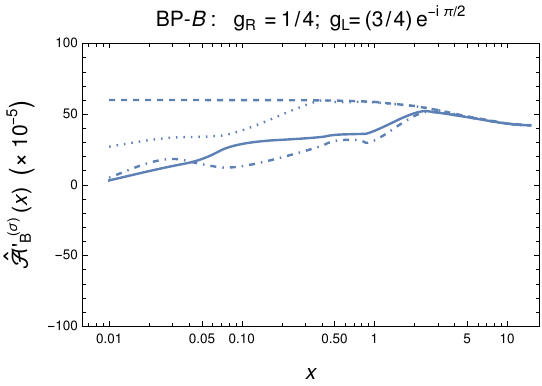}
  \end{center}    
  \caption{
    For BP-B,
    the $\langle  \sigma_0 v \rangle $ (in units of $M_\chi^2/\Lambda^4$) (left),
    $-\langle \Delta\hat\sigma_{01} v\rangle $ (in units of $M_\chi^4/\Lambda^6$) (middle),
    and $\AsymBPsigh$ (in units of $M_\chi^2/\Lambda^2$) (right), 
    from the multiple operator SC-2 $n\!\!=\!\!2$ contributions
    for $g_L=(3/4)e^{-i\pi/2}$, $g_R=1/4$, 
    with UV propagator correction included taking
    $(M_\chi/\Lambda,\hat\Gamma_\xi)$ to be 
    $(1/10,1/2)$ (solid),
    $(1/10,1/10)$ (dot-dashed),
    $(1/100,5)$ (dotted),
    and,
    without UV propagator correction (dashed).
    \label{sigvABPx.FIG} }
\end{figure}

In Fig.~\ref{sigvABPx.FIG} we show for BP-B
the same quantities now as a function of $x$,
for $g_L\!\!=\!\!(3/4)e^{-i\pi/2}$, $g_R\!\!=\!\!1/4$, $\tilde{g}\!\!=\!\!1$
for different choices of $(M_\chi/\Lambda,\hat\Gamma_\xi)$.  
We see that for $(M_\chi/\Lambda,\hat\Gamma_\xi)\! =\! (1/100,5)$ (so that $\tilde\Gamma_\xi\! =\! 1/2$ as before),
we find a similar shape as for the $(1/10,1/2)$ case
(with a factor of $1/10$ scaling of $x$), 
and the size of $\langle  \sigma_0 v \rangle $ and $-\langle \Delta\hat\sigma_{01} v\rangle $ are both higher by a factor of about 100,
leaving $\AsymBsigh$ to be about the same.
We can also see that for $(M_\chi/\Lambda,\hat\Gamma_\xi)\! =\! (1/10, 1/10)$,
we find that the behavior is largely
as for the $(1/10,1/2)$ case 
except for the $\langle  \sigma_0 v \rangle $ and $-\langle \Delta\hat\sigma_{01} v\rangle $ both being higher by a factor of about 5,
again leaving $\AsymBsigh$ to be about the same. 
These comparisons are without including the powers of $M_\chi/\Lambda$ scaling factors that have been pulled out as indicated above.

With the UV propagator correction included, we see from Fig.~\ref{sigvABPx.FIG} that
the $\langle \Delta\hat\sigma_{01}\, v\rangle $ is substantial only in the domain $x\in (0.1,10)$ for $M_\chi/\Lambda\!\! =\!\! 1/10$.
This domain shifts to smaller $x$ as $M_\chi/\Lambda$ decreases.  
These are in contrast to the case without the UV propagator correction (see the dashed curve), for which
$\langle \Delta\hat\sigma_{01}\, v\rangle $ keeps increasing as $x$ decreases. 
However, this growth in the $x \ll 1$ region is not to be trusted
since we see the UV propagator corrections modifies the behavior as the typical momenta become bigger than $\Lambda$,
bringing in sensitivity to the UV physics. 
This is particularly true here for SC-2 as the UV correction is resonant since it is in the s-channel.
The scattering baryon asymmetry with and without the UV correction
are in good agreement for $x \gtrsim 1$,
but differ for $x \lesssim 1$ where the latter keeps growing but the former turns around. 

Thus, if the first operator in Eq.~(\ref{modAops.EQ}) is also present in the effective theory
in addition to the operator in Eq.~(\ref{LIntVVMB.EQ}),
and if the kinematics allows (i.e. if $M_2\! > \! M_1\! +\! 2 M_U$),
the multiple operator scattering contribution obtained from crossing Fig.~\ref{chi2DDU-LONLO-modA.FIG} (left)
could be dominant as it is enhanced by a loop and $(\Lambda/M_\chi)^2$ factors 
compared to the single operator contribution.
If any of these conditions is not met, the single operator contribution of Sec.~\ref{singOpSig.SEC}
involving only the operator in Eq.~(\ref{LIntVVMB.EQ}) 
would remain as the only contribution to the scattering asymmetry.

\section{The $\Delta B = -2$ process $\Up\Dp\Dp \to \Up^c \Dp^c \Dp^c$}
\label{UDD2UcDcDc.SEC}

\begin{figure}
  \begin{center}
    \includegraphics[width=0.3\textwidth] {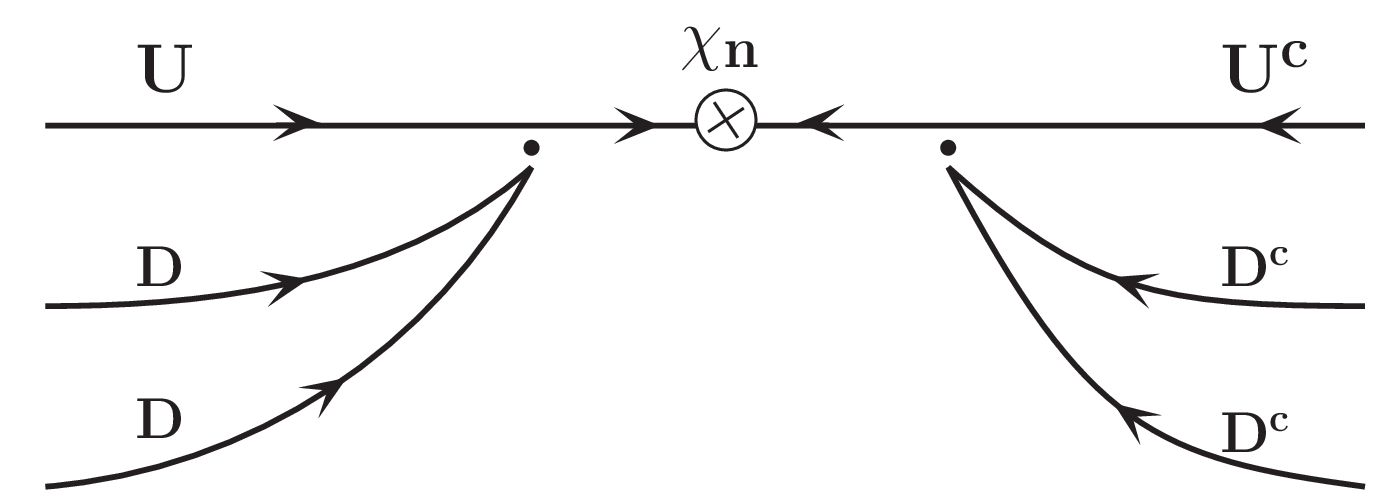} \hspace*{1cm}
    \includegraphics[width=0.3\textwidth] {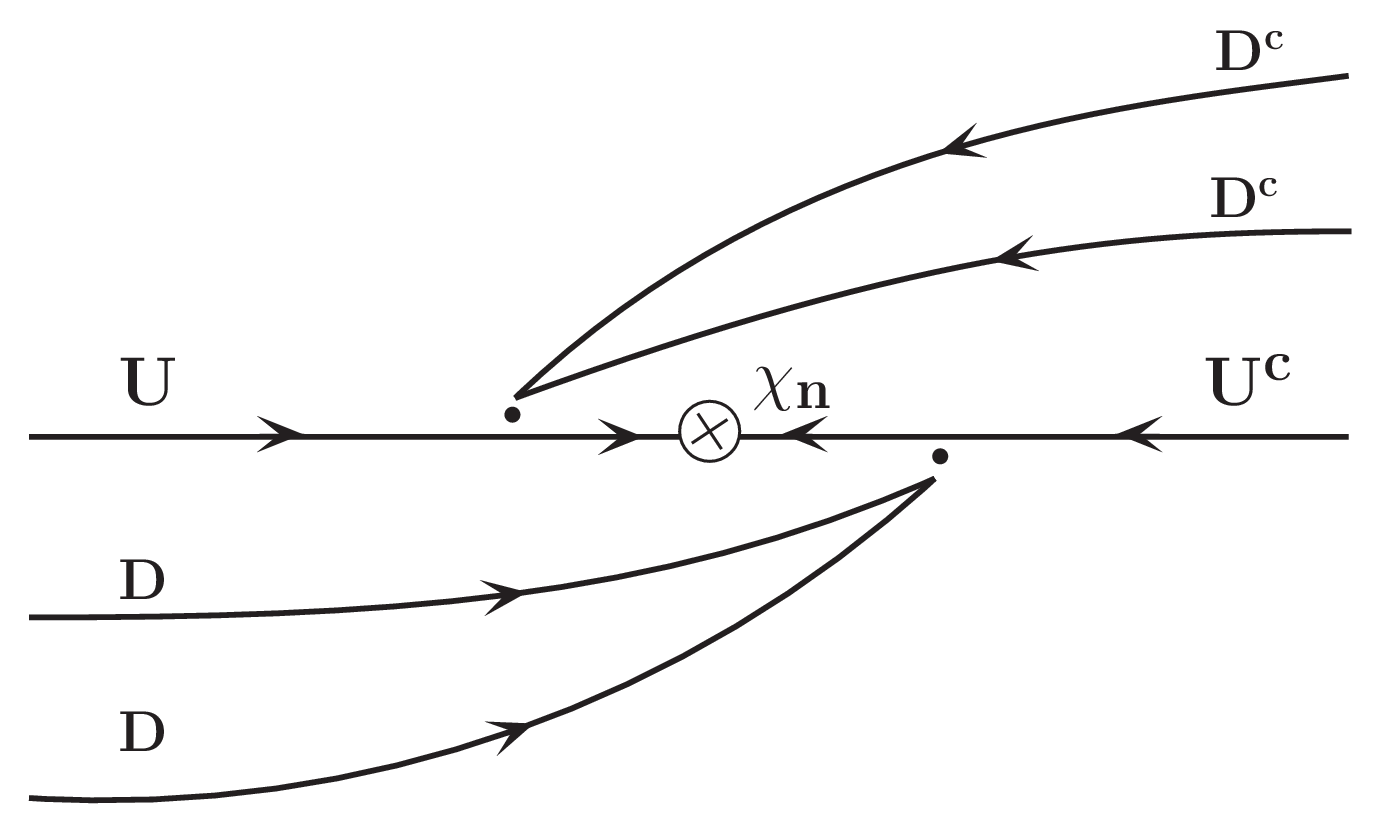}
    \caption{The S- and T-channel-like Feynman diagrams contributing to the $\Up\Dp\Dp \to \Up^c \Dp^c \Dp^c$ amplitude.
      \label{UDD2UcDcDcFD.FIG}
    }
  \end{center}
\end{figure}

The thermally averaged $\Up(q_U')\Dp(q_3)\Dp(q_4) \to \Up^c(q_U) \Dp^c(q_1) \Dp^c(q_2)$
scattering rate appears in the collision term (rhs) of the the BE.
It is sufficient to compute this rate at tree-level, i.e. at $\bigO(g^4)$, 
as the the other collision terms, for instance the $\DGm$ term, are at this order.
Here, we compute this contribution.
The Feynman diagram for this process is as shown in Fig.~\ref{UDD2UcDcDcFD.FIG}.
The left diagram has kinematics which is S-channel-like while the right diagram is T-channel-like,
and we call these contributions as $S$ and $T$ respectively.
(Another T-channel-like diagram similar to the right diagram but with one line from each vertex
is switched between the initial and final states is not shown.)   
The scattering rate is given as $|S+T|^2 = SS^* + TT^* + ST^* + TS^*$,
which we denote simply as $SS,~TT,~ST,~TS$ respectively for notational brevity.

We define a quantity for the $3\to 3$ process that has the units of a cross section (i.e. ${\rm GeV}^{-2}$) as
\beq
\sigzP \equiv M_\chi^3 \prod_{i=1}^{N_i} \left(\frac{1}{2E_{Q_i}}\right)
\int \, \prod_{f=1}^{N_f} \left(\frac{d^3 k_{Q_f}}{(2\pi)^3} \frac{1}{2E_{Q_f}} \right) (2\pi)^4 \delta^{(4)}(\underset{\ssty i}{\Sigma} p_{Q_i} - \underset{\ssty f}{\Sigma} k_{Q_f})\ |\ampM'_0|^2 \ ,
\label{sig0BBb.EQ}
\eeq
where 
$p_{Q_i}$ is an incoming $Q$ momentum and $k_{Q_i}$ an outgoing $Q$ momentum,
and here we have $N_i\!=\! 3$ and $N_f\!=\! 3$ since it is a $3\!\to\! 3$ process.
We write $p_n = q'_U + q_3 + q_4 = q_U + q_1 + q_2$,
and,
${\Mtn}^2 = E_n^2 - \pnr^{\!\!2}$ is the virtual mass (squared) of the $\Chi_n$.

We compute the matrix element mod-squared averaged over initial state spins and summed over final spins,
fold in phase-space and compute $\sigzP$ as in Eq.~(\ref{sig0BBb.EQ}).
We define the (dimensionless) $\sigzPh = M_\chi^2\, \sigzP$,
and from Eq.~(\ref{sig0BBb.EQ}) we write this in terms of the dimensionless $f_{\scriptscriptstyle \bar B B}^{nm}$ as  
\beq
\sigzPh = \left(\frac{M_\chi^8}{\Lambda^8}\right) \frac{1}{16} \sum_{n,m} f_{\scriptscriptstyle \bar B B}^{nm} \ . 
\eeq
We view the $\sigzP$ as an intermediate quantity,
and our final quantity of interest is really the thermally averaged scattering rate that we compute below. 

\begin{figure}
  \begin{center}
    \includegraphics[width=0.4\textwidth] {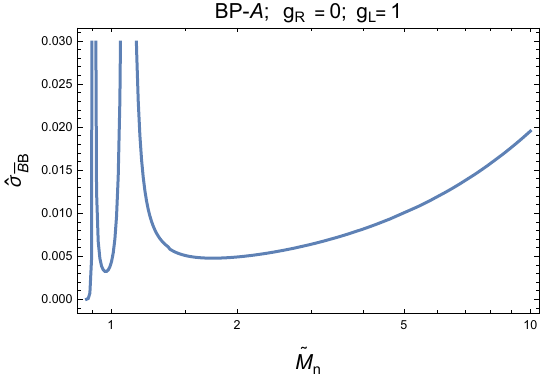}  \hspace*{0.1cm}
    \includegraphics[width=0.4\textwidth] {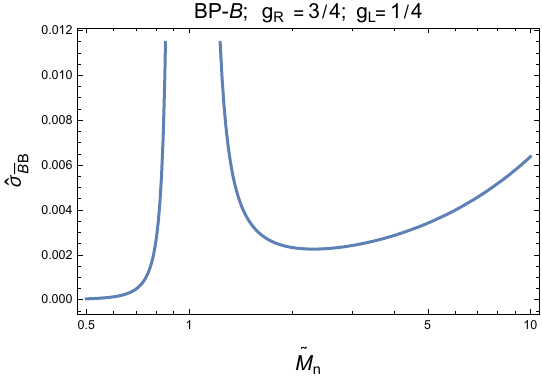}    
    \caption{In the CM frame, the (dimensionless) $\sigzPh(\Up\Dp\Dp \to \Up^c \Dp^c \Dp^c)$ for $\qUPr = 0$,
      for benchmark point BP-A (left) and BP-B (right). 
      \label{sigzPBBb.FIG}
    }
  \end{center}
\end{figure}

In this study, we include only the $SS$ contribution for $n\!\!=\!\!m$ as it is resonant and we expect it to be the most significant.
We show in Fig.~\ref{sigzPBBb.FIG}, for the benchmark points BP-A and BP-B,
the $\sigzPh$ in units of $(g M_\chi^2/(2 \Lambda^2))^4$, 
as a function of $\Mtn\! =\! \sqrt{p_n^2}$, in the CM frame with $\pnV\! =\! 0$,
for the sample value $\qUPr\! =\! 0$, taking $\hat\Gamma_{\!\Chi_n}\! =\! 10^{-6}$, and integrated over all the initial state angular variables. 
We see clearly the peaks due to the two poles at $\Mtn\! =\! M_{1,2}$ that are regulated by $\hat\Gamma_{\!\Chi_n}$.

For use in the collision term of the BE set in the early Universe,
we thermally average over the initial state 4-momenta $q'_U,q_3,q_4$ and define the
thermal average scattering rate as 
\beq
\GmzPTA \equiv \frac{1}{n_\chi^{(0)}} \int \prod_{i=1}^{N_i} \left(\frac{d^3 p_{Q_i}}{(2\pi)^3} f_{Q_i}^{(0)} \frac{1}{2E_{Q_i}} \right) \,  
\, \prod_{f=1}^{N_f} \left(\frac{d^3 k_{Q_f}}{(2\pi)^3} \frac{1}{2E_{Q_f}} \right) (2\pi)^4 \delta^{(4)}(\underset{\ssty i}{\Sigma} p_{Q_i} - \underset{\ssty f}{\Sigma} k_{Q_f})\ |\ampM'_0|^2 \ ,
\label{Gm0BBbTA.EQ}
\eeq
where
we have introduced a factor of $n_\chi^{(0)}$ in defining $\GmzPTA$ since we have an intermediate $\Chi$ exchanged (although offshell) in the process, 
and motivated by the (equilibrium) collision term having the $\GmzPTA n_\chi^{(0)}$ combination.
In terms of the thermally averaged
$\langle f_{\scriptscriptstyle \bar B B}^{nm}\rangle $
we obtain the thermally averaged dimensionless $UDD\to U^cD^cD^c$ scattering rate $\GmzPTAh\!=\!\GmzPTA/M_\chi$ as 
\beq
\GmzPTAh = \left(\frac{M_\chi^8}{\Lambda^8}\right) 
  \frac{1}{16} \sum_{n,m} \langle f_{\scriptscriptstyle \bar B B}^{nm}\rangle  \ . 
  \eeq

\begin{figure}[h]
  \begin{center}
    \includegraphics[width=0.4\textwidth] {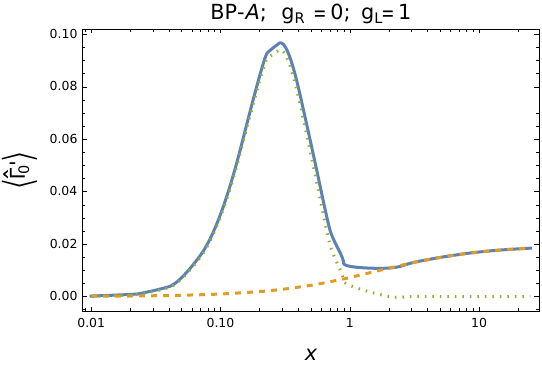}  \hspace*{0.1cm}
    \includegraphics[width=0.4\textwidth] {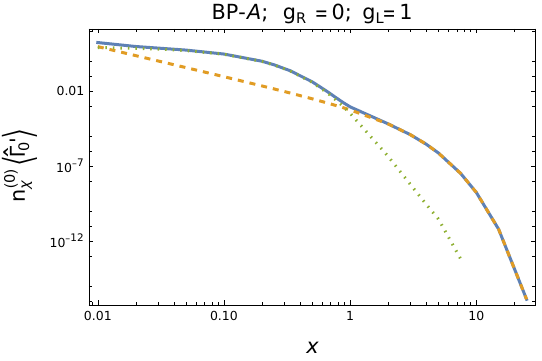}  \hspace*{0.1cm} 
    \includegraphics[width=0.4\textwidth] {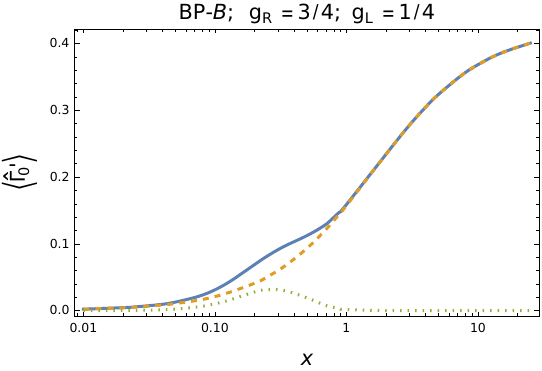}  \hspace*{0.1cm}
    \includegraphics[width=0.4\textwidth] {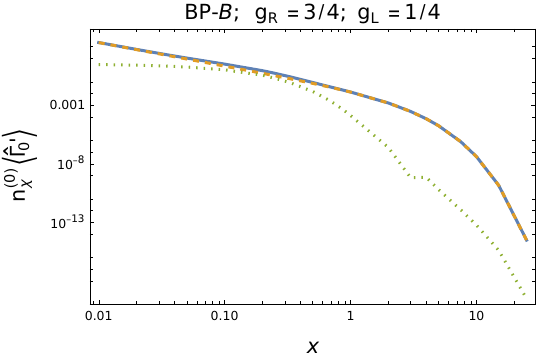}  \hspace*{0.1cm}     
    \caption{The (dimensionless) thermally averaged scattering rate for $\Up\Dp\Dp \to \Up^c \Dp^c \Dp^c$, for BP-A (top panel) and BP-B (bottom panel),
      showing the pole contribution (dashed), the nonpole (dotted), and the total rate (solid). 
      \label{UDD2UcDcDcGm.FIG}
    }
  \end{center}
\end{figure}

The very small $\hat\Gamma_{\!\Chi_n}$ makes the resonant $SS$ contribution very large for
$p_n^2\! \to\! M_n^2$ as seen in Fig.~(\ref{sigzPBBb.FIG}),
making the numerical evaluation unstable.
To get around this,
we obtain the integral due to the pole (narrow width approximation)
by replacing the $\Chi_n$ propagator by a $\delta(\Mtn-M_n)$ for the $SS$ contribution.
This allows us to analytically integrate over the $\Mtn$ variable, which sets $\Mtn\!=\! M_n$. 
The nonpole contribution we obtain by subtracting out the pole part from the integrand and numerically integrating
(cf. Appendix~\ref{dBBbInitMeas.SEC}). 
We compute the thermal average integrals numerically and
show the resulting (dimensionless) thermally averaged scattering rate $\GmzPTAh$ in Fig.~\ref{UDD2UcDcDcGm.FIG} for BP-A and BP-B,
in units of $(M_\chi/\Lambda)^8$.
We show the pole, nonpole, and total contributions. 
We also show the $\hat{n}_\chi^{(0)} \GmzPTAh$ in the same units, as these enter the rhs of the Boltzmann equations.  

We present some example values, setting $\hat\Gamma_n \!=\! 10^{-6}$,  
and computing the integrals numerically.
We find for $g_L\!=\! 1/4$, $g_R\!=\! 3/4$ we have $\sigzPh\!=\! 0.02$.
For $g_L\!=\! 1$, $g_R\!=\! 0$, $x\!=\! 1/4$, the (pole,~nonpole) values for BP-A
is $\GmzPTAh\!=\! (0.0023,0.09)$; 
while for BP-B we have for $g_L\!=\! 1/4$, $g_R\!=\! 3/4$, $x\!=\! 1/4$,
we have $\GmzPTAh\!=\! (0.05, 0.03)$.
The total thermally averaged scattering rate is the sum of the pole and nonpole pieces.
We can thus write with the couplings scaling,
the thermally averaged rate to be $\GmzPTA\!\approx \! 0.1\, g^4 (M_\chi^9/\Lambda^8)$, for $x\!=\! 1/4$, for both BP-A and BP-B.

\section{The Boltzmann Equations}
\label{nChinBBE.SEC}

Starting from a zero baryon number in the early Universe, we wish to compute the net baryon number generated at late time
and compare with the observed BAU.
We do this by tracking the net baryon number $n_B(t)$ as a function of cosmic time $t$ in the expanding Universe
using the Boltzmann equations (BEs) (see, for example, Refs.~\cite{Kolb:1990vq,Buchmuller:2000as,Grojean:2018fus},
and also the discussion in Refs.~\cite{Gopalakrishna:2022hwk,Gopalakrishna:2024qxk}).
$CPT$ and unitarity bring about relations between forward and inverse reaction matrix elements.
We discuss these relations in our new physics theory in Appendix~\ref{CPTUniBE.SEC}. 
We must take care to handle correctly the contributions of real onshell intermediate states (RIS) in scattering channels and related inverse channels. 
Doing so brings about a sign change in the $\nXz$ term relative to the $\nX$ term as pointed out in Ref.~\cite{Kolb:1979qa}.
We discuss these intricate aspects brought about by $CPT$ and unitarity considerations in Appendix~\ref{CPTUniBE.SEC}. 

Before we give the full set of BEs that follow from these considerations,
we list next the relevant decay and scattering thermally averaged rates that appear in the BE. 
We denote the thermally averaged decay rate for the process $\Chi\to UDD$ as $\GmTA$,
and that for the conjugate process $\Chi\to U^cD^cD^c$ as $\GmBTA$.
For the scattering processes, 
the relevant scattering cross sections are    
$\sigma_{\Chi\bar{Q}} \equiv \sigma(\Chi Q^c \to Q Q)$, 
$\sigma_{\Chi Q} \equiv \sigma(\Chi Q \to Q^c Q^c)$, 
with their inverse scattering channel cross sections being
$\sigma_{Q Q} \equiv \sigma(Q Q \to \Chi Q^c)$,
$\sigma_{\bar{Q}\bar{Q}} \equiv \sigma(Q^c Q^c \to \Chi Q)$ respectively, 
and the difference in the thermally averaged scattering rates is
$\DGmSigTA = \DSigvTA n_Q^{(0)}$ with $\Delta\sigma = \sigma_{\Chi\bar{Q}} - \sigma_{\Chi Q}$.
The thermally averaged LO scattering rate of the $\Delta B\!=\! 2$ process $UDD\to U^cD^cD^c$ is $\GmzPTA$, 
and that of the LO $2\to 4$ processes is $\GmzPSigUDTA$, $\GmzPSigDDTA$, $\GmzPSigUUTA$
for the $UD\to U^cD^cD^cD^c$, $DD\to U^cD^cD^cU^c$, $UU\to D^cD^cD^cD^c$ channels respectively.
We ignore the $4\to 2$ scattering processes in our analysis as we expect their contributions to be small. 
We write for the decay rates $\GmTA = \GmzTA + \DGmTA/2$ and $\GmBTA = \GmzTA - \DGmTA/2$, 
and similarly for the scattering rates 
$\GmSigXQBTA = \GmzSigXQTA + \DGmSigXQTA/2$ and $\GmSigXQTA = \GmzSigXQTA - \DGmSigXQTA/2$. 
We recall that the LO rates $\GmzTA$, $\GmzSigTA$, $\GmzPSigTA$ are the same for the process and conjugate process. 

The BE for $\nX$ is
\bea
\frac{d}{dt}n_\Chi + 3 H n_\Chi &=& - 2 \GmzTA \left[n_\Chi - n_\Chi^{(0)} \cosh{\frac{(\mu_U+2\mu_D)}{T}} \right] \nonumber \\ 
&&\hspace*{-1.5cm} - 2 \GmSigXDBTA \left[n_\Chi \cosh{\frac{\mu_D}{T}} - n_\Chi^{(0)} \cosh{\frac{(\mu_U+\mu_D)}{T}}\right]
- 2 \GmSigXUBTA \left[n_\Chi \cosh{\frac{\mu_U}{T}} - n_\Chi^{(0)} \cosh{\frac{2\mu_D}{T}}\right]  \nonumber \\
&\approx& - 2 \left(\GmzTA + \GmSigXDBTA + \GmSigXUBTA \right) (n_\Chi - n_\Chi^{(0)}) \ , \label{BEchiXUD.EQ}
\eea
where the $\mu_U$, $\mu_D$ are the chemical potentials of the $U$ and $D$, 
and for the last line we have taken $\cosh(...) \approx 1$ since $\mu_{U,D}/T \ll 1$, given that the baryon asymmetry is tiny. 
The BE for the $\Up,\Dp$ are 
\bea
\frac{d}{dt} n_U + 3 H n_U \!\!\!\!&=&\!\!\!\! \GmTA\, n_\Chi - \eU \eD^2 n_\Chi^{(0)} \GmBTA
+ \left(\eDB \GmSigXDBTA - \eU \GmSigXUTA \right) \nX - \left(\eU\eD \GmSigXDTA - \eDB^2 \GmSigXUBTA \right) \nXz \nonumber \\
&& \hspace*{-3cm} - \left( \eU\eD^2 \GmzPTA + \eD\eU \GmzPSigUDTA + \eU^2 \GmzPSigUUTA \right) \nXz
   + \left( \eUB\eDB^2 \GmzPTA + \eDB\eUB \GmzPSigUDTA + \eDB^2 \GmzPSigDDTA \right) \nXz, \label{UBE.EQ} \\
\frac{d}{dt} n_D + 3 H n_D \!\!\!\!&=&\!\!\!\! \GmTA\, n_\Chi - \eU \eD^2 n_\Chi^{(0)} \GmBTA
- \left(\eD \GmSigXDTA - \eUB \GmSigXUBTA - \eDB \GmSigXDBTA \right) \nX \nonumber \\
&& \hspace*{-3cm} + \left(\eUB\eDB \GmSigXDBTA - \eD^2 \GmSigXUTA - \eU\eD \GmSigXDTA \right) \nXz 
- \left( \eU\eD^2 \GmzPTA + \eD\eU \GmzPSigUDTA + \eD^2 \GmzPSigDDTA \right) \nXz \nonumber \\
&& \hspace*{-3cm} + \left( \eUB\eDB^2 \GmzPTA + \eDB\eUB \GmzPSigUDTA + \eDB^2 \GmzPSigDDTA + \eUB^2 \GmzPSigUUTA \right) \nXz,
\label{DBE.EQ}
\eea
where $\eU=e^{\mu_U/T}, \eD=e^{\mu_D/T}, \eUB=e^{-\mu_U/T}, \eDB=e^{-\mu_D/T}$.
The inverse channel thermal averages are written in terms of $n_\chi^{(0)}$
by using energy conservation in the statistical distribution functions~\cite{Kolb:1990vq}.
We obtain the BE for $\nUB, \nDB$ by taking $\eQ \lrarr \eQB$, $\GmTA \lrarr \GmBTA$, and $\GmSigXQTA \lrarr \GmSigXQBTA$, for $Q=\{\Up,\Dp\}$.
The thermally averaged scattering rate is
$\GmSigXQTA = \langle\sigma_{\Chi Q} v\rangle\, n_Q^{(0)}$ for $Q\!=\!\{U,D\}$,
and for its conjugate process is 
$\GmSigXQBTA = \langle\sigma_{\Chi \bar Q} v\rangle\, n_Q^{(0)}$. 
We obtain the BE for $\nDU \!=\! \nU \!-\! \nUB$ and $\nDD \!=\! \nD \!-\! \nDB$ by taking the difference of the corresponding BEs.

To obtain the baryon number density $n_B$ from the $\nDU,\nDD$ 
let us first start by considering the Dirac limit, i.e. with the Majorana masses set to zero. 
In this limit, baryon number is a good symmetry and coincides with $\chi$ number
since $B(\chi)\!=\!+1$ and $B(\bar\chi)\!=\!-1$ for the Dirac pair $\chi,\bar\chi$,
and we recall that $B(\Up)\!=\!B(\Dp)\!=\!1/3$.
We write
$N_{\Delta\Up}\!\equiv\! (N_{\Up}-N_{\bar \Up})$,
and $N_{\Delta\Dp}\!\equiv\! (N_{\Dp}-N_{\bar \Dp})$. 
The efficient QED processes $Q\bar Q \lrarr \gamma,\! \gamma\gamma$ etc. keep $Q\!=\!\{\Up,\Dp\}$ in thermal equilibrium,
implying a relation between the chemical potentials~\cite{Landau:1980mil} $\mu_Q + \mu_{\bar Q} = \mu_\gamma, 2\mu_\gamma$. 
We have $\mu_\gamma\!=\! 0$ since it is an EM neutral real gauge boson, 
which then implies $\mu_{\bar Q}\!=\! - \mu_Q$, consistent with EM charge conservation. 
Using these relations, we can write the relevant terms in the thermodynamic potential~\cite{Landau:1980mil} as      
$\Omega(T,V,\mu_i) \supset \mu_\chi N_{\chi} + \mu_{\bar\chi} N_{\bar\chi} + \mu_U N_{\Delta U} + 2 \mu_D N_{\Delta D}$, 
where we write $2 N_{\Delta D}$ for $N_{\Delta D_a} \!+\! N_{\Delta D_b}$, 
with the $a,b$, for instance, being different color indices.
When the $VV$ interaction is also in equilibrium, it ensures $\mu_\chi-(\mu_U+2\mu_D)\!=\!0$ 
and $\mu_{\bar \chi}\!=\!-\mu_\chi$.
(For such relations in other contexts, see for example, Refs.~\cite{Harvey:1990qw,Fridell:2021gag}.)
We note that changes in $N_{\Delta\Chi,\Delta U,\Delta D}$ are only due to the VV interaction which conserves
$N_{\Delta\chi}+(N_{\Delta U} + 2 N_{\Delta D})/3$. 
This conservation is enforced in the thermodynamics by including a chemical potential $\mu_B$
and writing $\Omega \supset \mu_B (N_{\Delta\Chi} + N_B)$,
where we have defined $N_B\! \equiv\! (N_{\Delta U} + 2 N_{\Delta D})/3$,
and setting $-(\partial\Omega/\partial\mu_B)_{T,V} = 0$. 
Matching the forms of $\Omega$ we have $\mu_\chi\!=\! \mu_B$, $\mu_Q \equiv \mu_U\!=\!\mu_D\!=\! \mu_B/3$. 
Returning to the Majorana mass nonzero case, $\chi$ number is not a good quantum number, and consequently $\mu_\chi\! =\! 0$ for the Majorana $\Chi$,
thus leaving $\Omega \supset \mu_B N_B$.
$N_B$ is nothing but baryon number, 
and the baryon number density $n_B\!=\!N_B/V$ is thus given as $\nB\!=\!(\nDU + 2\nDD)/3$.
If the VV interaction is in equilibrium, $\mu_\chi\!=\!0$ implies $\mu_B\! =\! 0$,
and the forward and inverse rates being equal keeps $n_B\!=\!0$ if we start with that initial condition. 
After the VV interaction decouples, there is no constraint on $N_B$,  
and if the forward and inverse rates are different in the microscopic physics,
in addition to $C$ and $CP$ violation as required by the Sakharov conditions, 
$n_B$ could build up to nonzero values. 
In the earlier sections we have shown that in our theory there is indeed such a rate asymmetry
enabling the possibility that $n_B$ could become nonzero. 
%
%
We track the build up of $n_B$ in the early Universe through its BE; 
correctly taking into account the change in sign due to the RIS contributions mentioned above,
from Eqs.~(\ref{UBE.EQ}) and (\ref{DBE.EQ}) we obtain the BE for $n_B$ to be 
\bea
\frac{d}{dt} n_B + 3 H n_B &=& \DGmTA \left( n_\Chi - n_\Chi^{(0)} \cosh{\frac{(\mu_U+2\mu_D)}{T}} \right)
- 2 \left(\GmzTA + 2\GmzPTA\right)\, n_\Chi^{(0)} \sinh{\frac{(\mu_U+2\mu_D)}{T}}
 \nonumber \\
  && \hspace*{-3cm} + \frac{5}{3} \DGmSigXDTA \left( n_\Chi \cosh{\frac{\mu_D}{T}} - n_\Chi^{(0)} \cosh{\frac{(\mu_U+\mu_D)}{T}} \right)
 - 2\!\times\!\frac{5}{3} \GmzSigXDTA \left(n_\Chi \sinh{\frac{\mu_D}{T}} + n_\Chi^{(0)} \sinh{\frac{(\mu_U+\mu_D)}{T}} \right)
  \nonumber \\
  && \hspace*{-3cm} + \DGmSigXUTA \left( n_\Chi \cosh{\frac{\mu_U}{T}} - n_\Chi^{(0)} \cosh{\frac{2\mu_D}{T}} \right)
   - 2 \GmzSigXUTA \left(n_\Chi \sinh{\frac{\mu_U}{T}} + n_\Chi^{(0)} \sinh{\frac{2\mu_D}{T}} \right)
   \nonumber \\
   && \hspace*{-3cm} -4 \GmzPSigUDTA\, n_\Chi^{(0)}\, \sinh{\frac{(\mu_U + \mu_D)}{T}} - 4 \GmzPSigDDTA\, n_\Chi^{(0)}\, \sinh{\frac{2\mu_D}{T}}
   -2 \GmzPSigUUTA\, n_\Chi^{(0)}\, \sinh{\frac{2\mu_U}{T}} \ . 
\label{BEnbXUD.EQ}
\eea

We can keep track of the nontrivial changes in number density over and above that due to the Hubble expansion
by changing variables from $n$ to a comoving variable $Y\!=\!n/s$, where $s$ is the entropy density
(see, for example, Refs.~\cite{Kolb:1990vq}, and also Ref.~\cite{Gopalakrishna:2024qxk} for a brief summary).
In this way, for each of the $\nX,\nXz,n_B$, we define a corresponding $Y_\Chi, Y_\Chi^{(0)}, Y_B$.
Also, since the observed BAU $\ll 1$, we can keep terms to first order in $Y_B$ or equivalently in $\muhh\!\equiv\! \mu/T$, 
taking $\cosh{\muhh}\! \approx\! 1$, $\sinh{\muhh}\! \approx\! \muhh$ for $\mu_U,\mu_D,\mu_B$. 
If $n_U^{(0)} \!\approx\! n_D^{(0)} \!\approx\! n_Q^{(0)}$ and $\mu_U \!\approx\! \mu_D \!\equiv\! \mu_Q$,
using $n=n^{(0)}e^{\mu/T}$ and $\mu_B \!=\! 3\mu_Q$ 
we obtain $n_B \!=\! 2 n_Q^{(0)} \sinh{(\hat{\hat\mu}_B/3)}$, which implies $\muhhB \!\approx\! (3/2) Y_B/Y_Q^{(0)}$ to first order in $\muhhB$,
where~\cite{Gopalakrishna:2024qxk}
$Y_Q^{(0)} = 45/(4\pi^4)\, (g_Q/g_{*S})\, (M_Q/T)^2 K_2(M_Q/T)$. 
Using these relations in Eqs.~(\ref{BEchiXUD.EQ}) and (\ref{BEnbXUD.EQ}) we obtain 
\bea
\frac{d}{dt}Y_\Chi
&\approx& - 2 \left(\GmzTA + \GmSigXDBTA + \GmSigXUBTA \right) (Y_\Chi - Y_\Chi^{(0)}) \ , \\
   \frac{d}{dt} Y_B &\approx& \left(\DGmTA + \frac{5}{3} \DGmSigXDTA + \DGmSigXUTA \right) \left( Y_\Chi - Y_\Chi^{(0)} \right)
 \nonumber \\ && \hspace*{-1cm}
   - 2 \left(\GmzTA + 2\GmzPTA\right)\, Y_\Chi^{(0)} \frac{3}{2} \frac{Y_B}{Y_Q^{(0)}} 
   - 2\!\times\! \left( \frac{5}{3} \GmzSigXDTA + \GmzSigXUTA \right) \left(\frac{1}{2} Y_\Chi + Y_\Chi^{(0)} \right) \frac{Y_B}{Y_Q^{(0)}}
 \nonumber \\ && \hspace*{-1cm}
   -4 \left( \GmzPSigUDTA + \GmzPSigDDTA + \frac{1}{2} \GmzPSigUUTA \right)\, Y_\Chi^{(0)}\, \frac{Y_B}{Y_Q^{(0)}} \ .
\label{BEYbYchi.EQ}
\eea
%

We see that the rhs of the BE (with the collision terms) involve the thermally averaged decay and scattering rates, which we
have spent effort in computing in this work.
Taking these rates as inputs, we solve the BE in Ref.~\cite{Gopalakrishna:2024qxk}.

\section{Conclusions}
\label{Concl.SEC}

In this work, we compute the baryon asymmetry generated in the decay and scattering processes 
involving the electromagnetically charge-neutral fermion $\chi$ and quark-like fermions $\Up,\Dp$,
in the effective theory developed in Ref.~\cite{Gopalakrishna:2022hwk}. 
In the theory, the Dirac fermion $\chi$ with Dirac mass $M_\chi$ is split into a pair of Majorana fermions $\Chi_n$
due to the presence of a baryon number violating $\chi$ Majorana mass. 
The $\chi,\Up,\Dp$ interaction is of the vector-vector (VV) form 
$(1/\Lambda^2)\, (\bar\chi\Gamma^\mu \Up)\,(\bar{\Dp^c}\Gamma_\mu \Dp)$. 
Being a dimension-6 operator, it is suppressed by the cutoff scale $\Lambda$ as shown.

For two representative choices of parameters, benchmark points BP-A and BP-B,
we compute the unitary matrix that diagonalizes the $\chi$ mass matrix and determine the
VV effective vertex for the mass eigenstates $\Chi_n$.
Both the benchmark points we choose are in the pseudo-Dirac limit in which the mass splitting between the $\Chi_n$ masses,
namely, $M_2 - M_1 \ll M_\chi$.
We show the dependence of the $\Chi_n$ mass eigenvalues $M_n$ on the Majorana mass and phase. 

We identify loop amplitudes for the $\Chi_n$ decay and scattering processes
such that the interference term between them and the tree-level amplitude is different for the process
and its conjugate process, leading to a baryon asymmetry. 
The phases in the Majorana mass, and in the VV interaction coupling act like weak phases
(opposite for process and conjugate process),
and the intermediate state $\Dp,\Up$ in a loop going on-shell produces a phase of $\pi/2$ that acts like a strong phase
(same for process and conjugate process).
These phases lead to $C$ and $CP$ violation, and along with the baryon number violation due to the Majorana mass, 
produces a baryon asymmetry. 

We refer to the loop processes that contain only the above VV effective operator as the single operator contributions. 
We show that if the $(1/\Lambda^2)\, (\bar\Chi\Gamma^\mu \Up)\,(\bar\Up\Gamma_\mu\Chi)$ effective operator
is also present in the effective theory, and if the kinematics allows it, 
additional loop contributions can be identified that contain both of these effective vertices,
that also produces a baryon asymmetry.
We refer to such contributions as multiple operator contributions.
In BP-A the baryon asymmetry is from the single operator contributions,
while in BP-B it is due to the multiple operator contributions. 

We write the Feynman diagrams for decay and scattering process,
and in the mass basis, compute the matrix elements, integrate over the loop momenta,
fold in the phase space, and compute the resulting baryon asymmetry.
The integrals we encounter are multi-scale and multi-dimensional,
and we expend considerable effort to make them computationally tractable.
We present our baryon asymmetry results in a general manner without picking a specific scale of new physics, i.e. $\Lambda$ or $M_\chi$,
but give scaling factors involving the $\Lambda$ and $r_{M\Lambda}\!\!=\!\! M_\chi/\Lambda$ that can be used to compute the actual numbers in any specific scenario
with its preference for the scale. 

We compute numerically the decay rate asymmetry between $\Chi_n \to \Up\Dp\Dp$ and $\Chi_n \to \bar{U}\bar{D}\bar{D}$,
for the benchmark points and for sample coupling choices.
For BP-A, we find the tree-level decay width
$\Gamma_{\Chi} \sim 10^{-6}\, \tilde{g}^2 g_{L,R}^2\, M_\chi^5/\Lambda^4$,
and from the single operator contributions, we find
the decay rate difference $\Delta\hat\Gamma_{01} \sim -10^{-12}\, \tilde{g}^4 g_{L,R}^4\, M_\chi^9/\Lambda^{8}$,
which yields the baryon asymmetry $\AsymB \sim 10^{-6}\, \tilde{g}^2 g_{L,R}^2\, M_\chi^4/\Lambda^4$.
For BP-B, we find 
$\Gamma_{\Chi} \sim 10^{-5}\, \tilde{g}^2 g_{L,R}^2\, M_\chi^5/\Lambda^4$,
and from the multiple operator contributions, we find 
$\Delta\hat\Gamma_{01} \sim -10^{-13}\, \tilde{g}^2 g_{L,R}^4\, M_\chi^7/\Lambda^{6}$,
and $\AsymBP \sim 10^{-8}\, g_{L,R}^2\, M_\chi^2/\Lambda^2$.
We show the dependence of the baryon asymmetry on the theory parameters in
Figs.~\ref{BPA-GmAsym.FIG},~\ref{GmAsymMRscan.FIG}, and ~\ref{BPB-Asym.FIG}. 

We use crossing relations on the decay matrix elements computed above to obtain the scattering matrix elements.
We then compute the thermally averaged scattering cross section difference between 
the process $\Chi_n \bar{Q} \to QQ$ and its conjugate process $\Chi_n Q \to \bar{Q}\bar{Q}$ (with $Q=\{\Up,\Dp\}$)
as a function of $x=M_\chi/T$, with $T$ being the temperature of the bath in which the $\Chi_n$ are interacting.
In the thermal spectrum, there is a probability for the $\Chi$ to have $p_\chi > \Lambda$,
and therefore UV physics effects can be sampled.
We capture this effect by having a propagating UV state $\xi$ (as in Ref.~\cite{Gopalakrishna:2022hwk})  
resolve the VV effective interaction, and compute the baryon asymmetry after including this correction.
We find that the scattering cross section asymmetry $\langle \Delta\sigma_{01} v\rangle $
is generated predominantly in the $0.01 \lesssim x \lesssim 10$ range (cf. Figs.~\ref{ABsigh.FIG} and~\ref{sigvABPx.FIG}),
indicating that our effective theory computation still gives a fairly good estimate of the size of the asymmetry generated in
our framework.
This also indicates to us that for a truly accurate scattering cross section result,
one better work with a specific UV completion. 

From our numerical computation,
for BP-A, for $x =1/2$,
from the single operator contribution,
we find
the thermally averaged LO cross section 
$\langle  \sigma_0 v \rangle  \sim 10\, g_{L,R}^2 M_\chi^2/\Lambda^4$,
and from the single operator contributions,
we find the thermally averaged cross section difference between the process and conjugate process 
$\langle \Delta\hat\sigma_{01}\, v\rangle  \sim -10^{-4}\, g_{L,R}^4 M_\chi^6/\Lambda^8$,
which yields the cross section asymmetry
$\AsymBsigh \sim 10^{-5}\, g_{L,R}^2 M_\chi^4/\Lambda^4$.
For BP-B, $x =1/2$,
from the multiple operator contribution,
we find
$\langle  \sigma_0 v \rangle  \sim 1.5 g_{L,R}^2 M_\chi^2/\Lambda^4$,
$\langle \Delta\hat\sigma_{01}\, v\rangle  \sim -5\times 10^{-4}\, g_{L,R}^4 M_\chi^4/\Lambda^6$,
and,
$\AsymBsigh \sim 4 \times 10^{-4}\, g_{L,R}^2 M_\chi^2/\Lambda^2$.
We vary the effective theory parameters about the benchmark points and show their effects on the baryon asymmetry generated.
We compute the tree-level thermally averaged scattering rate for the $\Delta B = -2$ three-to-three process
$\Up\Dp\Dp \to \Up^c \Dp^c \Dp^c$.
For instance, for $x\!=\! 1/4$, we find $\GmzPTA\!\sim \! 0.1\, g^4 (M_\chi^9/\Lambda^8)$ for both BP-A and BP-B.

We use $CPT$ and unitarity to relate the forward and inverse decay and scattering matrix elements.
Armed with these relations, we develop the Boltzmann equations for the $\Chi$ and baryon number number densities
as a function of cosmic time. 
The rates we compute in this work enter as collision term inputs on the rhs of the Boltzmann equations.  
To determine the preferred mass scale $M_\chi$ that leads to the observed BAU,
we numerically solve the Boltzmann equations
in Ref.~\cite{Gopalakrishna:2024qxk}. 
Knowing the scale would help us determine the prospects of observing such new physics
at future colliders, 
or in precision experiments such as the neutron-antineutron oscillation experiments
(see Refs.~\cite{Gopalakrishna:2022hwk,Gopalakrishna:2024qxk} and references therein).


\appendix




\section{Renormalization of the $\Chi_n \bar\Chi_m$ 2-point function}
\label{Chinm2ptFcn.SEC} 

For the Majorana fermions $\Chi_n$, we can write the effective Lagrangian in momentum space as
${\cal L}_2 \supset (1/2) \bar\Chi_m (\pslash - M_{mn}) \Chi_n$,
where we show only the bilinear part relevant to us here.
A hermitian $M$ ensures that the eigenvalues are real and the system is nondissipative in the $\{\Chi_1, \Chi_2\}$ sector.  
However, if intermediate particles can go onshell, i.e. the 2-point function can be cut, a dissipative piece arises, which we can encode by adding a non-hermitian piece
$i (\pslash + M)^{-1} M \Gamma/2$ to the ${\cal L}_2$ above.
We thus have a part that is singular (in the limit of zero width)
when intermediate particles in the cut are onshell giving a discontinuity in the imaginary piece,
and a regular principal value (PV) part. 
One renormalization and subtraction scheme we can adopt is an {\em onshell scheme} in which we pick counterterms
for the $\Chi_n \bar\Chi_m$ 2-point function to cancel the loop corrections so that the tree-level masses are the renormalized masses.
We take the counterterm Lagrangian as
${\cal L}_2^{\rm CT} \supset (1/2) \bar\Chi_m\, \delta\Sigma_{nm}\, {\Chi_n}$, with $\delta\Sigma$ hermitian.  
The nonhermitian piece due to the cut cannot be cancelled by the hermitian counterterm, and so survives without any subtraction.
For $n\!\!=\!\!m$, the non-hermitian piece implies $\Chi_n$ decay.
Related aspects in the onshell scheme are discussed, albeit in different contexts, for example, in Refs.~\cite{Denner:1991kt,Almasy:2009kn,Drauksas:2021xwz}. 

\begin{figure}
\begin{minipage}{6in}
  \centering
  \[
  \begin{array}{c}
    \raisebox{0\height}{\includegraphics[width=0.25\textwidth]{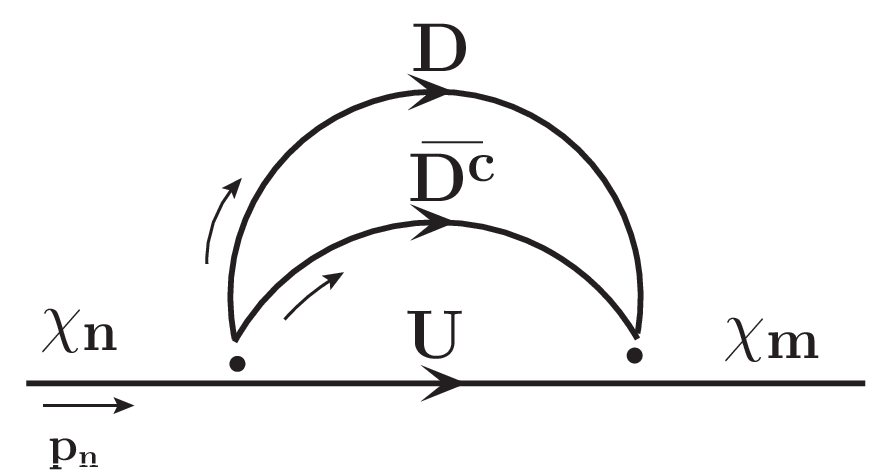}}
    \hspace*{1cm}
    \raisebox{0\height}{\includegraphics[width=0.25\textwidth]{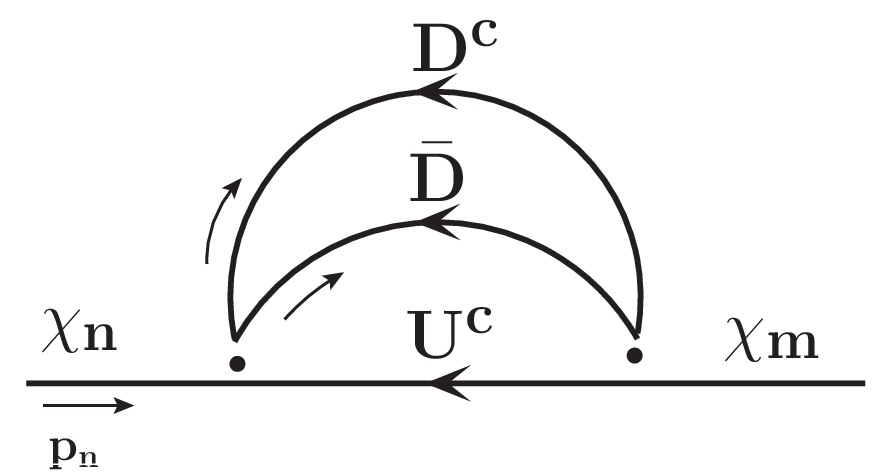}}
  \end{array}  
  \hspace*{-0.25cm} 
 \] \\
  \vspace*{-1.2cm}
  \beq \hspace*{0.5cm} \Omega_{nm}(p_n) \hspace*{3.75cm} \Omega^c_{nm}(p_n) \nonumber \eeq
  \vspace*{-0.75cm}
\end{minipage}
\caption{
  The two different loop contributions to the $\bar\Chi_m \Chi_n$ 2-point function $\Sigma_{nm}(p_n)$.
}
\label{Chinm2Pt.FIG}
\end{figure}

Due to the Majorana nature of the $\Chi_n$, the full 2-point function includes two different loop contributions, namely,
one from fields propagating in the loop, and another with their conjugate fields,
which we show diagrammatically in Fig.~\ref{Chinm2Pt.FIG}.
The full 2-point function is the sum of these, i.e. $\Sigma_{nm}(p) = \Omega_{nm}(p) + \Omega^c_{nm}(p)$.

We compute the baryon asymmetry in Sec.~\ref{oneOpAsymb.SEC} with a cut in the $\Chi_n \bar\Chi_m$ 2-point function
in the B$_1$ diagram in Fig.~\ref{chi2DDU-NLO.FIG},
or equivalently the B diagram contribution in Fig.~\ref{chi2DDU-LONLOCut.FIG} with $p'_n = p_n$. 
The cut is over the intermediate particles in $\Omega^c_{nm}$ for the process and $\Omega_{nm}$ for the conjugate process.
As mentioned, there is no counterterm for the discontinuity,
and we take the $\im(\Omega_{nm})$ and $\im(\Omega^c_{nm})$ without any subtraction. 
The $\im(\Omega_{nn}),\im(\Omega^c_{nn})$ (for $n\!\!=\!\!m$) with both legs on-shell yields the LO decay width $\Gamma_0,\Gamma_0^c$ respectively. 
The $\im(\Omega_{nm}),\im(\Omega^c_{nm})$ (for $n\!\!\neq\!\!m$) has the $\Chi_m$ offshell at $p_n^2 = M_n^2$
and is thus an intermediate state in some process, for instance in diagram B in our case. 
Although it does not contribute to the asymmetry, 
for completeness, we have computed the renormalization conditions and the subtraction for this
(cf. Appendix~\ref{Chinm2ptFcnDet.SEC}).

  As an aside, to glean further insight,
  we contrast our two Majorana fermions $\Chi_n$ case with the two Dirac fermions $\Psi_n$ case,
  with each containing two (different) Weyl fields $\psi_{n\alpha}$ and $\psi^c_{n\alpha}$.
  The 2-point functions $\bar\Psi_n \Psi_m$ now have bilinears of the form $\psi^c_n \psi_m + h.c.$ 
  In the Dirac case also, an identical argument can be used to cancel the real parts of $\Sigma_{11}, \Sigma_{22}$ at
  the on-shell subtraction scales $p_{1_0}, p_{2_0}$ and
  also $\Sigma_{12}$ at $p_{1_0}$.
  Now, the $\psi^c_1 \psi_2$ and $\psi^c_2 \psi_1$ bilinears are independent, allowing an independent counterterm to cancel the
  real parts of $\Sigma_{21}$ at $p_{2_0}$ also.
  Thus, in the Dirac case,
  all the $\PV((\Sigma_{mn}))$ can be canceled at the respective $p_{n_0}$.
  In contrast, in the Majorana case the $\Sigma_{12}$ and $\Sigma_{21}$ are not independent
  and therefore cannot both be canceled.

\section{Thermal Averaging Procedure}
\label{ThAvgProc.SEC}

Our interest ultimately is in the baryon asymmetry generated due to $\Chi_n$
decay and scattering taking place in a thermal bath of the hot early Universe at temperature $T$.
For a species in thermal equilibrium, the thermal average $\langle \Theta\rangle $ of any quantity $\Theta(\{p_i\},\{q_j\})$
with initial state momenta $\{p_i\}$ and final state momenta $\{q_j\}$ is given by 
\beq
\langle \Theta\rangle (\{q_j\}) =  \int \Pi_i \left(\frac{d^3 p_i}{(2\pi)^3} f(E_i) \right) \, \Theta(\{p_i\},~\{q_j\}) \ ,
\label{TADefn.EQ}
\eeq
with the equilibrium Boltzmann distribution function $f(E) = e^{(\mu-E)/T} \equiv e^{\mu/T} f^{(0)}(E)$ where
$\mu$ is the chemical potential, $f^{(0)}(E) = e^{-E/T}$, and, $E=\sqrt{{\bf p}^2 + M^2}$. 
For a species with $g$ degrees of freedom, we have
the number density $n = g\int d^3p/(2\pi)^3 f(E)$,
and we can write $n=n^{(0)} e^{\mu/T}$ with $n^{(0)}$ being the number density with $\mu=0$,
being given by $n^{(0)} = g\int d^3p/(2\pi)^3 f^{(0)}(E)$,
which we find to be
\beq
n^{(0)} = \frac{g M^2 T}{2\pi^2}\, K_2\left(\frac{M}{T}\right) \ , 
\eeq
where $K_2$ is the modified Bessel function of the second kind.
We note the expansions
$K_2(x) = 2/x^2 + {\cal O}(1)$ for $x\ll 1$, and $K_2(x) = \sqrt{\pi/(2x)} e^{-x} + {\cal O}(1/x)$ for $x\gg 1$,
which for $x\!\!=\!\! M/T$ give the relativistic and non-relativistic limits respectively as usual (see for example Ref.\cite{Dodelson:2003ft}).

We focus on thermal averages in scattering processes here since it is more involved given the two initial state particles. 
We take the initial state kinematical variables in the lab-frame, which in our early Universe setting can be identified with the standard cosmological frame.   
For the spin averaged scattering matrix element we expect a trivial dependence on $(c_n,\phi_n,\phi_i)$,
and perform those integrations, leaving $d^3 p_n d^3 k_i = 8\pi^2 d\pnr \pnr^{\!\!2} dc_i d\kir \kir^2$
over their usual domains $\pnr,\kir \in (0,\infty);\ c_i\in (-1,1)$.  

We define $x\equiv M_\chi/T$, and the scaled variables $\hat{M} = M/M_\chi$, $\hat{p}_r = p_r/M_\chi$, $\hat{E} = E/M_\chi$.
We define a distribution function normalized as a probability distribution function, namely, 
\beq
f_P^{(0)}(\hat{p}_r) = \frac{1}{n^{(0)}} \frac{(4 \pi)}{(2 \pi)^3} \hat{p}_r^2 f^{(0)}(\hat{p}_r)
= \frac{2 \pi^2 \hat{p}_r^2}{\hat{M}^2 K_2(\hat{M} x)}\, x e^{-\hat{E} x} \ ,
\label{feqProbpr.EQ}
\eeq
normalized as $\int_0^\infty d\hat{p}_r f^{(0)}_P(\hat{p}_r) = 1$.
We can equivalently write the energy probability distribution function
\beq
f_P^{(0)}(\hat{E}) = \frac{1}{n^{(0)}} \frac{(4 \pi)}{(2 \pi)^3} \hat{E} \sqrt{\hat{E}^2-\hat{M}^2} f^{(0)}(\hat{E})
= \frac{2 \pi^2 \hat{E} \sqrt{\hat{E}^2-\hat{M}^2}}{\hat{M}^2 K_2(\hat{M} x)}\, x e^{-\hat{E} x} \ ,
\label{feqProbE.EQ}
\eeq
which is normalized as $\int_{\hat{M}}^\infty d\hat{E} f^{(0)}_P(\hat{E}) = 1$.
Using the $f^{(0)}(E)/{n^{(0)}}$ that is normalized to have unit integral, we define another thermal average 
\beq
\langle \Theta\rangle _{\sssty\!P}(\{q_j\}) =  \int \Pi_i \left(\frac{d^3 p_i}{(2\pi)^3} \frac{f(E_i)}{n^{(0)}_i} \right) \, \Theta(\{p_i\},~\{q_j\}) \ ,
\label{TADefn.EQ}
\eeq
to be with the distribution function normalized as a probability distribution function.

The rest-frame decay width (see Eq.~(\ref{Gam0-3PSMSq.EQ})) can be written as
$\Gamma_0 = 1/(2M)\, \int d\Pi_3 |{\cal A}|^2$.
Since the decay matrix element ${\cal A}$, and the final-state phase-space element $d\Pi_3$ 
are individually Lorentz invariant,
if $\Chi$ has momentum $p=(E,{\bf p})$,
the generalization of the above rest-frame decay-width to a boosted frame is to replace the first factor as $1/(2M) \to 1/(2E)$.
In other words we have $\Gamma(E)/\Gamma_0 = M/E$, 
which clearly has the desirable physical property that the particle life-time $\tau = 1/\Gamma$ has a time-dilation factor of $E/M$ in the boosted frame.
In a thermal bath, the decaying $\Chi$ has a thermal energy distribution
$f^{(0)}_P(\hat{E})$ given in Eq.~(\ref{feqProbE.EQ}). 
with $\hat{E} = E/M_\chi$, and $\hat{M}=M/M_\chi$.
We can find the thermal average decay width as
\beq
\langle  \Gamma \rangle _{\sssty\!P} = \int_{\hat{M}}^\infty d\hat{E}\, f^{(0)}_P(\hat{E})\, \Gamma(\hat{E}) \ , 
\eeq
i.e. we have $\langle  \Gamma \rangle /\Gamma_0 = \langle  \hat{M}/\hat{E} \rangle $,
and
the ${\sssty P}$ subscript in $\langle ...\rangle _{\sssty\!P}$ reminding us of the $1/n_\chi^{(0)}$ in the distribution function, normalizing it as a probability distribution. 
Evaluating this, we obtain
\beq
\frac{\langle \Gamma\rangle _{\sssty\!P}}{\Gamma_0} = \frac{K_1(\hat{M} x)}{K_2(\hat{M} x)} \ ,
\label{reciETA.EQ}
\eeq
where the $K_n$ are the modified Bessel functions of the second kind.
We note the limits, 
$\langle  \Gamma \rangle _{\sssty\!P}/\Gamma_0 = \hat{M} x/2 + {\cal O}(x^3)$ for $x \ll 1$,
while
$\langle  \Gamma \rangle _{\sssty\!P}/\Gamma_0 = 1-3/(2\hat{M} x) + {\cal O}(1/x^2)$ for $x \gg 1$.
We give some numerical details on the behavior of the thermal distribution functions and thermal averages
(cf. Appendix~\ref{ThAvgProcDet.SEC}).

\section{$CPT$ and Unitarity Relations}
\label{CPTUniBE.SEC}

We start with the relations between matrix elements that follow from unitarity and the assumption of $CPT$ invariance,
following Ref.~\cite{Kolb:1979qa}. 

In the usual notation, we write the S-matrix as $\Smat\! =\! 1+i\Tmat$,
the nontrivial part of the S-matrix elements for any process $i\!\to\! j$ in terms of the T-matrix elements $\Tmat(i\to j)\! =\! \Tmat_{ji} = \matel{\Tmat}{j}{i}$, 
with $\{\ket{i}\}$ a complete set of states, 
and the amplitude $\ampM$ for this process defined via $\Tmat_{ji}\! =\! (2\pi)^4 \delta^4(p_i - p_j) \ampM(i\!\to\! j)$.   
Unitarity, namely $\Smat^\dagger \Smat\! =\! \Smat \Smat^\dagger\! =\! 1$, implies
$2 \im(\Tmat_{ii})\! =\! \sum_{j} \Tmat_{ji}^* \Tmat_{ji}$,
or equivalently
$2 \im(\ampM(i\!\to\! i))\! =\! \sum_{j} |\ampM(i\!\to\! j)|^2$.
Supplying kinematical factors, we obtain the optical theorem,
$\im(\ampM(i\!\to\! i))\! =\! 2 E_{cm} |{\bf p}| \sigma_{\rm total}$.
Furthermore, the above unitarity condition also implies
$\Tmat^\dagger \Tmat = \Tmat \Tmat^\dagger$ which leads to
$\sum_{j} |\ampM(i\!\to\! j)|^2\! =\! \sum_{j} |\ampM(j\!\to\! i)|^2$.

$CPT$ implies $\ampM(i\!\to\! j) = \ampM(\bar j\!\to\! \bar i)$ where $\bar i$ represents the CP conjugate state of the state $i$.
Combining $CPT$ and the above unitarity relation, we obtain
$\sum_{j} |\ampM(i\!\to\! j)|^2\! =\! \sum_{\bar j} |\ampM(\bar i\!\to\! \bar j)|^2$.
Since the sum over the complete set of states $j$ already includes also over $\bar j$,
there is no distinction between summing over $j$ and $\bar j$, 
and so we can write this as
\beq
\sum_{j} |\ampM(i\!\to\! j)|^2\! =\! \sum_{ j} |\ampM(\bar i\!\to\! j)|^2 \ .
\label{MijSqUni.EQ}
\eeq

If the sum above includes a sum over the initial and final state spins as appropriate for unpolarized observables,
the distinction between $CP$ transformed state $\bar j$ and the $C$ transformed state $j^c$ is unimportant.
For this reason, we use $\bar Q$ and $Q^c$ interchangeably as convenient.
Also, in our case,
for the Majorana $\Chi_n$ we have $\Chi_n^c = \Chi_n$, and so we prefer to write in the notation of $j^c$ rather than $\bar j$. 

Consider for instance the 1-particle $\Chi_n\!=\!\{\Chi_1,\Chi_2\}$ sector.
Given that $\Chi_n$ can decay to the 3-particle final states $B\!\equiv\! UD\bar D^c\!\sim\! UDD$ and to $\bar B\! \equiv\! \bar U \bar D D^c\! \sim\! U^cD^cD^c\! =\! B^c$,
the complete set of states $\Chi_n,B,B^c,...$ including the 3-particle states and other n-particle states not shown.
Only including all n-body final states preserves unitarity.
However, to leading order in perturbation theory, we can restrict to the 1- and 3-particle sectors $i=\{\Chi_n, B,B^c\}$ as we do below.
From the perspective of the 1-particle $\Chi_n$ sector, the $\Chi_n \to B,B^c$ decays appear as a dissipative effect, 
and in perturbation theory, the leading contribution to this dissipation is due to
the discontinuity in the imaginary part of the $\ampM(\Chi_n \to \Chi_n)$ forward amplitude
due to cuts from onshell $B,B^c$ intermediate states.
We have used this aspect to compute the $\Chi_n$ decay width.

We next obtain a relation between the inverse decay and (forward) decay matrix elements relevant for the rhs of the BE.
$CPT$ and unitarity as encoded in Eq.~(\ref{MijSqUni.EQ}) with $i=UDD$ implies
$\sum_{j} |\ampM(UDD\!\to\! j)|^2\! =\! \sum_{ j} |\ampM(U^cD^cD^c\!\to\! j)|^2$.
Explicitly putting in the $j$ we obtain
\bea
&& |\ampM(UDD\!\to\! \Chi_n)|^2 + |\ampM'(UDD\!\to\! UDD)|^2 + |\ampM'(UDD\!\to\! U^cD^cD^c)|^2 = \nonumber\\
&& \hspace{1cm} |\ampM(U^cD^cD^c\!\to\! \Chi_n)|^2 + |\ampM'(U^cD^cD^c\!\to\! UDD)|^2 + |\ampM'(U^cD^cD^c\!\to\! U^cD^cD^c)|^2 \ ,
\label{CPTunitr.EQ}
\eea
where by $\Chi_n$ we mean a sum over both $\Chi_{1,2}$,
and we restrict to the 1- and 3-body final states as appropriate for $\bigO(g^4)$ in the VV interaction effects we are after, and omit higher n-particle cuts.
Here, we have explicitly separated out the contribution due to a real (onshell) intermediate state (RIS)~\cite{Kolb:1979qa} $\Chi_n$
in $|\ampM_{RIS}(UDD\!\! \to\!\! U^cD^cD^c)|^2 \!\equiv\! |\ampM(UDD\!\to\! \Chi_n)|^2$,
and in $\ampM'(UDD\!\! \to\!\! U^cD^cD^c)$ a pure scattering piece that should not include the RIS piece with an onshell $\Chi_n$, 
i.e. we have $|\ampM|^2 = |\ampM_{RIS}|^2 + |\ampM'|^2$. 
Using the $CPT$ relation $|\ampM'(UDD\!\to\! UDD)|^2 = |\ampM'(U^cD^cD^c\!\to\! U^cD^cD^c)|^2$
we obtain
\bea
&& \left( |\ampM'(UDD\!\to\! U^cD^cD^c)|^2 - |\ampM'(U^cD^cD^c\!\to\! UDD)|^2 \right) = \nonumber \\
&& \hspace{1cm}  - \left(|\ampM(UDD\!\to\! \Chi_n)|^2 - |\ampM(U^cD^cD^c\!\to\! \Chi_n)|^2\right) \ .
\label{RISscatRel.EQ}
\eea
The rhs above is nothing but the difference in the inverse decay matrix elements.
This shows how inverse decay and pure scattering matrix elements are related.
We use this relationship below. 
This can also be written as
\bea
&& \left( |\ampM'(UDD\!\to\! U^cD^cD^c)|^2 - |\ampM'(U^cD^cD^c\!\to\! UDD)|^2 \right) = \nonumber\\
&& \hspace{1cm} -\left(|\ampM_{RIS}(UDD\!\to\! U^cD^cD^c)|^2 - |\ampM_{RIS}(U^cD^cD^c\!\to\! UDD)|^2 \right) \ , \\
{\rm or,} && |\ampM(UDD\!\to\! U^cD^cD^c)|^2 = |\ampM(U^cD^cD^c\!\to\! UDD)|^2 \ . \nonumber
\eea

We are now in a position to see what these relations imply for the rhs of the BE 
focusing on the decay and inverse decay terms first (cf. Ref.~\cite{Gopalakrishna:2022hwk}), 
which are given as
\bea
\frac{d}{dt} n_B + 3 H n_B \!\! &=&\!\! \int [d\Pi] \left[ 
 \left(|\ampM(\Chi_n \!\to\! UDD)|^2 - |\ampM(\Chi_n \!\to\! U^cD^cD^c)|^2\right) f_\chi \right. \label{BEscRIS.EQ} \\ 
&&\!\!\!\! - \left(|\ampM(UDD\!\to\! \Chi_n)|^2 e^{3\mu_Q/T} - |\ampM(U^cD^cD^c\!\to\! \Chi_n)|^2 e^{-3\mu_Q/T}\right) f_\chi^{(0)}  \nonumber \\
 &&\!\!\!\! \left. -2 \left(|\ampM'(UDD\!\to\! U^cD^cD^c)|^2 e^{3\mu_Q/T} - |\ampM'(U^cD^cD^c\!\to\! UDD)|^2 e^{-3\mu_Q/T}\right) (f_Q^{(0)})^3 \right] \ , \nonumber
\eea
where the $f_\chi$ is the $\Chi$ distribution function, $f_\chi^{(0)}$ is the equilibrium $\Chi$ distribution,
$f_Q = e^{\mu_Q/T} f_Q^{(0)}$ and using energy conservation we have substituted $(f_Q^{(0)})^3 = f_\chi^{(0)}$ in the second line above in which the $\Chi$ is onshell,
the factor of 2 in the last line is due to the processes being $\Delta B = \pm 2$,
in the last line we cannot use the energy conservation trick as we did in the second line as the $\Chi_n$ is not onshell, 
and, 
the integral measure $[d\Pi]$ is shown symbolically with the actual form suppressed.
The integrals with the distribution functions lead to terms that are proportional to $n_\chi,n_\chi^{(0)}$. 
$CPT$ relates the inverse decay matrix element to the decay matrix element as
$|\ampM(UDD\!\to\! \Chi_n)|^2\! =\! |\ampM(\Chi_n \!\to\! U^cD^cD^c)|^2$,
and, 
$|\ampM(U^cD^cD^c\!\to\! \Chi_n)|^2\! =\! |\ampM(\Chi_n \!\to\! UDD )|^2$.
Using these relations, we notice that in the limit of $\mu_Q \to 0$,
the decay and inverse decay terms (first two lines on the rhs) assemble into a form that goes like $(n_\chi + n_\chi^{(0)})$ 
and does not go to zero as $n_\chi \to n_\chi^{(0)}$.
This signals that something is amiss; what comes to the rescue is that the pure scattering contribution (third line), 
which we see from Eq.~(\ref{RISscatRel.EQ}) has the RIS piece contributing to inverse decay with a negative sign,
and along with the above factor of 2, together ends up overturning the sign of the $n_\chi^{(0)}$ piece, giving in the end a $(n_\chi - n_\chi^{(0)})$ behavior as expected.

For working out the result for nonzero $\mu_Q$,
we write 
$|\ampM'(UDD\!\to\! U^cD^cD^c)|^2 = |\ampM'_{avg}|^2 + (1/2) \Delta |\ampM'|^2$ and
$|\ampM'(U^cD^cD^c\!\to\! UDD)|^2 = |\ampM'_{avg}|^2 - (1/2) \Delta |\ampM'|^2$,
and,
$|\ampM(UDD\!\to\! \Chi_n)|^2 = |\ampM^{RIS}_{avg}|^2 + (1/2) \Delta |\ampM^{RIS}|^2$ and
$|\ampM(U^cD^cD^c\!\to\! \Chi_n)|^2 = |\ampM^{RIS}_{avg}|^2 - (1/2) \Delta |\ampM^{RIS}|^2$. 
In terms of these, Eq.~(\ref{RISscatRel.EQ}) is $\Delta |\ampM'|^2 = - \Delta |\ampM^{RIS}_{avg}|^2$. 
The rhs of Eq.~(\ref{BEscRIS.EQ}) can then be written as
$$-4 \sinh{(3\mu_Q/T)} \int (f_Q^{(0)})^3\, |\ampM'_{avg}|^2  + 2 \cosh{(3\mu_Q/T)} \int f_\chi^{(0)}\, \Delta |\ampM^{RIS}_{avg}|^2 \ .$$
Since we are interested in a tiny baryon asymmetry, i.e. $\mu_Q/T \ll 1$, it suffices to work to leading order in $\mu_Q$,
and we can take in the above rhs, the LO scattering matrix element, i.e. $|\ampM'_{avg}|^2 = |\ampM'_0|^2$.
We can write the first term as $-4 \sinh{(3\mu_Q/T)}\, n_\chi^{(0)} \langle  \Gamma_0'\rangle $
in terms of the thermally averaged transition rate $\GmzPTA$ defined in Eq.~(\ref{Gm0BBbTA.EQ}) and computed earlier. 
The $CPT$ relation between the inverse decay and decay matrix elements given above is   
$\Delta |\ampM^{RIS}|^2 = - \Delta |\ampM'|^2$. 
Using these results,
the contributions from decay, inverse decay, and $3\to 3$ scattering, 
finally leads to the first two terms on the rhs of the Boltzmann equation in Eq.~(\ref{BEnbXUD.EQ}). 
 
Using the $CPT$ and unitarity relations between the matrix elements above, 
we write down the Boltzmann Equations (BE) for the $\Chi$ number density $\nX$,
and the number densities of $\Chi,\Up,\Dp$, namely, $n_\Chi,n_\Up,n_\Dp$, and also for $\bar\Up,\bar\Dp$ namely $n_{\bar\Up}, n_{\bar\Dp}$. 
From these we obtain the BE for the baryon number density $n_B$, 
continuing our earlier discussion on the BE in Ref.~\cite{Gopalakrishna:2022hwk}. 

The rhs of the BE (collision terms) contain various thermally averaged decay and scattering rates. 
In our theory, the relevant rates are
$\GmTA$ for the $\Chi \to UDD$ decay process,
$\GmSigXDBTA$ for the $\Chi \bar{D} \to U D$ scattering process SC-1,
$\GmSigXUBTA$ for the $\Chi \bar{U} \to D D$ scattering process SC-2.
Their conjugate process rates are
$\GmBTA$ for the $\Chi \to UDD$ decay process,
$\GmSigXDTA$ for the $\Chi \bar{D} \to U D$ scattering process SC-1,
$\GmSigXUTA$ for the $\Chi \bar{U} \to D D$ scattering process SC-2.
The corresponding rate at LO are $\GmzTA$, $\GmzSigXDBTA$, $\GmzSigXUBTA$.
The $\Delta B=\pm 2$ rates at LO are 
$\GmzPTA$ for the $UDD\to U^cD^cD^c$ process,
$\GmzPSigUDTA$ for the $UD\to D^cU^cD^cD^c$ process,
and, 
$\GmzPSigDDTA$ for the $DD\to U^cU^cD^cD^c$ process.
There is no asymmetry at LO and so the same rates apply for the conjugate processes at LO.
The thermally averaged rates for the inverse processes of these are also required. 

A similar analysis is applicable for the scattering contribution on the rhs of the BE.
Consider first the SC-2 channel. 
The $CPT$ and unitarity relation in Eq.~(\ref{MijSqUni.EQ}) with $i=DD$,
and $j$ restricted to 2 and 4 particle states as appropriate for $\bigO(g^4)$ effect we are interested in, 
implies, 
\bea
&& |\ampM(DD\!\to\! \Chi_n U^c)|^2 + |\ampM'(DD\!\to\! U^cD^cD^cU^c)|^2 = \nonumber \\
&& \hspace{1cm}  |\ampM(D^cD^c\!\to\! \Chi_n U)|^2 + |\ampM'(D^cD^c\!\to\! UDDU)|^2 \ ,
\label{CPTunitrSC.EQ}
\eea
after we cancel common terms on both sides by using the relations 
$|\ampM'(DD\!\to\! DD)|^2 = |\ampM'(D^cD^c\!\to\! D^cD^c)|^2$ from CPT,
and,
$|\ampM'(DD\!\to\! UDDU^c)|^2 = |\ampM'(D^cD^c\!\to\! U^cD^cD^cU)|^2$ from $CPT$ and crossing symmetry.
We have $|\ampM(DD\!\to\! \Chi_n U^c)|^2 = |\ampM_{RIS}(DD\!\to\! U^cD^cD^cU^c)|^2$, 
where the RIS piece is when the $\Chi_n$ goes onshell in the scattering channel,
and similarly for its conjugate.
We can equivalently write Eq.~(\ref{CPTunitrSC.EQ}) as
$|\ampM_{RIS}(DD\!\to\! U^cD^cD^cU^c)|^2 + |\ampM'(DD\!\to\! U^cD^cD^cU^c)|^2 = |\ampM_{RIS}(D^cD^c\!\to\! UDDU)|^2 + |\ampM'(D^cD^c\!\to\! UDDU)|^2$,
or, 
$|\ampM(DD\!\to\! U^cD^cD^cU^c)|^2\!\! =\!\! |\ampM(D^cD^c\!\to\! UDDU)|^2$
combining the RIS piece and the pure scattering piece.
Using the $CPT$ relations $\ampM(DD\!\!\to\!\! \Chi U^c)\!\! =\!\! \ampM(\Chi U\!\!\to\!\! D^c D^c)$ and
$\ampM(DD\!\!\to\!\! \Chi U^c)\!\! =\!\! \ampM(\Chi U\!\!\to\!\! D^c D^c)$,
and defining
$\Delta\ampMsigSq\!\! =\!\! (|\ampM(\Chi U^c\!\!\to\!\! D D)|^2 - |\ampM(\Chi U\!\!\to\!\! D^c D^c)|^2)$
and
$\Delta\ampMPsigSq\!\! =\!\! (|\ampM'(D^cD^c\!\to\! UDDU)|^2 - |\ampM'(DD\!\to\! U^cD^cD^cU^c)|^2)$
we obtain
$\Delta\ampMPsigSq\!\! =\!\! -\Delta\ampMsigSq$.
This relates the pure scattering piece on the lhs to the RIS piece on the rhs,
and the negative sign in conjunction with a factor of 2 from being a $\Delta B = \pm 2$ scattering process,
overturns the sign of the inverse scattering piece on the rhs of the BE.
Similarly, for SC-1 the relevant scattering channel is
$U^cD^c\!\to\! DDDU$ and $UD\!\to\! D^cD^cD^cU^c$ and the RIS piece of these give
$U^cD^c\!\to\! \Chi_nD$ and $UD\!\to\! \Chi_nD^c$ respectively. 
Taking into account a nonzero $\mu_Q$, we obtain on the rhs of the BE the form
$\ampM'(QQ\!\! \to ...)\, e^{2\mu_Q/T}$ and $\ampM'(Q^cQ^c\!\! \to ...)\, e^{-2\mu_Q/T}$,
and proceeding along similar lines to the earlier inverse decay case,
we define the thermally averaged scattering rate $\GmzPSigTA$ as in Eq.~(\ref{Gm0BBbTA.EQ}) 
but now with $N_i\!=\!2$ and $N_f\!=\!4$ since the $QQ \to Q^cQ^cQ^cQ^c$ process is $2\!\to\! 4$, and $|{\ampM'}^{(\sigma)}_0|^2$ being the matrix element.  
Putting these together, we obtain the form of the scattering terms in the rhs of the Boltzmann equation  
in Eq.~(\ref{BEnbXUD.EQ})  
as the contribution from the scattering and inverse scattering processes,
where the RIS pieces have overturned the sign of the inverse scattering terms.

\section{UV Completion Corrections}
\label{VVintUV.SEC}

As the momentum transfer in the effective VV interaction of Eq.(\ref{LIntVVMB.EQ}) increases to approach $\Lambda$,
the point-like contact interaction will start to become resolved by propagating states in the UV completion. 
This issue is encountered, for instance, in the scattering channel when the Mandelstam $s,t,u$ approaches $\Lambda^2$.
For instance, in computing the thermal average cross section in Eq.~(\ref{TASC12.EQ}),
momenta $p \gtrsim \Lambda$ are sampled from the thermal spectrum.
Another aspect is the fact that we rely on loop corrections to obtain a CP violating effect leading to a baryon asymmetry,
and loop momenta being unrestricted, also resolve the effective vertex.
Due to these reasons, in a strict sense, an effective theory approach is not fully accurate and one will have to work with a specific UV completion.

This issue is not so pressing for the $\Chi$ decay channel as the momenta are limited by $M_\chi$
and the UV completion contributions appear only for momenta $\gtrsim \Lambda$.
This is true for loop momenta as well since we place cuts on the intermediate state propagators to obtain a
baryon asymmetry. In any case we study the effects of UV physics on the decay processes as well to confirm this expectation. 

Even for scattering processes, our effective theory analysis should be adequate for obtaining a reasonable estimate of the baryon asymmetry generated for the following reasons: 
(a) the probability of sampling momenta $p \sim \Lambda$ in the thermal spectrum for $T\sim M_\chi$ is not very large for $M_\chi/\Lambda \ll 1$, and
one can get a fairly good estimate using an effective theory; 
(b) the baryon number violation in our setup is due to the $\Chi$, and any contributions from propagating UV physics states
excited in the large momentum regions in the loops should not nullify the effects we have identified;
(c) we take differences of loop amplitudes between a process and conjugate process to obtain the baryon asymmetry,
which should have dominant contributions from the $\Chi$ sector.
Thus, although the results in the effective theory may not be fully accurate, it should capture the dominant contributions.
A big advantage is that it captures the behavior of a whole class of UV completions, 
which is the spirit of our effective theory analysis.
We quantify next how good an approximation the effective theory computation of the thermal average cross section is
by comparing it to the result with corrections from an example UV completion included.  

We compute the $\TAsigv$ in the effective theory and compare it with that obtained by including corrections from
the state that resolves the effective vertex in the UV completion as the momentum transfer approaches $\Lambda$.
To do this, for illustration, we pick the {\it UV completion A} discussed in Ref.~\cite{Gopalakrishna:2022hwk},
in which the vertex is resolved as the heavy state $\xi$ with mass $M_\xi$ and decay width $\Gamma_\xi$
begins to propagate for $p \gtrsim M_\xi$.
The $\xi$ propagator is $1/(p^2 - M_\xi^2 + i M_\xi \Gamma_\xi)$.  
We focus on the scattering channels SC-1 and SC-2 of Sec.~\ref{genBAsymScat.SEC}. 
SC-1 involves $\xi$ exchange in the t-channel, while SC-2 is in the s-channel.
Deviations from a contact VV interaction due to the propagating $\xi$ in $\sigma v$ can be captured by the replacement at each vertex 
$(1/\Lambda^2) \to (1/\Lambda^2) f_\xi(p)$, taking $\Lambda = M_\xi$,
where the correction factor due to a propagating $\xi$ is $f_\xi(p) = 1/(1-\tilde{p}^2-i \tilde\Gamma_\xi)$
with $\tilde{p} = p/\Lambda$ and $\tilde\Gamma_\xi = \Gamma_\xi/\Lambda$.
We have $f_\xi(p) \to 1$ as $\Lambda\to \infty$, in other words, the effective theory is recovered in the limit of $r_{M\Lambda} \equiv M_\chi/\Lambda \ll 1$.

For the s-channel scattering process (i.e. for SC-2),
we subtract the on-shell $\xi$ (resonance) contribution as this is included in the $\xi$ decay contribution
and should not also be included in the scattering channel
(for a detailed discussion of this aspect, see Ref.~\cite{Kolb:1979qa}).
We compare two alternate schemes for the subtraction:
(a) use the principal value (PV) of the $\xi$ propagator, and, 
(b) cut-out the contribution under a total window width of $5\tilde\Gamma_\xi$ centered at $\tilde{M}_\xi \equiv M_\xi/\Lambda = 1$.
We implement (a) by defining the (finite) PV part as $ 1/(p^2-M^2+i\epsilon) = {\rm PV}(1/(p^2-M^2)) - \pi i \delta(p^2 - M^2)$
isolating the singular (on-shell) part as the second term. 
We use the representation $\delta(p^2 - M^2) = \lim_{\epsilon\to 0} [ (\epsilon/\pi)/((p^2-M^2)^2 + \epsilon^2)]$,
and connect with a physical interpretation by taking $\epsilon \equiv M_\xi \Gamma_\xi$.
We thus obtain
$f^{\rm (PV)}_\xi(p) = 1/(1-\tilde{p}^2-i \tilde\Gamma_\xi) - i \tilde\Gamma_\xi/((1-\tilde{p}^2)^2 + \tilde\Gamma_\xi^2)$.
We see that our choice of the delta function ensures that the PV is real.
We view $f^{\rm (PV)}_\xi(p)$ as a (quadratic) function of the 3-momentum radial component $p_r$, 
and since $p_r \geq 0$ we pick up only the positive pole contribution.

We thus obtain UV propagator correction factors in the LO scattering amplitudes given by 
$f_{\xi\,(t)} = f_\xi (p_n - q_2)$ (for SC-1),
and, 
$f_{\xi\,(s)} = f_\xi (p_n + k_i)$ (for SC-2). 
In the LO cross sections the correction factors are   
$f_{\xi\,(t)}^{(00)} = |f_\xi (p_n - q_2)|^2 = 1/((1-\tilde{t})^2 + \tilde\Gamma_\xi^2)$ (for SC-1),
and 
$f_{\xi\,(s)}^{(00)} = |f_\xi (p_n + k_i)|^2 = {\rm PV}(1/((1-\tilde{s})^2))$ (for SC-2),
where
$(\tilde{s},\tilde{t}) = (s,t)/\Lambda^2$ for the Mandelstam variables $s=(p_n + k_i)^2$ and $t=(p_n - q_2)^2$,
and
$\tilde\Gamma_\xi = \Gamma_\xi/\Lambda$ with $\Gamma_\xi$ being the $\xi$ decay width,
which we also parametrize as the dimensionless $\hat\Gamma_\xi \equiv \Gamma_\xi/M_\chi$.



\pagebreak

\setcounter{partsection}{1}

\vfill
\custompartsection{Technical Details}
\label{techDet.PART}

Some technical details and basic formalism aspects are in the appendices below.  
Details of the computation of the matrix elements for our decay and scattering processes are in Appendix~\ref{MatElcomp.SEC}.
Our choice of independent phase-space and loop-momentum variables, and,
the phase space and loop integral measures for decay and scattering processes
are in Appendix~\ref{12U34UPVars.SEC}.
The matrix elements and phase space measures discussed above are combined to obtain the
decay and scattering rate asymmetries in Appendix~\ref{GmSigDet.SEC}. 
Implications of $CPT$ and unitarity relations on the matrix elements of forward and inverse processes
from a perturbation theory perspective in our new physics theory is discussed in Sec.~\ref{CPTUniDet.APP}. 
Some numerical details on the behavior of the thermal distribution functions and the thermal averages are in Appendix~\ref{ThAvgProcDet.SEC},
and on the UV propagator corrections are in Appendix~\ref{VVintUVDet.SEC}. 

\vfill

\pagebreak

\section{The Decay and Scattering Matrix Elements}
\label{MatElcomp.SEC}

Here we provide various technical details of computing the matrix elements.
We fall back on some first principles calculations to
handle carefully the two identical $\Dp$ in the interaction,
and the Majorana nature of the $\Chi_n$.
We start by discussing the normalization of the $\Dp\Dp\Up$ final state.
Following this, we compute the matrix element,
which we split into two fermion-line parts, the $\Chi\Up$ line part,
and another $\Dp\Dp$ line part in the final state and loops, and discuss them separately.
We then put them together to compute the tree-loop interference term in the
mod-squared of the matrix element, needed for the baryon asymmetry computation.
We give the details of how we renormalize (subtract) the $\bar\Chi_m\Chi_n$ 2-point function part of diagram B, 
to be consistent with working in the mass basis. 

In computing the matrix elements, we project onto a color-singlet combination of the $\Dp,\Dp,\Up$ involved,
consistent with the structure of the effective interaction,
and normalize the states appropriately as we discuss in Appendix~\ref{DDUstate.SEC}.
In Appendix~\ref{DDM12.SEC} we show how we handle the two identical $\Dp$ in the final state in computing the matrix element,  
and in Appendix~\ref{DDLoop.SEC} we show the corresponding aspects in computing the loop diagram.
We describe in detail our method of computing the tree-loop interference term in Appendix~\ref{LONLOintContr.SEC}.

\subsection{The $\Dp\Dp\Up$ state}
\label{DDUstate.SEC}

In computing the matrix element, we project onto a color-singlet combination of the 3-body $\Up\Dp\Dp$ final state as 
\beq
\ket{as{{\bf q}_1},bt{{\bf q}_2},cr{{\bf q}_U}} =
\frac{\epsilon^{abc}}{\sqrt{6}} \frac{1}{\sqrt{2}}\, \sqrt{2E_{\bf q_U}} \sqrt{2E_{\bf q_2}} \sqrt{2E_{\bf q_1}} \, {\hat{a}_{{\bf q}_U}}^{cr\, \dagger} {a_{{\bf q}_2}^{bt}}^\dagger {a_{{\bf q}_1}^{as}}^\dagger \ket{0} \ ,
\label{DDUstate.EQ}
\eeq
where $a,b,c$ denote the color indices, $s,t,r$ the spin indices, and, $q_1,q_2,q_U$ the momenta of the $\Dp,\Dp,\Up$ respectively,
we specialize to the $\epsilon^{abc}$ here
for having a color singlet final state with the $1/(\sqrt{2}\sqrt{6})$ providing the correct normalization,
and the fermionic creation operators anti-commute.

We can write the color-singlet state as a wave packet by introducing into Eq.~(\ref{DDUstate.EQ})
a weight $K^{str}$ for the spin helicities, and $\phi({{\bf q}_U},{{\bf q}_1},{{\bf q}_2})$ for the momenta.
We pay particular attention to the two identical particles, namely $\Dp\Dp$ (i.e. $\widebar{\Dp^c} \Dp$) in the final-state,
but distinguished by spin or color labels. 
Focusing on this sector for now, 
some example combinations of the two spins are: 
$K^{st} = \delta^{st}$ is a symmetric same spin combination;
$K^{st} = 1, \forall \{s,t\}$, democratically combines all spins; 
$K^{st} = \epsilon^{st}$ is an antisymmetric different spin combination.
As required for two fermions, we must get a minus sign under the interchange of the two $\Dp$ particles,
i.e. under $as{{\bf q}_1} \leftrightarrow bt{{\bf q}_2}$, 
which requires that $K^{st} \phi({\bf q_1},{\bf q_2}) \epsilon^{ab}$ combined must yield this minus sign.
Since the last (i.e. the color part) is already antisymmetric and will yield a minus sign under interchange,
the $K^{st} \phi({{\bf q}_1},{{\bf q}_2})$ part must yield a plus sign.
We note in particular that the simple state obtained by taking
$K^{st} = 1, \forall \{s,t\}$, and $\phi({{\bf q}_1},{{\bf q}_2}) = 1$ as in Eq.~(\ref{DDUstate.EQ})
is a fine $\Dp\Dp$ state with the correct minus sign being obtained
on interchange of the two particles simply on account of the anticommutation property of the $a^\dagger$. 

\subsection{The tree-level decay matrix element}
\label{chi2DDUMXULODet.SEC}

The $\MLOXU^\mu_n$ is given as
\beq
\MLOXU^\mu_n \equiv \bar{u}(q_U) \GVBn^\mu u(p_n) \ , 
\label{ampLOXUn.EQ} 
\eeq
The mod-squared of the LO matrix element for the process,
averaged over initial state spins and summed over final state spins and colors is given by
\beq
|\ampALOn|^2 = \frac{1}{4 \Lambda^4} \frac{1}{2} (\MLOXUSq)^{\mu\tau}_{nn} \, C_{00}\, (\MLOotSq)_{\mu\tau} \ , 
\label{ampALOn.EQ}
\eeq
where we have defined $\MLOotSq_{\mu\tau} \equiv \MLOot_\mu \MLOot_\tau^\dagger$.
(We note here that although written as $|...|^2$, this is not necessarily a real quantity due to the different Lorentz indices involved.)
The replacements for obtaining the mod-squared amplitude for the conjugate process have already been given above.

\subsubsection{The $\Dp\Dp$ final state part $\MLOot$}
\label{DDM12.SEC}

Here we describe the computation of $\MLOot$, the portion of the matrix element involving the $\Dp$ final state,
encountered in Eq.~(\ref{ampA0n.EQ}).
We pay careful attention to the presence of more than one identical $\Dp$ which lead to multiple contractions in the
matrix elements.

We start by being general in the effective interaction
and include scalar-scalar (SS) and vector-vector (VV) Lorentz structures
where for the VV interaction of Eq.~(\ref{LIntVVMB.EQ}) we have
$\widetilde{\Gamma} = \widetilde{G} = \tilde{g} \gamma^\mu $ and $\Gamma = G^n_{V\mu} =  \gamma_\mu \left(\ghLn P_L + \ghRn  P_R \right) $,
while for the SS interaction we have a similar coupling structure but without the $\gamma_\mu$. 
Here we explain details involving the contraction of the field bilinear $\overline{\Dp^c} \, \widetilde\Gamma \Dp_b $
appearing in the interaction,
with the two identical particle $\Dp\Dp$ (final) state. 
We use this to compute the $|M_{12}|^2$ appearing in the decay width $\Gamma(\Chi_n \to \widebar{\Dp^c} \Dp \Up)$, 
discussed in Sec.~\ref{ABdec.SEC}, 
where two identical $\Dp$ particles appear in the final state since $\widebar{\Dp^c} \Dp \sim \Dp \Dp$. 

\begin{figure}
  \begin{center}
    \includegraphics[width=0.3\textwidth] {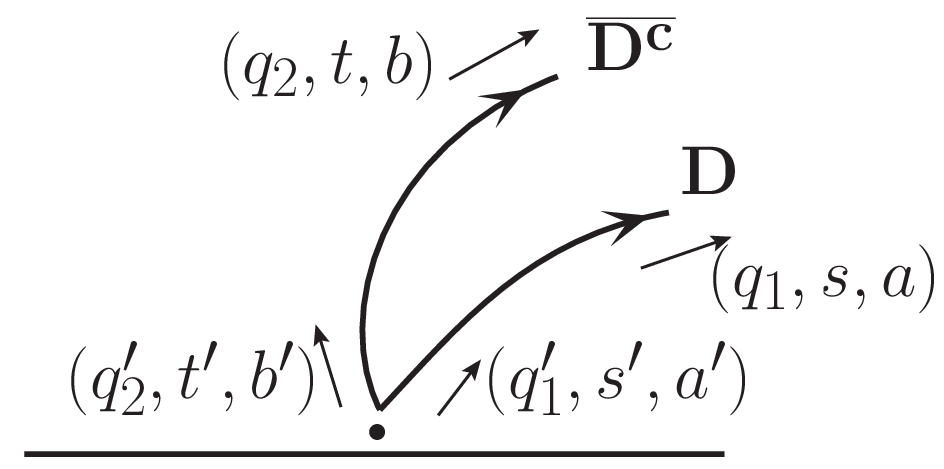}
    \hspace*{0.25cm}
    \includegraphics[width=0.3\textwidth] {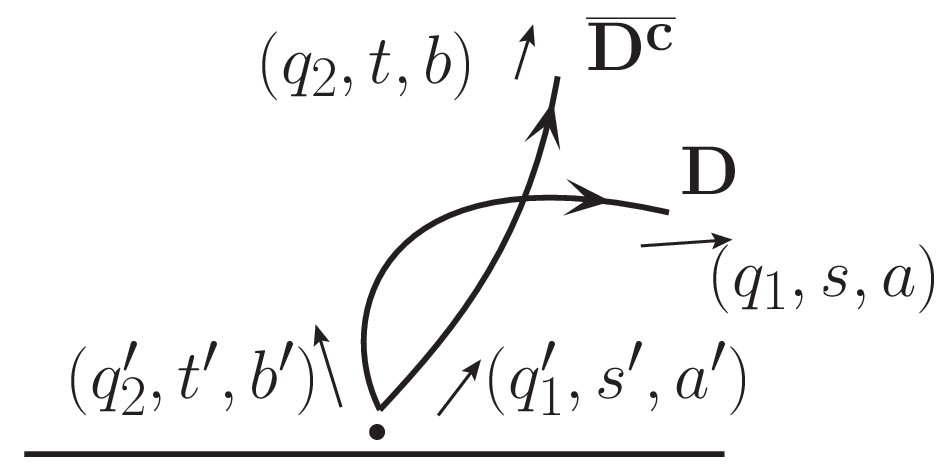}
    \caption{The left and right diagrams show the two contractions of the $\overline{D}\, \bar{\widetilde\Gamma} \, D^c$ part of the interaction
      with the $\Dp\Dp$ final state. The momentum flow, spin, and color indices are shown as $(q,s,a)$ respectively.  
      \label{chi2DDU-DDfs.FIG}
    }
  \end{center}
\end{figure}

We evaluate the contraction of the field bilinear $\overline{\Dp^c} \, \widetilde\Gamma \Dp_b$
with the 2-$\Dp$ state as in Eq.~(\ref{DDUstate.EQ}), for now taking the simple weighting
$K^{st} = 1, \forall \{s,t\}$, and $\phi({{\bf q}_1},{{\bf q}_2}) = 1$.
Since the two $\Dp$ are in the final state, we evaluate the contraction with the hermitian conjugate of
$\overline{\Dp^c} \, \widetilde\Gamma \Dp$, namely
$\overline{\Dp} \, \widebar{\widetilde\Gamma} \Dp^c$,
with $\widebar{\widetilde\Gamma} \equiv \gamma^0 \widetilde\Gamma^\dagger \gamma^0$. 
Thus, one of the things we need to evaluate the decay amplitude $\Gamma(\Chi_n \to \widebar{\Dp^c} \Dp \Up)$ is the matrix element
$M_{12}^{st} \equiv \matel{\overline{\Dp}_{a'} \, \widebar{\widetilde\Gamma} \Dp_{b'}^c}{btq_2,asq_1}{0}/\sqrt{2}$. 
Now,
$
\sum_{a',b'} \overline{\Dp}_{a'}(x) \, \widebar{\widetilde\Gamma} \Dp_{b'}^c(x) \epsilon^{a'b'}
\supset \int \sum_{a',b',s',t'} [dq'_1] [dq'_2] {a_{{\bf q}_1'}^{a's'}}^\dagger {a_{{\bf q}_2'}^{b't'}}^\dagger \epsilon^{a'b'}
\bar{u}^{s'}_{q_1'} \widebar{\widetilde\Gamma} v^{t'}_{q_2'} e^{i(q'_1+q'_2)x} + ...
$.
We thus have the factor
$\matel{{a_{\bf q_1}^{as}} {a_{\bf q_2}^{bt}} {a_{\bf q_1'}^{a's'}}^\dagger {a_{\bf q_2'}^{b't'}}^\dagger}{0}{0}$ in $M_{12}^{st}$,
which yields two contractions (due to the two identical $\Dp$), to give
\vspace{-0.25cm}
\begin{center}
$
-\delta({{\bf q}'_1 - {\bf q}_1}) \delta({{\bf q}'_2 - {\bf q}_2}) \delta^{a'a} \delta^{s's} \delta^{b'b} \delta^{t't} 
+ \delta({{\bf q}'_1 - {\bf q}_2}) \delta({{\bf q}'_2 - {\bf q}_1}) \delta^{a'b} \delta^{s't} \delta^{b'a} \delta^{t's}
$.
\end{center}
\vspace{-0.25cm}  
The two contractions are shown in Fig.~\ref{chi2DDU-DDfs.FIG}.
Using the antisymmetry in the color indices, which flips the sign on the second contraction (thus removing the relative minus sign),
we obtain\footnote{We do not include the factor of
  $(i \epsilon^{abc}/\Lambda^2)$ from the interaction vertex here but take it into account elsewhere.
  Also, we include elsewhere the factor $e^{i(q_1+q_2)x} /\sqrt{6}$.
  We choose to include only the $1/\sqrt{2}$ factor from the final state normalization here. 
}
\beq
M_{12}^{st} = -\frac{1}{\sqrt{2}} (\bar{u}^{s}_{q_1} \widebar{\widetilde\Gamma} v^{t}_{q_2} + \bar{u}^{t}_{q_2} \widebar{\widetilde\Gamma} v^{s}_{q_1} )  \ .
\label{M12stsqGen.EQ}
\eeq
Taking the mod-square of this and summing over $s,t$, we have $|M_{12}|^2 = \sum_{s,t} |M_{12}^{st}|^2 $, which is,   
\beq
|M_{12}|^2 =
\frac{1}{2} \left\{
 {\rm Tr}[\widetilde{\Gamma} (\qslash_1+M) (\widebar{\widetilde\Gamma}-\widetilde\Gamma^c) (\qslash_2-M) ]
 + (q_1 \leftrightarrow q_2)
 \right\} \ ,
\label{M12sqGen.EQ}
\eeq
where
$\Gamma^c \equiv C \Gamma^* C$,
and we have used the spinor identity
\beq
\bar{v}_{q_1}^s \Gamma u_{q_2}^t = \bar{v}_{q_2}^t (-\bar\Gamma^c) u_{q_1}^s \ ,
\label{uvSpId.EQ}
\eeq
and similarly for $u,v$ flipped. We note that $\widebar{\bar\Gamma} = \Gamma$ and $\left(\widebar{\bar\Gamma}\right)^c = \Gamma^c$.
In particular, these properties apply to $\widetilde\Gamma$ also. 

We first consider the $SS$ interaction with $\widetilde{\Gamma} = \tilde{g}_L P_L + \tilde{g}_R P_R$. 
For this, we find $\widebar{\widetilde\Gamma} = \widetilde\Gamma^c$, and Eq.~(\ref{M12sqGen.EQ}) then yields $|M_{12}^{SS}|^2 = 0$.
In addition to the discussion
in our earlier work  
showing that the $SS$ interaction cannot be written down, if one were to persist,
we have now shown that it would not cause this decay with the two identical $\Dp$ in the final state.
It should be remembered that this result has been obtained for $K^{st} = 1, \forall \{s,t\}$, and $\phi({{\bf q}_1},{{\bf q}_2}) = 1$. 

Continuing with the $SS$ interaction, we may still ask if contracting with a 2-$\Dp$ state with some other
$K^{st}$ and $\phi({{\bf q}_1},{{\bf q}_2})$ may be nonzero.
To address this question, we now generalize the above result.
Due to the presence of the $K^{st}$, writing the spinor products of Eq.~(\ref{M12stsqGen.EQ}) summed over the spins as a ${\rm Tr}(...)$
will not work so straightforwardly.
To facilitate this, we introduce the corresponding helicity projectors and then sum unrestricted over the spins,
leading to ${\rm Tr}(...)$ but now with the projectors. 
The helicity projectors for $u$-spinors is $\hat P_s({\bf q}) \equiv (1/2)(1 \pm \hat{q}^i \Sigma^i)$ for $s=\{1,2\}$,
and for $v$-spinors they are $\hat P'_s({\bf q}) \equiv (1/2)(1 \mp \hat{q}^i \Sigma^i)$ (i.e. $\hat P'_{1,2} = \hat P_{2,1}$),
with $\Sigma^i = \bmat \sigma^i & 0 \\ 0 & \sigma^i \emat$, and $\hat{q}^i$ is the unit 3-vector along ${\bf q}$.
These projectors satisfy $\hat P_s({\bf q}) \hat P_t({\bf q}) = \delta^{st} \hat P_s({\bf q})$. 
Introducing these projectors and denoting $\hat P_s({\bf q}_i) \equiv P_{si}$ and summing unrestricted over $s',t'$
allows us to generalize Eq.~(\ref{M12stsqGen.EQ}) as
\beq
|M_{12}^{st}|^2 =  \sum_{s',t'} |K^{st}|^2 |\phi({{\bf q}_1},{{\bf q}_2})|^2 \frac{1}{2}  |(\bar{u}^{s'}_{q_1} P_{s1} \widebar{\widetilde\Gamma} P'_{t2} v^{t'}_{q_2}
+ \bar{u}^{t'}_{q_2} P_{t2} \widebar{\widetilde\Gamma} P'_{s1} v^{s'}_{q_1} )|^2
\ .
\label{M12stsqGenPst.EQ}
\eeq
Since now the sum over $s',t'$ is unrestricted, we can write this as a ${\rm Tr}(...)$.
Using the notation $P_{s1} \Gamma P'_{t2} \equiv \Gamma_{s1t'2}$ we have
\bea
|M_{12}^{st}|^2 =  |K^{st}|^2 |\phi({{\bf q}_1},{{\bf q}_2})|^2
\frac{1}{2} \left\{
  {\rm Tr}[\widetilde\Gamma_{t'2s1} (\qslash_1 + M) (\widebar{\widetilde\Gamma}-\widetilde\Gamma^c)_{s1t'2} (\qslash_2 - M)] \right. \nonumber \\ \left. 
   +  {\rm Tr}[\widetilde\Gamma_{s'1t2} (\qslash_2 + M) (\widebar{\widetilde\Gamma}-\widetilde\Gamma^c)_{t2s'1} (\qslash_1 - M)]
\right\}
 \ , 
\label{M12stsqGenPstTr.EQ}
\eea
where we have used the identity $(\overline{\Gamma_{t2s'1}})^c = (\bar\Gamma^c)_{s1t'2}$.
Eq.~(\ref{M12stsqGenPstTr.EQ}) is the generalization we seek.
For the $SS$ interaction, we have $\widetilde\Gamma^c = \widebar{\widetilde\Gamma}$, and from Eq.~(\ref{M12stsqGenPstTr.EQ})  
we see that $|M_{12}^{st}|^2 = 0$ even for a general $K^{st}$. Again, even if we persisted and considered the $SS$ interaction,
it would not induce a decay with $\Dp^s\Dp^t$ in the final state for any orientations of the spins. 

We consider next the $VV$ interaction for $K^{st} = 1, \forall \{s,t\}$, and $\phi({{\bf q}_1},{{\bf q}_2}) = 1$. 
For the $VV$ interaction we have $\widetilde{\Gamma} \equiv \widetilde{G}_V^\mu = \tilde{g} \gamma^\mu$, 
and we must have $\tilde{g}_L = \tilde{g}_R \equiv \tilde{g}$. 
For this we have $\widebar{\widetilde\Gamma} = - \widetilde\Gamma^c = \bar{\widetilde{G}}_V^\mu = {\tilde{g}}^* \gamma^\mu$,
and from Eq.~(\ref{M12stsqGen.EQ}), we have
\beq
    \Mot^{st}_\mu  = - \frac{1}{\sqrt{2}} (\bar{u}^{s}_{q_1} \GtVB_\mu v^{t}_{q_2} + \bar{u}^{t}_{q_2} \GtVB_\mu v^{s}_{q_1} )
\ .
\label{M12stsqVV.EQ}
\eeq
Similarly, for the conjugate process, we find
\beq
    \Mcot^{st}_\mu  = - \frac{1}{\sqrt{2}} (\bar{u}^{s}_{q_1} \GtV_\mu v^{t}_{q_2} + \bar{u}^{t}_{q_2} \GtV_\mu v^{s}_{q_1} )
\ .
\label{Mc12stsqVV.EQ}
\eeq
The mod-squared of this (for different Lorentz indices $\mu,\tau$) summed over $s,t$,
is $|{M_{12}}|^2_{\mu\tau} = \sum_{s,t} {M_{12}^{st}}_\tau^\dagger {M_{12}^{st}}_\mu $, which, from Eq.~(\ref{M12sqGen.EQ}), is
\bea
\MotSq_{\mu\tau} &=&
 \left\{
 {\rm Tr}[ \GtV_\tau (\qslash_1+M) \GtVB_\mu (\qslash_2-M) ]
 + (q_1 \leftrightarrow q_2)
 \right\} \ , \nonumber \\
 &=&
 |\tilde{g}|^2
\left\{
 {\rm Tr}[\gamma_\tau (\qslash_1+M) \gamma_\mu (\qslash_2-M) ]
 + (q_1 \leftrightarrow q_2)  \right\} \\
 &=&
 2 |\tilde{g}|^2 {\rm Tr}[\gamma_\tau (\qslash_1+M) \gamma_\mu (\qslash_2-M) ] \ ,
\label{M12sqVV.EQ}
\eea
where the last line is due to the fact that the $(q_1 \leftrightarrow q_2)$ term gives the same contribution
since the trace of an odd number of $\gamma$-matrices is zero.
We show at the end of Appendix~\ref{DDLoop.SEC} that this can be thought of equivalently as the cut of a loop diagram,
and introduce the notation $\MotSq = \Lot$ to capture this equivalence. 
For the conjugate process, we similarly have 
\bea
\McotSq_{\mu\tau} &=&
 \left\{
 {\rm Tr}[ \GtVB_\tau (\qslash_1+M) \GtV_\mu (\qslash_2-M) ]
 + (q_1 \leftrightarrow q_2)
 \right\} \ , \nonumber \\
 &=&
  \MotSq_{\mu\tau} \ .
\label{Mc12sqVV.EQ}
\eea

\subsection{The single operator tree-loop interference terms}
\label{LONLOintDet.SEC}

We present here the details involved in the computation of the tree-loop interference term
for the decay and scattering rate difference. 

\subsubsection{The $\MLONLOXU$ matrix element part}
\label{LONLOintContr.SEC}

We obtain here the contractions leading to the tree-loop interference terms
for the decay rate of $\Chi_n \to \Dp_a \overline{\Dp^c}_b \Up_c$
depicted in Fig.~\ref{chi2DDU-LONLOCut.FIG}.
For this purpose we write the mode expansion of the fermion fields as 
\beq
\psi(x) = \int d^3p \frac{1}{\sqrt{2E_{\bf p}}} \left(u_p a_{\bf p} e^{-ipx} + v_p b_{\bf p}^\dagger e^{ipx} \right) \ , \quad
\Chi(x) = \int d^3p \frac{1}{\sqrt{2E_{\bf p}}} \left(u_p c_{\bf p} e^{-ipx} + v_p c_{\bf p}^\dagger e^{ipx} \right) \ ,
\label{psiChiModeExpn.EQ}
\eeq
where $\psi=(D,U)$ are Dirac fields, while $\Chi$ is a Majorana field.
The charge-conjugate field $\psi^c(x) = C \psi^*(x)$ has $a_{\bf p} \leftrightarrow b_{\bf p}$,
while $\Chi$ is self-conjugate, i.e. $\Chi^c(x) = \Chi(x)$. 
We use the VV interaction of Eq.~(\ref{LIntVVMB.EQ}) to obtain the tree-loop interference term.
 
\begin{figure}
  \begin{center}
    \includegraphics[width=0.5\textwidth] {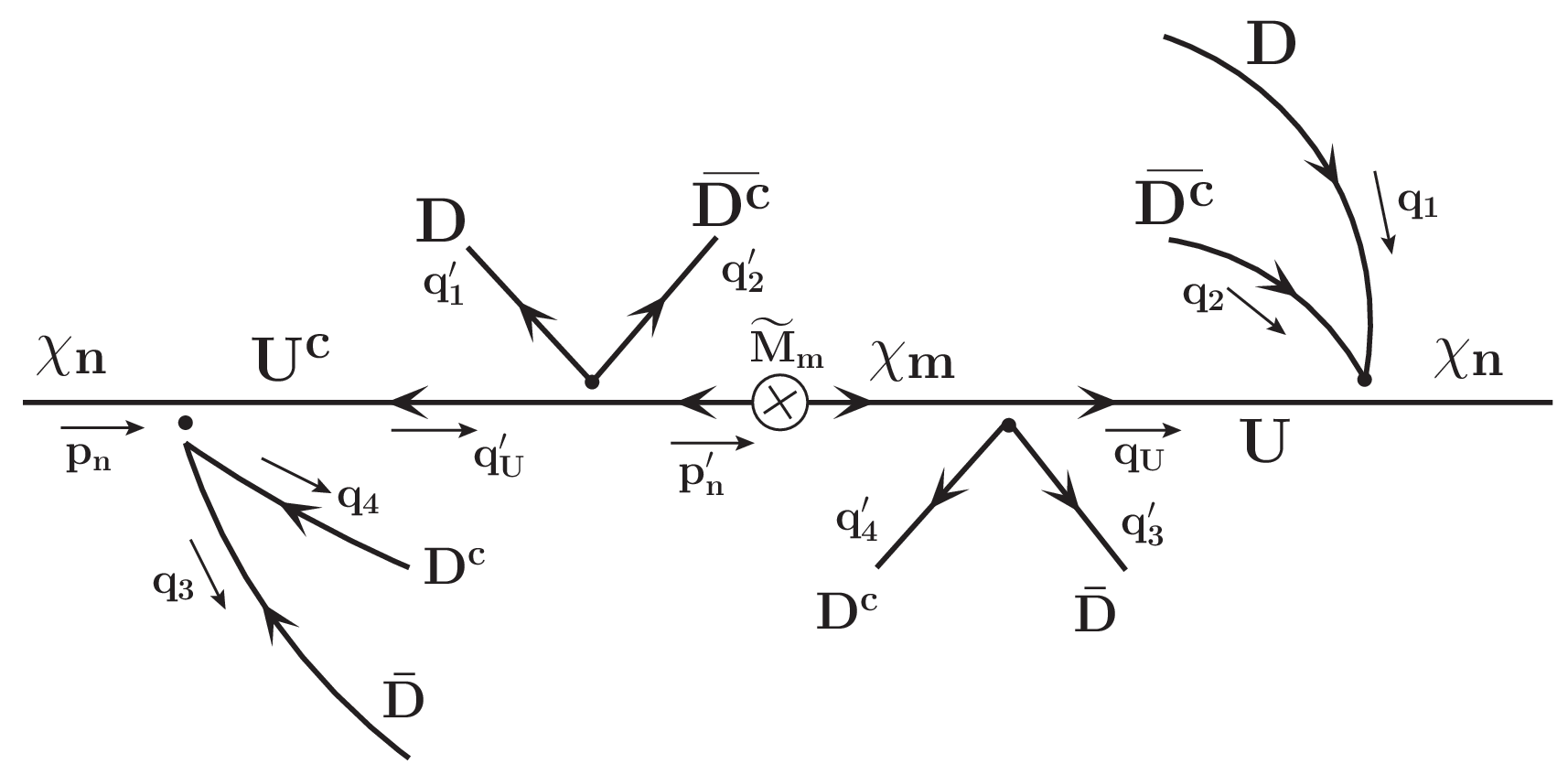}
    \caption{From this primitive diagram, the four different tree-loop interference processes A,~B,~C,~D can be generated by
      contracting the $q_i$ with the various $q'_j$ as explained in the text.
      The direction of the $q'_i$ momentum is such as to be in line with the momentum $q_j$ with which it is contracted. 
      \label{LONLOcontrGen.FIG}
    }
  \end{center}
\end{figure}

We start by deriving an expression for the $\chi-U$ fermion line in the tree-loop interference term
in coordinate space, which is obtained from
\beq
\MLONLOXU \sim \bra{0} c_n {\cal T}\left( (\widebar{U^c} \GVn^\tau \Chi_n) \, (\bar{U} \GVBm^\sigma \Chi_m) \, (\widebar{\Chi_m} \GZBm^\mu U^c) \, (\widebar{U^c} \GZn^\nu \Chi_n) \right)  c_n^\dagger \ket{0}
\eeq
where ${\cal T}$ is the time-ordering symbol, and each $(...)$ is at a different space-time point that is integrated over. 
Plugging in the mode expansion of Eq.~(\ref{psiChiModeExpn.EQ}) into this
gives us the momentum space diagram shown in Fig.~\ref{LONLOcontrGen.FIG}.
This gives the expression for $\MLONLOXU$ in momentum space given in Eq.~(\ref{ampA01hatn.EQ}).
For the conjugate process,
all fermion-line arrows are flipped in the diagram, and we have
\beq
\McLONLOXU \sim \bra{0} c_n {\cal T}\left( (\widebar{\Chi_n} \GZBn^\tau U^c) \, (\widebar{U^c} \GZm^\sigma \Chi_m) \, (\widebar{\Chi_m} \GVm^\mu U) \, (\bar{U} \GVBn^\nu \Chi_n) \right)  c_n^\dagger \ket{0} \ , 
\eeq
which yields a similar expression to that shown in Eq.~(\ref{ampA01hatn.EQ}) for $\MLONLOXU$
but with the change $G_V \leftrightarrow \bar{G}_\Lambda$. 

We obtain next the expression for the $\Dp$-loop in the tree-loop interference term.
What we need for this is 
\beq
\bra{0} a_1 a_2 {\cal T} \left( (\bar{\Dp} \bar{\widetilde{G}}_V^\sigma \Dp^c) (\bar{\Dp} \bar{\widetilde{G}}_V^\mu \Dp^c) \right) b_4^\dagger b_3^\dagger \ket{0}
\eeq
where $a_i$ is short for $a_{{\bf q}_i}$ etc. for the $\Dp$ particle,
and where, again, each $(\bar{D}\cdot D^c)$ field bilinear is at a different
space-time point that is integrated over.
We have the particular choice of the final $\bra{0} a_2 a_1$ and the initial $b_4^\dagger b_3^\dagger \ket{0}$
since our focus here is on the contribution for the process (and not the conjugate process)
that is relevant for generating the baryon asymmetry,
which are themselves to be obtained via the VV interactions, i.e. from 
\beq
-\bra{0} {\cal T} \left( (\widebar{\Dp^c_2} \widetilde{G}_V^\tau \Dp_1) (\bar{\Dp}_{3'} \bar{\widetilde{G}}_V^\sigma \Dp^c_{4'})
    (\bar{\Dp}_{1'} \bar{\widetilde{G}}_V^\mu \Dp^c_{2'}) (\widebar{\Dp^c_4} \widetilde{G}_V^\nu \Dp_3)  \right) \ket{0}
\eeq
where we have shown this in momentum space with $D_i \equiv D_{q_i}$.
This is shown diagrammatically in Fig.~\ref{LONLOcontrGen.FIG}.
The overall minus sign is due to the labeling of the momenta as in the figure. 

Putting together the color factors associated with these vertices we have
\beq
C_{01}^{(i)} = (\epsilon^{\alpha_4 \alpha_3 c'} \epsilon^{\alpha_2' \alpha_1' c'} \epsilon^{\alpha_4' \alpha_3' c} \epsilon^{\alpha_2 \alpha_1 c})/(\sqrt{6})^2\ ,
\label{clrFacGen.EQ}
\eeq
where the color index $\alpha_i$ is associated with $q_i$ etc., 
and the $1/\sqrt{6}$ factor is from the color-part of the normalization of the $\Dp\Dp\Up$ final state (see Eq.~(\ref{DDUstate.EQ})).

From the two bilinears $(\bar{\Dp}_{3'} \bar{\widetilde{G}}_V^\sigma \Dp^c_{4'}) (\bar{\Dp}_{1'} \bar{\widetilde{G}}_V^\mu \Dp^c_{2'})$,
we need two $a^\dagger$'s and two $b$'s to give us a nonzero contraction.
There are four such possibilities, namely,
$$-\bra{0} a_2 a_1 \left(  b_{3'} b_{4'} a^\dagger_{1'} a^\dagger_{2'} +  a^\dagger_{3'} a^\dagger_{4'} b_{1'} b_{2'} +
a^\dagger_{3'} b_{4'} a^\dagger_{1'} b_{2'} +  b_{3'} a^\dagger_{4'} b_{1'} a^\dagger_{2'} \right) b_4^\dagger b_3^\dagger \ket{0} \ , $$
where the momenta associated with $a^\dagger,b^\dagger$ flow outward while the momenta associated with the $a,b$ flow inward. 
These amount to attaching the different $q_i$ to the various $q'_j$ in Fig.~\ref{LONLOcontrGen.FIG}.
The direction of the $q'_i$ momentum is such as to be in line with the momentum $q_j$ with which it is contracted.
These four possibilities with the adjacent $a$'s and $b$'s contracted
(i.e. the momenta on them identified, and so too with the associated color indices) 
correspond to the processes A,~B,~C,~D respectively shown diagrammatically in Fig.~\ref{chi2DDU-LONLOCut.FIG}
where we have not shown the diagram corresponding to the fourth term above (i.e. process D)
which is just a flip of process C about the horizontal fermion-line,
i.e. process D corresponds to interchanging $q_{1'} \leftrightarrow q_{2'}$ and $q_{3'} \leftrightarrow q_{4'}$ in process C.
Any time there is a cross of the contracted lines, we get a $(-1)$ factor associated with an anticommutation of these fermionic operators.
All such $(-1)$ factors are multiplied to give us an overall sign of that term.
Note that contracting non-adjacent operators, in effect, gives 
a negative sign if it corresponds to doing an odd number of anticommutations.
There are totally four possibilities within each term including the adjacent and non-adjacent contractions, which can be viewed
as twists of the various legs in the diagrams.
Each contraction also has the associated color factor. 
Including the spinor factors, we have more fully,
\bea
-\bra{0} a_2 a_1\, \bar{v}_2 \widetilde{G}_V^\tau u_1 \left(
 b_{3'} b_{4'} a^\dagger_{1'} a^\dagger_{2'} \ \bar{v}_{3'} \bar{\widetilde{G}}_V^\sigma {u}_{4'}\ \bar{u}_{1'} \bar{\widetilde{G}}_V^\mu {v}_{2'}  +
a^\dagger_{3'} a^\dagger_{4'} b_{1'} b_{2'} \ \bar{u}_{3'} \bar{\widetilde{G}}_V^\sigma {v}_{4'}\ \bar{v}_{1'} \bar{\widetilde{G}}_V^\mu {u}_{2'} +
 \right.  \nonumber \\ \left. 
  a^\dagger_{3'} b_{4'} a^\dagger_{1'} b_{2'} \ \bar{u}_{3'} \bar{\widetilde{G}}_V^\sigma {u}_{4'}\ \bar{u}_{1'} \bar{\widetilde{G}}_V^\mu {u}_{2'}  +
  b_{3'} a^\dagger_{4'} b_{1'} a^\dagger_{2'} \ \bar{v}_{3'} \bar{\widetilde{G}}_V^\sigma {v}_{4'}\ \bar{v}_{1'} \bar{\widetilde{G}}_V^\mu {v}_{2'} 
  \right) b_4^\dagger b_3^\dagger\, \bar{u}_4 \widetilde{G}_V^\nu v_3 \ket{0} \ .
  \eea
  The initial-spin averaged total decay rate has all of the color and spin indices summed over.
  It turns out that in all cases the overall sign associated with twists within each process is undone by the color index structure.
  Thus all four twists come with the same overall sign after including color also,
  although the trace structure could be different.

  In the way of an example, consider the first term, and contract the adjacent indices associated with the $a,a^\dagger$ and $b,b^\dagger$,
  i.e. we have $1\!\!=\!\! 1', 2\!\!=\!\! 2', 3\!\!=\!\! 3', 4\!\!=\!\! 4'$,
  and since no fermion contraction lines cross, the overall sign is +1.
  The spinor factors with these contractions are
  $-\bar{v}_2 \widetilde{G}_V^\tau u_1\ \bar{v}_{3} \bar{\widetilde{G}}_V^\sigma {u}_{4}\ \bar{u}_{1} \bar{\widetilde{G}}_V^\mu {v}_{2}\
  \bar{u}_4 \widetilde{G}_V^\nu v_3$,
  which upon summing over all spins yields
  $-{\rm Tr}[\widetilde{G}_V^\tau (\qoslash + M_D) \bar{\widetilde{G}}_V^\mu (\qtslash - M_D)]\,
  {\rm Tr}[\bar{\widetilde{G}}_V^\sigma (\qfslash + M_D) \widetilde{G}_V^\nu (\qthslash - M_D)]$,
  as given in Eq.~(\ref{LDABCD.EQ}). 
  From Eq.~(\ref{clrFacGen.EQ}), the associated color factor with this is
  $C_{01}^A = \epsilon^{\alpha_4 \alpha_3 c'} \epsilon^{\alpha_2 \alpha_1 c'} \epsilon^{\alpha_4 \alpha_3 c} \epsilon^{\alpha_2 \alpha_1 c}/(\sqrt{6})^2$
  with all these indices summed over,
  which yields $C_{01}^A = 2\delta^{c'c} 2\delta^{c'c}/(\sqrt{6})^2 = 12/(\sqrt{6})^2 = 2$ as noted below Eq.~(\ref{LDABCD.EQ}).
  It turns out that all four twists yields the same overall sign (after including the signs due to the anticommutations and the color factor)
  and the same Tr[...] expression, explaining the factor of 4 for the $L_D^A$ contribution in Eq.~(\ref{LDABCD.EQ}).
  While writing the spinor products as ${\rm Tr}[...]$ we sometimes make use of the spinor identity in Eq.~(\ref{uvSpId.EQ}). 
  
  A similar computation for all the four processes and twists therein gives the expressions in Eq.~(\ref{LDABCD.EQ}),
  with the color-factors shown below that equation.
  This explicitly tells us also that the expressions for processes C and D are identical.

  Turning now to the $\Dp$-loop expressions for the conjugate process,
  we notice that the changes are only of the form $u_p \leftrightarrow v_p$,
  which could  potentially change the sign of the $M_D$ terms in the ${\rm Tr}[...]$ expressions,
  and are therefore reflected only in terms with odd powers of $M_D$.
  But such terms have the trace of an odd number of $\gamma$ matrices, which is zero.
  Therefore we conclude that the $\Dp$-loop expressions for the conjugate process are the same
  as for the process, i.e. $L^{c\,(i)}_D = L_D^{(i)}$ given in Eq.~(\ref{LDABCD.EQ}).

\subsubsection{The $\Dp\Dp$ loop function in diagram A$_1$}
\label{DDLoop.SEC}

Here we compute the $\Dp\Dp$ Loop Function $L_{34}^{\sigma\nu}$ appearing in diagram A$_1$ in
the loop contribution to the decay amplitude for $\Chi_n \to \widebar{\Dp^c} \Dp \Up$, 
discussed in Sec.~\ref{ABdec.SEC}, and shown in Fig.~\ref{chi2DDU-NLO.FIG}. 
We handle carefully the two identical $\Dp$ particles appearing in the loop.

\begin{figure}
  \begin{center}
    \includegraphics[width=0.32\textwidth] {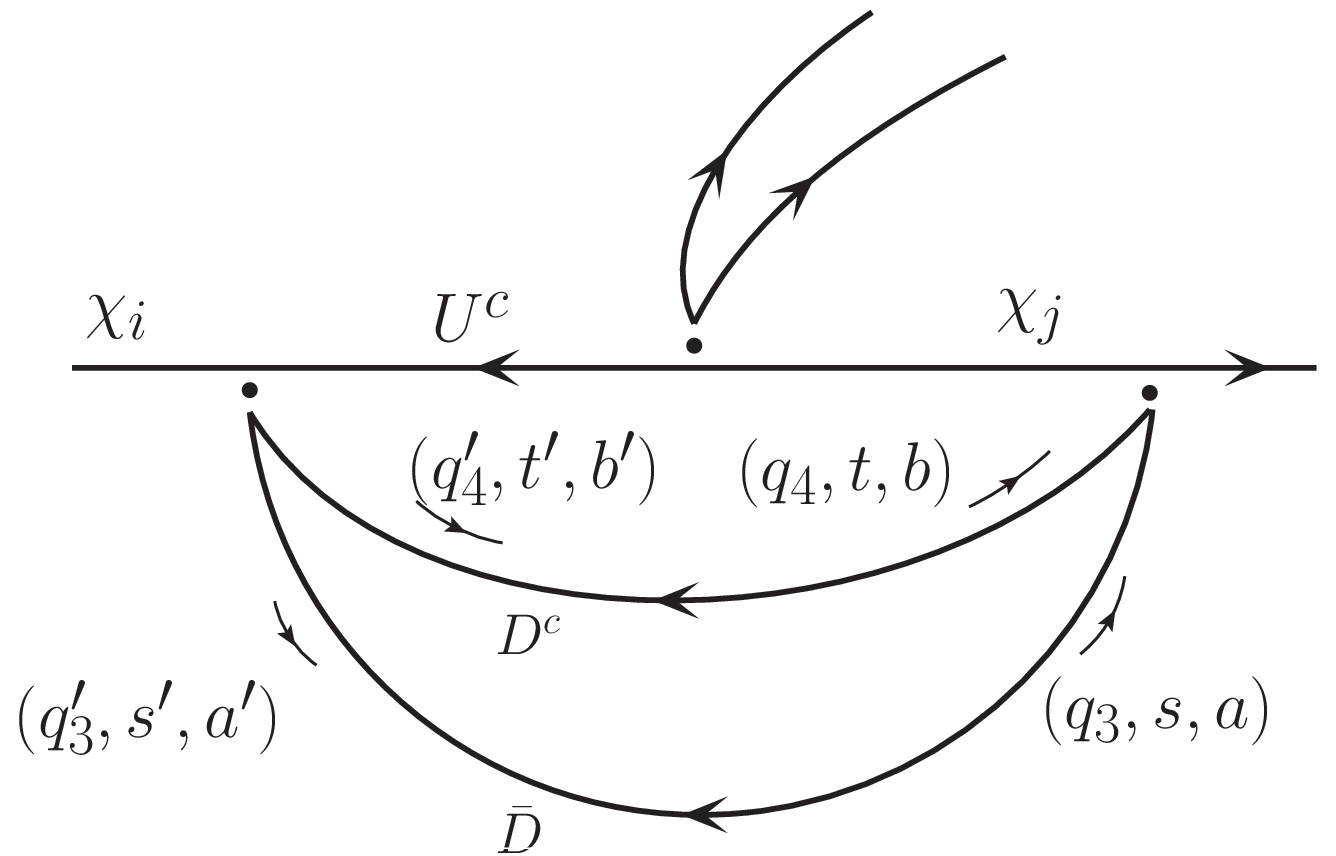}
    \hspace*{0.25cm}
    \includegraphics[width=0.32\textwidth] {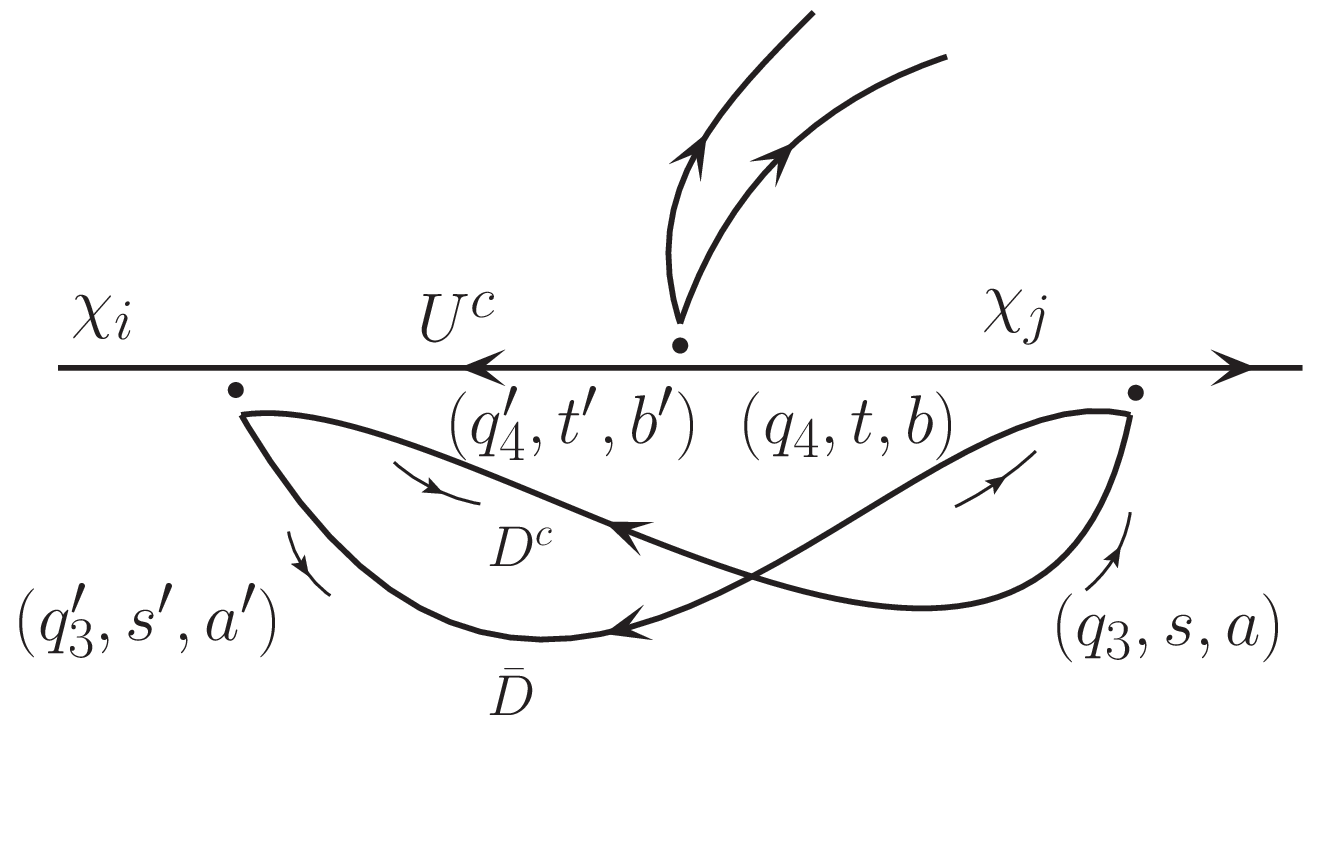}
    \caption{The first (left) and second (right) contraction for the loop $\Dp\Dp$ loop, 
      showing the routing of momentum, the spin index, and the color index, as $(q,s,a)$ respectively.
      \label{LO-NLO-DDloop.FIG}
    }
  \end{center}
\end{figure}

For the
general Lorentz structure of the interaction alluded to in Appendix~\ref{DDM12.SEC},
the loop function in coordinate space is 
$$
L_{34} \propto \matel{{\cal T}\ (\bar{D}_{a} 
  \bar{\widetilde{\Gamma}} D^c_{b} \epsilon^{abc})_y (\bar{D}^c_{b'} 
  \widetilde{\Gamma} D_{a'} \epsilon^{a'b'c'})_w }{0}{0},
$$
where ${\cal T}$ is the time-ordering symbol,
and the two operator bilinears are evaluated at different space-time points $y,w$.
The $D,D^c$ fields when contracted yield two propagators $\Delta_{yw}$ connecting the points.
From the usual mode expansion for the fields, in momentum space, we find the term
$$
L_{34} \propto \matel{{b_{\bf q_3}^{as}} {b_{\bf q_4}^{bt}} {b_{\bf q_4'}^{b't'}}^\dagger {b_{\bf q_3'}^{a's'}}^\dagger}{0}{0}
 (\bar{v}_{q_3}^{s} 
 \bar{\widetilde{\Gamma}} u_{q_4}^{t}) (\bar{u}_{q'_4}^{t'} 
 \widetilde{\Gamma} v_{q'_3}^{s'}) \epsilon^{abc} \epsilon^{a'b'c'} \ .
$$
From the first factor it is evident that two contractions are possible, which is due to the two $\Dp$ particles being identical. 
The two contractions yield
$$
\delta({{\bf q}_4 - {\bf q}'_4}) \delta({{\bf q}_3 - {\bf q}'_3}) \delta^{aa'} \delta^{ss'} \delta^{bb'}  \delta^{tt'} 
- \delta({{\bf q}_3 - {\bf q}'_4}) \delta({{\bf q}_4 - {\bf q}'_3}) \delta^{ab'} \delta^{st'} \delta^{ba'} \delta^{ts'} \ ,
$$
and the relative minus sign in the second term is from two $b$'s being (anti)commuted before contracting. 
These contractions are diagrammatically shown in Fig.~\ref{LO-NLO-DDloop.FIG}. 
Enforcing the $\delta$ functions and summing over $s,t$,
we get\footnote{The
  integration over the loop momenta $q_3,q'_U$ is implied but is not explicitly written down in our discussion here.
  We have also not included here the $i/\Lambda^{2}$ factor, nor the denominators of the two fermion propagators,
  nor a factor of $\epsilon^{abc} \epsilon^{abc'}$.
  All of these will be included elsewhere.}
\bea
L_{34} &=& \sum_{s,t}
(\bar{u}_{q_4}^{t} \widetilde{\Gamma} v_{q_3}^{s} \bar{v}_{q_3}^{s} \bar{\widetilde{\Gamma}} u_{q_4}^{t}  + 
\bar{u}_{q_3}^{s} \widetilde{\Gamma} v_{q_4}^{t} \bar{v}_{q_3}^{s} \bar{\widetilde{\Gamma}} u_{q_4}^{t} ) 
\label{L34uvgen.EQ} \\
  &=&
          {\rm Tr}[\widetilde{\Gamma} (\qslash_3 - M) (\bar{\widetilde{\Gamma}}-\widetilde{\Gamma}^c) (\qslash_4 + M) ]  \, 
\label{L34gen.EQ}
\eea
where we have used the spinor identity of Eq.~(\ref{uvSpId.EQ}) and  
note that an extra minus sign has flipped back the sign of the second term in Eq.~(\ref{L34uvgen.EQ})
due to interchanging the color indices $ab$. 

In particular, for the $VV$ interaction of Eq.~(\ref{LIntVVMB.EQ}), we get
\bea
    \Ltf^{\sigma\nu} &=&
    2\, {\rm Tr}[ \GtV^\nu (\qslash_3 - M) \GtVB^\sigma (\qslash_4 + M) ]
    = 2 |\tilde{g}|^2 \, {\rm Tr}[ \gamma^\nu (\qslash_3 - M) \gamma^\sigma (\qslash_4 + M) ]   \ .
     \label{L34VV.EQ}
\eea
We note that the RHS is in fact symmetric under $(q_3 \leftrightarrow q_4)$ interchange 
since the only term which breaks this symmetry, namely the term linear in $M_D$,
is zero since the trace of an odd number of $\gamma$-matrices is zero.
Similarly, for the conjugate process we find
\bea
    \Lctf^{\sigma\nu} &=&
     2\, {\rm Tr}[ \GtVB^\nu (\qslash_3 - M) \GtV^\sigma (\qslash_4 + M) ]  
   =  \Ltf^{\sigma\nu} \ .
           \label{Lc34VV.EQ}
\eea
          
We note the similarity of the above $\Ltf$ with the $\MotSq$ in Eq.~(\ref{M12sqVV.EQ}).
Comparing the direction of momenta and making the translation $q_3 \to -q_1$, $q_4 \to -q_2$,
and $(\GtVB) \leftrightarrow \GtV$,
we obtain $\Lot$
and see that the mapping is exact.
We thus can think of the $\Dp\Dp$ final state equivalently as a cut of the loop process $\MotSq \equiv \Lot$.

\subsubsection{The $\Chi_n \bar\Chi_m$ 2-point function counterterms}
\label{Chinm2ptFcnDet.SEC}

We have already discussed in Appendix~\ref{Chinm2ptFcn.SEC}
the important aspects of the renormalization and subtraction of the $\Chi_n \bar\Chi_m$ 2-point function.
Although not explicitly needed for our calculation of the baryon asymmetry in Sec.~\ref{oneOpAsymb.SEC}, for completeness,
we present next the counterterms and the subtraction conditions for the hermitian part, i.e. the non-dissipative or PV part, of the 2-point function
to demonstrate the subtractions that keep us consistently in the mass basis when loop corrections generate off-diagonal terms.
We can split the hermitian counterterm $\delta\Sigma_{nm}$ into individual counterterms and write
$\delta\Sigma_{nm}\!\! =\!\! \delta\Omega_{nm} + \delta\Omega^c_{nm}$,
and we can try to pick each piece to cancel the loop contributions above at the subtraction point $p_{0}$.

For $n\!\!=\!\!m$, the 2-point function $\Sigma_{nn} (p_n)$ has both legs the same particle and onshell. 
We impose a renormalization condition such that the loop contribution to the hermitian part of the 2-point function is
canceled by a counterterm at the on-shell subtraction points $p_n\!\! =\!\! p_{n_0}$ (for $n\!\!=\!\!1,2$), with
$p_{1_0}^2,p_{2_0}^2\!\! =\!\! M_{1,2}^2$,
leaving $M_n\!\! =\!\! M_{1,2}$ uncorrected and as the renormalized masses, consistent with being in the mass-basis.
We can take 
$\delta\Omega_{nn}\!\! =\!\! - \PV(\Omega_{nn}(p_{n_0}))$ and $\delta\Omega^c_{nn}\!\! =\!\! - \PV(\Omega^c_{nn}(p_{n_0}))$,
which implies $\delta\Sigma_{nn}\!\! =\!\! - \PV(\Sigma_{nn}(p_{n_0}))$. 
With this subtraction, there is no loop contribution to the PV part
for $nm\!\! =\!\! \{11,22\}$ at the subtraction point $\{p_{1_0},p_{2_0}\}$ respectively
as they are entirely canceled by the counterterms.

For $n\!\!\neq\!\! m$, with the $\Chi_n$-leg on-shell but the $\Chi_m$-leg necessarily off-shell.
Due to the Majorana nature of $\Chi_n$, we can show generally that
$\bar\Chi_m \Sigma_{nm} \Chi_n\!\! =\!\! \bar\Chi_n (C\gamma^0 \Sigma_{nm}^T \gamma^0 C) \Chi_m$ (no sum on $n,m$),
and therefore we must have the relation $\Sigma_{mn}\!\! =\!\! C\gamma^0 \Sigma_{nm}^T \gamma^0 C$.
In our theory with the VV interaction, we find that $\Omega_{mn}\!\! =\!\! C\gamma^0 \Omega_{nm}^{cT} \gamma^0 C$,
and the above relation is satisfied in an interesting way.
Thus, for Majorana fermions, $\delta\Sigma_{nm}$ and $\delta\Sigma_{mn}$
cannot be independently chosen to simultaneously cancel both $\PV(\Sigma_{21}(p_{2_0}))$ and $\PV(\Sigma_{12}(p_{1_0}))$.
We can at best cancel one of them, say $\PV(\Sigma_{12}(p_{1_0}))$ by the counterterm choice
$\delta\Sigma_{12}\!\! =\!\! - \PV(\Sigma_{12}(p_{1_0}))$,
suggesting $\delta\Omega_{12}\!\! =\!\! - \PV(\Omega_{12}(p_{1_0}))$ and $\delta\Omega^c_{12}\!\! =\!\! - \PV(\Omega^c_{12}(p_{1_0}))$. 
Owing to the earlier relation, this then fixes $\delta\Sigma_{21}$ also,
which individually fixes $\delta\Omega_{21}\!\! =\!\! - C\gamma^0 \PV({\Omega_{12}^c}^{\!\!T}(p_{1_0}))\gamma^0C\!\! =\!\! - \PV(\Omega_{21}(p_{1_0}))$ and
$\delta\Omega^c_{21}\!\! =\!\! - C\gamma^0 \PV({\Omega}^T_{12}(p_{1_0}))\gamma^0C\!\! =\!\! - \PV(\Omega^c_{21}(p_{1_0}))$ implying
$\delta\Sigma_{21}\!\! =\!\! - \PV(\Sigma_{21}(p_{1_0}))$. 

Finally, we can write the renormalized 2-point function as 
\beq
  \Omega_{nm}^{(ren)}(p) = \Omega_{nm}(p) - \Omega_{nm_s} \ ,  
\eeq
where the subtraction term for $n\!\! = \!\! m$ is $\Omega_{nn_s}\!\! =\!\! \PV(\Omega_{nn}(p_{n_0}))$ with $p_{n_0}\!\! =\!\! M_n^2$,
while for $n\!\! \neq \!\! m$ it is $\Omega_{nm_s}\!\! =\!\! \PV(\Omega_{nm}(p_{1_0}))$ with $p_{1_0}\!\! =\!\! M_1^2$.
  These now include both the discontinuity and the regular pieces of the 2-point function, 
  with a subtraction for the regular part but none for the discontinuity.
  We similarly define the subtracted $\Omega_{nm}^{c\, (ren)}(p)$.  

We can write the $\Omega_{mn}$ as  
$\Omega_{mn}\!\! =\!\! J^{\mu\nu}(M_\Up) L_{\mu\nu}(M_\Dp)$,
where the $J$ is the $\Chi\Up$ fermion line, and the $L$ is the $\Dp$ loop,
and similarly $\Omega^c_{mn}\!\! =\!\! J^{c\mu\nu}(M_\Up) L^c_{\mu\nu}(M_\Dp)$.   
  We can define dimensionless versions of the above loop functions that are functions of the scaled (hatted) masses and momenta defined in
  Sec.~\ref{oneOpAsymb.SEC}, namely $\hat{p}_i\!\! \equiv\!\! p_i/M_n$ and $\hat{M}_i\!\! \equiv\!\! M_i/M_n$. 
  We thus write  
  $\Omega_{nm}\!\! =\!\! (M_n^5/\Lambda^4) \hat\Omega_{nm}$
  with $\hat\Omega_{nm}\!\! =\!\! \hat{J}_{nm}^{\mu\nu}(\hat{M}_\Up) \hat{L}_{nm\mu\nu}(\hat{M}_\Dp)$,
  where
  the $1/\Lambda^4$ follows from the 2 VV interactions, implying the $M_n^5$ scaling for the $\Omega_{nm}$ since it has mass dimension +1.
  We then have 
  \beq
  \Omega_{nm}^{(ren)}(p)
  = (M_n^5/\Lambda^4)\, \left[\hat{J}_{nm}(p/M_n, M_\Up/M_n) \hat{L}_{nm}(p/M_n, M_\Dp/M_n) - (M_1/M_n)^5 \hat{J}_{nm_s} \hat{L}_{nm_s} \right] \ ,
 \eeq
 where we have defined
  $\hat{J}_{nm_s}\! \equiv\! \PV(\hat{J}_{nm}(p_{1_0}/M_1, M_\Up/M_1))$,
  and
  $\hat{L}_{nm_s}\! \equiv\! \PV(\hat{L}_{nm}(p_{1_0}, M_\Dp/M_1))$.
  We similarly define 
  $\Omega_{nm}^c\!\! =\!\! (M_n^5/\Lambda^4) \hat\Omega_{nm}^c$
  with $\hat\Omega_{nm}^c\!\! =\!\!  \hat{J}_{nm}^{c\mu\nu} \hat{L}^c_{\mu\nu} $,
  and write an analogous expression for $\Omega_{nm}^{c\, (ren)}(p)$.

\subsection{Contributions from diagrams without arrow clash}
\label{noArrowClash.SEC}

Here we discuss contributions to the baryon asymmetry from diagrams without a clash in the arrow depicting baryon number charge flow,
and show that these do not generate any rate difference. 

\subsubsection{Single operator contribution}
\label{noArClSingOp.SEC}
\begin{figure}
  \begin{center}
    \includegraphics[width=0.65\textwidth]{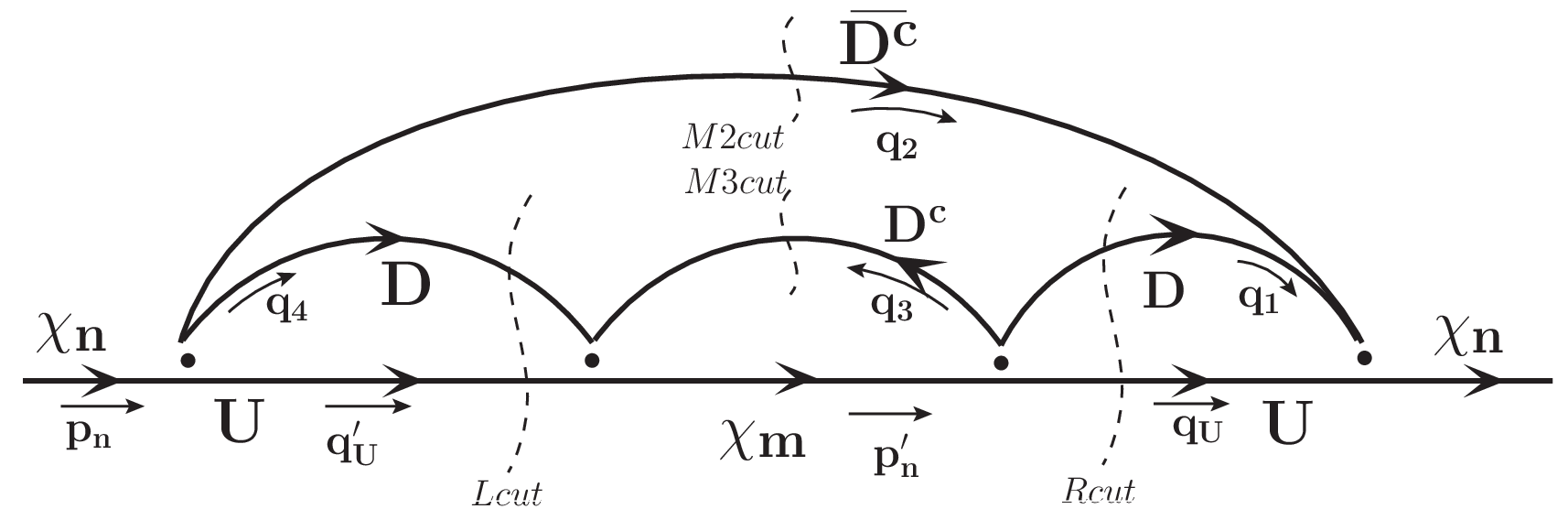}
  \end{center}    
  \caption{
    The diagram representing the tree-loop interference term 
    that does not have a clash of fermion number flow (i.e. no arrow clash), 
  where the arrows on the fermion lines show the direction of baryon number flow.
  The dashed line shows the cut propagators.
\label{chi2DDU-LONLO-diagJ.FIG} }
\end{figure}

Consider for example, the diagram shown in Fig.~\ref{chi2DDU-LONLO-diagJ.FIG}, and also similar diagrams with the $\Dp^c$ (also $\widebar{\Dp^c}$) and $\Up$ interchanged.
Although it may appear to contribute, upon a more careful analysis, we find that it does not.
We explain the reasons below.
The figure shows the diagram that corresponds to the tree-loop interference contribution $\ampA_{01} = \ampA_1 \ampA_0^*$ for the process,
where $\ampA_0$ is the LO amplitude and $\ampA_1$ is the loop amplitude.
The corresponding contribution $\ampA_{01}^c$ for the conjugate process is got by reversing all the arrows
changing all fields to conjugate fields and picking up the appropriate couplings.
The dashed lines in the figure shows the cuts we make in the diagram.
The ones labeled $Rcut$ together with $M2cut$ give the 3-body phase space for $\ampA_{01}$, where the part of the diagram to the right of these cuts
is nothing but $\ampA_0^*$ and to the left of these cuts $\ampA_1$.
As explained earlier, the ``weak phase'' (which flips in $\ampA_{01}$ vs. $\ampA_{01}^c$) is provided by complex couplings,
and the ``strong phase'' (which does not flip) by making the cuts labeled $Lcut$ and $M3cut$ in $\ampA_1$.
Taking into account the $i$'s in the vertices, propagators and the cuts, overall, we find an $i$ that remains and acts as the strong phase.

We recall from Sec.~\ref{genAB.SEC} that the baryon asymmetry is given by 
$\AsymB \propto\, 2\, {\rm Re}(\ampA_{01} - \ampA_{01}^c) = 2\, {\rm Re} (\Delta \ampA_{01}) = \Delta \ampA_{01} + \Delta \ampA_{01}^*$.
We realize that $\ampA_{01}^*$ corresponds to interchanging the interpretations of $Lcut$ and $Rcut$,
i.e. viewing the $Lcut$ together with $M2cut$ as giving the 3-body phase-space with the piece to the left of it being
$\ampA_0$ and the right of if being $\ampA_1^*$,
and the $Rcut$ together with $M3cut$ as contributing the strong phase to $\ampA_1^*$.
This different interpretation amounts to interchanging $q_1 \leftrightarrow q_4$ and $q_U \leftrightarrow q_U'$.
We therefore add this interchanged contribution and consider
$2 {\rm Re}(\ampA_{01}) = \ampA_{01}(q_1, q_U, q_4, q_U') + \ampA_{01}(q_1 \leftrightarrow q_4, q_U \leftrightarrow q_U') \equiv \ampA_{01}^{(L\leftrightarrow R)}$.
We take the difference between this and the corresponding amplitude for the conjugate process and obtain
$\Delta \ampA_{01}^{(L\leftrightarrow R)}$.

We split up the $\Delta \ampA_{01}^{(L\leftrightarrow R)}$ into different Dirac traces where we pick either the $\pslash$ or the $M$
in each of the four fermion propagator numerators in the $\Chi-\Up$ fermion line
(the $\Dp$ loop yields another Dirac-trace which we include in full in each of them),
namely,
(i) $M_n, M_m, M_U$ independent contribution, 
(ii) $M_m M_U$ contribution, 
(iii) $M_n M_U$ contribution,
(iv) $M_U M_U$ contribution.
The rest are all zero in the difference.
Given the remaining strong phase $i$ that we identified, we seek the Im part of each of them, 
which has two possibilities
${\rm Re}({\rm couplings}){\times}{\rm Im}({\rm Tr}[...])$, or, ${\rm Im}({\rm couplings})\times{\rm Re}({\rm Tr}[...])$.
The Im(Tr[...]) yields four 4-momenta contracted with the $\epsilon^{\mu\nu\sigma\tau}$,
which evaluated in the proper frame (see Sec.~\ref{12U34UPVars.SEC}) picks up the $0^{th}$ component ($E$) of one of the momenta involved
times a scalar triple product (of the other 3-momenta involved). 
This is odd in the azimuthal angle $\phi_{3}$ or $\phi_{U'}$
(we set the azimuthal angles $\phi_1$ and $\phi_U$ to zero without loss of generality as we are interested in the
spin-averaged decay rate of the unpolarized $\Chi_n$ in its rest frame),
and when integrated over these azimuthal angles, the decay rate difference between the process and conjugate process
due to these $\epsilon^{....}$ terms is zero.
Another way to see this is to consider this in the $\Chi_n$ rest frame.
By 3-momentum conservation, the three final-state 3-momenta lie in a plane and their scalar triple product is zero.
Now, the remaining ${\rm Im}({\rm couplings})\times{\rm Re}({\rm Tr}[...])$ contributions we find are all zero also,
where (ii) and (iii) become zero only in the $L \leftrightarrow R$ contribution discussed above.
Thus, in summary, the diagram of Fig.~\ref{chi2DDU-LONLO-diagJ.FIG} does not contribute to $\AsymB$.
We therefore focus on the diagrams with an arrow clash as detailed in the main text. 

\subsubsection{Multiple operator contribution}
\label{noArClMulOp.SEC}
We consider next diagrams $A_1^{\prime(2),(3)}$ in Fig.~\ref{chi2DDU-NLO-modA.FIG}
and show that no asymmetry arises from these,
as we could anticipate from there being no arrow clash in these diagrams. 
Putting the cut propagators on-shell we find that the decay rate difference between the process and conjugate process
$\Delta {\Gamma}^{(2),(3)}_{01} \propto \Delta {\cal A}^{\prime (2),(3)}_{01}$
can be written in terms of couplings times a combination of the two integrals
\beq
\Delta \hat{\cal I}'_{a,b} \equiv \int \frac{d^4 q_3}{(2\pi)^4} \ p_n^\mu q_1^\nu q_2^\sigma q_3^\tau \epsilon_{\mu\nu\sigma\tau} \left\{ p_n \cdot (q_1 - q_2)\, ,\ q_3 \cdot (q_1 - q_2)  \right\}  \ .
\eeq
These integrals involve Lorentz invariant dot-products of momenta, and each dot-product can be evaluated in a convenient frame,
as already discussed in Sec.~\ref{oneOpAsymb.SEC}.
Much of the techniques of that section apply here by realizing that the following simplifications in the variables of that section help us get results here:
identify $q_U' \equiv q_U$, identify the $*$-frame and $\SSt$-frame, integration is now only over $q_3$ (and {\em not} also over $q_U'$),
and applying the cuts discussed above imply that $\qthr$ is not independent anymore. 
Making these changes, we find that the integrands of these master integrals are proportional to $\sin{\phi_3^*}$,
and therefore when integrated over $\phi_3^*$ to get the total rate, these integrals are zero.
The same conclusion can be reached by considering the dot products in the $\Chi_n$ rest-frame. 
The first dot product is a scalar triple product of $\qoV,\qtV,\qthV$
that is odd with respect to flipping the direction of $\qthV$.
This is sufficient to show that $\Delta \hat{\cal I}'_{a}$ is zero.
Now for showing $\Delta \hat{\cal I}'_{b}$ is zero,
we observe that the integrand is odd with respect to flipping $\qthV,\qoV,\qtV$ simultaneously, and thus the integral is zero
after the integration over phase space is also done.
We thus find no integrated rate asymmetry from diagrams $A_1^{\prime(2),(3)}$.

\section{The Phase-space and Loop Integration Variables}
\label{12U34UPVars.SEC}

Here we arrive at our choice of independent variables for the 3-body phase-space momenta
and the loop momenta
in the decay process $\Chi_n(p_n) \to \Dp(q_1) \Dp(q_2) \Up(q_U)$,
and write down the phase-space and loop-integration measures. 
Our notation is that $q_i$ is a 4-vector, while ${\bf q}_i$ is the corresponding 3-vector
with 3-magnitude (radial component) ${q_i}_r = |{\bf q}_i|$.
An on-shell particle has $E_i = \sqrt{{q_i}^2_r + M_i^2}$.
All final state particles are on-shell;
the $\Dp,\Up$ particles in loops are also on-shell
given our focus on computing the baryon asymmetry as explained in the text.

\subsection{The 3-body phase-space variables and measure (for $\Chi$ decay)}
\label{3PS12Vars.SEC}

We start by writing the 3-body phase-space integral of the matrix element as a cut of the corresponding loop diagram as 
\beq
\int d\Pi_3\, |{\cal A}|^2 = \int \frac{1}{2} \frac{d^4q_U}{(2\pi)^4} \frac{d^4q_1}{(2\pi)^4} \frac{d^4q_2}{(2\pi)^4} (2\pi)^3 \delta^{\rm cut}_{12U}\,
(2\pi)^4 \delta^4 (p_n - (q_1 + q_2 + q_U)) |{\cal A}|^2 \ , \quad
\label{ampPSMEDefn.EQ}
\eeq
where we put the three propagators on-shell by replacing them with
\footnote{
  The $i$ is omitted in the expressions for $d\Pi_3$ here but is included elsewhere in the text. }
$(-2\pi i)^3 \delta^{\rm cut}_{12U}$
where $\delta^{\rm cut}_{12U} \equiv \delta(q_U^2 - M_U^2) \delta(q_1^2 - M_1^2) \delta(q_2^2 - M_2^2)$,
where for the decay channel, $M_1 = M_2 = M_D$. 
We include a factor of $1/2$ to account for the two identical $\Dp$ in the final state. 
We deal with the kinematical aspects of the phase-space here. 

At the outset, we choose to enforce the overall energy-momentum conserving $\delta$-function in Eq.~(\ref{ampPSMEDefn.EQ})
by performing the $d^4q_2$ integral, which sets $q_2 = p_n - (q_1 + q_U)$, 
and giving the form of Eq.~(\ref{3PS1U.EQ}).
This leaves 8 integration variables,
with the measure
\vspace{-0.25cm}
\begin{center}
  $d^4q_U d^4q_1 = dq_U^0 dc_U d\phi_U d\qUr \qUr^2 dq_1^0 dc_1 d\phi_1 d\qor \qor^2$,
\end{center}
\vspace{-0.25cm}
where $c_{i}\equiv \cos{\theta_{q_i}}$ and $\phi_i \equiv \phi_{q_i}$.
We have
$\delta(q_i^2 - M_i^2) = \delta({q_i^0}^2 - E_i^2) = \delta({q_i^0} - E_i)/|2 E_i|$
where $E_i^2 = {q_i}_r^2 + M_i^2$ for each of the $q_U,q_1$, picking the positive root for a physical on-shell particle. 

The Lorentz invariance of each factor allows us to take the $d^4q_U$ and $d^4q_1$ in {\em different} frames.
A convenient frame to consider is what we call as the *-frame in which $\pnV^* = \qUV^*$.
In this frame, 3-momentum conservation implies $\qtV^* = -\qoV^*$, i.e. these are back-to-back.
The advantage of this frame is that the end-points of $\qor^{\!*}$ are independent of the angular variables.

For a given 4-vector in the lab-frame, to obtain the corresponding 4-vector in the *-frame, we make a Lorentz transformation as follows. 
We first form the 4-vector $k_B \equiv (p_n-q_U) = (E_B,\kBV)$.
The boost rapidity $\eta$ that takes $k_B$ to $k_B^* \equiv (M_B,{\bf 0})$, i.e. to the $k_B$ rest-frame,
is given by $t_\eta \equiv \tanh\eta = \kBr/E_B$.
The required Lorentz transformation is a boost with this rapidity in the direction of the unit-vector $\kBVhat = \kBV/\kBr$.
More explicitly, consider any lab-frame 4-vector $p=(E_p,\pV)$ for which we seek the corresponding *-frame 4-vector $p^*=(E_p^*,\pV^*)$. 
We first decompose as $\pV = \pparV + \pperpV$, where $\pparV = (\pV\!\cdot\!\kBVhat)\, \kBVhat$ and $\pparr = |\pparV|$, and, $\pperpV = \pV - \pparV$.
Then, $\pV^* = \pparV^* + \pperpV^*$ with $\pparV^* = \pparr^{\!*}\, \kBVhat$, and  
\beq
\bmat E_p^* \\ \pparr^{\!*} \emat = {\cal B}_\eta \bmat E_p \\ \pparr \emat 
      {\rm where}\ {\cal B}_\eta \equiv \bmat c_\eta & -s_\eta \\ -s_\eta & c_\eta \emat \ ; \quad \pperpV^* = \pperpV \ ,  \nonumber
\eeq
with $s_\eta = \sinh\eta = t_\eta/\sqrt{1-t_\eta^2}$ and $c_\eta = \cosh\eta = 1/\sqrt{1-t_\eta^2}$.
The inverse transformation that takes a *-frame 4-vector to the lab-frame is obtained the same way, but with the rapidity $-\eta$. 

Using the Lorentz invariance of each measure as mentioned above, 
we take the $d^4q_U \delta(q_U^2 - M_U^2)$ in the lab-frame (i.e. the $\Chi$ rest-frame for decays),
and the $d^4q^*_1 \delta({q^*_1}^2 - M_1^2)$ and the remaining $\delta({q^*_2}^2 - M_2^2)$ in the *-frame.
The $dq_U^0$ and $d{q^*_1}^0$ integrals can be performed enforcing the first two $\delta$-functions, leaving
\beq
d\Pi_3\, = \frac{1}{(2\pi)^5} \frac{1}{2} dc_U d\phi_U d\qUr \frac{\qUr^2}{2 E_U} d\phi^*_1 dc^*_1 d\qorSt \frac{\qorSt^2}{2 E^*_1}   \delta({q^*_2}^2 - M_2^2) \ .
\eeq
We write the remaining $\delta-$function as $\delta({q^*_2}^2 - M_2^2) \equiv \delta(f(\qorSt))$, viewing the argument now as a function of $\qorSt$,
i.e. $f(\qorSt) = {q^*_2}^2 - M_2^2 = {E_2^*}^2 - \qtrSt^2 - M_2^2$,
with $E_2^* = E_n^* - E_U^* - E_1^*(\qorSt)$ and $\qtrSt = \qorSt$, owing to $\pnVSt = \qUVSt$. 
We have $\delta(f(\qorSt)) = \delta(\qorSt-{\qorSt}_z)/|df/d\qorSt|$ where ${\qorSt}_z$ is the zero of $f$, and the denominator is evaluated at the zero. 
We find that
${\qorSt}_z = (M_B/2)\{ [1+(M_1^2-M_2^2)/M_B^2]^2 - 4 M_1^2/M_B^2 \}^{1/2}$, 
$|df/d\qorSt| = 2(\qorSt/E_1^*)(E_n^* - E_U^*)$, 
and
$M_B^2 = (E_n^* - E_U^*)^2 = (p_n-q_U)^2$ being a Lorentz invariant quantity can be computed in any frame.
We can then perform the $d\qorSt$ integral enforcing the last $\delta-$function which sets ${\qorSt} = {\qorSt}_z$,
and we denote henceforth the zero simply as $\qorSt$.
We finally have
\beq
d\Pi_3 = \frac{1}{(2\pi)^5} \frac{1}{2} dc_U d\phi_U d\qUr d\phi^*_1 dc^*_1 \frac{\qUr^2}{2 E_U} \frac{\qorSt}{4 (E_n^* - E_U^*)} \ ,
\label{dPi3meas.EQ}
\eeq
in terms of the five independent 3-body phase-space variables $c_U,\phi_U,\qUr,c^*_1,\phi^*_1$.

For the decay process, in the $\Chi$ rest-frame, the matrix element will be independent of $\phi_U$ and the $d\phi_U$ integral just yields $2\pi$,
and we can include this factor and conveniently set $\phi_U = 0$.
The limits of integration for the independent variables are straightforward, except for the $\qUr$, which is bounded above.
Clearly, the minimum value of $\qUr$ is $0$, i.e. $\qUr^{(min)} = 0$.
We can find the maximum value $\qUr^{(max)}$ straightforwardly in the restframe of $\Chi$ i.e. with $\pnr=0$, as follows.  
From the Lorentz invariance of $M_B^2$, we can also write $M_B^2 = (p^*_n-q^*_U)^2 = (q_1^* + q_2^*)^2 = (E_1^* + E_2^*)^2$
where we have used 4-momentum conservation and the fact that $\qoVSt+\qtVSt=0$.
We also have $M_B^2 = (E_n^* - E_U^*)^2 = (p_n-q_U)^2 = M_n^2 + M_U^2 -2(E_n E_U - \pnr\qUr c_{nU})$ with $c_{nU} = \pnV\!\cdot\!\qUV/(\pnr\qUr)$,
and for $\pnr=0$, we see that $\qUr^{(max)}$ is attained when $M_B$ takes its minimum value,
which is for $\qorSt=\qtrSt=0$, and is $M_B^{(min)} = (M_1+M_2)$.
From this we find $\qUr^{(max)} = (M_n/2)\{ [1+(M_U^2-(M_1+M_2)^2)/M_n^2]^2 - 4 M_U^2/M_n^2 \}^{1/2}$. 
Thus, the $\qUr$ domain of integration is $\qUr \in (0,\qUr^{(max)})$, while the other integrations are over the usual full range.

\subsection{The loop variables and measure (for $\Chi$ decay)}
\label{loop34Vars.SEC}

This analysis closely parallels
that of the 3-body phase-space variables discussed in Sec.~\ref{3PS12Vars.SEC}.

\subsubsection{Two-loop integral measure}
\label{dPi3Pmeas.SEC}
We have the two-loop integral measure (such as for Fig.~\ref{chi2DDU-LONLOCut.FIG}) with three cut propagators given by\footnote{
  The $i$ is included elsewhere in the text.}
\beq
d\Pi'_3 = \frac{1}{2} \frac{d^4q'_U}{(2\pi)^4} \frac{d^4q_3}{(2\pi)^4} \frac{d^4q_4}{(2\pi)^4} (2\pi)^3 \delta^{\rm cut}_{34U} \,
(2\pi)^4 \delta^4 (p_n - (q_3 + q_4 + q'_U)) \ ,
\eeq
with 
$\delta^{\rm cut}_{34U} = \delta({q'_U}^2 - M_U^2) \delta(q_3^2 - M_3^2) \delta(q_4^2 - M_4^2)$ with $M_3=M_4=M_D$. 
We have shown also the overall 4-momentum conserving $\delta-$function,
which we enforce at the outset by performing the $d^4q_4$ integral, which sets $q_4 = p_n - q_3 - q'_U$,
and yields
$d\Pi'_3 = (1/2)\, (1/(2\pi)^5) d^4q_3\, d^4q'_U\, \delta^{\rm cut}_{34U}$.

We follow a procedure identical to the 3-body phase-space discussion in Sec.~\ref{3PS12Vars.SEC}, but with the following changes:
$q_U\to q'_U$, $q_1\to q_3$, $q_2\to q_4$, $M_1\to M_3$, $M_2\to M_4$, 
forming $k'_B \equiv (p_n-q'_U) = (E'_B,\kBPV)$, and analogous to the *-frame earlier as the rest-frame of $k_B$,
we now construct the $\SSt$-frame as the rest-frame of the $k'_B$, i.e. ${k'_B}^\SSt = (M'_B,{\bf 0})$, with ${M'_B}^2 = {k'_B}^2$.
In the $\SSt$-frame the $\qthV^\SSt = -\qfV^\SSt$, i.e. these are back-to-back.
The boost rapidity that takes us from the lab-frame to the $\SSt$-frame is now $t'_\eta \equiv \tanh\eta' = \kBPr/E'_B$, 
and the corresponding $p^\SSt = (E_p^\SSt,\pV^\SSt)$ with $\pV^\SSt = \pparV^\SSt + \pperpV^\SSt$ is constructed in an analogous way.
We use the Lorentz invariance of each factor in the integration measure and take $q'_U$ in the lab-frame, and $q_3^\SSt$ in the $\SSt-$frame.
We finally write the measure analogously as
\beq
d\Pi'_3 = \frac{1}{(2\pi)^5} \frac{1}{2} dc'_U d\phi'_U d\qUPr dc^\SSt_3 d\phi^\SSt_3 \frac{\qUPr^2}{2 E'_U} \frac{\qthrSSt}{4 (\EnSSt - \EUPSSt)} \ ,
\label{dPi3Primeas.EQ}
\eeq
in terms of the five independent loop variables $c'_U,\phi'_U,\qUPr,c^\SSt_3,\phi^\SSt_3$ for integration over the on-shell loop momenta.
The domain of integration for these variables are the usual full range,
except for $\qUPr \in (0,\qUPr^{(max)})$ with $\qUPr^{(max)}  = \qUr^{(max)}$, as before, in the $\Chi$ rest-frame.
We find ${M'_B}^2 = M_n^2 + M_U^2 -2(E_n E'_U - \pnr\qUPr c'_{nU})$ with $c'_{nU} = \pnV\!\cdot\!\qUPV/(\pnr\qUPr)$ giving $c'_{nU} = c'_U$ for us. 
We also find $\qthrSSt = (M'_B/2) [(1-2M_D^2/{M'_B}^2)^2 - 4 M_D^4/{M'_B}^4]^{1/2}$. 

For each numerical integration sampling point, using the Lorentz transformations given,
we transform the $q_1^*$ and the $q_3^\SSt$ to the lab-frame and compute all 4-momentum dot-products using the lab-frame momenta $\{p_n,q_U,q_1,q_2,q'_U,q_3,q_4\}$.
We also compute the $\epsilon^{\mu\nu\sigma\tau} {p_i}_\mu {p_j}_\nu {p_k}_\sigma {p_l}_\tau$ with the $p_{i,j,k,l}$ being any of these lab-frame momenta. 
Using the so computed (Lorentz invariant) dot-products, we evaluate the matrix element,
fold in the $d\Pi_3\, d\Pi'_3$ measures and obtain the required integrals numerically.

\subsubsection{One-loop integral measure}
\label{dPi2Pmeas.SEC}
Here we write down the one-loop integral measure applicable to the diagram such as the
$\ampA_{01}^{\prime(1)}$ depicted in Fig.~\ref{chi2DDU-LONLO-modA.FIG} (left).
Appealing to the Lorentz invariance of the loop measure, we write this in the $*$-frame as 
\beq
d\Pi_2' = \frac{1}{(2\pi)^2} dc_3^* d\phi_3^* \frac{\qthrSt}{4 (E_n^* - E_U^*)} \ ,
\label{dPi2UXmeas.EQ}
\eeq
where
$\qthrSt = (M_B/2) \{ [1-(M_3^2 + M_4^2)/M_B^2]^2 - 4 M_3^2 M_4^2/M_B^4 \}^{1/2} $
with $M_{3,4} = q_{3,4}^2$. 

We point out a subtlety in the maximum value of $\qUr$ in Fig.~\ref{chi2DDU-LONLO-modA.FIG}.
Since the left-cut also has to be enforced, 4-momentum conservation at the left-most vertex puts an additional constraint,
namely, we must now also have $\qUr \leq \qUrt^{(max)}$
where $\qUrt^{(max)} = (M_n/2)\{ [1+(M_U^2-(M_m+M_U)^2)/M_n^2]^2 - 4 M_U^2/M_n^2 \}^{1/2}$,
in addition to $\qUr \leq \qUr^{(max)}$ as earlier for the right-cut. 
We must thus take the stricter upper-bound of these, i.e. we have the
$\qUr$ domain of integration for $\ampA_{01}^{\prime(1)}$ now given as
$\qUr \in \left(0,{\rm min}\{\qUr^{(max)},\qUrt^{(max)}\} \right)$, while the other integrations are over the usual full range.

\subsection{The 2-body phase-space variables and measure (for $\Chi$ scattering)}
\label{2PS12Vars.SEC}

Our choice of the final-state phase-space variables has similarities with the considerations of Sec.~\ref{3PS12Vars.SEC},
but now for scattering it is a 2-body phase-space, unlike being 3-body phase-space earlier.   
One change for scattering is that instead of the final state momentum $q_U$ we now have the initial state momentum $k_i$.
Another change is that in decay, we could analyze the process in the $\Chi$ rest-frame without any loss of generality,
but in scattering, we have to generalize this to have arbitrary $\pnr$, i.e. we no longer take the lab-frame to be the $\Chi$ rest-frame here.
We make these generalizations in our implementation of $d\Pi_2$ for our scattering processes SC-1 and SC-2 discussed in Sec.~\ref{genBAsymScat.SEC}. 

Similar to the procedure of Sec.~\ref{loop34Vars.SEC}, we now take the incoming momenta $\pnV$ in terms of the independent $(\pnr,c_n,\phi_n)$,
and $\kiV$ in terms of the independent variables $(\kir,c_i,\phi_i)$.
For unpolarized $\Chi$ scattering the matrix element will have no $\phi_n$ dependence and we can integrate it trivially to get a factor of $2\pi$,
and conveniently set $\phi_n=0$.

The Lorentz invariance of $d\Pi_2$ allows us to compute it in any frame, and we choose to compute it in the
center-of-momentum (CM) frame, which is the rest-frame of $k'_A = p_n+k_i$.\footnote{
This has parallels to the *-frame in Sec.~\ref{3PS12Vars.SEC}, which was the rest-frame of $k_B$.}
We now write 
\beq
d\Pi_2 = \frac{1}{(2\pi)^2} dc_1^{CM} d\phi^{CM}_1 \frac{\qor^{CM}}{4M_A}\, \left[\frac{1}{2}\right]_{SC-2} \ ,
\label{dPi2SC.EQ}
\eeq
where $\theta_1 \equiv \theta_{q_1}$, $\phi_1 \equiv \phi_{q_1}$ with $c_1 \equiv \cos{\theta_1}$, $M_A^2 = {k'_A}^2$.
We have $M_A = E_{CM} = E^{CM}_n + E_i^{CM}$, 
and 
$\qor^{CM} = (M_A/2)\, \{ [1+(M_2^2-M_1^2)/M_A^2]^2 - 4 M_2^2/M_A^2 \}^{1/2}$,
and since $M_A$ is frame-independent, we also have
$M_A^2 = M_n^2 + M_i^2 + 2(E_n E_i - \pnr \kir c_{ni})$, and without loss of generality we choose $\pnV$ to be along the z-axis, and therefore, $c_{ni} = c_i$.
We include an extra factor of $1/2$ only for SC-2 due to two identical $\Dp$ in the final state.

\subsection{The loop variables and measure (for $\Chi$ scattering)}
\label{loop34PriVars.SEC}

Our implementation of the loop integration measure in scattering  
closely parallels the decay case in Sec.~\ref{loop34Vars.SEC}, and we discuss here how we adapt this for scattering.
As mentioned in Sec.~\ref{2PS12Vars.SEC} we generalize to have the scattering $\pnr \neq 0$ here too.

\subsubsection{Two-loop measure}

We construct the $\SSt$ frame exactly as discussed before in Sec.~\ref{loop34Vars.SEC}.
But using the Lorentz invariance of the measure, we choose to now take as the independent variable
$\qUPV$ variables in the $\SSt$ frame (instead of the lab frame earlier).
As before we take $\qthV$ independent variables also in the $\SSt$ frame.
We can then write the loop integration measure for the scattering channels as
\beq
d\tilde\Pi'_3 = \frac{1}{(2\pi)^5} \frac{1}{2} d\cUPSSt d\phiUPSSt d\qUPrSSt dc^\SSt_3 d\phi^\SSt_3 \frac{\qUPrSSt^2}{2 \EUPSSt} \frac{\qthrSSt}{4 (\EnSSt - \EUPSSt)} \ , 
\eeq

The domain of integration for these variables are the usual full range, except for $\qUPrSSt \in (0,\qUPrSSt^{(max)})$.
Now, $\qUPrSSt^{(max)}$ is determined as follows.
We first compute the Lorentz invariant
${M_B'}^2 = (p_n - q_U')^2 = (\EnSSt - \EUPSSt)^2 = M_n^2 + M_U^2 + 2(\qUPrSSt^2 - \EnSSt \EUPSSt)$.
But we also have from 4-momentum conservation $M_B' = \EthSSt + \EfoSSt$.
We realize that $\qUPrSSt$ attains its maximum value when $\qthrSSt=0$, i.e. when $M_B' = 2 M_D$,
and we obtain
$\qUPrSSt^{(max)} = (M_n^2/(4 M_D)) \{ [1+(M_U^2 - 4 M_D^2)/M_n^2 ]^2 - 4 M_U^2/M_n^2 \}^{1/2}$.

Given the above independent momenta in the $\SSt$ frame, we transform all the loop momenta to the lab-frame
using the boost transformation that we have already given in Sec.~\ref{loop34Vars.SEC}.
With all the momenta now available in the lab-frame, we compute the 4-momentum dot-products,
and the matrix element in terms of those. 
We then fold in the loop integration measure discussed here and obtain the required integrals numerically. 

\subsubsection{One-loop measure}
\label{dPi2PMASC2.SEC}
For the multiple operator contribution in SC-2, we have $p_n+k_i=q_3+q_4$ as can be seen from the crossed Fig.~\ref{chi2DDU-NLO-modA.FIG}.
Thus, it would be most convenient to pick the independent loop momenta in the CM frame as the $\qthV$ and $\qfV$ are back-to-back in this frame.
We follow our analysis in Sec.~\ref{2PS12Vars.SEC} and write analogous to Eq.~(\ref{dPi2SC.EQ}) the one-loop measure for this case as 
\beq
d\Pi'_2 = \frac{1}{(2\pi)^2} dc_3^{CM} d\phi^{CM}_3 \frac{\qthr^{CM}}{4M_A} \ ,
\eeq
where $q_3=q_m$ is the $\Chi_m$ momentum.

\section{The Decay and Scattering Rates}
\label{GmSigDet.SEC}

Here we give details on the tree-loop interference terms 
putting together the matrix element and phase-space measure discussed above
in Appendices~\ref{MatElcomp.SEC} and \ref{12U34UPVars.SEC} respectively,
and about the numerical evaluation of the decay and scattering rate differences. 

\subsection{The tree-level decay rate}
\label{chi2DDUGmLODet.SEC}

To obtain the total decay width, we fold in the 3-body phase-space element and integrate.
We evaluate the 3-body phase-space element equivalently as a cut of a loop diagram as
\beq
d\Pi_3 \equiv \frac{1}{2} \frac{d^4q_U}{(2\pi)^4} \frac{d^4q_1}{(2\pi)^4} (2\pi)^3 \delta^{\rm cut}_{12U} \ , \quad
 {\rm where}\ \delta^{\rm cut}_{12U} \equiv \delta(q_U^2 - M_U^2) \delta(q_1^2 - M_D^2) \delta(q_2^2 - M_D^2) \ ,
\label{3PS1U.EQ}
\eeq
and we have a factor of 1/2
due to the two identical $\Dp$ in the final state.
We
explicitly show in Appendices~\ref{DDM12.SEC}~and~\ref{DDLoop.SEC}
that $\MotSq = \Lot$, the latter being the corresponding loop expression. 
We take $q_1,q_U$ as the independent momenta, and 
the full details on our choice of independent 3-body phase-space variables and integrations over them are given in
Appendix~\ref{3PS12Vars.SEC}.
As explained there, we have integrated over the $d^4 q_2/(2\pi)^4$ and enforced the overall energy-momentum conserving
$\delta$-function $ (2\pi)^4 \delta^4 (p_n - (q_1 + q_2 + q_U))$.

We have $\MLOotSq_{\tau\mu} = (\MLOotSq_{\mu\tau})|_{M_D \to (-M_D)}$. 
Because the trace of an odd number of $\gamma$ matrices is zero, we have the $M_D$ dependence like
  $(\MLOotSq_{\tau\mu}) \supset -M_D^2 (...)$,
which
implies that $\MLOotSq_{\tau\mu} = \MLOotSq_{\mu\tau}$ and therefore $(\MLOotSq_{\mu\tau})^* = (\MLOotSq_{\mu\tau})$.
We also have $\MLOotSq_{\mu\tau} = \McLOotSq_{\mu\tau}$ (cf. Appendix~\ref{DDM12.SEC}).  
We can also show $\McLOXUSq_{\mu\tau} = \MLOXUSq_{\tau\mu} = {\MLOXUSq_{\mu\tau}}^{\!\!\!\!*}$. 
From these relations, we can infer that $|\ampALOn|^2 = |\ampAcLOn|^2$,
which implies $\Gamma_0^n = \Gamma_0^{cn}$, 
showing explicitly that the LO decay rate is the same for the process and the conjugate process and that there is no baryon asymmetry at LO.
The total decay width is the sum of the widths for the process and the conjugate process,
and to obtain the leading contribution to the baryon asymmetry,
it is sufficient to take for the total width $\Gamma_{\Chi_n} \approx \Gamma_0^n + \Gamma_0^{cn} = 2 \Gamma_0^n$,
since there is no asymmetry at LO. 

For the numerical computation of the integrals,
we scale all dimensionful quantities by $M_n$
and define dimensionless momenta $\hat{p}_i \equiv p_i/M_n$ and dimensionless masses $\hat{M}_i \equiv M_i/M_n$,
strip off the couplings
and define (dimensionless) coupling-independent traces and integrals
\bea
    (\Mhzzn_{\lambda\lambda'})_{nn}^{\mu\tau} &=&
          {\rm Tr}[\gamma^\mu P_\lambda (\hat\pslash_n + \hat{M}_n) \gamma^\tau P_{\lambda'} (\hat\qslash_U + \hat{M}_U)] \ , \label{Mh00def.EQ} \\
           {\IhLOn}_{\lambda\lambda'} &=& \int
        d\hat\Pi_3\,
   \frac{1}{2} (\Mhzzn_{\lambda\lambda'})_{nn}^{\mu\tau} \, C_{00}\, (\Lhot)_{\mu\tau} \ ,  
          \label{I0Mh00n.EQ}
\eea
where $P_\lambda$ is the chirality projection operator with $\lambda,\lambda' = \{L,R\}$.
Including the couplings we have 
$\MhLOXUSq = \sum_{\lambda\lambda'} \hat{g}^*_\lambda \hat{g}_{\lambda'} \Mhzzn_{\lambda\lambda'}$ with hatted momenta and masses,
and,
$|\tilde{g}|^2\, \Lhot = \MhotSq$ is
with hatted momenta and masses.
Including the couplings, we define the dimensionless
\bea          
f_{00}^n &=& |\tilde{g}|^2\left(|\ghLn|^2 {\IhLOn}_{LL} + |\ghRn|^2 {\IhLOn}_{RR} + \ghLn^* \ghRn {\IhLOn}_{LR} + \ghRn^* \ghLn {\IhLOn}_{RL} \right) \ , \nonumber \\
&=& |\tilde{g}|^2\left[ ( |\ghLn|^2 + |\ghRn|^2)\, {\IhLOn}_{LL} + (\ghLn^* \ghRn + \ghRn^* \ghLn)\, {\IhLOn}_{LR} \right] \ .
\label{f0nIh.EQ} 
\eea
In terms of this we write the LO decay rate as in Eq.~(\ref{Gm0f00.EQ}). 

We find $\Mhzzn_{RR} = (\Mhzzn_{LL})^*$, $\Mhzzn_{RL} = \Mhzzn_{LR} = (\Mhzzn_{LR})^*$,  
which imply ${\IhLOn}_{RR} = ({\IhLOn}_{LL})^*$, ${\IhLOn}_{RL} = {\IhLOn}_{LR} = ({\IhLOn}_{LR})^*$.
We also find that $(\Mhzzn_{LL} - \Mhzzn_{RR})^{\mu\tau} = 2 i \epsilon^{\mu\tau\nu\sigma} {p_n}_\nu {q_U}_\sigma$, and when contracted with the
symmetric $\Lhot_{\mu\tau}$, the RHS gives zero, which implies that ${\IhLOn}_{LL} = {\IhLOn}_{RR}$,
and since the symmetric piece is real, these integrals are real.
Therefore, $f_{00}^n$ is real as required.

\subsection{The single operator decay rate difference}
\label{oneOpAsymbDet.SEC}

Computing the $\MNLOXUi$ of Eq.~(\ref{ampANLOn.EQ}) we find 
\beq
{\MNLOXUi}^{\sigma\mu\nu}_{nm} \equiv \bar{u}(q_U) \GVBm^\sigma ({\pslash^{\prime (i)}_n} + M_m) \GZBm^\mu (\qslash'_U + M_U) \GZn^\nu u(p_n) \nonumber \ . 
\eeq
Next, we consider the computation of $\Ltf$ of Eq.~(\ref{ampANLOn.EQ}). 
As an example, consider the contribution from diagram A$_1$.
For this, the $L^{34}_{(A_1)}$   
is derived in Appendix~\ref{DDLoop.SEC}, and is
\beq
    \Ltf^{\sigma\nu} 
    = 2 |\tilde{g}|^2 \, {\rm Tr}[ \gamma^\nu (\qslash_3 - M_D) \gamma^\sigma (\qslash_4 + M_D) ]   \ .
\label{L34VVcp.EQ}
\eeq
The $M^{12}_{1(A_1)}$ is the same as the $M_0^{12}$ of the LO amplitude computed earlier. 
The color factor for this is $C_1^{(A_1)} = \epsilon^{a'b'c} \epsilon^{abc'} \epsilon^{a'b'c'} = 2 \epsilon^{abc}$.
The final state $|M_{12}|^2$ can be equivalently thought of as a cut of the corresponding loop giving $L_{12}$,
as already discussed above.
This equivalence is very useful in writing the tree-loop interference term 
as cuts of a loop process and we use this technique below (cf. Fig.~\ref{chi2DDU-LONLOCut.FIG}) in computing the $\Delta\hat\Gamma_{01}$.
We give the details of the loop integration measure $d\Pi'_3$ in Sec.~\ref{dPi3Pmeas.SEC}.

As an example, we highlight the following aspects of diagram A.  
The color factor is $$C_{01}^{(A)} = \epsilon^{abc} \epsilon^{a'b'c} \epsilon^{abc'} \epsilon^{a'b'c'}/(\sqrt{6})^2 = 12/(\sqrt{6})^2 = 2 \ ,$$ 
using $\epsilon^{a'b'c} \epsilon^{a'b'c'} = 2\delta^{cc'}$ and $\epsilon^{abc} \epsilon^{abc} = 6$.
The $\MotSq_{\mu\tau}$ is given earlier
(recall $M^{12}_{1(A_1)} = M_0^{12}$, as mentioned earlier),
and $\Ltf_{\sigma\nu}$ in Eq.~(\ref{L34VVcp.EQ}). 
We can show using an identical argument as earlier that
  $\Ltf_{\sigma\nu}^* = \Ltf_{\sigma\nu}$, and, $\Ltf_{\sigma\nu} = \Lctf_{\sigma\nu}$.
In a similar way, we compute these factors for the other diagrams also, but do not give the details here. 

We obtain
the decay rate difference as      
\bea
\Delta\hat\Gamma^{n}_{01} &=& \frac{1}{2M_n} \frac{1}{16 \Lambda^8}
\int d \Pi_3\, d\Pi'_3\,
\frac{1}{2}\,
\sum_{m,(i)}\, 2\, {\rm Im}\left[\frac{1}{2} {\MLONLOXUi}^{\sigma\mu\nu\tau}_{nm}\right] \,
\frac{C_{01}^{(i)} (\MNLOoti\Ltfi)_{\mu\sigma\nu} \MLOot^\dagger_\tau}{[{p'_n}^{\!\!(i)\, 2} - M_m^2]}  \ ,
\label{Gam01-3PS.EQ}
\eea
and using this we can compute $\AsymB$ from Eq.~(\ref{ABDGm01Gm0.EQ}).
But before we do this, we present a slightly different but equivalent method that we use to compute the $\Delta\hat\Gamma^{n}_{01}$,
in the form of Eq.~(\ref{Gm01cuts.EQ}).

The computation of the $\Dp$-loop in Fig.~\ref{chi2DDU-LONLOCut.FIG} is detailed in Appendix~\ref{LONLOintContr.SEC}, with 
the momenta ${p'_n}^{\!\!(A)} = p_n - (q_1 + q_2 + q_3 + q_4)$, ${p'_n}^{\!\!(B)} = p_n$, ${p'_n}^{\!\!(Cij)} = p_n - (q_i + q_j)$, 
and for diagram D we get the same contribution as diagram C, i.e. $L_D^D = L_D^C$.
In each process, either the $q_1-q_2$ legs or the $q_3-q_4$ legs can be twisted independently (since the 2$\Dp$ are identical)
to give us four different sub-diagram; diagrams $A$ and $B$ have these equal, giving an overall factor of 4 as seen above,
while for diagram C (and D) each of these are split up into the $L_D^{Cij}$ as given above.
Although each of these subdiagrams has a negative sign due to anti-commuting two fermionic operators,
in each case the anti-symmetry of the two color indices involved undoes this minus sign. 
The $\Dp$-loop factor for the conjugate process are the same, i.e. $L^{c\,(i)}_D = L_D^{(i)}$ of Eq.~(\ref{LDABCD.EQ})
as argued in Appendix~\ref{LONLOintContr.SEC}.
Furthermore, since there are no $\gamma^5$ in any of the ${\rm Tr}[...]$ in $L^{(i)}_D$, they are all real, 
and the phase in $\hat\ampA_{01}$ is entirely due to the phase in $\MLONLOXU$.

We recall that the expression for the ${\MLONLOXUi}$
is given in Eq.~(\ref{ampA01hatn.EQ}) (see also Appendix~\ref{LONLOintContr.SEC}). 
In the tree-loop interference term for the conjugate process, the analogous factor to this,
namely $\McLONLOXU$, is obtained by making the change $\GVB \leftrightarrow \GZ$ in $\MLONLOXU$ above
(see also footnote \ref{GVB2BZcorr.FN} on p.\pageref{GVB2BZcorr.FN}).
This implies $\McLONLOXU = \MLONLOXU^*$, and that $\hat\ampA^c_{01} = \hat\ampA_{01}^*$,
which proves the claim we made in writing Eq.~(\ref{DGm01Ahn.EQ}). 

As earlier, for the numerical computation of the integrals, here too we scale all dimensionful quantities by $M_n$,
and take the dimensionless $\MhLONLOXUnmi$ and $(\hat{L}^{(i)}_D)$ as the corresponding quantities
but with hatted momenta and masses, namely, $\hat{p}_i \equiv p_i/M_n$ and $\hat{M}_i \equiv M_i/M_n$.
We factor out the couplings, and expand out the
${\rm Tr}[(...)({\hat\pslash'_n} + \hat{M}_m)(...)(\hat\pslash_n + \hat{M}_n)(...)]$ structure
in the $\MhLONLOXUi$ (cf. Eq.~(\ref{M01XUdefn.EQ}))
into separate trace-terms indexed by $k=\{0,1,2,3,4\}$ and the individual chiralities $\lambda=\{L,R\}$, as
\beq
\MhLONLOXUnmi^{\sigma\mu\nu\tau} = \sum_{k,\lambda}
G^{nm}_{k\lambda}  \hat{{\cal M}}_{k\lambda(i) nm}^{\sigma\mu\nu\tau} \ ,
\label{Mh01nmghLR.EQ}
\eeq
in terms of the coupling-independent traces $\hat{{\cal M}}_{k\lambda}$ (the ${\sigma\mu\nu\tau}$, $nm$, and $(i)$ are suppressed on the LHS below for notational brevity)
\bea
\hat{{\cal M}}_{0} &\equiv& {\rm Tr}[\gamma^\sigma \hat\pslash'_n \gamma^\mu (\hat{\qslash}'_U + \hat{M}_U) \gamma^\nu \hat\pslash_n \gamma^\tau (\hat{\qslash}_U + \hat{M}_U)]
\ , \nonumber \\
    \hat{{\cal M}}_{1\lambda} &\equiv& \hat{M}_n \hat{M}_m {\rm Tr}[\gamma^\sigma \gamma^\mu \hat{\qslash}'_U \gamma^\nu \gamma^\tau \hat{\qslash}_U P_{\lambda}] \ , \quad
    \hat{{\cal M}}_{2\lambda} \equiv \hat{M}_n \hat{M}_m \hat{M}_U^2 \, {\rm Tr}[\gamma^\sigma \gamma^\mu \gamma^\nu \gamma^\tau P_{\lambda}] \ , \nonumber \\
\hat{{\cal M}}_{3\lambda} &\equiv& \hat{M}_m \hat{M}_U
    {\rm Tr}[\gamma^\sigma \gamma^\mu (\gamma^\nu \hat\pslash_n \gamma^\tau \hat{\qslash}_U + \hat{\qslash}'_U \gamma^\nu \hat\pslash_n \gamma^\tau) P_{\lambda}] \ , \nonumber \\
\hat{{\cal M}}_{4\lambda} &\equiv& \hat{M}_n \hat{M}_U
    {\rm Tr}[\gamma^\sigma \hat\pslash'_n \gamma^\mu (\hat{\qslash}'_U \gamma^\nu \gamma^\tau P_{\lambda} + \gamma^\nu \gamma^\tau P_\lambda \hat{\qslash}_U )] \ ,
\label{MasTrDefn.EQ}    
\eea
where for $k=0$ we are to omit $\lambda$,
and,
the coupling combinations\footnote{
We can also define the coupling combinations
$G^{nm}_{1\pm} \equiv G^{nm}_{1R} \pm G^{nm}_{1L}$ and 
$G^{nm}_{2\pm} \equiv G^{nm}_{2L} \pm G^{nm}_{2R}$,
since $f_{01}^{nm}$ picks up the combinations
$2[ \Im{(G_{1+})} (\hat{{\cal M}}_{1+}) + \Im{(G_{2+})} (\hat{{\cal M}}_{2+})
  + \Re{(G_{1-})} (\hat{{\cal M}}_{1-}) + \Re{(G_{2-})} (\hat{{\cal M}}_{2-})]$,
where we have defined: 
$\hat{{\cal M}}_{1+} \equiv \Re{(\hat{{\cal M}}_{1R})} = \Re{(\hat{{\cal M}}_{1L})}$,
$\hat{{\cal M}}_{2+} \equiv \Re{(\hat{{\cal M}}_{2L})} = \Re{(\hat{{\cal M}}_{2R})}$,
$\hat{{\cal M}}_{1-} \equiv \Im{(\hat{{\cal M}}_{1R})} = - \Im{(\hat{{\cal M}}_{1L})}$,
$\hat{{\cal M}}_{2-} \equiv \Im{(\hat{{\cal M}}_{2L})} = - \Im{(\hat{{\cal M}}_{2R})}$.
The $\hat{\cal I}_{\pm}$ definitions are in analogy with these $\hat{{\cal M}}_{\pm}$ definitions. 
}
\bea
G^{nm}_0 &=& \hat{g}_{Rn} \hat{g}_{Ln} \hat{g}^*_{Rm} \hat{g}^*_{Lm} \ , \nonumber \\  
G^{nm}_{1R} &=& \hat{g}^2_{Ln} \hat{g}^{*2}_{Lm} \ ,  G^{nm}_{1L} = \hat{g}^2_{Rn} \hat{g}^{*2}_{Rm} \ , \quad
G^{nm}_{2R} = \hat{g}^2_{Rn} \hat{g}^{*2}_{Lm} \ , G^{nm}_{2L} = \hat{g}^2_{Ln} \hat{g}^{*2}_{Rm} \ ,\nonumber  \\
G^{nm}_{3R} &=& \hat{g}_{Ln} \hat{g}_{Rn} \hat{g}^{*2}_{Lm}  \ , G^{nm}_{3L} = \hat{g}_{Ln} \hat{g}_{Rn} \hat{g}^{*2}_{Rm} \ , \nonumber  \\ 
G^{nm}_{4R} &=& \hat{g}_{Rn}^2 \hat{g}^*_{Lm} \hat{g}^*_{Rm} \ , G^{nm}_{4L} = \hat{g}_{Ln}^2 \hat{g}^*_{Lm} \hat{g}^*_{Rm} \ .
\label{GklnmDefn.EQ}
\eea
We define dimensionless coupling-independent integrals of the above traces folded in with the $\Dp$-loop functions as
\bea
\hat{\cal I}^{(01)}_{k\lambda\, nm} &\equiv& \int
   d\hat\Pi_3\,
   d\hat\Pi'_3\,
    \frac{1}{2}\, \sum_{(i)} \frac{1}{2} \hat{{\cal M}}_{k\lambda(i)nm}^{\sigma\mu\nu\tau} \,
     \frac{C^{(i)}_{01} {(\hat{L}^{(i)}_D)}_{\mu\tau\nu\sigma}}{[({{\hat{p}}_n}^{\prime (i)})^2 - \hat{M}_m^2]}
    \ , \label{I01nm.EQ}
    \eea
    where
    we define ${(L^{(i)}_D)} \equiv |\tilde{g}|^4 {(\hat{L}^{(i)}_D)}$, now written with hatted momenta and masses.
We include the diagram B contribution, i.e. for $(i)\!\!=\!\!(B)$,
only for $m\!\!\neq\!\! n$ (i.e. $nm\!\!=\!\!12$ for $\Chi_1$ decay, and $nm\!\!=\!\!21$ for $\Chi_2$ decay)
for the reasons already explained on p.\pageref{diagBsub.PG}. 

In terms of these integrals and couplings we define 
\bea
f_{01}^{nm} = |\tilde{g}|^4  \sum_{k,\lambda}\, {\rm Im} \left[ G^{nm}_{k\lambda} \hat{\cal I}^{(01)}_{k\lambda\, nm} \right]  \ ,
\label{f1nmMIgG.EQ}
\eea
where for $k=0$ we are to omit $\lambda$.
We can write Eq.~(\ref{Gm01cuts.EQ}) now as in Eq.~(\ref{DGm01f01.EQ}). 

An analysis of the superficial degree of divergence suggests that only $\hat{\cal I}^{(01)}_{0\, nm}$ could be potentially UV divergent,
assuming that the $1/\Lambda^2$ in the operators is resolved in a gauge invariant way in the UV completion.
In the extreme high-energy limit, the masses can be ignored, and in this limit, when summed over $m$,
the structure of the $G^{nm}_0$ is such that the unitarity of the $\Umat$ ensures that this goes to zero,
analogous to the GIM-cancellation in the SM.
This implies that the $f_{01}$ is finite with no UV divergences present.
Furthermore, using the unitarity of $\Umat$ we can show that $G_0^{nm}$ is always real $\forall (n,m)$.
Also, $\hat{\cal M}_0$ is always real and therefore $k=0$ does not contribute to $f_{01}^{nm}$. 
All other $k$ have at least one $M_n$ or $M_m$ factor, which indicates that at least one Majorana mass insertion is required
to generate a nonzero $\AsymB$, as we expect.

We compute the traces in Eqs.~(\ref{LDABCD.EQ})~and~(\ref{MasTrDefn.EQ}) leading to dot-products of the 4-momenta involved.
Since these dot-products and the 3-body phase-space and loop integration measures are Lorentz invariant,
we can evaluate them in any convenient frame.
We discuss in Appendix~\ref{12U34UPVars.SEC} full details on the frames we use, the choice of independent momenta in them,
the limits of integration,
and the phase-space and loop integration measures.
As explained in full detail there,
we sample the 3-body phase-space independent momenta of the $(q_1, q_2, q_U)$-sector in the $*$-frame
in which ${\bf p}_n^* = {\bf q}_U^*$, implying ${\bf q}_2^{*} = -{\bf q}_1^{*}$.
We choose this frame, since in this frame, the 3-momentum radial-component upper limit of integration does not depend on the angular variables,
simplifying the sampling. 
For a similar reason, we sample the independent loop momenta of the $(q_3,q_4,q_U')$-sector in the $\SSt$-frame
in which ${\bf p}_n^{\SSt} = {{\bf q}'_U}^{\!\!\SSt}$, implying ${\bf q}_4^{\SSt} = -{\bf q}_3^{\SSt}$.
For an integration sample point of the independent momenta picked in the respective frames,
we compute all the other 4-momenta using 4-momentum conservation and on-shell conditions,
and the Lorentz-invariant phase-space and loop integration-measures in those frames.
Our choice of the five independent 3-body phase space variables and the measure are as explained in Appendix~\ref{3PS12Vars.SEC}, 
and of the five independent loop variables and the loop measure are as in Appendix~\ref{loop34Vars.SEC}.  

We transform all of the 4-momenta to the $\Chi_n$ rest-frame (lab-frame), and evaluate all 4-momentum dot-products in this frame.
Using these dot-products, we compute the integrand of Eq.~(\ref{I01nm.EQ}),
fold-in the above integration measures, and numerically integrate to obtain the $\hat{\cal I}^{(01)}_{k\lambda\, nm}$. 
Before discussing the numerical results, we make the following observations. 

We find that the ${\rm Im}(\hat{\cal M}_{k\lambda})$ are proportional to $\epsilon^{\sigma\mu\nu\tau}$ contracted with the various 4-momenta
involved, and since they are all Lorentz invariant, we can evaluate them in a convenient frame, namely, either the $*$-frame or the $\SSt$-frame.
These dot-products have one $0^{\rm th}$ component $(E)$ and a scalar triple product of the other three 3-momenta
that is odd in the loop momenta azimuthal angles $\phi_3, \phi'_U$, and when we integrate over these angles, we get zero.
Thus all the $\hat{\cal I}^{(01)}_{k\lambda\, nm}$ are real, and therefore from Eq.~(\ref{f1nmMIgG.EQ}) we note that
only the ${\rm Im}(G^{nm}_{k\lambda})$ are picked up in generating the $\AsymB$. 
\label{intsReal.PG}

We consider here some limits in which we expect $\AsymB$ to be zero and see if our formalism is consistent with that.
We know from the Sakharov conditions that the generation of a nonzero $\AsymB$ requires $C$ and $CP$ violation.
We found the following conditions for these to be good symmetries.
First, we found that $C$-invariance is present if and only if $\ghLn^* = \ghRn$.
If this condition holds, the following are true for $f_{01}^{nm}$ in Eq.~(\ref{f1nmMIgG.EQ}): 
(a) $G_0^{nm}$ is real, and since $\hat{\cal M}_0$ is real, $k=0$ does not contribute in $f_{01}^{nm}$, and, 
(b) for $k=\{1,2,3,4\}$, $G_{kR} = G_{kL}^*$ and since $\hat{\cal M}_{kR} = \hat{\cal M}_{kL}^*$, 
we have $(G_{kL} \hat{\cal M}_{kL} + G_{kR} \hat{\cal M}_{kR})$ is real, and therefore these also do not contribute to $f_{01}^{nm}$.
Thus, $f_{01}^{nm} = 0$ in the $C$-invariance limit, i.e. $\AsymB = 0$ in this limit.
Next, we found that $CP$-invariance is present if and only if $\ghLn, \ghRn$ are real, in which case all the $G^{nm}_{k\lambda}$ are real.
In this limit, $k=0$ does not contribute to $f_{01}$ since $\hat{\cal M}_0$ is real.
For all other $k$, $\hat{\cal M}_{kR} + \hat{\cal M}_{kL}$ is real and therefore these do not contribute to $f_{01}$ either,
and we have $\AsymB = 0$ in the $CP$-invariance limit.
Next, we consider the ``Dirac limit'', i.e. the $\tilde{M}_{L,R} = 0$,
whence $\Umat_{an} = ((i,1),(-i,1))/\sqrt{2}$,
where the $i$ in the first column ensure that the mass eigenvalues are positive. 
In $G_{k\lambda}$, this $i$ always appears twice, giving $\pm 1$.
So the phase dependence is potentially only due to $\phi_L$, the only physical phase present in the Dirac limit. 
$k=0$ does not contribute in $f_{01}$ since $G_{0}$ and $\hat{\cal M}_0$ are real. 
For the other $k$, the $G_{1L},G_{2L}$ are insensitive to this phase,
the $G_{2L}$ and $G_{2R}$ are conjugate with respect to $\phi_L$ and this phase does not contribute when summed over $L,R$,
the $G_{3}, G_{4}$ give zero with this $\Umat$ when summed over the $m$ (which can be done since $M_m$ are degenerate in the Dirac limit).  
Thus, we see that $\AsymB = 0$ in the Dirac limit. 
As mentioned already, it is reassuring that our formalism passes these nontrivial checks.

Following the method already discussed above, we use the $\Mhzzn$ and $\Lhot$ written in terms of dot-products of the 4-momenta involved. 
For each 4-momentum sample-point of the numerical integration,
we compute these dot-products and the phase-space measure,  
and numerically integrate to obtain the LO integrals ${\cal I}_0^n$ of Eq.~(\ref{I0Mh00n.EQ}).  
We then fold-in the couplings to obtain the $f_{00}^n$ from Eq.~(\ref{f0nIh.EQ}).
We then compute the LO decay width $\Gamma_0^n$ from Eq.~(\ref{Gm0f00.EQ}), 
omitting the $(M_\chi^5/\Lambda^4)$ factor in our numerical calculation, which is to be included later. 
We obtain the tree-loop interference term by computing the dot-products for the traces in $\MhLONLOXUnmi$ in Eq.~(\ref{MasTrDefn.EQ})
and in the $\Dp$-loop function ${(\hat{L}^{(i)}_D)}$ in Eq.~(\ref{LDABCD.EQ}),
fold in the phase-space and loop measures,
and numerically integrate to obtain $\hat{\cal I}^{(01)}_{k\lambda\, nm}$ of Eq.~(\ref{I01nm.EQ}).  
We then include the couplings and compute the loop function $f_{01}^{nm}$ from Eq.~(\ref{f1nmMIgG.EQ}). 
We then obtain the tree-loop interference decay width difference $\Delta\hat\Gamma_{01}^{n}$ from Eq.~(\ref{DGm01f01.EQ}), 
and the baryon asymmetry $\AsymB$ from Eq.~(\ref{ABf01f00.EQ}),
omitting in our numerical calculation the $(M_\chi^9/\Lambda^8)$ factor in the former and the $(M_\chi/\Lambda)^4$ factor in the latter, which are to be included later. 

For the BP-A, single operator contributions, we find the following values for the integrals.
The LO integrals ${\IhLOn}_{\lambda\lambda'}$ are ($\times 10^{-5}$):
${\IhLOo}_{LL} = 0.0372$, ${\IhLOo}_{LR} = -0.035$, ${\IhLOt}_{LL} = 0.88$, ${\IhLOt}_{LR} = -0.61$.
The tree-loop interference integrals $\hat{\cal I}^{(01)}_{k\lambda\, nm}$ are ($\times 10^{-9}$): 
$\hat{\cal I}^{(01)}_{0\, 21} = -2.41$, 
$\hat{\cal I}^{(01)}_{1R\, 21} = -0.64$, 
$\hat{\cal I}^{(01)}_{2R\, 21} = -0.3$, 
$\hat{\cal I}^{(01)}_{3R\, 21} = 0.88$, 
$\hat{\cal I}^{(01)}_{4R\, 21} = 1.1$, 
$\hat{\cal I}^{(01)}_{4R\, 22} = 0.03$, 
and we have $\hat{\cal I}^{(01)}_{kL\, nm} = \hat{\cal I}^{(01)}_{kR\, nm}$.  
These integrals are determined with a
sampling error better than about a few percent
and integrals with values less than $10^{-11}$ are not shown.
As already discussed (see p.\,\pageref{intsReal.PG}), we find that the imaginary parts of these integrals are zero.  

We can trace the reason for the suppression in $\hat{f}_{00}$
to a 3-body phase-space suppression leading to a suppressed weight in the measure
due to $\qUr,\qorSt \sim {\cal O}(0.1)$ for the decay kinematics in BP-A. 
Similarly,
we can trace the reason for the suppressed $\hat{f}_{01}$ 
to a product of the suppression factor in the 3-body phase-space as above in the LO integrals,
and another $10^{-3}$ factor from the suppressed weight in the loop-integral measure
due to the cut of the loop-diagram picking out similar sizes of loop-momenta as phase-space momenta, i.e. suppressed as $\qUPr,\qthrSSt \sim {\cal O}(0.1)$.

For our parameter choices here, we find that the diagram-B contribution is about 75 times 
the contribution from diagrams A,~C,~D combined. One enhancement in diagram B is a relative color factor of 3,
while another relative factor of about 10 can be traced to the different Lorentz structures in Eq.~(\ref{LDABCD.EQ}).

\subsection{The multiple operator decay rate difference}
\label{multiOpAsymbDet.SEC}

We compute the multiple operator matrix element similar to the single operator contribution above. 
We give the one-loop measure $d\Pi_2'$ in Appendix~\ref{dPi2Pmeas.SEC},
and the 3-body phase-space measure $d\Pi_3$, as already discussed, in Appendix~\ref{3PS12Vars.SEC}.
The dimensionless hatted versions of these are obtained by scaling all the momenta and masses by $M_n$.
Requiring both the left cut and the right-cut in this case implies that the upper limit $\qUr^{(max)}$
is now modified as discussed in Appendix~\ref{dPi2Pmeas.SEC}.  

As mentioned, we find no asymmetry from diagrams $A_1^{\prime(2),(3)}$ in Fig.~\ref{chi2DDU-NLO-modA.FIG},  
as we could anticipate from there being no arrow clash in these diagrams.
This is shown in Appendix~\ref{noArClMulOp.SEC}. 

We give next the numerical values we obtain for the multiple-operator contribution in BP-B. 
The LO integrals ${\IhLOn}_{\lambda\lambda'}$ are ($\times 10^{-5}$):
${\IhLOo}_{LL} = 7.9$, ${\IhLOo}_{LR} = -0.85$, ${\IhLOt}_{LL} = 8.06$, ${\IhLOt}_{LR} = -0.72$.

As explained already, the smallness in the value of the rate asymmetry in this case is the result of suppressions due to    
the 3-body phase-space weight of Eq.~(\ref{dPi3meas.EQ}) (without the $1/(2\pi)^5$) of $\sim 10^{-3}$,
the loop measure weight of Eq.~(\ref{dPi2UXmeas.EQ}) (without the $1/(2\pi)^2$) of $\sim 10^{-2}$,
and,
the dimensionless $\ampA_{01}^{\prime(1)}$  (without the $1/\Lambda^6$) of $\sim 10^{-4}$ including the (imaginary part of) coupling factors.

\subsection{The single operator scattering rate difference}
\label{singOpSigDet.SEC}

The $d\Pi_2$ is the final-state 2-body phase-space element, 
and we give full details of our choice of independent variables in Sec.~\ref{2PS12Vars.SEC}.
As explained there, we appeal to the frame invariance of the phase space measure and take the $d\Pi_2$ independent variables,
namely $c_1 \equiv \cos{\theta}_1$ and $\phi_1$, in the center-of-momentum (CM) frame. 

We strip out the couplings and write in terms of the traces as in the decay case,
with the dimensionless hatted momenta and masses, i.e. $\hat{p_i} \equiv p_i/M_n$ and $\hat M_i\equiv M_i/M_n$, 
and compute integrals analogous to those in Eq.~(\ref{I0Mh00n.EQ}),
namely,
\beq
\IhLOnsig_{\lambda\lambda'} = \frac{1}{2\hat{E}_n 2\hat{E}_i} \int d\hat\Pi_2 \left[ -\frac{1}{4} (\Mhzzn_{\lambda\lambda'})_{nn}^{\mu\tau} \, C_{00}\, (\Lhot)_{\mu\tau} \right]_{(q_i\ {\rm cross})} \ .
\label{IhLOnsig.EQ}
\eeq
Unlike in the decay case, here we include the $1/(2\hat{E}_n 2\hat{E}_i)$ factors in the definition of the integrals since
this factor should be in the integrand of the thermal averaging (cf. Sec.~\ref{thAvgcsAB.SEC}). 

We fold back the couplings analogous to Eq.~(\ref{f0nIh.EQ})
to obtain 
\beq        
f_{00}^{n(\sigma)}
= |\tilde{g}|^2\left[ ( |\ghLn|^2 + |\ghRn|^2)\, \IhLOnsig_{LL} + (\ghLn^* \ghRn + \ghRn^* \ghLn)\, \IhLOnsig_{LR} \right] \ ,
\label{f0signIh.EQ} 
\eeq
and compute from this the LO cross sections in Eq.~(\ref{sig0vf00.EQ}). 

We move next to the tree-loop interference contribution in scattering.
In analogy with Eqs.~(\ref{Gam01-3PS.EQ})
we write
\beq
\Delta\hat\sigma_{01}^{n}\, v = \frac{1}{2E_n 2E_i}\, \frac{1}{16 \Lambda^8} \sum_m
 \int d\Pi_2\,
d\tilde\Pi'_3\,
  \frac{1}{2}
    \sum_{(i)}\, \left\{ 2\, {\rm Im}\left[-\frac{1}{4} {\MLONLOXUi}^{\sigma\mu\nu\tau}_{nm}\right] \,
    \frac{C_{01}^{(i)} (\MNLOoti\Ltfi)_{\mu\sigma\nu} \MLOot^\dagger_\tau}{[{p'_n}^{\!\!(i)\, 2} - M_m^2]} \right\}_{(q_i\ {\rm cross})} \ . 
    \label{Dsigv01.EQ}
\eeq
Here, $[d\tilde\Pi'_3]$ is the measure for integration over the loop momenta, and we give its details in Sec.~\ref{loop34PriVars.SEC}.
As earlier in Eq.~(\ref{I01nm.EQ}), we strip off the couplings and define dimensionless integrals
\bea
\hat{\cal I}^{(01)(\sigma)}_{k\lambda\, nm} &\equiv& \frac{1}{2\hat{E}_n 2\hat{E}_i}\, \int
   d\hat\Pi_2\,
   d\hat{\tilde\Pi}'_3\,
    \frac{1}{2} \sum_{(i)} \left\{ -\frac{1}{4} \hat{{\cal M}}_{k\lambda(i)nm}^{\sigma\mu\nu\tau} \,
    \frac{C^{(i)}_{01} {(\hat{L}^{(i)}_D)}_{\mu\tau\nu\sigma}}{[({{\hat{p}}_n}^{\prime (i)})^2 - \hat{M}_m^2]} \right\}_{(q_i\ {\rm cross})} \ .
    \label{I01nmsig.EQ}
\eea
We fold back in the couplings analogous to Eq.~(\ref{f1nmMIgG.EQ}) to obtain
\beq
f_{01}^{nm(\sigma)} = |\tilde{g}|^4  \sum_{k,\lambda}\, {\rm Im} \left[ G^{nm}_{k\lambda} \hat{\cal I}^{(01)(\sigma)}_{k\lambda\, nm} \right]  \ ,
\label{f1nmMIgGsig.EQ}
\eeq
for each of the SC-1 and SC-2. 

For obtaining the thermally averaged LO cross section, 
we first thermally average taking the $\IhLOnsig$ of Eq.~(\ref{IhLOnsig.EQ}) as the $\Theta$ of Eq.~(\ref{TADefn.EQ})
to obtain the thermally averaged integrals $\langle \IhLOnsig\rangle $.
In order to obtain the thermally averaged cross section difference and the thermally averaged baryon asymmetry,
we thermally average in the sense of Eq.~(\ref{TADefn.EQ}) using Eq.~(\ref{I01nmsig.EQ})
and setting $\Theta=\hat{\cal I}^{(01)(\sigma)}_{k\lambda\, nm}$
to obtain the corresponding thermally averaged integrals $\langle \hat{\cal I}^{(01)(\sigma)}_{k\lambda\, nm}\rangle $. 

We evaluate numerically the thermal averages of the integrals of Eqs.~(\ref{IhLOnsig.EQ}) and (\ref{I01nmsig.EQ}),
generating the independent 4-momenta samples in the various frames as discussed earlier. 
For illustration, we give here the values of the thermal average integrals for the benchmark point BP-A
laid out on p.\,\pageref{BPADefn.PG}.
The tree-level thermal average integrals $\langle {\IhLOn}_{\lambda\lambda'}^{(\sigma)}\rangle $ for $x\!\!=\!\! 1/2$
with the UV propagator correction factor present for $M_\chi/\Lambda\!\! =\!\! 1/10$ and $\hat\Gamma_\xi = 1/10$,
for (SC-1,~SC-2) respectively are:
$\langle {\IhLOo}_{LL}^{(\sigma)}\rangle  = (2.45, 84.1)$,
$\langle {\IhLOo}_{LR}^{(\sigma)}\rangle  = (-0.008, 1.1)$,
$\langle {\IhLOt}_{LL}^{(\sigma)}\rangle  = (2.45, 83.9)$,
$\langle {\IhLOt}_{LR}^{(\sigma)}\rangle  = (-0.007, 0.88)$.
The tree-loop interference thermal average integrals $\langle \hat{\cal I}^{(01)}_{k\lambda\, nm}\rangle $
for $x\!\!=\!\! 1/2$ for (SC-1,~SC-2) respectively are ($\times 10^{-3}$):
$\langle \hat{\cal I}^{(01)(\sigma)}_{0\, 21}\rangle  = (-3, -36)$, 
$\langle \hat{\cal I}^{(01)(\sigma)}_{1R\, 21}\rangle  = (-1.2, -23)$, 
$\langle \hat{\cal I}^{(01)(\sigma)}_{3R\, 21}\rangle  = (0.8, 17)$, 
$\langle \hat{\cal I}^{(01)(\sigma)}_{4R\, 21}\rangle  = (1.1, 22)$, 
We have shown only the $\lambda \!\!=\!\! R$ integral values,
since we find $\langle \hat{\cal I}^{(01)}_{kL\, nm}\rangle  = \langle \hat{\cal I}^{(01)}_{kR\, nm}\rangle $.
These integrals are determined with a
sampling error better than about 20\,\%, 
and integrals with values less than $10^{-4}$ are not shown.

For these parameter choices, with the UV propagator correction present,
the diagram B contribution of Fig.~\ref{chi2DDU-LONLOCut.FIG} (but crossed for scattering) is dominant,
being about 1000 times the contributions from diagrams A,C,D.
We trace this to a suppression from the $\Chi_m$ propagator of about a factor of 50 in diagram A,C,D, 
while it enhances the diagram B contribution by a factor of about 5,
thus largely explaining the relatively suppressed A,C,D contributions.

\subsection{The multiple operator scattering rate difference}
\label{multiOpSigDet.SEC}

We give details on the phase space variables and measure in Appendix~\ref{2PS12Vars.SEC},
and,
the loop integration variables and measure in Appendix~\ref{loop34PriVars.SEC}.
We generate the independent loop and phase-space momenta of the integration sampling points in the CM frame and compute all the
4-momentum dot-products in the CM frame. 

We evaluate the integrals numerically,
and for illustration we give for BP-B, some sample numerical values for the thermally averaged integrals. 
With the UV state propagator correction included with $M_\chi/\Lambda\!\! = \!\! 1/10$ and $\hat\Gamma_\xi\!\! = \!\! 1/2$,
the tree-level thermal average integrals $\langle {\IhLOn}_{\lambda\lambda'}^{(\sigma)}\rangle $, for $x\!\!=\!\! 1/2$,
for (SC-1,~SC-2) respectively are:
$\langle {\IhLOo}_{LL}^{(\sigma)}\rangle \! = \! (2.45, 15.7)$,
$\langle {\IhLOo}_{LR}^{(\sigma)}\rangle \! = \! (-1.1\times 10^{-3}, 0.027)$, 
$\langle {\IhLOt}_{LL}^{(\sigma)}\rangle \! = \! (2.45, 15.7)$,
$\langle {\IhLOt}_{LR}^{(\sigma)}\rangle \! = \! (-9\times 10^{-4}, 0.023)$.

Denoting the multiple operator scattering cross section difference as $\langle \Delta\hat\sigma'_{01}\, v\rangle $, 
we can compare, for example for the $(M_\chi/\Lambda,\hat\Gamma_\xi)\!\! =\!\! (1/10,1/2)$ case here
with the single operator contribution $\langle \Delta\hat\sigma_{01}\, v\rangle $ in the previous section. 
Including the $M_\chi/\Lambda$ scaling factors but suppressing the coupling dependence, we find, 
$\langle \Delta\hat\sigma'_{01}\, v\rangle /\langle \Delta\hat\sigma_{01}\, v\rangle \! \sim\! (2\pi^2) (\Lambda/M_\chi)^2 \langle \hat{f}_{01}^{'(\sigma)}\rangle /\langle \hat{f}_{01}^{(\sigma)}\rangle  \sim 10^3$.
Similarly, comparing the multiple operator to single operator scattering baryon asymmetry, for $M_\chi/\Lambda\!\! =\!\! 1/10$,
we find
$\AsymBPsigh/\AsymBsigh\! \sim\!
2 \pi^2  (\Lambda/M_\chi)^2
\langle \hat{f}_{01}^{'(\sigma)}\rangle /\langle \hat{f}_{01}^{(\sigma)}\rangle  \cdot
\langle \hat{f}_{00}^{(\sigma)}\rangle /\langle \hat{f}_{00}^{'(\sigma)}\rangle 
\! \sim\! 10^4$.

\subsection{The $\Delta B = -2$ $\Up\Dp\Dp \to \Up^c \Dp^c \Dp^c$ scattering rate}
\label{UDD2UcDcDcDet.SEC}

Taking the matrix element mod-squared averaged over initial state spins and summed over final spins,
and folding in phase-space we compute the $\sigzP$ in Eq.~(\ref{sig0BBb.EQ}),
by decomposing this computation into various integrals similar to those in Eq.~(\ref{I01nm.EQ}). 
These integrals are 
\beq
\IBBbklnmh \equiv \frac{1}{(2E'_U)(2E_3)(2E_4)} \int d\hat\Pi_3\,
    \sum_{(i)} \frac{1}{8} \hat{{\cal M}}_{k\lambda(i)nm}^{\sigma\mu\nu\tau} \,
     \frac{C^{(i)}_{\bar B B} {(\hat{L}^{(i)}_{\bar B B})}_{\mu\nu\tau\sigma}}{[({{\hat{p}}_n}^{\prime})^2 - \hat{M}_n^2 + i \hat{M}_n \hat{\Gamma}_n]\, [({{\hat{p}}_m}^{\prime})^2 - \hat{M}_m^2 + i \hat{M}_m \hat{\Gamma}_m]}
\label{IBBb.EQ}
\eeq
where
for the S-channel-like contribution we have ${p_n}^{\!\!\!\prime~(S)}\! =\! p_n$,
while for the T-channel-like contribution we have ${p_n}^{\!\!\!\prime~(T)}\! =\! q'_U - (q_1+q_2)$, 
the index is over $(i)\!\! =\!\! SS, ST, TS, TT$ contributions from the mod-square of the sum of the S- and T-channel-like contributions,
and the $\hat{{\cal M}}_{k\lambda(i)nm}$ are as in Eq.~(\ref{MasTrDefn.EQ}) with the associated couplings as in Eq.~(\ref{GklnmDefn.EQ}). 
The $DD$-loop part is given as
\beq
(\hat{L}^{\scriptscriptstyle (i)}_{\scriptscriptstyle \bar B B})_{\mu\nu\tau\sigma} = 4 C^{\scriptscriptstyle (i)}_{\scriptscriptstyle \bar B B} \
    {\rm Tr}[\gamma^\mu (\qoslash + M_D) \gamma^\nu (\qtslash - M_D)]\ {\rm Tr}[\gamma^\tau (\qfslash + M_D) \gamma^\sigma (\qthslash - M_D)] \ , 
\eeq
where the indices are as shown for $SS$,
exchanged as $\mu \leftrightarrow \sigma$ and $\nu \leftrightarrow \tau$ for $TT$,
$\mu \leftrightarrow \sigma$ for $ST$,
$\nu \leftrightarrow \tau$ for $TS$,
and,
the color factors are $C^{\scriptscriptstyle (SS),(TT)}_{\scriptscriptstyle \bar B B}\! =\! 6$ as in diagram-B, and $C^{\scriptscriptstyle (ST),(TS)}_{\scriptscriptstyle \bar B B}\! =\! 2$ as in diagram-A (cf. Eq.~(\ref{LDABCD.EQ})).
We then compute the analog of Eq.~(\ref{f1nmMIgG.EQ}), but now picking up the real part, as
\bea
f_{\scriptscriptstyle \bar B B}^{nm} = |\tilde{g}|^4  \sum_{k,\lambda}\,
   {\rm Re} \left[ G^{nm}_{k\lambda} \hat{\cal I}^{\scriptscriptstyle (\bar B B)}_{k\lambda\, nm} \right] \ .
\label{f1nmMIgGBBb.EQ}
\eea 

\subsubsection{Thermally averaged $UDD \to U^cD^cD^c$ scattering rate}
\label{dBBbInitMeas.SEC}
Here we review aspects of the initial state, the center-of-momentum (CM) momentum variables, and thermal averaging 
in the $UDD \to U^cD^cD^c$ process of Sec.~\ref{UDD2UcDcDc.SEC}, 
and simplify the expression for its thermally averaged scattering rate $\GmzPTA$ in Eq.~(\ref{Gm0BBbTA.EQ}).
We exploit the parallels in kinematics with the initial state integration in thermal averaging
with the loop integration in the $\Chi_n$ decay tree-loop interference calculation in Sec.~\ref{ABdec.SEC},
and pick independent momentum variables as there. 

The measure for integration over the initial state momenta $q'_U, q_3, q_4$, 
in the notation $[dq] \equiv d^3q/(2\pi)^3\, 1/(2 E_q)$, is  
\beq
\GmzPTA = \frac{1}{n_\chi^{(0)}} \int [dq'_U][dq_3][dq_4]\frac{d^4\pCM}{(2\pi)^4} (\Pi_i f_{Q_i}^{(0)})
    \ |{\cal P}_\chi|^2 \ (2\pi)^4 \delta^{(4)}(\pCM - (q'_U+q_3+q_4))\, |\ampMPzt|^2\ d\Pi_3 \ , 
 \label{GmzPTADefn.EQ}
\eeq
where
$\Pchi\! =\! 1/({p'_n}^2 - M_n^2 + iM_n \Gamma_n)$ is the $\Chi_n$ propagator with $p'_n$ being the $\Chi_n$ momentum,
we define $\ampMPz = \ampMPzt\, \Pchi$,
the thermal (equilibrium) distribution for each of the initial $Q_i\!=\!\{U,D,D\}$ is $f_{Q_i}^{(0)} = e^{-E_i/T}$,
$d\Pi_3$ is the final state phase space measure as detailed earlier (cf. Sec.~\ref{dPi3Pmeas.SEC}),
and
we show an integration over the CM momentum $\pCM\! =\! q'_U + q_3 + q_4$ along with the energy-momentum conserving $\delta$-function,
the zeroth component of which enforces $\pCMz\!=\! E'_U+E_3+E_4$, which implies $\Pi_i f_{Q_i}^{(0)} = e^{-\pCMz/T}$.
In this work, we specialize to the $SS$ contribution discussed in Sec.~\ref{UDD2UcDcDc.SEC}, with $p_{CM}\! =\! p'_n\!=\! p_n$.

We obtain the $\Chi_n$ pole contribution 
by replacing $|\Pchi|^2 \to \pi/(M_n \Gamma_n)\ \delta({\pnz}^2 - E_n^2)$,
with $E_n\!=\! \sqrt{\pnr^2 + M_n^2}$.
This amounts to making the narrow width approximation for the $\Chi_n$. 
We use the representation
$\delta(x^2 - a^2) = \lim_{\epsilon \to 0}~[(\epsilon/\pi)/((x^2-a^2)^2 + \epsilon^2)]$ with $\epsilon = M_n \Gamma_n$.
We notice that the integral is singular and is regulated by $\Gamma_n$.
The $\delta$-function sets $\pnz\!=\! E_n$ putting $\Chi_n$ onshell, giving $\Pi_i f_{Q_i}^{(0)} = e^{-E_n/T} = f^{(0)}_\chi$.
We perform as earlier in Sec.~\ref{dPi3Pmeas.SEC} the $q_4$ integral to enforce the energy-momentum conserving $\delta$-function,
and follow the same procedure as there, leading to the $d\Pi_3'$.  
Carrying out the $\pnz$ integral to enforce the $\delta$-function, we obtain 
\beq
\GmzPTA_{\rm pole} = \int d\Pi_3 d\Pi'_3 \frac{1}{2\pi} \frac{\pi}{M_n \Gamma_n} \left| \ampMPzt \right|^2 \
      \left[(4\pi)\frac{d\pnrL {\pnrL}^2}{(2\pi)^3} \frac{1}{2 \EnL} \frac{f^{(0)}_\chi}{n_\chi^{(0)}} \right] \ , 
\eeq
where we are integrating over the CM-frame momentum in the lab-frame $\pnL$ with a trivial angular dependence.   
Owing to the Lorentz invariance of $d\Pi_3$, $d\Pi'_3$, $|\ampMPzt|^2$, we are free to evaluate these in convenient frames
(eg. $\Chi_n$ rest-frame, *-frame, $\SSt$-frame, cf. Sec.~\ref{ABdec.SEC}) 
and they would be independent of any boost of the CM-frame to $\pnL$;
the quantity in $[...]$ has a dependence on $\pnL$ (and on no other momenta)
and this factor can be integrated separately over $\pnL$, giving $\langle 1/(2E_n)\rangle _{\!P}$,
the $P$ subscript reminding us of the $1/n_\chi^{(0)}$ normalization making it like a probability distribution
(cf. Appendix~\ref{ThAvgProc.SEC}).
We can use here the result of Eq.~(\ref{reciETA.EQ}) (and shown in Fig.~(\ref{GmTAx.FIG})) to get $\langle \hat{M_n}/\hat{E}_n\rangle _{\!P}\! =\! K_1(\hat{M_n} x)/K_2(\hat{M_n} x)$. 
Thus, in the narrow width approximation, we have 
\beq
\GmzPTAh_{\rm pole} = \frac{1}{2 \hat{M}_n} \frac{K_1(\hat{M}_n x)}{K_2(\hat{M}_n x)} \alzPh \ ; \quad {\rm where}\ \alzPh = \frac{1}{2 \hat{M}_n \hat\Gamma_n} \ \int d\hat\Pi_3 d\hat\Pi'_3 \left| \ampMPzt \right|^2 \ ,
\label{GmzPTAPhNW.EQ}
\eeq
now written with dimensionless variables, with the $\alzPh$ being $x$ independent. 
We follow the same procedure as for the tree-loop interference term in Sec.~\ref{ABdec.SEC} and evaluate the integral above numerically. 

We obtain the nonpole contribution by subtracting out from
the integrand of Eq.~(\ref{GmzPTADefn.EQ}) the pole contribution, i.e. the $\delta$-function discussed above.
In other words, we take instead of $|\Pchi|^2$ in Eq.~(\ref{GmzPTADefn.EQ}) the subtracted form
$\PchiMSq \equiv |\Pchi|^2 - \pi/(M_n \Gamma_n)\ \delta({\pnz}^2 - E_n^2)$ using the representation of the delta function given above.
Using the subtracted propagator ensures that the integrand is nonsingular and leads to a more stable numerical integration
over $p_n^0$.
For the integration, we trade $p_n^0$ for $\Mtn$, with $dp_n^0 = d\Mtn \Mtn/\Etn$ where $\Etn=\sqrt{\pnr^2 + \Mtn^2}$, the $\Chi_n$ now being offshell with virtual mass $\Mtn$.
Also, owing to the zeroth component $\delta$-function (energy conservation),
we have here $\Pi_i f_{Q_i}^{(0)} = e^{-\Etn/T} \equiv \tilde{f}^{(0)}(\Etn)$.
The numerical evaluation of the integrals is similar to the pole contribution case above,
with the extra integration over $\Mtn$ now added, again taking the $\Chi_n$ momentum in the lab-frame ($\pnL$) as above.
Putting these together we have for the nonpole (remainder) contribution
\bea
\GmzPTA_{\rm nonpole} &=& \int d\Pi_3 d\Pi'_3 
  \frac{d\Mtn}{2\pi}2\Mtn\, \left|\ampMPzt \right|^2 \, \PchiMSq \, (4\pi)\frac{d\pnrL {\pnrL}^2}{(2\pi)^3} \frac{1}{2\EtnL} \frac{\tilde{f}^{(0)}(\EtnL)}{n_\chi^{(0)}} \ , \\
  &&\hspace*{-1cm} {\rm where}\
    \PchiMSq = \left|\frac{1}{\Mtn^2 - M_n^2 + iM_n \Gamma_n}\right|^2 - \frac{\pi}{M_n \Gamma_n}\ \frac{\epsilon}{\pi} \frac{1}{(\Mtn^2 - M_n^2)^2 + \epsilon^2} \ . \nonumber
\label{GmzPTARhNW.EQ}
\eea
We have checked that the numerically evaluated nonpole integrals are not very sensitive to the choices of $\Gamma_n$ and $\epsilon$,
and for numerical stability we conveniently choose the corresponding scaled variables to be $1/10$ and $1/100$ respectively. 

We compute the thermal average of Eq.~(\ref{IBBb.EQ})
by supplying the $d^3 \hat{p}_{Q_i}/(2\pi)^3~f_{Q_i}^{(0)}$ factors for the initial state
as required by Eq.~(\ref{Gm0BBbTA.EQ}) and integrate  
to obtain $\langle \hat{\cal I}^{\scriptscriptstyle (\bar B B)}_{k\lambda\, nm}\rangle $.

The $\IBBbklnmh$ values before thermal averaging, for $\Mtn\!=\! 1$, $\pnrL\!=\! 0$, $\qUPr\!=\! 0$, 
for BP-A are
$\IBBbolooh\! =\! 0.12$, $\IBBboltth\! =\! 0.14$,
$\IBBbtlooh\! =\! 0.1$, $\IBBbtltth\! =\! 0.12$,
which for $g_L\!=\! 1$, $g_R\!=\! 0$ gives $\sigzPh\!=\! 0.004$;
while for BP-B are
$\IBBbolooh\! =\! 1.7$, $\IBBboltth\! =\! 2$,
$\IBBbtlooh\! =\! 0.21$, $\IBBbtltth\! =\! 0.24$. 

We find the thermally averaged integral $\IBBbTAklnmh$
(pole,~nonpole) values for BP-A to be
$\IBBbTAolooh\! =\! (4.2\times 10^{-5}, 2.36)$, $\IBBbTAoltth\! =\! (0.15, 3.5)$,
$\IBBbTAtlooh\! =\! (3.8\times 10^{-5}, 3.1\times 10^{-2})$, $\IBBbTAtltth\! =\! (0.086, 0.046)$.
For BP-B we find  
$\IBBbTAolooh\! =\! (1.16, 2.5)$, $\IBBbTAoltth\! =\! (8.84, 3.7)$,
$\IBBbTAtlooh\! =\! (0.023, 6.2\times 10^{-4})$, $\IBBbTAtltth\! =\! (0.12, 9.3\times 10^{-4})$.

\section{CPT and Unitarity Relations in Perturbation Theory}
\label{CPTUniDet.APP}

\begin{figure}
  \begin{center}
    \includegraphics[width=0.6\textwidth]{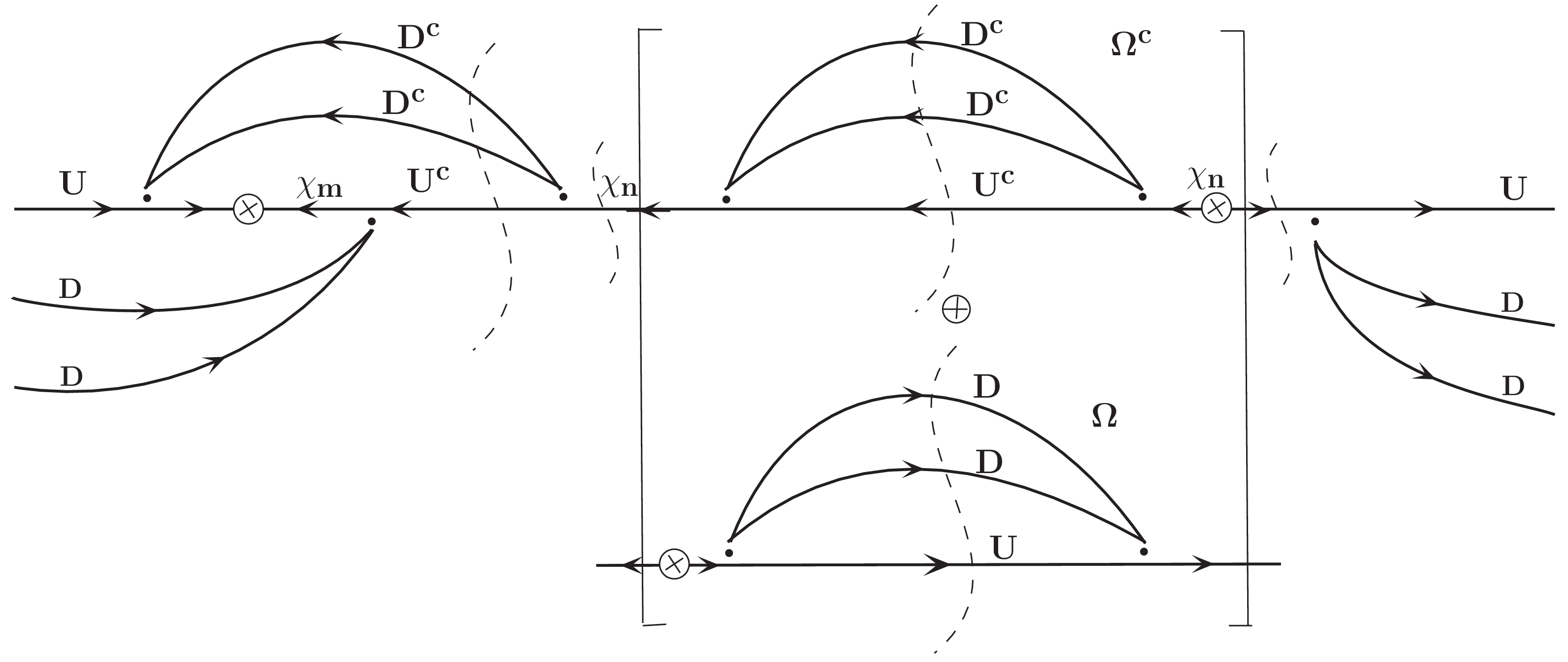} 
  \end{center}    
  \caption{The $DDU\to DDU$ forward scattering amplitude showing the diagram-A like tree-loop interference term  
    whose RIS piece yields the inverse decay contribution. The dashed curves show the cuts.  
 \label{DDU2DcDcUcRIS.FIG} }
\end{figure}

We discuss here how the content of Eq.~(\ref{RISscatRel.EQ}) works in perturbation theory, 
which is quite intricate.
In Sec.~\ref{ABdec.SEC} 
we computed the decay width difference with Feynman diagrams at order $\bigO(g^4)$
for the single operator contributions represented by diagrams A,B,C,D, 
which are the interference terms between the loop diagrams A$_1$,B$_1$,C$_1$,D$_1$ with the tree-level diagram.
Since in these diagrams the intermediate $UDD$ are anyway being cut to obtain the strong phase, 
we can obtain the inverse decay asymmetry by repositioning the sub-factors from there.
Eq.~(\ref{RISscatRel.EQ}) suggests to us that to obtain the inverse decay asymmetry,
we are to start with the $UDD \to U^cD^cD^c$ scattering channel and focus on diagrams in which an intermediate $\Chi$ can go on-shell
giving the required RIS piece.
Such an onshell $\Chi$ propagator puts a $1/\Gamma_\Chi$ factor, which at LO is $1/\bigO(g^2)$,
and therefore we must also look at scattering diagrams at $\bigO(g^6)$ since we can obtain a 
$\bigO(g^6)/\bigO(g^2)\!\! =\!\! \bigO(g^4)$ decay asymmetry contribution. 
An example of an $\bigO(g^6)$ contribution
is shown in Fig.~\ref{DDU2DcDcUcRIS.FIG} as the cut in the $UDD\to UDD$ forward amplitude,
where we start with diagram-A, move parts of the diagram around for it to be the interference term
between loop and tree diagrams of the $UDD\to U^cD^cD^c$ scattering channel  
and insert a $\Chi$ 2-point function $\Omega^c + \Omega$ (see Appendix~\ref{Chinm2ptFcn.SEC}) 
We show this for the process; the conjugate process is obtained by reversing all arrows in the diagram,
i.e. each of the fields is replaced with its conjugate, 
and the baryon asymmetry is of course given by the difference of these. 
The part to the right of the $U^cD^cD^c$ cut in $\Omega^c,\Omega$ is the tree-level scattering amplitude,
while the part to the left of this cut is the loop amplitude.
The RIS contribution is obtained when the $\Chi_n$ on the left goes onshell.
This RIS contribution differs from the pure scattering contribution by the additional factor shown in $[...]$
in the figure.
This factor is $i(\Omega + \Omega^c) i(\pslash_n+M_n)/(p^2 - M_n^2 + iM_n\Gamma_\Chi)$.
For the contribution to survive we must cut $\Omega,\Omega^c$ giving $i\hat\Omega,i\hat\Omega^c$,
or otherwise the counterterm cancels this onshell $\Chi$ contribution (see Appendix~\ref{Chinm2ptFcn.SEC}) 
For the onshell (RIS) contribution then this factor is $-i(\hat\Omega+\hat\Omega^c)/(iM_n\Gamma_\Chi))$.
Using the fact that the total width is $\Gamma_\Chi = \int(\hat\Omega+\hat\Omega^c)$,
this factor is $-\int ((...) \times (\hat\Omega+\hat\Omega^c) \times (...))/\int(\hat\Omega+\hat\Omega^c))$.
We take all integration measures above appropriately and do not show them explicitly.
The integral and Dirac trace in the numerator shown as $(...)$ is split by the $\delta$ functions of the RIS cut
into separate integral and Dirac trace factors,
one of which is exactly like the denominator, which cancels, giving -1 for the $[...]$ part of the figure.
This is the origin of the minus sign in Eq.~(\ref{RISscatRel.EQ}) now obtained perturbatively at $\bigO(g^4)$,
and proves the relation at this order. 
This minus sign in conjunction with the factor of 2 in the last line of Eq.~(\ref{BEscRIS.EQ})
coming from the considered scattering channel being $\Delta B\!\!=\!\! \pm 2$, 
gives the RIS piece to be -2 of the first line in this equation, which overturns the sign of the first line, as required. 

To understand the behavior of the scattering collision terms perturbatively, we consider the forward scattering amplitude similar to Fig.~\ref{DDU2DcDcUcRIS.FIG},
but now modified with an initial state $Q$ ($\Dp$ for SC-1 or $\Up$ for SC-2) and a same-kind final state $Q$ crossed to become intermediate states
and joined to get an additional loop.
The insertion of $(\Omega + \Omega^c)$ to obtain an $\bigO(g^6)$ contribution and the $\Chi_n$ RIS gives a $\bigO(g^2)$
denominator, and leads to a $\bigO(g^4)$ asymmetry.
Following a similar logic as in the inverse decay case earlier, the above inserted factor gives a minus sign,
now understood perturbatively in the scattering case.

\section{Thermal Averages - Numerical Details}
\label{ThAvgProcDet.SEC}

In Appendix~\ref{ThAvgProc.SEC} we give details on our method of computing thermal averages.
Here we give numerical details on the behavior of the distribution functions and the thermal averages with some example parameter choices. 

\begin{figure}[h]
  \begin{center}
    \includegraphics[width=0.4\textwidth] {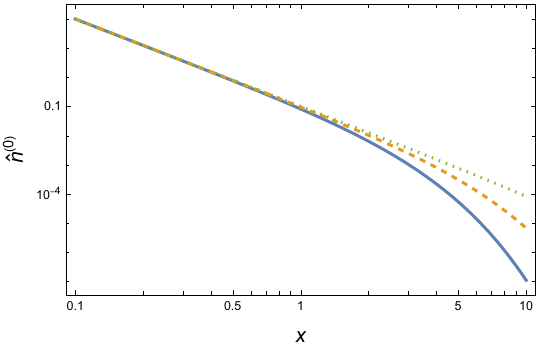}
    \caption{The dimensionless $\hat{n}^{(0)}$ vs. $x$ for $\hat{M}\!=\!1,1/3,1/10$ (solid, dashed, dotted). 
      \label{neq0.FIG}
    }
  \end{center}
\end{figure}

\begin{figure}[h]
  \begin{center}
    \includegraphics[width=0.4\textwidth] {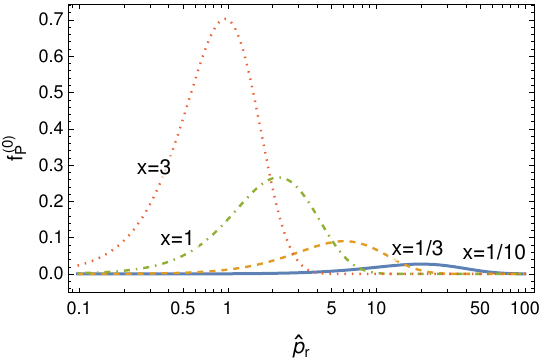}
    \includegraphics[width=0.4\textwidth] {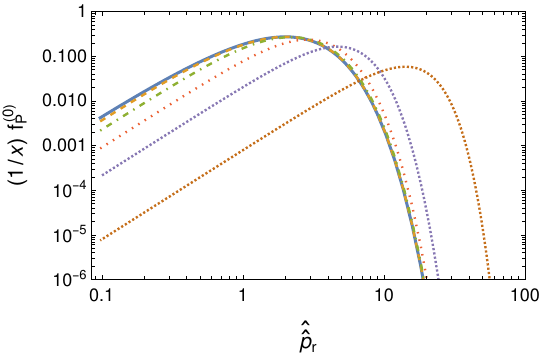}
    \caption{The probability distribution function $f_P^{(0)}$ vs. $\hat{p}_r$ (left), and $(1/x) f_P^{(0)}$ vs. $\hat{\hat{p}}_r$ (right).
      The right plot is for $x=\{1/10,1/3,1,3,10,100\}$ (solid to dotted). 
      \label{feqProb.FIG}
    }
  \end{center}
\end{figure}

In Fig.~(\ref{neq0.FIG}) we show the dimensionless $\hat{n}^{(0)}=n^{(0)}/M_\chi^3 = g/(2\pi)^2\, (\hat{M}^2/x)\, K_2(\hat{M} x)$ for various $\hat{M}$. 
In Fig.~(\ref{feqProb.FIG})~(left) we show the distribution function normalized as a probability distribution function
$f_P^{(0)}(\hat{p}_r)$ defined in Eq.~(\ref{feqProbpr.EQ}). 
In Fig.~(\ref{feqProb.FIG})~(right) we show the probability distribution function $(1/x) f_P^{(0)}$ as a function of
$\hat{\hat{p}}_r \equiv x \hat{p}_r = p_r/T$,
where we have $\int_0^\infty d\, {\hat{\hat{p}}}_r (1/x) f^{(0)}_P(\hat{\hat{p}}_r) = 1$. 

\begin{figure}
  \begin{center}
    \includegraphics[width=0.4\textwidth] {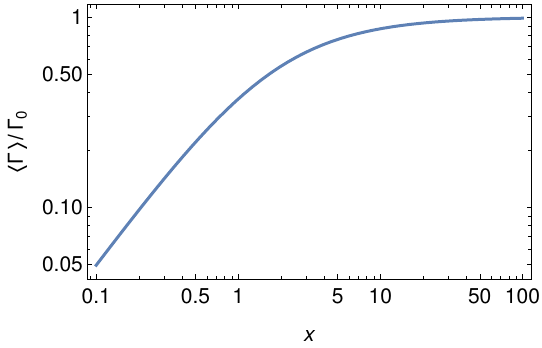}
    \caption{The ratio of the thermally averaged width to the zero temperature width as a function of $x=M/T$.
      \label{GmTAx.FIG}
    }
  \end{center}
\end{figure}

If the $\Chi$ is in (or close to) thermal equilibrium with the bath at temperature $T$,
we investigate next how the decay width is modified,
following the method detailed in Appendix~\ref{ThAvgProc.SEC}.
As usual, we define $x=M_\chi/T$,
and for notational ease, we drop the subscript $\Chi$ on quantities here, for instance, we denote the $\Chi_n$ width as $\Gamma$, its mass as just $M$, etc.
In Fig.~(\ref{GmTAx.FIG}) we show the $\langle  \Gamma \rangle _{\sssty\!P}/\Gamma_0$ vs. $x$, for $\hat{M} = 1$. 

\section{UV Completion Corrections - Numerical Details}
\label{VVintUVDet.SEC}

In Appendix~\ref{VVintUV.SEC} we discuss the manner in which we handle potential UV propagator corrections in our
thermally averaged rates. Here we provide more details on the behavior of these corrections for some sample parameter choices. 

\begin{figure}
  \begin{center}
  \includegraphics[width=0.32\textwidth]{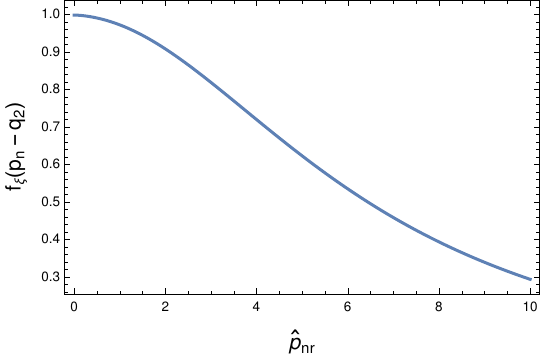}  
  \includegraphics[width=0.32\textwidth]{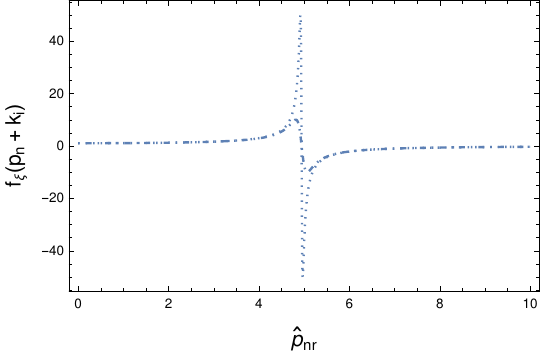} \\  
  \includegraphics[width=0.32\textwidth]{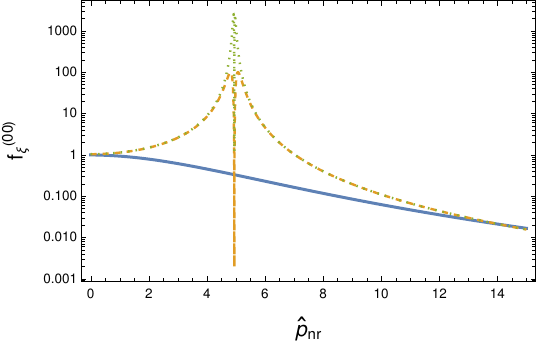}
  \includegraphics[width=0.32\textwidth]{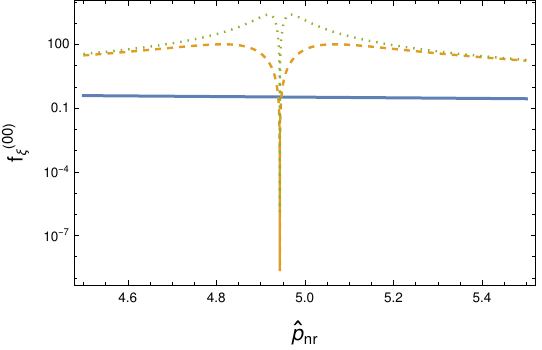}  
  \end{center}    
  \caption{
    The CM frame LO UV propagator correction $f_\xi$ for BP-A parameters, $M_\chi/\Lambda\!\! =\!\! 1/10$,
    for SC-1 (top left) and SC-2 (top right), 
    and $f_{\xi\,(s,t)}^{(00)}$ (bottom panel) for SC-1 (solid), and 
    SC-2 with $\hat\Gamma_\xi\!\! =\!\! 1/2$ (dashed) and $\hat\Gamma_\xi = 1/10$ (dotted); 
    the bottom right plot zooming in on the resonance region.
    \label{fxi00nCh.FIG} 
  }
\end{figure}

In Fig.~\ref{fxi00nCh.FIG} we show the LO $f_{\xi\,(s,t)}$ and $f_{\xi\,(s,t)}^{(00)}$ for the t-channel SC-1,
and the s-channel SC-2, in the CM frame for $r_{M\Lambda} = M_\chi/\Lambda\!\! =\!\! 1/10$. 
With our parametrization of the $\xi$ width, $\hat\Gamma_\xi = \Gamma_\xi/M_\chi = r_{M\Lambda} \tilde\Gamma_\xi$,
we present our results for sample choices of $\hat\Gamma_\xi$.  
For SC-2 we show the PV for $\hat\Gamma_\xi\!\! =\!\! 1/2, 1/10$,
which correspond to $\tilde\Gamma_\xi = 1/20, 1/100$ respectively, for $M_\chi/\Lambda\!\! =\!\! 1/10$. 
The resonant enhancement is clearly seen, with the peak being larger for a smaller value of $\hat\Gamma_\xi$.
For illustration, we have taken the mass parameters as in BP-A for $n=2$, 
and set the CM frame phase space angular variables to $c_1^{CM} = 1/2$, and $\phi_1^{CM} = \pi/6$. 

We include the $f_\xi$ factor and obtain the $\TAsigv$ by numerical integration as detailed in Sec.~\ref{genBAsymScat.SEC}. 
For illustration we evaluate the $\TAsigv$ for $x=1/3$ and $\tilde\Gamma_\xi = \Gamma_\xi/\Lambda = 1/10$. 
Compared to the effective theory value, we find the following $\TAsigv$ numbers for scheme (a) (with scheme (b) shown in parenthesis): \\
\indent for $r_{M\Lambda} = 1/10$, SC-1 is 66\% (70\%) lower, while the SC-2 is 200\% higher (33\% lower);\\
\indent for $r_{M\Lambda} = 1/25$, SC-1 is 33\% (33\%) lower, while SC-2 is 600\% (30\%) higher; \\
\indent for $r_{M\Lambda} = 1/50$, SC-1 is 10\% (13\%) lower, while SC-2 is 65\% (45\%) higher;\\
\indent for $r_{M\Lambda} = 1/100$, SC-1 is 3\% (4\%) lower, while SC-2 is 10\% (10\%) higher. \\
From this we learn that
the relative difference in the thermal average cross section between the effective theory and the {\it UV completion A} is at most an
${\cal O}(1)$ factor for $M_\chi/\Lambda < 1/10$,
and the agreement is better than to 10\% (10\%) for $r_{M\Lambda} < 1/100$.

For the tree-loop interference contribution of Fig.~\ref{chi2DDU-LONLOCut.FIG}, 
we include the UV state $\xi$ propagator correction by multiplying the $f_\xi(p)$ for each vertex, which yields
$f^{(01)}_{\xi} = f_\xi(p_n-q'_U)\, f_\xi(q'_U-p_n')\, f_\xi(p'_n-q_U)\, f_\xi(q_U - p_n)$,
which is equivalently, 
$f^{(01)}_{\xi} = f_\xi(p_n-q'_U)\, f_\xi(q'_U-p_n')\, f^*_\xi(q_U-p'_n)\, f^*_\xi(p_n-q_U)$, 
and we take the PV part of each factor.
For the UV correction factor for the scattering channels we apply the crossing relations in Eq.~(\ref{crRel.EQ}) on this.

\begin{figure}
  \begin{center}
    \includegraphics[width=0.32\textwidth]{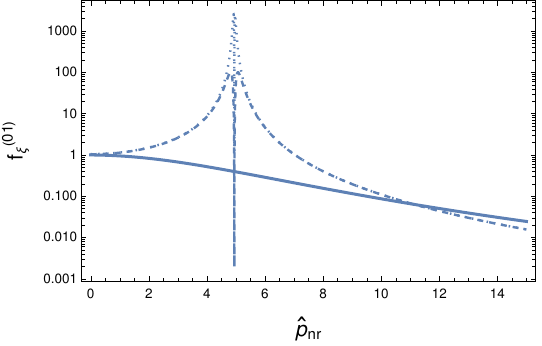}
    \includegraphics[width=0.32\textwidth]{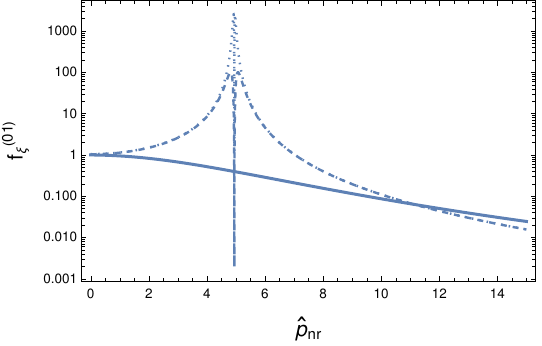}
    \includegraphics[width=0.32\textwidth]{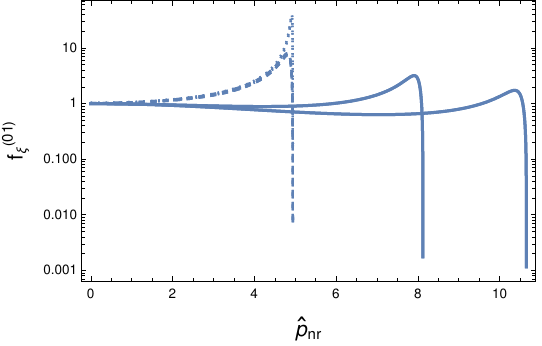}
  \end{center}    
  \caption{
    The $\pnrh$ dependence of the CM frame UV propagator correction factor $f_{\xi}^{(01)}$
    for the tree-loop interference contribution
    in diagram A (left), diagram B (middle) and diagram C (right), 
    for BP-A parameters,
    and, $M_\chi/\Lambda\!\! =\!\! 1/10$, for SC-1 (solid), and SC-2 with $\hat\Gamma_\xi\!\! =\!\! 1/2$ (dashed) and $\hat\Gamma_\xi = 1/10$ (dotted). 
    \label{fxi01nCh.FIG} 
  }
\end{figure}

In Fig.~\ref{fxi01nCh.FIG} we show the
$\pnrh$ dependence of the CM frame UV propagator correction factor $f_{\xi}^{(01)}$ for SC-1 and SC-2
for the tree-loop interference contribution in diagrams A,B,C of
(the crossed) Fig.~\ref{chi2DDU-LONLOCut.FIG}
for $M_\chi/\Lambda\!\! =\!\! 1/10$, where for SC-2 we show the PV for $\hat\Gamma_\xi\!\! =\!\! 1/2,1/10$ 
and the resonant enhancement is clearly seen.
The different curves for diagram $C$ is for the different subdiagrams $Cij$ discussed below Eq.~(\ref{LDABCD.EQ}).  
For illustration we have set the CM frame phase space angular variables to $c_1^{CM} = 1/2$, and $\phi_1^{CM} = \pi/6$,
and the loop variables to $c'_U=1/2$, $\phi'_U=\pi/6$, $\qUPr=0.15$, $c^\SSt_3=1/2$, $\phi^\SSt_3=\pi/6$. 

\begin{figure}
  \begin{center}
    \includegraphics[width=0.4\textwidth]{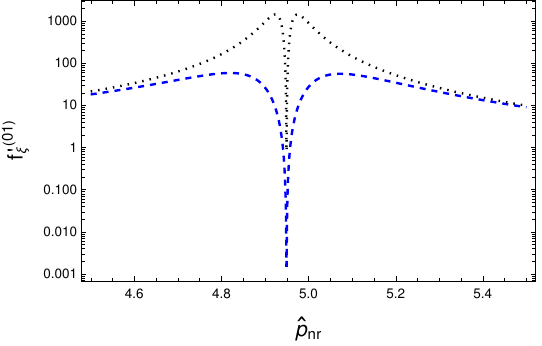}
  \end{center}    
  \caption{
    The $\pnrh$ dependence of the CM frame UV propagator correction factor $f_{\xi}^{'(01)}$
    in the multiple-operator tree-loop interference contribution for SC-2 with BP-B parameters,
    and, $M_\chi/\Lambda\!\! =\!\! 1/10$, $\hat\Gamma_\xi\!\! =\!\! 1/2$ (dashed) and $\hat\Gamma_\xi = 1/10$ (dotted). 
    \label{fxi01MA.FIG} 
  }
\end{figure}

For the mixed-operator tree-loop interference correction factor in Fig.~\ref{chi2DDU-LONLO-modA.FIG}, we have
the UV propagator correction factors product
$f^{\prime (01)}_{\xi} = f_\xi(q_3+q_U)\, f_\xi(q_1+q_2)\, f^*_\xi(q_1+q_2)$.
Again, for the scattering channels we apply the crossing relations on this.
In Fig.~\ref{fxi01MA.FIG} we show the
$\pnrh$ dependence of the CM frame UV propagator correction factor $f_{\xi}^{'(01)}$ for SC-2
in the multiple-operator tree-loop interference contribution of the (crossed) Fig.~\ref{chi2DDU-LONLO-modA.FIG} diagram
for $M_\chi/\Lambda\!\! =\!\! 1/10$, for $\hat\Gamma_\xi\!\! =\!\! 1/2,1/10$.
For this illustrative plot, we set the phase space momentum angular variables at $c_1^{CM} = 1/2$, $\phi_1^{CM} = \pi/6$,
and the loop momentum angular variables at $c_m=1/2$, $\phi_m=\pi/6$. 


\end{document}